\documentclass[a4paper,11pt]{article}
\usepackage[dvipsnames]{xcolor}
\usepackage[utf8]{inputenc}
\usepackage[bbgreekl]{mathbbol}
\usepackage{geometry}
\usepackage{multicol}

\usepackage{extarrows}
\usepackage{xcolor}
\usepackage{amsmath, array, amssymb, amsfonts,amsthm}
\usepackage{upgreek}
\usepackage{xfrac}
\usepackage[inline]{enumitem}
\setlist{nolistsep}
\usepackage[french, italian, english]{babel}
\usepackage[all]{xy}
\usepackage{txfonts}  
\usepackage{sectsty}
\usepackage{booktabs}
\usepackage{caption}
\usepackage{dsfont}
\usepackage{mathtools}
\usepackage{slashed}
\usepackage[makeroom]{cancel}
\usepackage[hidelinks]{hyperref}
\usepackage{textcomp}
\usepackage{multicol}
\usepackage[title]{appendix}
\usepackage[square,numbers,sort,compress,semicolon,merge]{natbib}
\let\cite\citep 
\usepackage{hyperref}
\hypersetup{colorlinks=true, urlcolor=blue, citecolor=blue, linktoc=page}

\usepackage{tikz}
\usepackage{tikz-cd}
\usetikzlibrary{cd}
\tikzcdset{
arrow style=tikz,
diagrams={>={Straight Barb[scale=0.8]}}
}
\usetikzlibrary{matrix,arrows,decorations.pathmorphing}

\DeclareMathAlphabet{\mathpzc}{OT1}{pzc}{m}{it}

\usepackage{fancybox}

\usepackage{mathrsfs}
\usepackage{scrextend}

\makeatletter
\renewcommand*\env@matrix[1][\arraystretch]{%
  \edef\arraystretch{#1}%
  \hskip -\arraycolsep
  \let\@ifnextchar\new@ifnextchar
  \array{*\c@MaxMatrixCols c}}
\makeatother

\newcommand{\defeq}{\vcentcolon=}
\newcommand{\rdefeq}{=\vcentcolon}

\newcommand\M{\mathcal{M}}

\newcommand\R{\mathcal{R}}
\newcommand\RR{\mathbb{R}}
\newcommand\CC{\mathbb{C}}
\newcommand\C{\mathcal{C}}

\newcommand\id{\textit{id}}

\newcommand\G{\mathcal{G}}
\renewcommand\H{\mathcal{H}}
\newcommand\Hl{\H_\text{\tiny loc}}

\renewcommand\S{\mathcal{S}}

\newcommand\U{\mathcal{U}}

\newcommand\K{\mathcal{K}}
\newcommand\J{\mathcal{J}}
\renewcommand\O{\mathcal{O}}

\newcommand\D{\mathcal{D}}

\newcommand\vphi{\varphi}

\renewcommand\u{\text{\bf u}}

\renewcommand\epsilon{\varepsilon}

\newcommand\rarrow{\rightarrow}

\newcommand\aut{\mathfrak{aut}}
\newcommand\diff{\mathfrak{diff}}

\newcommand\LieG{\mathfrak{g}}

\renewcommand\t{\tilde}
\newcommand\h{\widehat}
\renewcommand\b{\bar }
\newcommand\w{\wedge}
\renewcommand\d{\partial}
\newcommand\s{\sigma}
\newcommand\bs{\boldsymbol}
\newcommand \omegaf{\bs\omega_{\text{\tiny$0$}}}

\renewcommand\-{^{-1}}
\newcommand\Ad{\text{Ad}}
\newcommand\ad{\text{ad}}
\renewcommand\id{\text{id}}

\def\munderline#1#2{\color{#1}\underline{{\color{black}#2}}\color{black}}

\makeatletter

\newcommand{\Rmnum}[1]{\expandafter\@slowromancap\romannumeral #1@}
\makeatother

\makeatletter
\newcommand{\leqnomode}{\tagsleft@true\let\veqno\@@leqno}
\newcommand{\reqnomode}{\tagsleft@false\let\veqno\@@eqno}
\makeatother

\DeclareMathOperator{\Diff}{Diff}
\DeclareMathOperator{\Aut}{Aut}
\DeclareMathOperator{\Der}{Der}
\DeclareMathOperator{\Tr}{Tr}
\DeclareMathOperator{\vol}{vol}

\newtheorem{thm}{Theorem}

\newtheorem{prop}[thm]{Proposition}
\theoremstyle{definition}

\makeatother

\begin{document}


\title{Geometric relational framework \\
for general-relativistic gauge field theories}

\author{J. T. \textsc{François} $\,{}^{a,\, b,\, c,\,*}$ \and L. \textsc{Ravera} $\,{}^{d,\,e,\,f,\,\star}$ }

\date{}

\maketitle
\begin{center}
\vskip -0.6cm
\noindent
${}^a$ Department of Mathematics \& Statistics, Masaryk University -- MUNI.\\
Kotlářská 267/2, Veveří, Brno, Czech Republic.\\[2mm]
 
${}^b$  Department of Philosophy -- University of Graz. \\
 Heinrichstraße 26/5, 8010 Graz, Austria.\\[2mm]
 
${}^c$ Department of Physics, Mons University -- UMONS.\\
 Service \emph{Physics of the Universe, Fields \& Gravitation}.\\
20 Place du Parc, 7000 Mons, Belgium.
\\[2mm]

${}^d$ DISAT, Politecnico di Torino -- PoliTo. \\
Corso Duca degli Abruzzi 24, 10129 Torino, Italy. \\[2mm]

${}^e$ Istituto Nazionale di Fisica Nucleare, Section of Torino -- INFN. \\
Via P. Giuria 1, 10125 Torino, Italy. \\[2mm]

${}^f$ \emph{Grupo de Investigación en Física Teórica} -- GIFT. \\
Universidad Cat\'{o}lica De La Sant\'{i}sima Concepci\'{o}n, Concepción, Chile. \\[2mm]

\vspace{1mm}

${}^*$ {\small{francois@math.muni.cz}} \qquad \quad ${}^\star$ {\small{lucrezia.ravera@polito.it}}
\end{center}

\vspace{1mm}


\begin{abstract}
We remind how relationality arises as the core insight of general-relativistic gauge field theories from the articulation of the generalised hole and point-coincidence arguments. 
Hence, a compelling case for a manifestly relational framework ensues naturally. 
We propose our formulation for such a framework, based on a significant development of the dressing field method of symmetry reduction.  

We first develop a version for the group $\Aut(P)$ of automorphisms of a principal bundle $P$ over a manifold $M$, as it is the most natural and elegant, and as $P$ hosts all the  mathematical structures relevant to general-relativistic gauge field theory. 
Yet, as the standard formulation is local, on $M$, we then develop the relational framework for local field theory. 
It manifestly implements the generalised point-coincidence argument, whereby the physical field-theoretical degrees of freedoms co-define each other and define, coordinatise, the physical spacetime itself.
Applying the framework to General Relativity, we obtain relational Einstein equations, encompassing various notions of ``scalar coordinatisation" à la Kretschmann-Komar and Brown-Kucha\v{r}.
\end{abstract}


\vspace{-5mm}

\clearpage

\tableofcontents

\section{Introduction}  
\label{Introduction}  

A key moment in the development of General Relativity (GR) was the realisation by Einstein of the meaning of diffeomorphism covariance in the theory: the fact that spacetime points, or regions, are defined relationally, via field values coincidences, and so are, by extension, the physical degrees of freedoms (d.o.f.) of those fields. 
The~conclusion results from the articulation of the famous ``hole argument" and ``point-coincidence argument". 
This~essential physical insight, \emph{relationality}, at the heart of general-relativistic physics is curiously often overlooked. 
Which unfortunately leads to misleading statements spreading frictionlessly in the modern technical literature.

In section \ref{Relationality in general-relativistic gauge field theory}, we therefore propose to briefly recapitulate the logic behind the relational picture in general-relativistic gauge field theory (gRGFT), detailed in~\cite{JTF-Ravera2024c}.
We start with reminding the logic as it arises in GR, 
and then also argue that a similar line of arguments makes a strong case for the notion that the gauge principle in gauge field theory (GFT) is a way to encode the relational character of gauge physics. 
For this, one has to admit that the internal d.o.f. of gauge fields are probing an enriched spacetime whose points are not structureless: such a space is  described by the geometry of a fiber bundle $P$ (bundle geometry being widely recognised as the foundation of classical GFT). 
We articulate the dialectics arising from requiring ``gauge invariance" under both passive and active gauge transformations on a bundle. 
In such a space, we argue that relationality results from the conjunction of an ``internal hole argument" and an ``internal point-coincidence argument". 
Finally, bringing GR and GFT together, we thus conclude that the paradigmatic core of the framework of gRGFT is the relational character of the physics it describes, and that it is  encoded in the requirement of covariance under the diffeomorphism and gauge groups, its local covariance group.
\medskip

Therefore, the current default formalism  of gRGFT is manifestly covariant and tacitly relational. 
It is certainly a worthy endeavor to search for an invariant and manifestly relational formulation of gRGFT, which would have many avantages: its observables and fundamental degrees of freedom (d.o.f.) would be easily identified, its field equations would have a well-posed Cauchy problem, its  quantization would possibly be more easily achieved, etc.

In this paper, we propose such a formulation based on the dressing field method (DFM) of symmetry reduction \cite{GaugeInvCompFields, Francois2014, Francois2021, Berghofer-et-al2023, Francois2023-a}  -- see also~\cite{Zajac2023}. 
The latter being best understood in terms of the  geometry of field space, we will give our account of this geometry, \emph{twice over}. 
\medskip

First, in section \ref{Geometry of field space}, we shall consider what we call the ``global field space", i.e. the space of fields (differential forms) on a finite-dimensional principal bundle $P$ over a base manifold $M$, with structure Lie group $H$. 
This global field space will be described as an infinite-dimensional principal bundle with structure group the  automorphism group $\Aut(P)$ of $P$ which contains (so to speak) both diffeomorphisms of its base manifold $M$ and the subgroup of vertical automorphisms, also known as (a.k.a.) the gauge group $\H$ of $P$. 
We further frame integration on $P$ as an operation on what we call the ``associated bundle of regions" of $P$. 
Then, in section \ref{The dressing field method}, we  develop the DFM for  $\Aut(P)$: in a nutshell, the method amounts to a systematic algorithm to build \emph{basic} objects on the global field space, and on the bundle of regions. 
In the latter case, a notion of field-dependent regions of $P$ arises naturally. 

The standard formulation of field theory is not done on $P$, but rather on its base manifold $M$, in what one may call ``bundle coordinate patches": field theory is thus done non-intrinsically, up-to bundle coordinate changes (a.k.a. ``gauge choices"). 
Yet the bundle $P$ is the natural space hosting the mathematical structures relevant to field theory. 
In that respect, it may seem that a more fundamental formulation of gRGFT on $P$ is lacking.
It is as if the only available, state  of the art, formulation of GR was its coordinate tensor calculus version, rather than its formulation via intrinsic differential geometric methods. 
When such a bundle formulation of gRGFT is available, our formalism will allow its relational formulation. This is one of the motivation for pursuing it to the extent we do. 

Another motivation is that it gives a (much simpler) template for the formalism as it applies to local field theory. 
The latter requires to elaborate on the bundle geometry of what we call the ``local field space", i.e. fields  on $M$ that are the local representatives of global objects living on $P$.
The~structure group of the local field space is the group $\Diff(M)\ltimes \Hl$ -- with $\Hl$ the local gauge group of internal gauge theory -- usually understood to be the covariance group of gRGFT. 
We develop this bundle geometry in section \ref{Local field theory}, also framing integration as an operation on the associated bundle of regions of $M$. 
The DFM then implies to build basic objects on the local field space and on its associated bundle of regions. The physical relational spacetime arises from the notion of dressed integrals. 
The formalism is applied to obtain a manifestly relational and invariant reformulation of a gRGFT.  
We illustrate it on GR and GR coupled to (scalar) electromagnetism. 
In the conclusion \ref{Conclusion}, we take stock of our results and sketch further developments of our program. 
Appendices complete the main text.

\section{Relationality in  general-relativistic gauge field theory}  
\label{Relationality in  general-relativistic gauge field theory}  

The statement that physics is relational may appear too obvious to mention.  
At an elementary level it may be understood as a simple kinematical proposition, meaning that physical objects evolve with respect to (w.r.t.) each other.
General-relativistic physics adds two fundamental refinements.
The first is that the very definition of physical objects has to be relational, so that they \emph{co-define}, and evolve w.r.t., each others.
Secondly, it  takes this refinement so seriously so as to insist that there are no physical entities that can influence others without being influenced in return: as it is often phrased ``nothing can act upon without being acted upon". 
This implies not only that no physical structure may constitute an absolute reference (i.e. a fixed background)
but also that these are unnecessary. They arise only as  limit cases of more fundamental dynamical entities. This is known as background independence. 

It is usually much less widely appreciated that the physics of gauge field theory (GFT) shares this fundamental \mbox{relational} character, 
so that it is a core insight of the union of the two: 
general-relativistic gauge field theory (gRGFT).
In this global framework, relationality emerges from the requirement of 
heuristic symmetry \mbox{principles}, the general covariance and gauge principles, understood as principles of ``democratic epistemic access" to Nature, according to which there are no privileged situated viewpoints to contemplate the law of Physics. 
The dialectics between the generalised hole argument and point-coincidence argument is key to the outcome. 
The precise logic is laid out in detail in \cite{JTF-Ravera2024c}. As~it~serves as motivational background for this work, we briefly review it below.

\paragraph{General-relativistic physics}
The principle of general covariance of field equations requires that laws of physics be indifferent to the choice of coordinates, which implies that the equations describing these laws, the field equations, be tensorial. 
In a first step of analysis, it 
means that the physics described is that of a spacetime and its (field) content faithfully modelled by a differentiable manifold $M$ and  fields $\upphi$ defined on it, satisfying tensorial field equations $\bs E(\upphi)$: The geometrical entities  $(M, \upphi)$ are coordinate-invariant, and model the objective -- i.e. viewpoint independent -- structure of spacetime and its field content.

But a second step of analysis is required. 
As the automorphism group of $M$, diffeomorphisms $\Diff(M)$, are also automorphisms of any \emph{natural bundles} over $M$ and their spaces of sections, i.e. of all geometric objects on $M$, in particular of the space of tensors. 
The field equations are then $\Diff(M)$-covariant, so that $\Diff(M)$ is  an automorphism group of the solution space 
  $\S \defeq \{ \upphi  \, |\, \bs E(\upphi)=0 \}$, which is furthermore  \emph{foliated into orbits}: any solution $\upphi \in \S$ has a  $\Diff(M)$-orbit $\O_\upphi \subset \S$.\footnote{The action of $\Diff(M)$ is a priori not free, a solution may have Killing symmetries, ${\sf K}_\upphi\defeq \{\psi \in \Diff(M)\, |\, \psi^*\upphi=\upphi\} \neq \id_M$.}
This has fundamental consequences, stemming from articulating Einstein's famous hole argument and point-coincidence argument  \cite{Norton1993, Stachel1989, Stachel2014, Giovanelli2021}. 

The hole argument highlights the consequence of the collision between $\Diff(M)$-covariance of $E(\upphi)=0$ and the view that $(M, \phi)$ faithfully represent a physical state of affair. 
It is standard to phrase it in terms of solutions $\upphi, \upphi' \in \O_\upphi$, i.e. $\upphi'=\psi^*\upphi$,  s.t.  $\psi$ is a compactly supported diffeomorphism whose support $D_\psi \subset M$ is the ``hole":  Manifestly, such a situation raises an issue with the Cauchy problem, and in particular with the initial value problem, i.e. with determinism. 
To avoid these, two options are available: 
One may either renounce general covariance (GC) of the field equations -- which was envisaged by Einstein in the 1913-1915 period \cite{Norton1993} -- or one may  conclude that all solutions within the same $\Diff(M)$-orbit $\O_\upphi$ represent the \emph{same} physical state. 
This one-to-many correspondence between a physical state and its mathematical descriptions in general/$\Diff(M)$-covariant theories means they are unable to physically distinguish between $\Diff(M)$-related solutions of $\bs E=0$, and consequently make no physical distinction between $\Diff(M)$-related points of $M$. 
In other words, spacetime and its field content are not described by $(M, \phi)$ only, but by its $\Diff(M)$-class. 

This fact, far from being a drawback of the formalism, encodes the essential insight of GR physics, which Einstein identified through his famous point-coincidence argument: 
it is the apparently obvious  observation that physical interactions -- thus all measurements -- happen as  spacetime coincidences of the objects involved, and that the description of such coincidences is $\Diff(M)$-invariant. 
We may write this statement of the $\Diff(M)$-invariance of point-wise mutual \emph{relations} $\R$ among the fields in the  collection $\{\upphi\}$   symbolically as
\begin{equation}
\begin{aligned}
\label{PC-arg}
\R :  \S \times M\  &\ \rarrow \ \   \S \times M /\sim \  \ \xrightarrow{\simeq} \  \text{Relational spatiotemporal physical d.o.f.,} \\
\big(\upphi, x \big) \  &\  \mapsto  \ \big(\upphi, x \big) \sim \big(\psi^*\upphi, \psi\-(x)\big)   \ \mapsto\   \R \big(\upphi; x \big) =\R \big(\psi^*\upphi; \psi\-(x)\big),
\end{aligned} 
\end{equation}
where $\S \times M /\sim$ is the quotient of $ \S \times M$ by the equivalence relation $\big(\upphi, x \big) \sim \big(\psi^*\upphi, \psi\-(x)\big)$, i.e. it is the space of these equivalence classes.\footnote{We shall meet such spaces  later, in   section \ref{Associated bundle of regions}, understood as  ``associated bundles" to the field space  $\Phi$ seen as an infinite-dimensional principal bundle.}
The point-coincidence argument not only dissolves the apparent indeterminism issue raised by the hole argument, but taken to its logical conclusion it may be read both ways:   
\eqref{PC-arg} can be understood to mean first that  physical spacetime points  are  \emph{defined}, or \emph{individuated}, as relational coincidences of distinct physical field-theoretical d.o.f., and then, second, that these d.o.f. are not instantiated within the individual, mathematical, fields $\{\upphi\}$ but by the \emph{relations} among them.  

In summary, the ultimate (ontological) consequence of the (epistemic) GC principle, resulting from the conjunction of the hole and point-coincidence arguments, is the relationality of general-relativistic physics, which we may state as follows:
 Spacetime is relationally defined via its field content, 
 and fields are relationally defined, and evolve,  w.r.t.  each other. 
Relationality of physics is thus what is \emph{tacitly} encoded by the $\Diff(M)$-covariance of a general-relativistic \mbox{theory}.\footnote{We may add that relationality is also a priori a feature of solutions of $\bs E(\upphi)=0$ with Killing symmetries 
${\sf K}_\upphi \subset \Diff(M)$,
the subgroup ${\sf K}_\upphi$ encoding further physical properties of a solution: e.g.  signaling that $\upphi$ has a privileged class of observers. This holds in particular for homogeneous metric solutions, $\upphi=g$, among which Minkowski solution $g =\eta$. Thus, as Kretschmann hinted at, Special Relativity (SR) in fact enjoys $\Diff(M)$-covariance. What makes it special is that its field equation $\bs E(g)=\text{Riem}(g)=0$ implies that the metric decouples from other fields and has frozen dynamics (no d.o.f.), making it a background structure. The solution $g=\eta$ has a Killing group, the Poincaré group ${\sf K}_\eta = IS\!O(1,3)$, distinguishing  geodesic (inertial) observers as privileged.}
The  diagram  Fig.\ref{Diag1}  below summarises the logic leading to the relational picture in GR. 
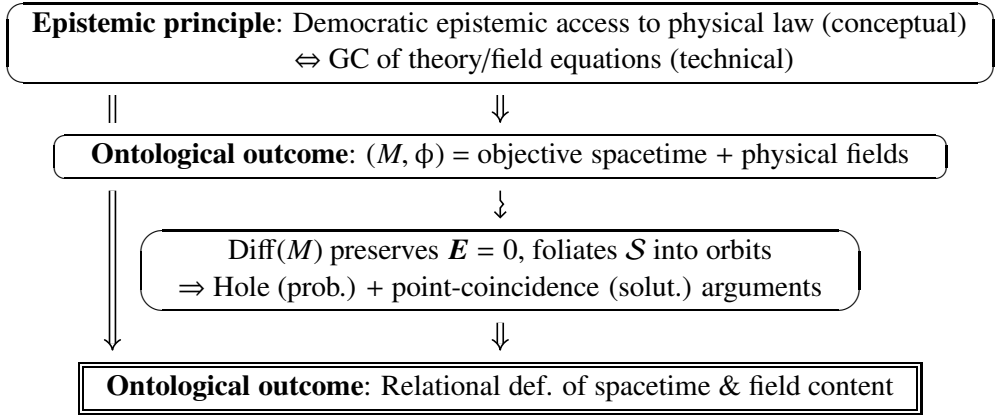
\begin{figure}[h!]
\centering
\begin{tikzcd}[row sep=1em]
\ovalbox{\parbox{\dimexpr\linewidth-20\fboxsep-200\fboxrule\relax}{\centering {\bf Epistemic principle}: Democratic epistemic access to physical law (conceptual) \\ $\qquad\quad \Leftrightarrow$ GC of theory/field equations (technical)}} 
\arrow[d, Rightarrow] \arrow[d, equal,  start anchor={[xshift=-5.10cm]}, end anchor={[xshift=-5.10cm]}] \\
\ovalbox{\parbox{\dimexpr\linewidth-25\fboxsep-250\fboxrule\relax}{\centering {\bf Ontological outcome}:  $(M, \upphi)$ $ = $ objective spacetime $+$ physical fields }}
\arrow[d, rightsquigarrow] \arrow[dd, Rightarrow,  start anchor={[xshift=-5.10cm]}, end anchor={[xshift=-5.10cm]}] \\
\ovalbox{\parbox{\dimexpr\linewidth-20\fboxsep-450\fboxrule\relax}{\centering  $\Diff(M)$ preserves $\bs E=0$, foliates $\S$ into orbits \\ $ \Rightarrow$ Hole (prob.) $+$ point-coincidence (solut.) arguments}}
\arrow[d, Rightarrow] \\
\doublebox{\ \text{ {\bf Ontological  outcome}: 
 Relational def. of spacetime \& field content   }\  } 
\end{tikzcd}
\caption{Relationality in the general-relativistic framework.} \label{Diag1}
\end{figure}

\noindent 
An all but similar analysis can be carried through for gauge field theory (GFT).

\paragraph{Gauge field physics} 
The gauge principle (GP) requires that the laws of physics be indifferent to the choice of ``gauge" representatives of the fields under consideration, which implies that the field equations for describing these laws be gauge-tensorial.  
Classical GFT being based on the geometry of fiber bundles, one easily identifies changes of gauge representatives as changes of principal bundle coordinates, a.k.a.  \emph{local gluings}, or \emph{passive gauge transformations}.
The GP can thus be understood, in first analysis, to imply that gauge field physics describes the structure and dynamics of an \emph{enriched spacetime} and its field content, faithfully represented by a smooth principal fiber bundle $P$ and  fields $\upphi$ defined on it, satisfying gauge-tensorial equations $\bs E(\upphi)=0$: The geometrical objects $(P, \upphi)$ are  bundle-coordinate invariant, and model the objective -- i.e. viewpoint independent -- structure of the enriched spacetime and its gauge-field content.   

Now, the group of \emph{vertical automorphisms} $\Aut_v(P)$ of the bundle, isomorphic to its gauge group $\H$, is an automorphism group of the geometric objects defined on $P$, notably connections (i.e. gauge potentials) and tensorial forms (field strengths, matter fields and their covariant derivatives): it defines their \emph{active gauge transformations}.
The field equations are then $\Aut_v(P)$-covariant, and $\Aut_v(P)$ is thus an automorphism group of the space of solutions $\S\defeq \{\upphi \,|\, \bs E (\upphi)=0\}$, which it foliates into orbits: any solution $\upphi\in \S$ has a $\Aut_v(P)$-orbit $\O_\upphi \subset \S$.\footnote{There again, the action of $\Aut_v(P)$ is a priori not free: some solutions may have gauge Killing symmetries, ${}^{\text{gt}}{\sf K}_\upphi \neq \id_P$.} 
From this fact, one may articulate an \emph{internal} hole argument and an \emph{internal} point-coincidence argument. 

The internal hole argument stresses the incompatibility between $\Aut_v(P)$-covariance of $\bs E (\upphi)=0$ and 
the view that $(P, \upphi)$ faithfully represent the physical structure of an enriched spacetime and its field content. It can be expressed thus:
The existence of two solutions $\upphi, \upphi' \in \O_\upphi$, i.e. $\upphi'=\psi^*\upphi$,  s.t.  $\psi$ is a compactly supported vertical automorphism whose support $D_\psi \subset P$ is the ``internal" hole,   manifestly raises an issue with the Cauchy problem and determinism. 
There are again two possibilities to deal with this: 
One may either drop the requirement of gauge-covariance of the theory (abandon the GP), which is unadvisable given  the empirical success of the GFT framework,
or conclude that all solutions within the same $\Aut_v(P)$-orbit $\O_\upphi$ represent a single  physical state. 
This is the well-know fact that in GFT there is a one-to-many correspondence between a physical state and its mathematical descriptions. 
A gauge field theory is thus unable to physically distinguish between $\Aut_v(P)$-related solutions of the field equations $\bs E(\upphi)=0$, and consequently cannot  distinguish $\Aut_v(P)$-related points within fibers of $P$ either. 
One must then conclude that the ennriched spacetime and its field content is not described by $(P, \upphi)$, but by its $\Aut_v(P)$-class.

The \emph{active} gauge-covariance under  $\Aut_v(P)$ is not  a mere redundancy, it encodes a fundamental physical insight, established by the internal point-coincidence argument, which is this:
Only the \emph{relative} values of the internal  d.o.f. of the fields at a point $p\in P$  have  physical meaning, and the description of these point-wise coincidences is $\Aut_v(P)$-invariant.\footnote{For example, in QED, only the \emph{relative phase} of the electromagnetic (EM) potential and charged field is meaningful. A stark illustration of this is the Aharonov-Bohm (AB) effect.}
The internal point-coincidence argument
dissolves any appearance of  indeterminism arising from (active) gauge symmetry of the theory and the associated internal hole argument.
Taking it a step further, it can be understood to mean that points of the physical internal structure of spacetime  are   \emph{defined} via coincidences of distinct internal physical field-theoretical d.o.f., and that these d.o.f. do not belong to the individual mathematical fields $\{\upphi\}$ per se, but are instantiated as \emph{internal relations} among them.

Summarising, the ontological consequence of the  GP, resulting from the internal hole and point-coincidence arguments, is the relationality of GFT physics, which one may state thus:
The internal structure of spacetime is relationally defined via its field content, and internal d.o.f. are relationally defined and evolve w.r.t. each other.
Relationality of gauge field physics is thus \emph{tacitly} encoded by the  $\Aut_v(P)$-covariance of a gauge field theory.\footnote{
Notice that relationality is also enjoyed in particular by solutions with gauge Killing symmetries 
${}^{\text{gt}}{\sf K}_\upphi \subset \Aut_v(P)$ -- the latter encoding futher physical properties of the solution. }
\medskip

The lessons from the general-relativistic and gauge field-theoretic frameworks obviously carry over to their union, which we now consider.

\paragraph{General-relativistic gauge field theory} 
Taken together, the GC principle and the GP are the starting points of the framework of general-relativistic gauge field theory (gRGFT): They must be understood as principles of democratic epistemic access to Nature, technically implemented as the requirement that the field equations representing the laws of physics must be general covariant and gauge-covariant, i.e. tensorial and gauge-tensorial.

In first analysis, this means that the general-relativistic gauge physics describes the structure and dynamics of an enriched spacetime, with structureful points, and its field content. 
The enriched spacetime is described by a principal bundle $P$ with structure group $H$, acting freely (and transitively on fibers, i.e. $H$-orbits), so that the space of fibers is a manifold $P/H = M$  (the base manifold) and one has the projection map $\pi : P\rarrow M$.  
Thus, the geometric structure that is to describe the spatiotemporal d.o.f. is seen to emerge as a quotient space of the geometric structure describing  the totality of the elementary d.o.f. (spatiotemporal $+$ internal). 
Physical fields are described by geometric objects $\phi$ on $P$, satisfying tensorial and gauge-tensorial field equations $\bs E(\upphi)=0$. 
The geometric structure $(P, \upphi)$ is coordinate and bundle-coordinate invariant, and models the objective, viewpoint independent, structure of the enriched relativistic spacetime and its gauge field content. 

Now, the maximal group of transformations of a bundle $P$ is its group of automorphisms $\Aut(P)$:  the subgroup of $H$-equivariant diffeomorphisms, preserving the fibration structure. 
It thus induces~smooth diffeomorphisms of the space of fibers $P/H= M$:
there is a surjection  $\t\pi: \Aut(P) \rarrow \Diff(M)$. 
The group of vertical automorphisms induces the identity transformation on $M$ -- i.e. it is the kernel of $\t \pi$ -- and is contained as a normal subgroup, $\Aut_v(P) \triangleleft \Aut(P)$.\footnote{One may define vertical diffeomorphisms of $P$, $\Diff_v(P)\defeq \big\{ \psi\in \Diff(P)\, |\, \pi \circ \psi =\pi \big\}$, so that $\Aut_v(P) \subset \Diff_v(P)$, which also induce $\id_M \in \Diff(M)$ on $M$. But, since these are not $H$-equivariant, they are not natural morphisms (arrows) in the category of principal bundles. 
Still, they induce \emph{generalised} gauge transformations, so are relevant for GFT. See \cite{Francois2023-b, Francois2023-a}. We will  encounter this structure again in the next section.}
We~have the short exact sequence (SES) of groups
\begin{align}
\label{SES-P}
\id_P\rarrow \Aut_v(P) \simeq \H \xlongrightarrow{\triangleleft} \Aut(P)  \xlongrightarrow{\t \pi}  \Diff(M) \rarrow  \id_M.
\end{align}
Naturally, $\Aut(P)$ acts as an automorphism group on the space of geometric objects on $P$, notably connections and tensorial forms: its action on objects of those spaces defines their combined active gauge transformations and diffeomorphism transformations. 
The field equations of a general-relativistic gauge theory are thus $\Aut(P)$-covariant, and $\Aut(P)$, as an automorphism group of the space of solutions $\S \defeq \{\upphi\, |\, \bs E(\upphi)=0\}$, foliates it into orbits so that a solution $\upphi \in \S$ has an $\Aut(P)$-orbit $\O_\upphi \subset \S$. 
This set the stage for the \emph{generalised} hole and point-coincidence arguments. 

The generalised hole argument establishes the clash between $\Aut(P)$-covariance of the field equations $\bs E(\upphi)=0$ and the initial notion that $(P, \upphi)$ faithfully represents an enriched general-relativistic spacetime and its gauge field content.
Indeed, the possibility of having two solutions 
$\upphi, \upphi' \in \O_\upphi$, i.e. $\upphi'=\psi^*\upphi$,  s.t.  $\psi$ is a compactly supported automorphism whose support $D_\psi \subset P$ is the ``bundle hole",  is an issue for the Cauchy problem and determinism.
The empirical success of the framework, preventing us to abandon general covariance and gauge-covariance, forces to admit that physical d.o.f. have to be $\Aut(P)$-invariant, yielding the revision:
The  structure and dynamics of the enriched general-relativistic spacetime and its gauge field content are modelled by the $\Aut(P)$-class of 
$(P, \upphi)$.

Again,  $\Aut(P)$-covariance, far from being a mere redundancy, encodes the fundamental physical insight brought forth by the generalised point-coincidence argument:
Only the relative values of the fields at a point $p\in P$  have a physical meaning, and the description of these point-wise coincidences is invariant under automorphisms of $P$.
The~$\Aut(P)$-invariance of point-wise mutual relations $\R$ between the fields  $\{\upphi\}$ we shall write  
 \begin{equation}
 \begin{aligned}
 \label{gen-PC-arg}
 \R :  \S \times P\  &\ \rarrow \ \   \S \times P / \sim \  \ \xrightarrow{\simeq} \  \text{Relational physical d.o.f.,} \\
 \big(\upphi, p \big) \  &\  \mapsto  \ \big(\upphi, p \big) \sim \big(\psi^*\upphi, \psi\-(p)\big)   \ \mapsto\   \R \big(\upphi; p \big) =\R \big(\psi^*\upphi; \psi\-(p)\big),
 \end{aligned} 
\end{equation}
where $\S \times P /\sim$ is the quotient of  $ \S \times P$ by the equivalence relation $\big(\upphi, p \big) \sim \big(\psi^*\upphi, \psi\-(p)\big)$, for $\psi \in Aut(P)$ -- See section \ref{Associated bundle of regions}. 
Taking the generalised point-coincidence argument to its logical conclusion implies that \eqref{gen-PC-arg}  can be understood to mean both that points of the physical enriched spacetime are   \emph{defined} via coincidences of distinct physical field-theoretical d.o.f., and that  the latter are not the individual mathematical fields $\{\upphi\}$ per se, but the \emph{relational} d.o.f. established between them.

The key ontological consequence of the epistemic general covariance and gauge principles, reached via the articulation of the generalised hole and point-coincidence arguments, is the \emph{relationality} of general-relativistic gauge physics:
The enriched spacetime is relationally defined via its field content, and
all physical d.o.f. are relationally defined and evolve w.r.t. each other. 
The  diagram  Fig.\ref{Diag3} below summarises the logic.
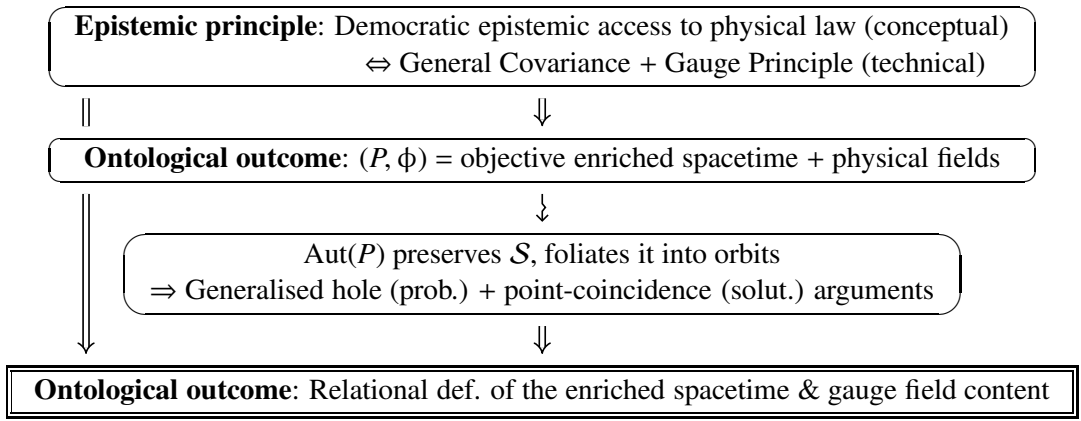
\begin{figure}[h!]
\centering
\begin{tikzcd}[row sep=1em]
\ovalbox{\parbox{\dimexpr\linewidth-20\fboxsep-200\fboxrule\relax}{\centering {\bf Epistemic principle}: Democratic epistemic access to physical law (conceptual) \\  \hspace{3.5cm}$ \Leftrightarrow$ General Covariance $+$ Gauge Principle (technical)}} 
\arrow[d, Rightarrow] \arrow[d, equal,  start anchor={[xshift=-6cm]}, end anchor={[xshift=-6cm]}] \\
\ovalbox{\parbox{\dimexpr\linewidth-24\fboxsep-170\fboxrule\relax}{\centering {\bf Ontological outcome}:  $(P, \upphi)$ $ = $ objective enriched spacetime $+$ physical fields}}
\arrow[d, rightsquigarrow] \arrow[dd, Rightarrow,  start anchor={[xshift=-6cm]}, end anchor={[xshift=-6cm]}] \\
\ovalbox{\parbox{\dimexpr\linewidth-20\fboxsep-340\fboxrule\relax}{\centering  $\Aut(P)$ preserves $\S$, foliates it into orbits \\ $ \Rightarrow$ Generalised hole (prob.) $+$ point-coincidence (solut.) arguments}}
\arrow[d, Rightarrow] \\
\doublebox{\ \text{ {\bf Ontological  outcome}: 
 Relational def. of the enriched spacetime \& gauge field content    }\  } 
\end{tikzcd}
\caption{Relationality in  general-relativistic gauge field theory.} \label{Diag3}
\end{figure}

The above logic, establishing relationality as the conceptual core of gRGFT, would naturally extended to any (empirically successful) field theory resting on covariance/invariance under  local symmetries, as these may point to an underlying geometric structure modelling  objective physical entities.
For example,  it could apply to models of supersymmetric theory, such as supergravity or strings models, as well as higher gauge theory --  relying respectively on supergeometry, higher geometry, or higher supergeometry -- if ever these were to make relevant empirical contact with fundamental physics.
\bigskip

In the standard mathematical formulation of gRGFT, relationality of physics is tacit, encoded in the manifest invariance, or covariance, of the theory under its symmetries ($\Diff(M)$, $\Aut_v(P)\simeq \H$, and $\Aut(P)$). 
It can thus be easily overlooked, which may lead to a number of unfortunate misconceptions. 
For instance, the notion that ``boundaries" break $\Diff(M)$ or $\H$-gauge symmetries, which has the same conceptual structure as a hole argument. This notion evaporates once it is recognised that a physical boundary is relationally defined, and is invariant (under $\Diff(M)$, $\Aut_v(P)\simeq \H$, or $\Aut(P)$). 
This and similar misconceptions, together with the various  countermeasures put forward to solve the alleged issue, would be avoided had one a framework in which both relationality and strict invariance are manifest.
Such a reformulation of gRGFT would also give ready access to the observables of a theory. 

In the following, we propose such a framework, whose technical touchstone is the \emph{dressing field method} (DFM). It is best formulated within the bundle geometry of the space of fields of a general-relativistic gauge field theory. 
In the next section \ref{Geometry of field space}, we give our account of this bundle geometry, before 
describing the DFM for the gRGFT framework in section \ref{The dressing field method}. 
Its adaptation to field theory is the object of section \ref{Local field theory}.

\section{Geometry of global field space}  
\label{Geometry of field space}  

In this section we elaborate on the geometry of field space as an infinite-dimensional fiber bundle, extending  standard notions defined in the finite-dimensional context to the infinite-dimensional setting \cite{Frolicher-Kriegl1988, Kriegl-Michor1997}.

\subsection{Field space as a principal bundle}  
\label{Field space as a principal bundle}  

We are interested in  gauge field theories based on a principal bundle $P(M,  H)$ with structure group $H$. For a realistic description of spinorial matter fields, the principal bundles considered must be of the form $P=Q \times OM$ with $H=G\times S\!O(r,s)$, where $Q(M, G)$  is a $G$-bundle over $M$ and $OM$ is the orthonormal frame bundle of $M$.\footnote{$P=Q \times OM$ is bundle over $M$, \emph{not} $M \times M$: it is a \emph{pullback}, or \emph{fibered product} in the category of principal bundles.} 
Therefore, our field space  $\Phi$ is made of tensors or differential forms taking values in various representations $V$ of $H$, $\Omega^\bullet(P) \otimes V$. 
For~example (Ehresmann or Cartan) connection 1-forms and their curvature 2-forms ($V=$ Lie$H$), or $V$-valued tensorial 0-forms (describing e.g. charged spinors) and their covariant derivative 1-forms.
A collection of such objects will be noted as $\phi \in \Phi$, i.e. it is a single point in field space. 

 The group $\Aut(P)$  has a natural  action by pullback on $\Phi$:
  Given $\phi \in \Phi$ and $\psi \in \Aut(P)$ we have  $\phi^\psi := \psi^*\phi$, where the left-hand side is a notation for the pullback on the right-hand side.
  This is a right-action since it is well-known that the pullback satisfies $(f\circ g)^*=g^* \circ f^*$ for any two smooth maps $M \xrightarrow{g}N \xrightarrow{f} Q$.\footnote{This is clear on forms, a.k.a. contravariant tensors. On vector fields and more general covariant tensors, the pullback action is the pushforward by the inverse: $\psi^* \defeq {\psi^{-1}}_*$. So, on general (mixed) tensors on $P$, $\psi^*$ is indeed a well-defined right action of $\Aut(P)$. } 
 We therefore write
 \begin{equation}
  \begin{aligned}
  \label{Right-action-on-Phi}
  \Phi \times\Aut(P) &\rarrow \Phi,  \\
               (\phi, \psi) &\mapsto R_\psi \phi:= \psi^*\phi,
\end{aligned}
 \end{equation}              
with indeed, for another  $\psi' \in\Aut(P)$: $R_{\psi'} R_\psi \phi:= {\psi'}^*\psi^*\phi= (\psi \circ \psi')^*\phi=:R_{\psi \circ \psi'} \phi$. 
The field space is fibered by the action of $\Aut(P)$, the fiber through a point $\phi$ being its orbit $\mathcal O(\phi)$ under automorphisms. 
We denote the set of orbits, or moduli space, by $\Phi/\Aut(P)=: \M$. 
Under adequate restrictions of either of $\Phi$ or $\Aut(P)$, the field space $\Phi$ can be understood as an infinite-dimensional principal fiber bundle over the base $\M$ with structure group $\Aut(P)$:\footnote{In particular, this requires that all points $\phi$ have trivial stability groups, meaning that we a priori reject Killing symmetries of $\phi$.} 

 \begin{equation}
  \begin{aligned}
  \Phi &\xrightarrow{\pi} \M,  \\
   \phi &\mapsto \pi(\phi)=: [\phi].
\end{aligned}
 \end{equation}     
The projection $\pi$ is such that (s.t.) $\pi \circ R_\psi= \pi$. 
The fiber over a point $[\phi] \in \M$, $\pi\-([\phi])=\mathcal O(\phi)$,  is diffeomorphic to the structure group $\Aut(P)$ as a manifold.
We pass on the local bundle structure of $\Phi$: since the space of orbits is not manageable operationally, one usually does not work on it. For details we refer the reader to \cite{Francois2023-a}.

\subsubsection{Natural transformation groups} 
\label{Natural transformation groups}  

As an infinite-dimensional manifold, $\Phi$ has a diffeomorphism group $\bs{\Diff}(\Phi)$, but \emph{as a principal bundle} its maximal transformation group is its group of \emph{automorphisms}
\begin{align}
    \bs{\Aut}(\Phi) := \big\{ \, \Xi \in \bs{\Diff}(\Phi) \, |\, \Xi \circ R_\psi = R_\psi \circ \Xi  \, \big\},
\end{align}
whose elements  preserve the fibration structure and thus project naturally as elements of $\bs{\Diff}(\M)$. 

The subgroup of \emph{vertical diffeomorphisms}
\begin{align}
    \bs{\Diff}_v(\Phi) \defeq \left\{ \, \Xi \in \bs{\Diff}(\Phi)\, |\, \pi \circ \Xi = \pi \, \right\}
\end{align}
induces the identity transformation on $\M$: 
Since these are motions along fibers, to $\Xi \in \bs{\Diff}_v(\Phi)$ must correspond a unique $\bs\psi: \Phi \rarrow \Aut(P)$ s.t. $\Xi(\phi)=R_{\bs\psi(\phi)} \phi := [\bs\psi(\phi)]^* \phi$: i.e.  $\bs{\Diff}_v(\Phi) \simeq C^\infty\big(\Phi, \Aut(P)\big)$. 
Notice that the map composition law for $\bs{\Diff}_v(\Phi)$ gives rise to a peculiar composition operation for $C^\infty\big(\Phi, \Aut(P)\big)$: For $\Xi, \Xi' \in \bs{\Diff}_v(\Phi)$ to which correspond $\bs\psi, \bs\psi' \in C^\infty\big(\Phi, \Aut(P)\big)$, one finds
$
\Xi' \bs \circ\,\! \Xi \,(\phi) = R_{\bs\psi' \left( \Xi(\phi) \right)}\, \Xi(\phi) = R_{\bs\psi' \left( \Xi(\phi) \right)}\, R_{\bs\psi(\phi)}\phi = R_{\bs\psi(\phi) \circ \bs\psi' \left( \Xi(\phi) \right) } \phi.
$
Thus we have, 
\begin{align}
\label{twisted-comp-law}
 \Xi' \bs \circ\,\! \Xi \in  \bs{\Diff}_v(\Phi) \ \ \text{ corresponds to }\ \ \bs\psi \circ \big( \bs\psi' \bs \circ R_{\bs \psi}\big) \in C^\infty\big(\Phi, \Aut(P)\big). 
\end{align}
Remark the distinction between the composition law $\bs \circ$ of maps on $\Phi$, and the composition law $\circ$ of maps on $P$. 

This is an example of the composition law of the infinite-dimensional group of bisections of a Lie groupoid \cite{Mackenzie2005, Schmeding-Wockel2015, Maujouy2022}. As a mater of fact, what we have been discussing above can be reframed in the groupoid framework as  follows: 
One defines the generalised action groupoid $\bs\Gamma  \rightrightarrows \Phi$ with $\bs \Gamma = \Phi \rtimes \Big(C^\infty\big(\Phi, \Aut(P)\big) \simeq \bs\Diff_v(\Phi)\Big)$, and with source and target maps 
$s: \bs\Gamma \rarrow \Phi$, $(\phi, \bs\psi \simeq \Xi) \mapsto \phi$,  and 
$t: \bs\Gamma \rarrow \Phi$, $(\phi, \bs\psi \simeq \Xi) \mapsto \Xi(\phi)=\bs\psi(\phi)^*\phi$.
The associative composition law: $g \circ f \in \bs\Gamma$ for $f,  g\in \bs\Gamma$ is defined whenever $t(f)=s(g)$.
 It generalises the action groupoid $\b{\bs\Gamma} =  \Phi \rtimes \Aut(P) \rightrightarrows \Phi$ associated with the right action of $\Aut(P)$ on $\Phi$. The group of bisections $\mathcal B(\bs\Gamma)$ of $\bs\Gamma$ is the set of sections of  $s$ --  maps $\sigma : \Phi \rarrow \bs\Gamma$, $\phi \mapsto \s(\phi)=(\phi, \bs\psi \sim \Xi)$  s.t. $s\circ \s =\id_\Phi$ -- such that $t \circ \s: \Phi\rarrow \Phi$ is invertible: thus $t \circ \s \in \bs\Diff(\Phi)$. We have indeed, for  $\s \in \mathcal B(\bs\Gamma)$, that $t \circ \s=\Xi \, (=\bs\psi^*) \in \bs\Diff_v(\Phi)\simeq C^\infty\big(\Phi,  \Aut(P) \big)$. 
The group law of bisections is defined as 
$\big(\s_2 \star \s_1\big)(\phi) \defeq \s_2\big( (t \circ \s_1 (\phi) \big) \circ \s_1(\phi)$. 
After composition with the target map $t$ on the left, this reproduces precisely the peculiar group law \eqref{twisted-comp-law}. 
Then, slightly abusing the terminology, we may refer to $C^\infty\big(\Phi,  \Aut(P) \big)$ as the group of bisections of the generalised action groupoid $\bs\Gamma$ associated to $\Phi$. 
In the rest of this paper though, we will simply, and accurately, refer to it as the group of generating maps of $\bs\Diff_v(\Phi)$.

This  generalises the subgroup of \emph{vertical automorphisms}
\begin{align}
    \bs{\Aut}_v(\Phi) := \big\{ \,\Xi \in \bs{\Aut}(\Phi) \, |\, \pi \circ \Xi = \pi \, \big\}=\bs\Diff_v(\Phi)  \cap \bs\Aut(\Phi),
\end{align}
isomorphic to the \emph{gauge group} 
\begin{align}
 \label{GaugeGroup}
 \bs{\Aut}(P) := \big\{ \,\bs\psi : \Phi \rarrow  \Aut(P) \, |\,  \bs\psi(\phi^\psi) =\psi^{-1} \circ \bs\psi(\phi) \circ \psi \,  \big\} 
\end{align}
via $\Xi(\phi)=R_{\bs\psi(\phi)}\, \phi$ still. 
The equivariance of elements $\bs\psi$ of $\bs\Aut(P)$ implies that to $ \Xi' \bs \circ\,\! \Xi \in  \bs{\Aut}_v(\Phi)$ \mbox{corresponds} $\bs\psi' \circ  \bs\psi \in\bs\Aut(P)$: i.e. the composition operation $\bs\circ$ in $ \bs{\Aut}_v(\Phi)$ translates to  the usual  composition operation $\circ$ of the group $\bs\Aut(P)$. 

We have that $\bs{\Aut}(\Phi)$ is in the normaliser of  $\bs{\Diff}_v(\Phi)$ and, since a group is a subgroup of its normaliser, we get:  $N_{\bs\Diff(\Phi)}\big(\bs\Diff_v(\Phi) \big) \supset \langle \bs\Diff_v(\Phi) \cup \bs\Aut(\Phi) \rangle$. 
We have the special case $N_{\bs\Diff(\Phi)}\big(\bs\Aut_v(\Phi) \big) = \bs\Aut(\Phi)$, i.e.  $ \bs\Aut_v(\Phi) \triangleleft  \bs\Aut(\Phi)$, which gives the short exact sequence (SES)
 \begin{align}
 \label{SESgroup}
\id_\Phi\rarrow \bs\Aut(P) \simeq \bs{\Aut}_v(\Phi)  \xlongrightarrow{\triangleleft}   \bs{\Aut}(\Phi) \longrightarrow \bs{\Diff}(\M) \rarrow \id_\M,
 \end{align}
where the image of each arrow is in the kernel of the next.  
One often encounters in the literature the notion of ``\emph{field-dependent}" gauge transformations and diffeomorphisms. The proper mathematical embodiment of this notion is the group $\bs\Diff_v(\Phi) \simeq C^\infty\big(\Phi, \Aut(P) \big)$. Nonetheless, stricto sensu, the gauge transformations on $\Phi$ are defined via the action of the subgroup $\bs{\Aut}_v(\Phi) \simeq \bs{\Aut}(P)$. 
The structure group $\Aut(P)$ supplies the notion of  ``\emph{field-independent}" gauge transformations and diffeomorphisms.
The linearization of \eqref{SESgroup} gives a SES of Lie algebras defining the Atiyah Lie algebroid of the principal bundle $\Phi$. 
\medskip

To study the differential structure of $\Phi$, let us first recall the following general results.
Consider the manifolds $M$, $N$ and their tangent bundles $TM$, $TN$, together with a diffeomorphism $\psi: M \rarrow N$ and the flow $\upphi_\tau:N \rarrow N$ of a vector field $X \in \Gamma(TN)$, s.t. $X_{|\upphi_0}=\tfrac{d}{d\tau} \upphi_\tau |_{\tau=0} \in T_{\upphi_0}N$. One defines the flow
 \begin{align}
 \label{Psi-conjugation}
\vphi_\tau:= \psi\- \circ \upphi_\tau \circ \psi : M \rarrow M
\end{align}
of a vector field $Y \in \Gamma(TM)$  related to $X$ as
 \begin{align}
 \label{Psi-relatedness}
Y:=  \tfrac{d}{d\tau} \left( \psi\- \circ \upphi_\tau \circ \psi \right)\big|_{\tau=0}= (\psi\-)_*\, X \circ \psi.
 \end{align}
We have the composition of maps $M \xrightarrow{\psi} N \xrightarrow{X} TN \xrightarrow{(\psi\-)_*} TM$ resulting in the above vector field $Y: M \xrightarrow{} TM$.
So,~$X\in \Gamma(TN)$ and $Y\in \Gamma(TM)$ are $\psi$-related \eqref{Psi-relatedness} when their flows are $\psi$-\emph{conjugated} \eqref{Psi-conjugation}. Furthermore,  $\psi$-relatedness is a morphism of Lie algebras, that is $[Y, Y']=  (\psi\-)_* [X, X'] \circ \psi$.

As a variation of the above, suppose $\phi$ is some tensor field on $M$ and $\mathfrak L_X \phi$ its Lie derivative along $X \in \Gamma(TM)$ with flow $\varphi_\tau$, for  $\psi \in \Diff(M)$ we have 
\begin{equation}
\begin{aligned}
\label{lemma1}
\psi^*\big( \mathfrak L_X \phi \big) &=  \psi^* \tfrac{d}{d\tau} \varphi^*_\tau \phi \big|_{\tau=0} =\tfrac{d}{d\tau} (\varphi_\tau \circ \psi )^* \phi \big|_{\tau=0} 
						     =\tfrac{d}{d\tau} (\varphi_\tau \circ \psi )^* \ (\psi\-)^*\psi^* \phi \big|_{\tau=0} \\
						     &=\tfrac{d}{d\tau} ( \psi\- \circ \varphi_\tau \circ \psi )^* \ \psi^* \phi \big|_{\tau=0} \\
						     & =: \mathfrak L_{[(\psi\-)_*X \circ \psi]} \, (\psi^* \phi). 
\end{aligned}
\end{equation}

\subsection{Differential structure} 
\label{Differential structure}  

As a manifold, $\Phi$ has a tangent bundle $T\Phi$, a cotangent bundle $T^\star\Phi$, or more generally a space of forms $\Omega^\bullet(\Phi)$. 
Considering these structures in turn, it will be important to distinguish  the pushforward and pullback on $P$ and $\Phi$: we  reserve $*$ to denote these operations on $P$, and $\star$ for their counterparts on $\Phi$.  

\subsubsection{Tangent bundle and subbundles} 
\label{Tangent bundle and subbundles}  

Sections of the tangent bundle  $\mathfrak X : \Phi \rarrow T\Phi$ are vector fields on $\Phi$, we note $\mathfrak X \in \Gamma(T\Phi)$. They form a Lie algebra under the bracket of vector field $[\ , \, ]: \Gamma(T\Phi) \times \Gamma(T\Phi) \rarrow \Gamma(T\Phi)$. 
We may write a vector field at $\phi \in \Phi$ as $\mathfrak X_{|\phi} =\tfrac{d}{d\tau}  \Psi_\tau(\phi) \, \big|_{\tau=0}$, with  $\Psi_\tau \in \bs\Diff(\Phi)$ its flow s.t.  $\Psi_{\tau=0}(\phi)=\phi$.
As derivations of the algebra of functions $C^\infty(\Phi)$ we write: $\mathfrak X = \mathfrak X(\phi) \tfrac{\delta}{\delta \phi} $, where $\tfrac{\delta}{\delta \phi}$ denotes  the functional differentiation w.r.t. $\phi$, and $\mathfrak X(\phi)$ are the functional components. 
The Lie bracket in the Lie algebra of $\bs\Diff(\Phi)$ is \emph{minus} the bracket in $\Gamma(T\Phi)$:  
$\bs\diff(\Phi)\defeq \big(\Gamma(T\Phi), -[\ , \ ]_{\text{\tiny{$\Gamma(T\Phi)$}}} \big)$.

 The pushforward by the projection is $\pi_\star: T_\phi\Phi \rarrow T_{\pi(\phi)}\M=T_{[\phi]}\M$. 
The pushforward by the right action of $\psi \in \Aut(P)$ is  $R_{\psi\star}: T_\phi\Phi \rarrow T_{\psi^*\phi}\Phi$.  
 In general $R_{\psi\star} \mathfrak X_{|\phi} \neq \mathfrak X_{|\psi^*\phi}$, 
 meaning that a generic vector field ``rotates" as it is pushed vertically along fibers.  
So, in general, $\pi_\star \mathfrak X$ is not a well-defined vector field on the base $\M$:  at $[\phi]\in \M$  the vector  obtained would vary depending on where on the fiber over $[\phi]$ the projection is taken.

The Lie subalgebra of \emph{right-invariant} vector fields, which do not rotate as they are pushed vertically,  is
 \begin{align}
 \label{Inv-vector-fields}
\Gamma_{\text{\!\tiny{inv}}}(T\Phi):=\left\{\mathfrak X \in \Gamma(T\Phi)\, |\, R_{\psi\star} \mathfrak X_{|\phi}=  \mathfrak X_{|\psi^*\phi}  \right\}.
 \end{align}
 They have well-defined projections on $\M$:  For $\mathfrak X \in \Gamma_{\text{\!\tiny{inv}}}(T\Phi)$, we have $\pi_\star \mathfrak X_{|\psi^*\phi}=\pi_\star R_{\psi\star} \mathfrak X_{|\phi} = (\pi \circ R_\psi)_\star \mathfrak X_{|\phi} = \pi_\star \mathfrak X_{|\phi} =: \mathfrak Y_{|[\phi]}  \in T_{[\phi]}\M$.  Then, $\pi_\star \mathfrak X =: \mathfrak Y \in \Gamma(T\M)$ is a well-defined vector field. 
 The defining property of invariant vector fields implies that their flows are automorphisms of $\Phi$: we have
 \begin{align}
R_{\psi\star} \mathfrak X_{|\phi}=\tfrac{d}{d\tau} R_\psi \Psi_\tau (\phi)\, \big|_{\tau=0} \quad \text{and} \quad  \mathfrak X_{|R_\psi \phi}=\tfrac{d}{d\tau}  \Psi_\tau (R_\psi \phi)\, \big|_{\tau=0},
 \end{align}
which  implies $R_\psi \circ \Psi_\tau = \Psi_\tau \circ R_\psi$. 
The Lie subalgebra $\Gamma_{\text{\!\tiny{inv}}}(T\Phi)$ is thus the Lie algebra of  $\bs\Aut(\Phi)$:
  \begin{align}
  \label{LieAlg-Aut}
  \bs\aut(\Phi)=\big(\Gamma_{\text{\!\tiny{inv}}}(T\Phi); -[\ , \ ]_{\text{\tiny{$\Gamma(T\Phi)$}}}  \big).
  \end{align}
The latter should not be confused with the Lie algebra $\aut(P)$ of  the structure group $\Aut(P)$:
  \begin{align}
\aut(P)=\big(\Gamma_{\text{\!\tiny{inv}}}(TP); -[\ , \, ]_{\text{{\tiny $\Gamma(TP)$}}}\big). 
  \end{align}
We may use the notation $[X,Y]_{\text{{\tiny $\aut$}}}:=-[X , Y]_{\text{{\tiny $\Gamma(TP)$}}}$ when useful.
\smallskip

The \emph{vertical tangent bundle} $V\Phi := \ker \pi_\star$ is a canonical  subbundle of the tangent bundle $T\Phi$.
Vertical vector fields are elements of $\Gamma(V\Phi):=\left\{  \mathfrak X \in \Gamma(T\Phi)\, |\, \pi_\star \mathfrak X=0 \right\}$. 
Since $V\Phi$ is a subbundle,  $\Gamma(V\Phi)$ is a Lie ideal of $\Gamma(T\Phi)$. 
Indeed,  since $\pi_\star: \Gamma(T\Phi) \rarrow \Gamma(T\M)$ is a Lie algebra morphism, we have, for $\mathfrak X \in \Gamma(V\Phi)$ and $\mathfrak Y \in \Gamma(T\Phi)$: $\pi_\star[\mathfrak X, \mathfrak Y]=[\pi_\star \mathfrak X, \pi_\star \mathfrak Y ]=[0, \pi_\star \mathfrak Y]=0$, i.e. $[\mathfrak X, \mathfrak Y] \in \Gamma(V\Phi)$.

We now consider the vertical vector fields induced by the respective actions of $\aut(P)$, $\bs\aut(P)$ and $\bs\diff_v(\Phi)$. 
 A~\emph{fundamental} vertical vector field at $\phi \in \Phi$ generated by  $X=\tfrac{d}{d\tau} \psi_\tau \big|_{\tau=0} \in \aut(P)$ with flow $\psi_\tau \in \Aut(P)$ is: 
 \begin{align}
 \label{Fund-vect-field}
 X^v_{|\phi} := \tfrac{d}{d\tau} R_{\psi_\tau} \phi\, \big|_{\tau=0} = \tfrac{d}{d\tau}  \psi^*_\tau \phi \, \big|_{\tau=0} =: \mathfrak L _X \phi, 
 \end{align} 
The  Lie derivative on $P$ is also given by the Cartan formula $ \mathfrak L _X=[\iota_x, d]=\iota_Xd + d\iota_X$, with $d$ the de Rham exterior derivative on $P$. It is a degree 0 derivation of the algebra $\Omega^\bullet(P)$ of forms on $P$, since $\iota_X$ is of degree $-1$ and $d$ is of degree $1$. 
Manifestly, fundamental vector fields satisfy $\pi_\star X^v\equiv 0$, since 
$\pi_\star X^v_{|\phi} =\tfrac{d}{d\tau} \pi \circ R_{\psi_\tau} \phi\, \big|_{\tau=0} = \tfrac{d}{d\tau} \pi(\phi)\, \big|_{\tau=0}$. 
One shows (see Appendix \ref{Lie algebra (anti-)isomorphisms}) that the map $|^v :\aut(P) \rarrow \Gamma(V\Phi)$, $X \mapsto X^v$, is a Lie algebra morphism: i.e. $([X,Y]_{\text{{\tiny $\aut$}}})^v=(-[X, Y]_{\text{{\tiny $\Gamma(TP)$}}})^v=[X^v, Y^v]$.
The pushforward by the right-action of $\Aut(P)$ on  a fundamental vertical vector field is:
\begin{equation}
\begin{aligned}
\label{Pushforward-fund-vect}
 R_{\psi\star} X^v_{|\phi} :=&\, \tfrac{d}{d\tau} R_\psi \circ R_{\psi_\tau} \phi \,  \big|_{\tau=0} 
                                         =  \tfrac{d}{d\tau} R_{\psi_\tau \circ \psi} \phi \, \big|_{\tau=0} =  \tfrac{d}{d\tau} R_{\psi_\tau \circ \psi} \, R_{\psi^{-1} \circ \psi} \, \phi \, \big|_{\tau=0}  \\
                                         =&\,  \tfrac{d}{d\tau} R_{(\psi\- \circ \psi_\tau \circ \psi)}  \, R_{\psi} \, \phi \, \big|_{\tau=0} =   \tfrac{d}{d\tau} R_{(\psi\- \circ \psi_\tau \circ \psi)}  \, \psi^*\phi \, \big|_{\tau=0}   \\
                                         =&\!: \left(  (\psi\-)_* \, X \circ \psi \right)^v_{|\psi^*\phi}. 
\end{aligned}
\end{equation}
Therefore,  fundamental vector fields generated by $\aut(P)$ are  not right-invariant. 

On the other hand,  the fundamental vector fields induced by $\bs\aut(P)$, the Lie algebra of the gauge group $\bs\Aut(P)$, are right-invariant. To $\bs\psi_\tau \in \bs\Aut(P)$ corresponds  $\bs X =\tfrac{d}{d\tau}\, \bs\psi_\tau\, \big|_{\tau=0} \in \bs\aut(P)$. Given the definition  \eqref{GaugeGroup} of the gauge group, whose elements transformation property is  $R^\star_\psi \bs\psi = \psi\- \circ \bs\psi \circ \psi $, by  \eqref{Psi-conjugation}-\eqref{Psi-relatedness} we have
 \begin{align}
 \label{LieAlg-GaugeGroup}
 \bs\aut(P):=\left\{\, \bs X: \Phi \rarrow \aut(P)\  |\ R^\star_\psi \bs X = (\psi\-)_*\, \bs X \circ \psi \,  \right\}.
 \end{align}
 This transformation property can also be written as: $\bs X(\phi^\psi)= \bs X(\psi^*\phi) =  (\psi\-)_*\, \bs X(\phi) \circ \psi $. 
Observe that the infinitesimal version is given by the Lie derivative on $\Phi$ along the corresponding fundamental vector field:
 \begin{align}
 \label{inf-equiv-gauge-Lie-alg}
\bs L_{X^v} \bs X = X^v(\bs X) = \tfrac{d}{d\tau} \, R^\star_{\psi_\tau} \bs X \, \big|_{\tau=0} = \tfrac{d}{d\tau} \, (\psi\-_\tau)_*\, \bs X \circ \psi_\tau  \, \big|_{\tau=0} 
				=: \mathfrak L_X \bs X =[X, \bs X]_{\text{{\tiny $\Gamma(TP)$}}} =[\bs X, X]_{\text{{\tiny $\aut$}}}. 
 \end{align}
A fundamental vector field generated by $\bs X \in \bs\aut(P)$ is
  \begin{align}
 \bs X^v_{|\phi} := \tfrac{d}{d\tau} R_{\bs\psi_\tau(\phi)} \phi\, \big|_{\tau=0} = \tfrac{d}{d\tau}  (\bs\psi_\tau(\phi))^* \phi \, \big|_{\tau=0} =: \mathfrak L _{\bs X} \phi.
 \end{align}
 Its pushforward by the right-action of $\Aut(P)$ is
  \begin{align}
 R_{\psi\star} \bs X^v_{|\phi} :=&\, \tfrac{d}{d\tau} R_\psi \circ R_{\bs\psi_\tau(\phi)} \phi \,  \big|_{\tau=0} 
                                                =  \tfrac{d}{d\tau} R_{(\psi\- \circ \bs\psi_\tau(\phi) \circ \psi)}  \, R_{\psi} \, \phi \, \big|_{\tau=0} =   \tfrac{d}{d\tau} R_{\bs\psi_\tau(\psi^*\phi)}  \, \psi^*\phi \, \big|_{\tau=0}
                                         \rdefeq \bs X^v_{|\psi^*\phi}. 
 \end{align}
Furthermore, one shows  (see Appendix \ref{Lie algebra (anti-)isomorphisms}) that the ``verticality map" $|^v : \bs\aut(P) \rarrow \Gamma_{\text{\!\tiny{inv}}}(V\Phi)$, $\bs X \mapsto \bs X^v$, is a Lie algebra \emph{anti}-morphism: i.e. $([\bs X, \bs Y]_{\text{{\tiny $\aut$}}})^v=(-[\bs X, \bs Y]_{\text{{\tiny $\Gamma(TP)$}}})^v=-[\bs X^v, \bs Y^v]$.
 Therefore, since the Lie subalgebra of right-invariant vertical vector fields is the Lie algebra of the group $\bs\Aut_v(\Phi)$, we have 
  \begin{align}
  \label{Gauge-Lie-alg-morph}
 \bs\aut(P)  \simeq \bs\aut_v(\Phi)=\big(\Gamma_{\text{\!\tiny{inv}}}(V\Phi); -[\ , \ ]_{\text{\tiny{$\Gamma(T\Phi)$}}}  \big).
  \end{align}
We can thus write the  infinitesimal version of \eqref{SESgroup}, i.e. the SES describing the Atiyah Lie algebroid of the  bundle~$\Phi$:
\begin{align}
 \label{Atiyah-Algebroid}
0\rarrow \bs\aut(P)  \simeq \bs\aut_v(\Phi)  \xlongrightarrow{|^v}   \bs{\mathfrak{aut}}(\Phi) \xlongrightarrow{\pi_\star} \bs{\diff}(\M) \rarrow 0,
\end{align}
A splitting of this SES, i.e. the datum of a map $ \bs{\mathfrak{aut}}(\Phi) \rarrow \bs\aut(P)$ -- or equivalently of a map $\bs{\diff}(\M) \rarrow  \bs{\mathfrak{aut}}(\Phi)$ -- which would allow to decompose a (right-invariant) vector field on $\Phi$ as a sum of a gauge element and a vector field on $\M$, is supplied by a choice of Ehresmann connection $1$-form on $\Phi$. 

Finally, consider the Lie algebra of the group of vertical diffeomorphisms $\bs\Diff_v(\Phi)$:
\begin{align}
\label{Lie-Alg_diff_v}
\bs\diff_v(\Phi)\defeq \left\{ \bs X_{|\phi}^v =\tfrac{d}{d\tau} \Xi_\tau(\phi) \big|_{\tau=0}=\tfrac{d}{d\tau} R_{\bs\psi_\tau(\phi)} \phi \big|_{\tau=0} \in \Gamma(V\Phi) \right\},
\end{align}
 where  $\Xi_\tau \in \bs\Diff_v(\Phi)$, and $\bs\psi_\tau \in  C^\infty\big(\Phi, \Aut(P) \big)$ is the flow of $ \bs X: \Phi \rarrow \Gamma_{\text{\!\tiny{inv}}}(TP)\simeq \aut(P)$. 
 Therefore, we have that $\bs\diff_v(\Phi) \simeq C^\infty\big(\Phi, \aut(P) \big)$. 
The pushforward of $\bs X^v$ by the action of $\Aut(P)$ is the same as for a fundamental vector field \eqref{Pushforward-fund-vect},
 \begin{align}
 \label{Pushforward-fund-vect_bis}
 R_{\psi\star} \bs X^v_{|\phi} = \left(  (\psi\-)_* \, \bs X \circ \psi \right)^v_{|\psi^*\phi}. 
 \end{align}
The map $|^v : C^\infty\big(\Phi, \aut(P) \big) \rarrow \Gamma(V\Phi)$ is  a Lie algebra morphism, yet the bracket on $C^\infty\big(\Phi, \aut(P) \big)$  extends the bracket in $\aut(P)$  taking into account the $\Phi$-dependence of its elements. 
Indeed, for $\bs X^v, \bs Y^v \in \bs\diff_v(\Phi)$ one has
\begin{equation}
\begin{aligned}
 \label{extended-bracket1}
\big[ \bs X^v, \bs Y^v\big]_{\text{\tiny{$\Gamma(T\Phi)$}}}  &= \{-[  \bs X,  \bs Y]_{\text{{\tiny $\Gamma(TP)$}}}\}^v +  [ \bs X^v(\bs Y)]^v -  [\bs Y^v(\bs X)]^v \\
			    &= \left\{  [\bs X, \bs Y]^{\phantom{-}}_{\text{\tiny{$\aut$}}} +  \bs X^v(\bs Y) - \bs Y^v(\bs X)  \right\}^v \rdefeq \{\bs X, \bs Y\}^v.  
\end{aligned}
\end{equation}
The result is proven in  \cite{Francois2023-b}  for  the finite-dimensional case. 
The bracket on $C^\infty\big(\Phi, \aut(P) \big) $ is
\begin{align}
 \label{extended-bracket2}
 \{\bs X, \bs Y\} :=  [\bs X, \bs Y]^{\phantom{-}}_{\text{\tiny{$\aut$}}} +  \bs X^v(\bs Y) - \bs Y^v(\bs X).
\end{align}
From this follows naturally that,
\begin{align}
 [\bs L_{\bs{ X}^v}, \bs L_{ \bs{ Y}^v }]  = \bs L_{[ \bs{ X}^v,  \bs{ Y}^v]}= \bs L_{\{ \bs{ X},  \bs{ Y}\}^v }.     \label{Lie-deriv-ext-bracket}
\end{align}
The relation \eqref{extended-bracket1} is the infinitesimal version of \eqref{twisted-comp-law},  \eqref{extended-bracket2}  reflecting the bisection composition law in $C^\infty \left(\Phi, \Aut(P) \right)$. 
  
Echoing our discussion of the action groupoid $\bs\Gamma$, let us briefly recall the Lie algebroid picture of the above \cite{Mackenzie2005, Crainic-Fernandez2011}. Associated to the action of $\aut(P)$ on $\Phi$, i.e. to the Lie algebra morphism $\alpha = |^v: \aut(P) \rarrow \Gamma(V\Phi) \subset \Gamma(T\Phi)$, is the action (or transformation) Lie algebroid $\bs A=\Phi \rtimes \aut(P) \rarrow \Phi$ with anchor $\rho : \bs A \rarrow V\Phi \subset T\Phi$, $(\phi, X) \mapsto \alpha(X)_{|\phi}=X^v_{|\phi}$, inducing the Lie algebra morphism $\t \rho: \Gamma(\bs A) \rarrow \Gamma(V\Phi) \subset \Gamma(T\Phi)$. 
The space of sections $\Gamma(\bs A)=\{ \Phi \rarrow \bs A, \phi \mapsto \big(\phi, \bs X(\phi)\big) \}$ is naturally identified with the space $C^\infty\big(\Phi , \aut(P) \big)$,\footnote{Sections of $\Gamma(\bs A)$ are just the graphs of elements of $C^\infty\big(\Phi ,\aut(P) \big)$.} so~that $\t \rho=\alpha=|^v$.  
The~Lie~algebroid bracket $[\ \,,\  ]_{\text{\tiny{$\Gamma(\bs A)$}}}$ is 
uniquely determined by the Leibniz condition $[\bs X, f \bs Y]_{\text{\tiny{$\Gamma(\bs A)$}}}=f[\bs X, \bs Y]_{\text{\tiny{$\Gamma(\bs A)$}}} + \t \rho(\bs X) f \cdot \bs Y$, for $f\in C^\infty(\Phi)$,
and the requirement that  $[\ \,,\ ]_{\text{\tiny{$\Gamma(\bs A)$}}} = [\ \, , \  ]_{\text{\tiny{$\aut(P)$}}}$ on constant sections.
It is found to be: $[\bs X, \bs Y]_{\text{\tiny{$\Gamma(\bs A)$}}} = [\bs X, \bs Y  ]_{\text{\tiny{$\aut(P)$}}} + \t \rho (\bs X) \bs Y - \t\rho(\bs Y) \bs X$.
The action Lie algebroid bracket indeed reproduces \eqref{extended-bracket2} above. 

Quite intuitively, the action Lie algebroid $\bs A$ is the limit of the action Lie groupoid $\bs\Gamma$ when the target and source maps are infinitesimally close. Hence $C^\infty\big(\Phi,\aut(P) \big)\simeq \Gamma(A)$ is the Lie algebra of the group (of bisections) $C^\infty\big(\Phi, \Aut(P)\big)$, as we have shown  explicitly above. 

Remark that Lie algebroid brackets appear at two levels in the geometry just exposed. 
First,~as previously mentioned, the SES \eqref{Atiyah-Algebroid} defines the (transitive) Atiyah Lie algebroid of $\Phi$ as a principal bundle. So, the bracket in $\bs\aut(\Phi)$ is a Lie algebroid bracket. This would be true for any choice of structure group for $\Phi$ -- i.e. an internal gauge group $\Aut_v(P)\simeq \H$, or $\Diff(M)$ as done in \cite{Francois2023-a}.
Its restriction to $\bs\aut_v(\Phi)$  trivialises as the Lie bracket of the Lie algebra of the gauge group of $\Phi$. In the case at hand, with structure group $\Aut(P)$ for $\Phi$, this Lie algebra is $\bs\aut(P)$: so the Lie algebra bracket is the Lie algebroid bracket of $P$.
The (action Lie algebroid) bracket \eqref{extended-bracket2}  in $C^\infty\left(\Phi, \aut(P) \right) \simeq \bs\Diff_v(\Phi) \supset \bs\aut(P)\simeq \bs\aut_v(\Phi)$  \emph{extends} the bracket of the Lie algebra of the gauge group of $\Phi$; in our case, it extends the Atiyah Lie algebroid bracket $[\ \,,\, ]_{\text{\tiny{$\aut$}}}$ of $P$.  
\medskip
 
To the best of our knowledge, the first introduction of a bracket of the type \eqref{extended-bracket2} is to be found in 
Bergmann \& Komar \cite{Bergmann-Komar1972} (for $\phi=g_{\mu\nu}$ and $C^\infty\big(\Phi, \Diff(M) \big)$) and Salisbury \& Sundermeyer \cite{Salisbury-Sundermeyer1983} -- one may look up equations (3.1)-(3.2) in \cite{Bergmann-Komar1972} and equation (2.1) in  \cite{Salisbury-Sundermeyer1983}. It was later reintroduced by Barnich \& Troessaert in  \cite{Barnich-Troessaert2009},  equation (8), for the study of asymptotic symmetries of GR.  This  bracket 
also appears more recently in the covariant phase space literature, e.g. in \cite{Freidel-et-al2021, Freidel-et-al2021bis, Gomes-et-al2018, Chandrasekaran-et-al2022, Speranza2022}.  
It has been interpreted as an action Lie algebroid bracket first in \cite{Barnich2010}. 
In the next section we will show that it is  also a special case of the Frölicher-Nijenhuis bracket of vector-valued forms.

\medskip

Finally, we state the following result (proven in Appendix  \ref{Lie algebra (anti-)isomorphisms}), key to the geometric definition of general vertical  and gauge transformations on field space. The pushforward by a vertical diffeomorphism $\Xi \in \bs\Diff_v(\Phi)$, to which corresponds $\bs \psi \in C^\infty\big(\Phi, \Aut(P) \big)$,
is  a map $\Xi_\star: T_\phi\Phi \rarrow T_{\Xi(\phi)}\Phi=T_{\bs\psi^*\phi}\Phi$. 
For a generic $\mathfrak X \in \Gamma(T\Phi)$ it is  
\begin{equation}
\begin{aligned}
\label{pushforward-X}
 \Xi_\star \mathfrak X_{|\phi} &= R_{\bs\psi(\phi) \star} \mathfrak X_{|\phi} + \left\{   \bs\psi(\phi)\-_* \bs d \bs\psi_{|\phi}(\mathfrak X_{|\phi})   \right\}^v_{|\,\Xi(\phi)}  \\
 					    &=R_{\bs\psi(\phi) \star} \left( \mathfrak X_{|\phi} +  \left\{  \bs d \bs\psi_{|\phi}(\mathfrak X_{|\phi})  \circ \bs\psi(\phi)\-  \right\}^v_{|\phi} \right). 
\end{aligned}
\end{equation}
The proof holds the same for  $\Xi \in \bs\Aut_v(\Phi)\sim\bs \psi \in \bs\Aut(P)$. 
This~relation can be used to obtain the formula for repeated pushforwards:  e.g. to obtain the result for $(\Xi' \bs \circ \Xi)_\star \mathfrak X_{|\phi}$, per \eqref{twisted-comp-law},  one only needs to substitute $\bs\psi \rarrow \bs\psi \circ \big( \bs\psi' \bs \circ R_{\bs \psi}\big)$. In case  $\Xi, \Xi' \in \bs{\Aut}_v(\Phi)$, one substitutes  $\bs\psi \rarrow \bs\psi'  \circ  \bs\psi$.

\subsubsection{Differential forms and their derivations} 
\label{Differential forms and their derivations}  

Consider the space of forms $\Omega^\bullet(\Phi)$  with the graded Lie algebra of its derivations 
$\Der_\bullet \big(\Omega^\bullet(\Phi) \big)=\bigoplus_k \Der_k \big(\Omega^\bullet(\Phi) \big)$  whose graded bracket is $[D_k, D_l]=D_k \circ D_l - (-)^{kl} D_l \circ D_k$, with $D_i \in \Der_i\big(\Omega^\bullet(\Phi) \big)$.  

The de Rham complex of $\Phi$ is $\big( \Omega^\bullet(\Phi); \bs d  \big)$ with $\bs d \in \Der_1$ the de Rham (exterior) derivative, which is nilpotent -- $\bs d ^2 =0=\sfrac{1}{2}[\bs d, \bs d]$ -- and defined via the Koszul formula.  The exterior product $\w$ is defined on scalar-valued forms as usual, so that  
$\big( \Omega^\bullet(\Phi, \mathbb K), \w, \bs d \big)$ is a differential graded algebra. The exterior product can also be defined on the space $\Omega^\bullet(\Phi, \sf A)$ of forms with values in an algebra $(\sf A, \cdot)$, using the product in $\sf A$ instead of the product in $\mathbb K$. 
So~$\big( \Omega^\bullet(\Phi, \sf A), \w, \bs d \big)$ is a again a differential graded algebra.\footnote{On the other hand, an exterior product cannot be defined on $\Omega^\bullet(\Phi, \bs V)$ where $\bs V$ is merely a vector space.}

One may define vector field-valued differential forms $\Omega^\bullet(\Phi, T\Phi)=\Omega^\bullet(\Phi) \otimes T\Phi$. Then, the subalgebra of 
\emph{algebraic} derivations is defined as $D_{|\Omega^0(\Phi)}=0$; they have the form $\iota_{\bs K} \in \Der_{k-1}$ for $\bs K \in \Omega^{k}(\Phi, T\Phi)$, with $\iota$ the inner product. 
It generalises the inner contraction of a form on a vector field:
For $  \omega  \otimes  \mathfrak X \in \Omega^\bullet(\Phi, T\Phi)$ we have $\iota_{\bs K}( \omega  \otimes  \mathfrak X)  := \iota_{\bs K} \omega\otimes  \mathfrak X = \bs\omega \circ \bs K \otimes  \mathfrak X$.
The \emph{Nijenhuis-Richardson bracket} (or \emph{algebraic} bracket) is defined by
\begin{align}
\label{NR-bracket} 
 [\bs K, \bs L]_{\text{\tiny{NR}}}:=\iota_{\bs K} \bs L -(-)^{(k-1)(l-1)} \, \iota_{\bs L} \bs K
 \end{align}
 and makes the map $\iota: \Omega^\bullet(\Phi, T\Phi) \rarrow \Der_\bullet \big(\Omega^\bullet(\Phi) \big)$,  $\bs K \mapsto  \iota_{\bs K}$, 
a graded Lie algebra morphism:
\begin{align}
\label{NR-bracket-id}
[\iota_{\bs K}, \iota_{\bs L}]=\iota_{[\bs K, \bs L]_{\text{\tiny{NR}}}}. 
\end{align}
The \emph{Nijenhuis-Lie derivative} is the map  
\begin{equation}  
\label{NL-derivative}
  \begin{aligned}    
  \bs L:=[\iota, \bs d] : \Omega^\bullet(\Phi, T\Phi) &\rarrow \Der_\bullet \big(\Omega^\bullet(\Phi) \big) \\
  					\bs K &\mapsto  \bs L_{\bs K}:= \iota_{\bs K} \bs d - (-)^{k-1} \bs d  \iota_{\bs K}.      
  \end{aligned}
  \end{equation}     
We have $\bs L_{\bs K} \in \Der_k$ for $\bs K \in \Omega^{k}(\Phi, T\Phi)$.  It generalises the Lie derivative along vector fields, $\bs L_\mathfrak{X} \in \Der_0$. 
It is such that $[\bs L_{\bs K}, \bs d]=0$. 
Given $\bs K = K \otimes \mathfrak X \in \Omega^k(\Phi, T\Phi)$ and   $\bs J = J \otimes \mathfrak Y \in \Omega^l(\Phi, T\Phi)$, the  \emph{Frölicher-Nijenhuis bracket}  is 
  \begin{align}
  \label{FN-bracket}
  [\bs K, \bs J]_{\text{\tiny{FN}}}= K \w J \otimes [\mathfrak X, \mathfrak Y]  + K \w \bs L_\mathfrak{X} J \otimes \mathfrak Y
  														  -  \bs L_\mathfrak{Y} K \w J \otimes \mathfrak X
														  + (-)^k \big(  dK \w \iota_\mathfrak{X} J \otimes \mathfrak Y
														                       + \iota_\mathfrak{Y} K \w dJ \otimes \mathfrak X   \big).
  \end{align}
It makes the Nijenhuis-Lie derivative  a  morphism of graded Lie algebras: 
\begin{align}
\label{NL-deriv-morph}
[\bs L_{\bs K}, \bs L_{\bs J}]=\bs L_{[\bs K, \bs J]_{\text{\tiny{FN}}}}.
\end{align} 
The following relations hold: 
\begin{equation}
  \begin{aligned}
  \label{Relations-L-iota}
[\bs L_{\bs K}, \iota_{\bs J}] &= \iota^{}_{ [\bs K, \bs J]_{\text{\tiny{FN}}} } - (-)^{k(l-1)} \bs L_{(\iota_{\bs K} \bs J)},  \\
[\iota_{\bs J}, \bs L_{\bs K}] &= \bs L_{(\iota_{\bs K} \bs J)} +(-)^k \, \iota^{}_{ [\bs J, \bs K]_{\text{\tiny{FN}}} }. 
  \end{aligned}
  \end{equation}
  For a  systematic exposition of the above notions (in the finite-dimensional setting), see \cite{Kolar-Michor-Slovak} Chapter II, section 8.
  \medskip
  
The Frölicher-Nijenhuis (FN) bracket  reproduces the  bracket \eqref{extended-bracket1} as a special case.
Indeed, specialising  \eqref{FN-bracket} in degree $0$, for  $\bs f = f \otimes \mathfrak X$ and $\bs g = g \otimes \mathfrak Y \in \Omega^0(\Phi, T\Phi)$ we get
    \begin{align}
    \label{FN-bracket-special}
     [\bs f, \bs g]_{\text{\tiny{FN}}} &= f \w g \otimes [\mathfrak X, \mathfrak Y]  + f \w \bs L_\mathfrak{X} g \otimes \mathfrak Y
  														  -  \bs L_\mathfrak{Y} f \w g \otimes \mathfrak X, \notag \\
						&= f \w g \otimes [\mathfrak X, \mathfrak Y] +  [\bs f, \bs{dg}]_{\text{\tiny{NR}}}  -  [\bs{g}, \bs{df}]_{\text{\tiny{NR}}},
    \end{align}
 and 
   \begin{align}
   \label{FN-NL-der}
   [\bs L_{\bs f}, \iota_{\bs g}] &= \iota^{}_{ [\bs f, \bs g]_{\text{\tiny{FN}}} }. 
    \end{align}      
The map  $|^v : C^\infty\big(\Phi, \aut(P)\big) \rarrow  \Gamma(V\Phi)$,  $\bs{ X} \mapsto \bs{  X}^v$, allows to see $\bs{ X}^v \in \diff_v(\Phi)$ as a (vertical) vector-valued 0-form on $\Phi$:  $ \bs{ X}^v \in \Omega^0(\Phi, V\Phi) \subset \Omega^\bullet(\Phi, T\Phi)$. 
The Nijenhuis-Richardson and Frölicher-Nijenhuis brackets thus  apply: 
for $\bs{ X}^v, \bs{  Y}^v \in \Omega^0(\Phi, V\Phi)$ we find
\begin{equation}
\label{NR-bracket-field-dep-diff}
\begin{split}
 [\bs{ X}^v, \bs d \bs{ Y}^v ]_{\text{\tiny{NR}}} &= \{ \iota_{\bs{ X}^v} \bs{d}\bs{ Y} \}^v = \{ \bs{ X}^v(\bs{ Y})\}^v, \\
  [\bs d \bs{ X}^v,  \bs{ Y}^v ]_{\text{\tiny{NR}}} &= -[\bs{ Y}^v, \bs d \bs{ X}^v ]_{\text{\tiny{NR}}} = -\{ \iota_{\bs{ Y}^v} \bs{d}\bs{ X} \}^v = - \{ \bs{ Y}^v(\bs{ X})\}^v, 
  \end{split}
\end{equation}
 so that the FN bracket for 0-forms \eqref{FN-bracket-special} is
 \begin{align}
 \label{FN-bracket-field-dep-diff=BT-bracket}
   [\bs{ X}^v,  \bs{ Y}^v ]_{\text{\tiny{FN}}} 
     =  \big(  [\bs{ X}, \bs{ Y}]_{\text{{\tiny $\aut$}}}  +  \bs{ X}^v(\bs{ Y}) -   \bs{ Y}^v(\bs{ X}) \,\big)^v   = \{ \bs{ X},  \bs{ Y}\}^v. 
 \end{align}
 Then we have the following special cases of identities \eqref{NR-bracket-id}, \eqref{Relations-L-iota}, and \eqref{NL-deriv-morph} among derivations in $\Der^\bullet$: 
 \begin{align}
 [\iota_{\bs{ X}^v}, \iota_{\{\bs{d Y}\}^v}] &= \iota^{}_{ [\bs{ X}^v, \bs d \bs{ Y}^v ]_{\text{\tiny{NR}}} } = \iota^{}_{ \{ \iota_{\bs{ X}^v} \bs{d}\bs{ Y} \}^v },
 \label{identity-iota}  \\
  [\bs L_{\bs{ X}^v}, \iota_{ \bs{ Y}^v }] &= \iota^{}_{[\bs{ X}^v,  \bs{ Y}^v ]_{\text{\tiny{FN}}}}, \label{identity-L-iota} \\
 [\bs L_{\bs{ X}^v}, \bs L_{ \bs{ Y}^v }] &= \bs L_{[\bs{ X}^v,  \bs{ Y}^v ]_{\text{\tiny{FN}}}} 
 										    = \bs L_{\{ \bs{ X},  \bs{ Y}\}^v }.     \label{NL-derr-ext-bracket}
 \end{align}
 As expected, \eqref{NL-derr-ext-bracket} reproduces \eqref{Lie-deriv-ext-bracket}.
 The above reproduces as special cases various identities  derived heuristically in the covariant phase space literature, e.g.  \cite{Freidel-et-al2021, Freidel-et-al2021bis, Chandrasekaran-et-al2022}.

 \medskip

\paragraph{Remarkable forms}  The action by pullback of   $\Aut(P)$ on a form $\bs \alpha \in \Omega^\bullet(\Phi)$ defines its \mbox{\emph{equivariance}}, $R^\star_\psi \alpha$. 
\mbox{The~action} by pullback of   $\bs\Diff_v(\Phi)\simeq C^\infty\big( \Phi, \Aut(P)\big)$, which we write $\bs\alpha^{\bs\psi}\defeq \Xi^\star \bs\alpha$, defines  \emph{vertical transformations},
while the action by pullback of $\bs\Aut_v(\Phi)\simeq \bs\Aut(P)$  defines  \emph{gauge transformations}. 

We write a generic form at $\phi\in \Phi$ as
\begin{align}
\bs\alpha_{|\phi} = \alpha\big(\! \w^\bullet \!\bs d\phi_{|\phi}; \phi \big),
\end{align}
where $\bs d\phi \in \Omega^1(\Phi)$ is the basis 1-form on $\Phi$ 
 and $\alpha(\ ;\ )$ is the functional expression of $\bs\alpha$, alternating multilinear in the first arguments and with arbitrary $\phi$-dependence  in the second argument  (in physics, often polynomial). The~equivariance and  vertical transformation of $\bs\alpha$ are: 
 \begin{equation}
\begin{aligned}
R^\star_\psi \bs\alpha_{|\phi^\psi} &= \alpha\big(\! \w^\bullet \! R^\star_\psi \bs d\phi_{|\phi^\psi};\,  R_\psi \phi \big) = \alpha\big(\! \w^\bullet \! R^\star_\psi \bs d\phi_{|\phi^\psi}; \, \phi^\psi \big), 
\quad  \text{for } \ \psi \in \Aut(P), \\
\bs\alpha^{\bs\psi}_{|\phi} \defeq \Xi^\star \bs\alpha^{}_{|\Xi(\phi)} &= \alpha\big(\! \w^\bullet \! \Xi^\star \bs d\phi^{}_{|\Xi(\phi)};\  \Xi( \phi) \big) = \alpha\big(\! \w^\bullet \! \Xi^\star \bs d\phi_{|\phi^{\bs\psi}};\,  \phi^{\bs\psi} \big),
\quad \text{for } \  \Xi \in \bs \Diff_v(\Phi) \sim \bs\psi \in C^\infty\big( \Phi, \Aut(P)\big).    \label{GT-general}
\end{aligned}
\end{equation}
The infinitesimal equivariance and  vertical transformations are given by the (Nijenhuis-)Lie derivative along the elements of $\Gamma(V\Phi)$ generated  respectively by $\aut(P)$ and $C^\infty\big( \Phi, \aut(P)\big)$:
\begin{align}
\bs L_{X^v}\bs\alpha = \tfrac{d}{d\tau}  R^\star_{\psi_\tau} \bs\alpha \big|_{\tau=0}  \quad  \text{ with } \ X \in \aut(P),  
 \qquad \quad  \bs L_{\bs X^v}\bs\alpha = \tfrac{d}{d\tau}  \Xi_\tau^\star \bs\alpha \big|_{\tau=0}\quad  \text{ with } \ \bs X \in C^\infty\big( \Phi, \aut(P)\big). \label{GT-inf-general}
\end{align}
There are forms of particular interest whose gauge transformations need not be computed explicitly, but  rather are read from the forms special properties.

\medskip

First, \emph{equivariant} forms are those whose equivariance is controlled by either representations of the structure group, or by 1-cocycles for the action of the structure group. 
\emph{Standard equivariant} forms are valued in representations $(\rho, \bs V)$ of the structure group $\Aut(P)$ and s.t.: 
\begin{align}
\Omega_\text{eq}^\bullet(\Phi, \rho) \defeq \left\{ \, \bs\alpha \in \Omega^\bullet(\Phi, \bs V)\,|\, R^\star_\psi\bs\alpha_{|\phi^\psi}=\rho(\psi)\-\bs\alpha_{|\phi}\, \right\}. 
 \end{align} 
The infinitesimal version of the equivariance property is $\bs L_{X^v}\bs\alpha=-\rho_*(X) \bs\alpha$ for $X \in \aut(P)$.

The \emph{twisted equivariant} forms \cite{Francois2019_II} have equivariance  controlled by a 1-cocycle for the action of $\Aut(P)$ on $\Phi$, 
 i.e. a map:
 \begin{equation}
\begin{aligned}
\label{cocycle}
C: \Phi \times \Aut(P) &\rarrow G, \quad \text{$G$ some Lie group (possibly infinite-dimensional).}   \\
	(\phi, \psi) &\mapsto C(\phi; \psi) \qquad \text{s.t.} \quad C(\phi; \psi'\circ \psi) =C(\phi; \psi') \cdot C(\phi^{\psi'}; \psi). 
 \end{aligned} 
  \end{equation}
 Manifestly,   $\phi$-independent 1-cocycles are  group morphisms,  i.e. 1-cocycles are generalisations of  representations. 
 From the 1-cocycle property \eqref{cocycle} follows that $C(\phi; \id_M)=\id_G=C(\phi^\psi; \id_M)$, thus that $C(\phi; \psi)\-\!= C(\phi^\psi; \psi\-)$. 
Given   a $G$-space $\bs V$, one defines twisted equivariant forms as 
 \begin{align}
 \label{C-eq-form}
 \Omega^\bullet_\text{eq}(\Phi, C) \defeq \left\{ \bs\alpha \in \Omega^\bullet(\Phi, \bs V)\, | \, R^\star_\psi \bs\alpha_{|\phi^\psi} = C(\phi  ; \psi)\- \bs\alpha_{|\phi}\right\}.
 \end{align} 
 The property \eqref{cocycle} ensures compatibility with the right action: $R_{\psi'}^\star R^\star_{\psi} = R^\star_{\psi' \circ\,\psi}$. The infinitesimal equivariance  is $\bs L_{X^v}\bs\alpha=-a(X; \phi) \bs\alpha$, where $a(X, \phi):= \tfrac{d}{d\tau}\, C(\phi, \psi_\tau) |_{\tau=0}$ is a 1-cocycle for the action of $\aut(P)$~on~$\Phi$:
  \begin{equation}
 \begin{aligned}
 \label{inf-cocycle}
a: \Phi \times \aut(P) &\rarrow \mathfrak g, \quad \text{$\mathfrak g$ the Lie algebra of $G$.}   \\
	(\phi, X) &\mapsto a(X; \phi) \qquad \text{s.t.} \quad X^v\! \cdot a(Y; \phi)- Y^v\!\cdot a(X; \phi)+ \big[a(X; \phi)\,,\ a(Y; \phi)\big]_{\mathfrak g}=a([X,Y]_{\text{{\tiny $\aut$}}}; \phi). 
 \end{aligned} 
   \end{equation}
 The infinitesimal relation \eqref{inf-cocycle} ensures compatibility with the right action: $[\bs L_{X^v}, \bs L_{X^v}]=\bs L_{[X^v, Y^v]}=\bs L_{([X, Y]_{\text{{\tiny $\aut$}}})^v}$. 
 Observe that it is a non-Abelian generalisation of the Wess-Zumino (WZ) consistency condition for  anomalies $a(X; \phi)$.
 The WZ consistency condition being reproduced for $G$ Abelian. 
 
 The subspace of \emph{invariant} forms are those whose  equivariance is trivial, and \emph{horizontal} forms are those vanishing on vertical vector field:
 \begin{equation}
  \begin{aligned}
 \Omega_\text{inv}^\bullet(\Phi)&=\left\{ \, \bs\alpha \in \Omega^\bullet(\Phi)\,|\, R^\star_\psi \bs\alpha=\bs\alpha \,\right\}, \  \text{infinitesimally }\  \bs L_{X^v}\bs\alpha=0, \\
 \Omega_\text{hor}^\bullet(\Phi)&=\left\{ \, \bs\alpha \in \Omega^\bullet(\Phi)\,|\, \iota_{X^v}\bs\alpha=0 \right\}.
  \end{aligned}
   \end{equation}
   
 A form which is both equivariant and horizontal is said \emph{tensorial}. We have thus standard tensorial forms  
 \begin{align}
 \label{tens-forms}
\Omega_\text{tens}^\bullet(\Phi, \rho)\defeq \left\{ \, \bs\alpha \in \Omega^\bullet(\Phi, \bs V)\,|\, R^\star_\psi\bs\alpha=\rho(\psi)\-\bs\alpha,\, \text{ \& }\ \iota_{X^v}\bs\alpha=0\, \right\}. 
\end{align} 
Similarly, the space of \emph{twisted tensorial} forms is
 \begin{align}
 \label{twisted-tens-forms}
\Omega_\text{tens}^\bullet(\Phi, C)\defeq \left\{ \, \bs\alpha \in \Omega^\bullet(\Phi, \bs V)\,|\, R^\star_\psi\bs\alpha=C(\phi; \psi)\-\bs\alpha,\, \text{ \& }\ \iota_{X^v}\bs\alpha=0\, \right\}. 
\end{align} 
In either case, we have  $\Omega_\text{tens}^0(\Phi)=\Omega_\text{eq}^0(\Phi)$. 

Let us recall the well-known fact that the de Rham derivative $\bs d$ does not preserve the space of tensorial forms (horizontality is lost). 
This is a reason for the introduction of a notion of \emph{connection} on $\Phi$ so as to define a \emph{covariant derivative} on the space of tensorial forms. 
As we will review in section \ref{Connections on field space} below,  for standard tensorial forms one needs an Ehresmann connection 1-form, while for twisted tensorial forms one needs a generalisation called \emph{twisted connection}  \cite{Francois2019_II}.  
\medskip

Finally, forms that  are both  invariant and horizontal are called  \emph{basic}: 
 \begin{align}
\Omega_\text{basic}^\bullet(\Phi)\defeq \left\{ \, \bs\alpha \in \Omega^\bullet(\Phi)\,|\, R^\star_\psi\bs\alpha=\bs\alpha\, \text{ \& }\ \iota_{X^v}\bs\alpha=0\, \right\}. 
\end{align} 
This space is preserved by $\bs d$, so 
$\big( \Omega^\bullet_\text{basic}(\Phi), \bs d \big)$ is a subcomplex of the de Rham complex of $\Phi$: the \emph{basic subcomplex}.  
Therefore, basic forms can also be defined as  Im$(\pi^\star)$ (hence their name): 
 \begin{align}
\Omega_\text{basic}^\bullet(\Phi)\defeq\left\{ \, \bs\alpha \in \Omega^\bullet(\Phi)\,|\, \exists\, \bs\beta \in \Omega^\bullet(\M) \text{ s.t. } \bs\alpha=\pi^\star\bs\beta \,  \right\}. 
\end{align} 
The cohomology of $\big( \Omega^\bullet_\text{basic}(\Phi), \bs d \big)$ is  the \emph{equivariant cohomology} of $\Phi$. As  $[\bs d, \pi^\star]=0$, it is isomorphic to the cohomology  $\big(\Omega^\bullet(\M), \bs d \big)$ of the base moduli space. Hence its importance, especially when it is unpractical (or impossible) to work concretely on $\M$, as it is the case in gauge field  theory. 

We stress that the analogue of $\Gamma_{\text{\!\tiny{inv}}}(T\Phi)$ for forms 
is not $\Omega^\bullet_\text{inv}(\Phi)$ 
but $\Omega_\text{basic}^\bullet(\Phi)$. Only  basic forms project to well-defined forms in $\Omega^\bullet(\M)$, containing only physical d.o.f.
In section \ref{The dressing field method}, we will detail a systematic method to build the basic version $\bs\alpha^b \in \Omega_\text{basic}^\bullet(\Phi)$ of a form $\bs\alpha \in \Omega^\bullet(\Phi)$.

\subsection{Vertical transformations and gauge transformations} 
\label{Vertical transformations and gauge transformations}  

As   previously seen, the  vertical transformation a form $\bs\alpha \in \Omega^\bullet(\Phi)$ is its  pullback by $\bs\Diff_v(\Phi)\simeq C^\infty\big( \Phi, \Aut(P)\big)$
: $\bs\alpha^{\bs\psi} \defeq \Xi^\star \bs\alpha$.
  The notation on the left-hand side is justified by the fact that the vertical transformation is expressed in term of the generating  element $\bs\psi \in C^\infty\big( \Phi, \Aut(P)\big)$ associated to $\Xi \in \bs\Diff_v(\Phi)$. 
  Performing two vertical transformations, using \eqref{twisted-comp-law}, one has 
  \begin{align}
  \label{2-Vert-Trsf}
  \big( \bs\alpha^{\bs\psi}\big)^{\bs\psi'} \defeq {\Xi'}^\star \Xi^\star \bs\alpha = \big( \Xi \bs \circ \Xi' \big)^\star \bs\alpha \rdefeq \bs\alpha^{ \bs\psi' \circ\, (\bs\psi \bs\circ R_{\bs\psi'}) }.
  \end{align} 
   For gauge transformations, defined by the action of $\bs\Aut_v(\Phi)\simeq \bs{\Aut}(P)$, whose elements have specific equivariance  $R^\star_\psi \bs\psi = \bs\psi \bs\circ R_\psi =  \psi\- \circ \bs\psi \circ \psi$, the previous expression simplifies:
     \begin{align}
  \label{2-Gauge-Trsf}
  \big( \bs\alpha^{\bs\psi}\big)^{\bs\psi'} = \bs\alpha^{ \bs\psi \,\circ\, \bs\psi' }.
  \end{align} 
 Infinitesimal  vertical transformations by  $\bs\diff_v(\Phi) \simeq C^\infty\big(\Phi, \aut(P) \big)$ are given by the Nijenhuis-Lie derivative: 
  \begin{align}
  \label{Inf-2-Vert-Trsf}
  \bs L_{\bs X^v} \bs\alpha = \left\{   \begin{matrix} \tfrac{d}{d\tau} \, \Xi_\tau^\star \bs\alpha\, \big|_{\tau=0} \\[3mm]
					                                       \hspace{-.2cm} [\iota_{\bs X^v}, \bs d]\, \bs \alpha \end{matrix} \right. \qquad \text{so} \qquad    
 [\bs L_{\bs{ X}^v}, \bs L_{ \bs{ Y}^v }] \, \bs\alpha = \bs L_{[\bs{ X}^v,  \bs{ Y}^v ]_{\text{\tiny{FN}}}} \, \bs\alpha = \bs L_{\{ \bs{ X},  \bs{ Y}\}^v }\, \bs\alpha.    
  \end{align}
 The second equation uses  \eqref{Lie-deriv-ext-bracket}/\eqref{NL-derr-ext-bracket}. It is the infinitesimal version of \eqref{2-Vert-Trsf}. 
 For infinitesimal gauge transformations, defined by the action of $\bs\aut_v(\Phi)\simeq \bs\aut(P)$, this reduces to 
 $[\bs L_{\bs{ X}^v}, \bs L_{ \bs{ Y}^v }] \, \bs\alpha = \bs L_{( -[\bs X,\bs  Y]_{\text{\tiny{$\aut$}}})^v}\, \bs\alpha = \bs L_{([ \bs X, \bs Y]_{\text{{\tiny $\Gamma(TP)$}}})^v}\, \bs\alpha$.

To get concrete expressions, one uses the duality between pullback and pushforward together with  \eqref{pushforward-X}:  For any $\mathfrak X,\mathfrak X', \ldots \in \Gamma(T\Phi)$   one has 
\begin{equation}
\label{GT-geometric}
\begin{aligned}
\bs\alpha^{\bs\psi}_{|\phi} (\mathfrak X_{|\phi}, \ldots) =  \Xi^\star \bs\alpha_{|\Xi(\phi)} (\mathfrak X_{|\phi}, \ldots)= \bs\alpha^{}_{|\Xi(\phi)} (\Xi_\star \mathfrak X_{|\phi}, \ldots) &= \bs\alpha^{}_{|\phi^{\bs\psi(\phi)}}\left( R_{\bs\psi(\phi) \star} \left( \mathfrak X_{|\phi} +  \left\{  \bs d \bs\psi_{|\phi}(\mathfrak X_{|\phi})  \circ \bs\psi(\phi)\-  \right\}^v_{|\phi} \right), \ldots\right)  \\
&=R_{\bs\psi(\phi)}^\star\, \bs\alpha_{|\phi^{\bs\psi(\phi)}} \left(  \mathfrak X_{|\phi} +  \left\{  \bs d \bs\psi_{|\phi}(\mathfrak X_{|\phi})  \circ \bs\psi(\phi)\-  \right\}^v_{|\phi}, \ldots \right).
\end{aligned}
\end{equation}
It is clear from \eqref{GT-geometric} that the vertical transformation of a form is controlled by its equivariance and verticality properties. 
In particular,  the vertical transformation of a tensorial form is simply given by its equivariance: 
\begin{equation}
\label{GT-tensorial}
\begin{split}
&\text{For }\ \bs\alpha \in \Omega_\text{tens}^\bullet(\Phi, \rho), \quad \bs\alpha^{\bs\psi} = \rho(\bs\psi)\- \bs\alpha. \\
&\text{For }\ \bs\alpha \in \Omega_\text{tens}^\bullet(\Phi, C), \quad \bs\alpha^{\bs\psi} = C(\bs\psi)\- \bs\alpha.
\end{split}
\end{equation}
In the second line we introduce the simplified notation $[C(\bs\psi)](\phi) \defeq C\big(\phi; \bs\psi(\phi)\big)$.  
The map $C(\bs\psi) : \Phi \rarrow G$ is twice dependent on the point $\phi\in \Phi$.
We stress that $\bs\alpha^{\bs\psi}=\Xi^\star \bs\alpha  \notin \Omega_\text{tens}^\bullet(\Phi, \rho)$ unless $\bs\psi \in \bs\Aut(P) \sim \Xi \in \bs\Aut_v(\Phi)$ -- see \cite{Francois2023-b} -- making \emph{gauge transformations} special indeed: they preserve the space of tensorial forms, while $\bs\Diff_v(\Phi)$ does not. 

The infinitesimal versions of \eqref{GT-geometric} is, by definition \eqref{Inf-2-Vert-Trsf}, 
\begin{align}
\label{Inf-GT-general}
\bs L_{\bs X^v} \bs\alpha = \tfrac{d}{d\tau} \, R_{\bs\psi_\tau}^\star \bs\alpha\, \big|_{\tau=0} + \iota_{\{\bs{dX}\}^v} \bs\alpha,
\end{align}
with $\bs X=\tfrac{d}{d\tau}\bs\psi_\tau \big|_{\tau=0}$. 
Notice that $\{\bs{dX}\}^v$ can be seen as an element of $\Omega^1(\Phi, V\Phi)$, so $\iota_{\{\bs{dX}\}^v}$ is an algebraic derivation (of degree 0) as discussed in section \ref{Differential forms and their derivations}. 
For $\bs\alpha \in \Omega^\bullet_\text{eq}(\phi)$, depending if it is standard or twisted equivariant, \eqref{Inf-GT-general} specialises to:
\begin{align}
\label{Inf-GT-eq}
\bs L_{\bs X^v} \bs\alpha = \left\{   \begin{matrix}   -\rho_*(\bs X)\, \bs\alpha + \iota_{\{\bs{dX}\}^v} \bs\alpha, \\[3mm]
						                              -a(\bs X)\, \bs \alpha + \iota_{\{\bs{dX}\}^v} \bs\alpha,
					  \end{matrix} \right.
\end{align}
where we  introduce the notation $[a(\bs X)](\phi)\defeq a\big(\bs X(\phi); \phi\big)$ for the linearised 1-cocycle. 
In particular,  for the pullback representation  $\rho(\psi)\-=\psi^*$, i.e. for  $\Omega^\bullet(P)$-valued (or tensor-valued) forms $\bs\alpha$, \eqref{Inf-GT-general} gives naturally: 
\begin{align}
\label{Inf-GT-diff-rep}
\bs L_{\bs X^v} \bs\alpha = \mathfrak L_{\bs X} \bs\alpha + \iota_{\{\bs{dX}\}^v} \bs\alpha.
\end{align}
This formula  clarifies the geometrical meaning of the so-called ``anomaly operator", $\Delta_{\bs X}$, featuring in the covariant phase space literature  \cite{Hopfmuller-Freidel2018, Chandrasekaran_Speranza2021, Freidel-et-al2021, Freidel-et-al2021bis, Speziale-et-al2023}: in our notations  
 $\Delta_{\bs X}\defeq \bs L_{\bs X^v} - \mathfrak L_{\bs X} - \iota_{\{\bs{dX}\}^v}$. 
This operator can only be non-zero on $\Phi$ in theories admitting background non-dynamical structures or fields ``breaking" $\Aut(P)$-covariance. Those fundamentally fail to comply with the core physical (symmetry) principles  of general-relativistic gauge field theory. 
We further elaborate on this point in section \ref{The dressing field method}.

The infinitesimal versions of \eqref{GT-tensorial}, for tensorial forms, are: 
\begin{align}
\label{GT-inf-tensorial}
\bs L_{\bs X^v} \bs \alpha = -\rho_*(\bs X) \bs \alpha \quad\text{and } \quad \bs L_{\bs X^v} \bs \alpha = -a(\bs X) \bs \alpha.
\end{align}
From the commutativity property \eqref{Lie-deriv-ext-bracket}/\eqref{NL-derr-ext-bracket} of the Nijenhuis-Lie derivative applied to a twisted tensorial form $\bs \alpha$,  
$[ \bs L_{\bs X^v}, \bs L_{\bs Y^v}] \, \bs \alpha=\bs L_{[\bs X^v, \bs Y^v]_{\text{{\tiny FN}}}}\,  \bs \alpha$ $= \bs L_{\{\bs X, \bs Y\}^v}\,  \bs \alpha$,  follows the relation for the infinitesimal 1-cocycle:
\begin{equation}
\label{BB}
\begin{aligned}
& \bs X^v\big( a(\bs Y; \phi) \big) -  \bs Y^v\big( a(\bs X; \phi) \big)    - a(\{\bs X, \bs Y\}; \phi)  + [a(\bs X; \phi), a(\bs Y; \phi)]_{\text{{\tiny $\LieG$}}} =0, \\
&\bs X^v\big( a(\munderline{red}{\bs Y}; \phi) \big) -  \bs Y^v\big( a( \munderline{red}{\bs X}; \phi) \big)    - a([\bs X, \bs Y]_{\text{{\tiny $\aut(P)$}}}; \phi)  + [a(\bs X; \phi), a(\bs Y; \phi)]_{\text{{\tiny $\LieG$}}}=0.  
\end{aligned}
\end{equation}
The second equation is obtained from the first using the FN bracket \eqref{FN-bracket-field-dep-diff=BT-bracket}/\eqref{extended-bracket1}: 
The notation $\munderline{red}{\bs Y},\munderline{red}{\bs X}$  means that the elements $\bs Y, \bs X$ are considered $\phi$-independent, so $\bs X^v, \bs Y^v$ pass through. 
Therefore,  \eqref{BB}  reproduces the defining  infinitesimal 1-cocycle property  \eqref{inf-cocycle}. 
 
To illustrate, let us consider the case of elements of the gauge group, $\bs\eta \in  \bs\Aut(P)$, and its Lie algebra, $\bs Y \in  \bs\aut(P)$. As 0-forms they are trivially horizontal, and their equivariance are specified by definition  \eqref{GaugeGroup}-\eqref{LieAlg-GaugeGroup}: they are thus tensorial, so we have 
 \begin{align}
 \bs \eta^{\bs \psi} = \bs\psi \- \circ \bs \eta \circ \bs \psi, \quad \text{and} \quad \bs Y^{\bs\psi}= (\bs\psi\-)_* \bs Y\circ \bs\psi.
 \end{align}
 The infinitesimal gauge transformation of  $\bs Y$ is then  $\bs L_{\bs X^v} \bs Y = \mathfrak L_{\bs X} \bs Y = [\bs Y, \bs X]_{\text{{\tiny $\aut(P)$}}}$, as expected from its  infinitesimal equivariance \eqref{inf-equiv-gauge-Lie-alg}. 
 
As a special case of \eqref{GT-tensorial}, or given their definition, basic forms are strictly gauge invariant: 
\begin{align}
\label{GT-basic}
\text{For }\ \bs\alpha \in \Omega_\text{basic}^\bullet(\Phi): \quad \bs\alpha^{\bs\psi} = \bs\alpha, \quad \text{so}\quad  L_{\bs X^v} \bs \alpha=0. 
\end{align}
 
Another important example is that of the basis 1-form $\bs d\phi  \in \Omega^1(\Phi)$, 
since its vertical transformation $\bs d\phi^{\bs\psi}\defeq \Xi^\star \bs d \phi$ features in the general formulae \eqref{GT-general}  for the vertical transformation of a generic form.
The equivariance and verticality properties of $\bs d\phi$ are given by definition: 
 \begin{align}
 \label{basis-1-form}
R^\star_\psi \bs d\phi \defeq \psi^* \bs d \phi, \qquad \text{and} \qquad \bs \iota_{X^v}\bs d\phi \defeq \mathfrak L_X \phi. 
 \end{align}
The verticality must reproduce the $\aut(P)$-transformation of the field $\phi$.
 It is then immediate that
 \begin{align}
  \label{GT-basis-1-form}
  \bs d\phi^{\bs\psi}   \defeq \Xi^\star \bs d \phi = \bs\psi^*   \big(  \bs d\phi   +   \mathfrak L_{ \bs d \bs\psi  \circ \bs\psi\-} \phi  \big).   
 \end{align}
This generalises a standard result of the covariant phase space literature (see e.g. \cite{DonnellyFreidel2016}). 
From \eqref{Inf-GT-diff-rep}, the linear version~is 
  \begin{align}
  \label{GT-inf-basis-1-form}
 \bs L_{\bs X^v} \bs d\phi   =  \mathfrak L_{\bs X} \bs d \phi +   \mathfrak L_{\bs d \bs X} \phi.
 \end{align}
 The same may be obtained via $ \bs L_{\bs X^v} \bs d\phi = \bs d (\iota_{\bs X^v} \bs d\phi) = \bs d \big( \mathfrak L_{\bs X} \phi \big)$.

\subsection{Connections on field space}  
\label{Connections on field space}  

As previously observed, the exterior derivative $\bs d$ does not preserve $\Omega^\bullet_{\text{tens}}(\Phi)$  of standard/twisted tensorial forms. To build a first order linear differential operator that does, the \emph{covariant derivative}, one needs to endow  $\Phi$ with  an adequate notion of connection 1-form.

\subsubsection{Ehresmann connections} 
\label{Ehresmann connections} 

A Ehresmann connection 1-form $\bs \omega \in \Omega^1_{\text{eq}}\big(\Phi, \aut(P) \big)$ on field space $\Phi$ is defined by the following two properties:
\begin{equation}
\label{Variational-connection}
\begin{aligned}
\bs\omega_{|\phi} \big( X^v_{|\phi}\big)&=X, \quad \text{for } X\in \aut(P), \\
R^\star_\psi \bs\omega_{|\phi^\psi} &= \psi\-_* \, \bs\omega_{|\phi}  \circ\, \psi. 
\end{aligned}
\end{equation}
Infinitesimally, the equivariance of the connection under $\aut(P)$ is
\begin{align}
\label{inf-equiv-connection}
\bs L_{X^v} \bs\omega = \tfrac{d}{d\tau}\, R^\star_{\psi_\tau} \bs\omega\, \big|_{\tau=0} = \tfrac{d}{d\tau}\,  {\psi_\tau\-}_* \, \bs\omega  \circ\, \psi_\tau \,\big|_{\tau=0} 
																   = [X, \bs \omega]_{\text{{\tiny $\Gamma(TP)$}}} = [\bs\omega, X]_{\text{{\tiny $\aut(P)$}}}.
\end{align}
The space of connection $\C$ is an affine space modelled on the vector space $\Omega^1_{\text{tens}}\big(\Phi, \aut(P)\big)$:  
For $\bs\omega, \bs\omega' \in \C$, we have that $\bs\beta\defeq \bs \omega' - \bs\omega \in \Omega^1_{\text{tens}}\big(\Phi, \aut(P)\big)$. Or, given $\bs\omega \in \C$ and $\bs\beta \in \Omega^1_{\text{tens}}\big(\Phi, \aut(P)\big)$, we have that $\bs\omega'=\bs\omega + \bs \beta \in \C$. 

A connection allows to define the horizontal subbundle $H\Phi \defeq \ker \bs\omega$ complementary to the vertical subbundle, $T\Phi=V\Phi \oplus H\Phi$. The horizontal projection is the map $|^h : T\Phi \rarrow H\Phi$, $\mathfrak X \mapsto \mathfrak X^h \defeq \mathfrak X - [\bs\omega(\mathfrak X)]^v$, as clearly $\bs\omega (\mathfrak X^h)=0$.

A covariant derivative associated to $\bs\omega$ is defined as $\bs D\defeq \bs d \bs \circ |^h :  \Omega^\bullet_\text{eq} \big(\Phi, \rho) \rarrow   \Omega^{\bullet+1}_\text{tens} \big(\Phi, \rho)$. On tensorial forms it has the algebraic expression $\bs D :  \Omega^\bullet_\text{tens} \big(\Phi, \rho) \rarrow   \Omega^{\bullet+1}_\text{tens} \big(\Phi, \rho)$, $\bs\alpha \mapsto \bs{ D\alpha}=\bs{d\alpha} +\rho_*(\bs\omega) \bs\alpha$, the sought after first order linear operator. 

The curvature 2-form is defined as $\bs\Omega \defeq \bs{d \omega} \bs \circ |^h$,  which implies  $\bs \Omega \in \Omega^2_\text{tens} \big(\Phi, \aut(P)\big)$. Algebraically, it is also given by Cartan structure equation
\begin{align}
\label{Curvature-Cartan-eq}
\bs\Omega = \bs{d \omega} +\tfrac{1}{2}[\bs \omega, \bs \omega]_{\text{{\tiny $\aut(P)$}}}.
\end{align}
The Bianchi identity $\bs{D\Omega}=\bs{d\Omega}+[\bs\omega, \bs \Omega]_{\text{{\tiny $\aut(P)$}}} \equiv0$ is an algebraic consequence. On tensorial forms, $\bs D \circ \bs D =\rho_*(\bs \Omega)$. 
To prove \eqref{Curvature-Cartan-eq}, the FN bracket  \eqref{FN-bracket-field-dep-diff=BT-bracket}/\eqref{extended-bracket1} plays a key role. Indeed,  one needs to show that both sides of the equality vanish on $ \bs X^v, \bs Y^v \in \bs\diff_v(\Phi)$ with $\bs X, \bs Y \in C^\infty\big(\Phi, \aut(P)\big)$. It is the case of the left-hand side since $(\bs X^v)^h\equiv0$; the right-hand side, using Koszul formula for $\bs d$ and  \eqref{extended-bracket1}, yields:
\begin{equation}
\begin{aligned}
\bs{d \omega}\big(\bs X^v, \bs Y^v \big) +[\bs \omega(\bs X^v), \bs \omega(\bs Y^v)]_{\text{{\tiny $\aut(P)$}}}
&=\bs X^v\big(\omega( \bs Y^v)\big) - \bs Y^v\big(\omega(\bs X^v)\big) - \bs \omega\big([\bs X^v, \bs Y^v]\big)    + [\bs X, \bs Y]_{\text{{\tiny $\aut(P)$}}}  \\
&= \bs X^v\big( \bs Y\big) - \bs Y^v\big(\bs X\big)    - \bs \omega\big(\{\bs X, \bs Y\}^v\big)  + [\bs X, \bs Y]_{\text{{\tiny $\aut(P)$}}}  \\
&=   [\bs X, \bs Y]_{\text{{\tiny $\aut(P)$}}} +  \bs X^v\big( \bs Y\big) - \bs Y^v\big(\bs X\big)   - \{\bs X, \bs Y\} \equiv 0.  
\end{aligned}
\end{equation}

Given the defining equivariance and verticality properties \eqref{Variational-connection} of a connection,  using \eqref{pushforward-X}/\eqref{GT-geometric} one shows that its  vertical transformation under $\bs\Diff_v(\Phi)\simeq C^\infty\big( \Phi, \Aut(P)\big)$ is
\begin{align}
\label{Vert-trsf-connection}
\bs\omega^{\bs\psi }\defeq \Xi^\star \bs\omega =\bs\psi\-_* \, \bs\omega  \circ \bs\psi + \bs\psi\-_*\bs{d\psi}. 
\end{align}
The formula is the same for its  gauge transformation under $\bs\Aut_v(\Phi)\simeq \bs\Aut(P)$. The difference between the two is seen only upon repeated transformations of each type, as  stressed in section \ref{Vertical transformations and gauge transformations}, see \eqref{2-Vert-Trsf}--\eqref{2-Gauge-Trsf}. 
Here again, we note that $\bs\omega^{\bs\psi}=\Xi^\star \bs\omega  \notin \C$ unless $\bs\psi \in \bs\Diff(M) \sim \Xi \in \bs\Aut_v(\Phi)$ -- see \cite{Francois2023-b} -- only  $\bs\Aut_v(\Phi)\simeq \bs\Aut(P)$ preserves $\C$.  
By \eqref{Inf-2-Vert-Trsf}, the $\bs\diff_v(\Phi)\simeq C^\infty\big( \Phi, \aut(P)\big)$ transformations of a connection are given by the Nijenhuis-Lie derivative,
\begin{align}
\label{Inf-Vert-trsf-connection}
\bs L_{\bs X^v} \bs\omega =\bs{dX}+ [\bs\omega, \bs X]_{\text{{\tiny $\aut(P)$}}}. 
\end{align}
Infinitesimal gauge transformations, under $\bs\aut_v(\Phi)\simeq \bs\aut(P)$, are given by the same relation, but can be written $\bs L_{\bs X^v} \bs\omega =\bs{DX}$ as $\bs X \in \bs\aut(P)$ is a tensorial 0-form. 
Similarly, the finite and infinitesimal  vertical transformations of the curvature (and gauge transformations, with the above caveat) are given, as special cases of \eqref{GT-tensorial} and \eqref{GT-inf-tensorial}, by:
\begin{align}
\label{GT-curvature}
\bs\Omega^{\bs\psi }\defeq \Xi^\star \bs\Omega =\bs\psi\-_* \, \bs\Omega  \circ \bs\psi, \qquad \text{so} \qquad \bs L_{\bs X^v} \bs\Omega =[\bs\Omega, \bs X]_{\text{{\tiny $\aut(P)$}}}. 
\end{align}

Equation \eqref{Vert-trsf-connection} allows to write the following useful lemma: For $\bs\alpha, \bs{D\alpha} \in \Omega^\bullet_\text{tens}(\Phi, \rho)$, 
we have on the one hand $\bs d\, \Xi^\star \bs\alpha = \bs d\big( \rho(\bs \psi)\- \bs\alpha \big)$. On the other hand, by $\Xi^\star \bs {D\alpha} =  \rho(\bs \psi)\- \bs{D\alpha}$, 
\begin{equation*}
\begin{aligned}
\Xi^\star \bs {d\alpha} &= \rho(\bs \psi)\- \bs{D\alpha} - \Xi^\star \big(\rho_*(\bs\omega)\bs\alpha \big) 
				   =  \rho(\bs \psi)\- \bs{d\alpha} +  \rho(\bs \psi)\-\,  \rho_*(\bs\omega) \bs\alpha  -  \rho_*(\bs\omega^{\bs \psi})\bs\alpha^{\bs \psi}
				   = \rho(\bs \psi)\- \bs{d\alpha}  -  \rho_*(\bs \psi\-_* \bs d\bs\psi) \rho(\bs\psi)\-\bs\alpha\\
				 &= \rho(\bs \psi)\- \left(  \bs{d\alpha} -  \rho_*( \bs d\bs\psi \circ \bs\psi\-) \bs\alpha \right). 
\end{aligned}
\end{equation*}
By naturality of the exterior derivative, $[\Xi^\star, \bs d]=0$, we  obtain the identity:
\begin{align}
\label{Formula-general-rep}
\bs d\big( \rho(\bs \psi)\- \bs\alpha \big) = \rho(\bs \psi)\- \left(  \bs{d\alpha} -  \rho_*( \bs d\bs\psi \circ \bs\psi\-)\, \bs\alpha \right).
\end{align}
In particular, for the pullback representation, $\rho(\psi)\-\!= \psi^*$ and $-\rho_*(X) =\mathfrak L_X$, this is:
  \begin{align}
\label{Formula-pullback-rep}
\bs d\big( \bs \psi^* \bs\alpha \big) = \bs \psi^* \left(  \bs{d\alpha} + \mathfrak L_{\bs d\bs\psi \circ \bs\psi\-} \bs\alpha \right). 
\end{align}
The latter appears in the covariant phase space literature, e.g. in \cite{DonnellyFreidel2016, Speranza2018}.\footnote{
It ought not to be confused with with \eqref{GT-basis-1-form} as, despite the superficial similarity, the two results are distinct geometric statements.}
\smallskip

\subsubsection{Twisted connections} 
\label{Twisted connections} 

Twisted equivariant/tensorial forms $\bs\alpha$ have values in a $G$-space $\bs V$, $G$ a (possibly infinite-dimensional) Lie group, 
and  their equivariance is  given by a 1-cocycle \eqref{cocycle}
 for the action of $\Aut(P)$ on $\Phi$,
$C: \Phi \times \Aut(P) \rarrow G$,   $(\phi, \psi) \mapsto C(\phi; \psi)$, such that
\begin{align}
\label{twist-equiv}
R^\star_\psi \bs\alpha = C(\phi; \psi)\- \bs\alpha, \qquad \text{with} \quad C(\phi; \psi'\circ \psi) =C(\phi; \psi') \cdot C(\phi^{\psi'}; \psi).
\end{align}
Their infinitesimal equivariance is  given by $\bs L_{X^v} \bs\alpha = -a(X, \phi) \bs \alpha$, with 
 $a(X, \phi):= \tfrac{d}{d\tau}\, C(\phi, \psi_\tau) |_{\tau=0}$ a 1-cocycle \eqref{inf-cocycle}  for the action of $\aut(P)$ on $\Phi$.  
If the target group of the 1-cocycle is the automorphism group itself, $G=\Aut(P)$, or a subgroup thereof, and $\bs V$ is a space of tensors of $P$, \eqref{twist-equiv} specialises to
\begin{align}
\label{twist-equiv-diff}
R^\star_\psi \bs\alpha = C(\phi; \psi)^* \bs\alpha, \qquad \text{with} \quad C(\phi; \psi'\circ \psi) =C(\phi; \psi') \circ C(\phi^{\psi'}; \psi).
\end{align}

A \emph{twisted connection} 1-form $\bs \varpi \in \Omega^1_{\text{eq}}\big(\Phi, \LieG \big)$ is defined by the two properties:
\begin{equation}
\label{Variational-twisted-connection}
\begin{aligned}
\bs\varpi_{|\phi} \big( X^v_{|\phi}\big)&= \tfrac{d}{d\tau} C(\phi; \psi_\tau) \,\big|_{\tau=0}=a(X, \phi) \ \in \LieG, \quad \text{for } X\in \aut(P), \\
R^\star_\psi \bs\varpi_{|\phi^\psi} &= \Ad_{C(\phi; \psi)\-} \, \bs\varpi_{|\phi} + C(\phi; \psi)\-\bs d C(\ \, ; \psi)_{|\phi}. 
\end{aligned}
\end{equation}
In the special case $G=\Aut(P)$, the equivariance of $\bs \varpi \in \Omega^1_{\text{eq}}\big(\Phi, \aut(P) \big)$  is 
\begin{align}
\label{Special-eq-twisted}
R^\star_\psi \bs\varpi_{|\phi^\psi} &= [C(\phi; \psi)\big)\-]_* \, \bs\varpi_{|\phi} \circ  C(\phi; \psi) +  [C(\phi; \psi)\big)\-]_*  \bs d C(\ \, ; \psi)_{|\phi},
\end{align}
The infinitesimal equivariance under $\aut(P)$ is
\begin{align}
\label{inf-equiv-twisted-connection}
\bs L_{X^v} \bs\varpi = \tfrac{d}{d\tau}\, R^\star_{\psi_\tau} \bs\varpi\, \big|_{\tau=0} = \bs d a(X; \ ) + [\bs\varpi, a(X;\ )]_{\text{{\tiny $\LieG$}}}.
\end{align}
Or, in the case  $G=\Aut(P)$, $ \bs L_{X^v} \bs\varpi= \bs d a(X; \ ) + [\bs\varpi, a(X;\ )]_{\text{{\tiny $\aut(P)$}}} $. 

The space of twisted connections $\b\C$ is an affine space modelled on the vector space $\Omega^1_{\text{tens}}\big(\Phi, \LieG\big)$: For $\bs\omega, \bs\omega' \in \C$, we have  $\bs\beta\defeq \bs \omega' - \bs\omega \in \Omega^1_{\text{tens}}\big(\Phi, \LieG\big)$. Or, given $\bs\omega \in \b\C$ and $\bs\beta \in \Omega^1_{\text{tens}}\big(\Phi, \LieG\big)$, we have that $\bs\omega'=\bs\omega + \bs \beta \in \b\C$. 

A \emph{twisted covariant derivative} is  defined as  $\b{\bs D} :  \Omega^\bullet_\text{eq} \big(\Phi, C) \rarrow   \Omega^{\bullet+1}_\text{tens} \big(\Phi, C)$, $\bs\alpha \mapsto \b{\bs D} \bs\alpha \defeq \bs{d\alpha} +\rho_*(\bs\varpi) \bs\alpha$. This is the first order linear operator adapted to twisted equivariant/tensorial forms.

The curvature 2-form of $\bs\varpi$ is  defined by the Cartan structure equation:
\begin{align}
\label{Twisted-curvature}
\b{\bs\Omega} \defeq \bs{d \varpi} +\tfrac{1}{2}[\bs \varpi, \bs \varpi]_{\text{{\tiny $\LieG$}}} \ \  \in  \Omega^2_\text{tens} \big(\Phi, \LieG \big).
\end{align}
It thus satisfies the Bianchi identity, $\b{\bs D} \b{\bs\Omega}=\bs d\b{\bs\Omega}+[\bs\varpi, \b{\bs \Omega}]_{\text{{\tiny $\LieG$}}}=0$. And it holds that $\b{\bs D} \circ \b{\bs D} =\rho_*(\b{\bs \Omega})$.
For $X^v, Y^v \in \Gamma(V\Phi)$ with $X, Y \in \diff(M)$, we have
\begin{equation}
\begin{aligned}
 \b{\bs\Omega}(X^v, Y^v)&= X^v\big(\omega(  Y^v)\big) -  Y^v\big(\omega( X^v)\big) - \bs \omega\big([ X^v,  Y^v]\big)   +[\bs \omega( X^v), \bs \omega( Y^v)]_{\text{{\tiny $\LieG$}}} \\
0&=  X^v\big( a(Y; \phi) \big) -  Y^v\big( a(X; \phi) \big)    - a([X, Y]_{\text{{\tiny $\aut(P)$}}}; \phi)  + [a(X; \phi), a(Y; \phi)]_{\text{{\tiny $\LieG$}}},  \label{C}
\end{aligned}
\end{equation}
which reproduces the infinitesimal 1-cocycle property  \eqref{inf-cocycle}. 

The vertical transformation under $\bs\Diff_v(\Phi)\simeq C^\infty\big( \Phi, \Aut(P)\big)$ of a twisted connection is found to be, using \eqref{Variational-twisted-connection} and \eqref{pushforward-X}/\eqref{GT-geometric}, 
\begin{align}
\label{Vert-trsf-twisted-connection}
\bs\varpi^{\bs\psi }\defeq \Xi^\star \bs\varpi =\Ad_{C(\bs\psi)\- } \, \bs\varpi  + C(\bs\psi)\-\bs{d} C(\bs\psi).
\end{align}
 In case $G=\Aut(P)$ this specialises to: $\bs\varpi^{\bs\psi }=C(\bs\psi)\-_* \, \bs\varpi  \circ C(\bs\psi) + C(\bs\psi)\-_*\bs{d} C(\bs\psi)$.
Gauge transformations under $\bs\Aut_v(\Phi)\simeq \bs\Aut(P)$ are given by the same formula, the difference showing upon repeated transformations of each type,  see \eqref{2-Vert-Trsf}--\eqref{2-Gauge-Trsf} in section \ref{Vertical transformations and gauge transformations}. 
Transformations of a twisted connection under $\bs\diff_v(\Phi)\simeq C^\infty\big( \Phi, \aut(P)\big)$  are given by the Nijenhuis-Lie derivative, 
\begin{align}
\label{Inf-Vert-trsf-twisted-connection}
\bs L_{\bs X^v} \bs\varpi =\bs{d} a(\bs X)+ [\bs\varpi, a(\bs X)]_{\text{{\tiny $\LieG$}}}. 
\end{align}
Infinitesimal gauge transformations, under $\bs\aut_v(\Phi)\simeq \bs\aut(P)$, are given by the same relation. 
The difference being seen upon iteration, as reflected by the commutation property  of the Nijenhuis-Lie derivative \eqref{Lie-deriv-ext-bracket}/\eqref{NL-derr-ext-bracket}. 

Finite and infinitesimal general vertical transformations of the curvature are given by, 
\begin{align}
\label{GT-twisted-curvature}
\b{\bs\Omega}^{\bs\psi }\defeq \Xi^\star \b{\bs\Omega} =\Ad_{C(\bs\psi)\-} \, \b{\bs\Omega}, \qquad \text{so} \qquad \bs L_{\bs X^v}\b{\bs\Omega} =[\b{\bs\Omega}, a(\bs X)]_{\text{{\tiny $\LieG$}}}. 
\end{align}
For  $G=\Aut(P)$, this is $\b{\bs\Omega}^{\bs\psi }\defeq \Xi^\star \b{\bs\Omega} =\bs\psi\-_* \, \b{\bs\Omega}  \circ \bs\psi$ and $\bs L_{\bs X^v}\b{\bs\Omega} =[\b{\bs\Omega}, \bs X]_{\text{{\tiny $\aut(P)$}}}$. 
These illustrate \eqref{GT-tensorial} and \eqref{GT-inf-tensorial}. The same relations hold for its gauge transformations, with the usual caveat.
When $\bs X^v, \bs Y^v \in \bs\diff_v(\Phi)$ with $\bs X, \bs Y \in C^\infty \big(\Phi, \aut(P)\big)$, using the definition  \eqref{FN-bracket-field-dep-diff=BT-bracket}/\eqref{extended-bracket1}  of the FN bracket, we have
\begin{equation}
\label{B}
\begin{aligned}
 \b{\bs\Omega}(\bs X^v, \bs Y^v)&= \bs X^v\big(\omega(  \bs Y^v)\big) - \bs Y^v\big(\omega(\bs X^v)\big) - \bs \omega\big([ \bs X^v,  \bs Y^v]\big)   +[\bs \omega( \bs X^v), \bs \omega( \bs Y^v)]_{\text{{\tiny $\LieG$}}},\\
0&=  \bs X^v\big( a(\bs Y; \phi) \big) -  \bs Y^v\big( a(\bs X; \phi) \big)    - a(\{\bs X, \bs Y\}; \phi)  + [a(\bs X; \phi), a(\bs Y; \phi)]_{\text{{\tiny $\LieG$}}}, \\
0&=\bs X^v\big( a(\munderline{red}{\bs Y}; \phi) \big) -  \bs Y^v\big( a( \munderline{red}{\bs X}; \phi) \big)    - a([\bs X, \bs Y]_{\text{{\tiny $\diff(M)$}}}; \phi)  + [a(\bs X; \phi), a(\bs Y; \phi)]_{\text{{\tiny $\LieG$}}}. 
\end{aligned}
\end{equation}
This reproduces \eqref{BB}, where the notation of the last line was first used.

\subsection{Associated bundles, bundle of regions of $P$ and integration}
\label{Associated bundles}  

Given a principal fiber bundle, it is  standard that one can build an associate bundle (over the same base) via each representation of the structure group. This can be generalised by replacing  representation by 1-cocycles for the action of the structure group  \cite{Francois2019_II}. Below we review these constructions in our case, where the principal bundle is $\Phi$ with structure group $\Aut(P)$.

Given a representation space $(\rho,  \bs V)$ of $\Aut(P)$, consider the direct product space $\Phi \times \bs V$, with the two natural projections: $\pi_\Phi : \Phi \times \bs V \rarrow \Phi$ and  $\pi_{\bs V} : \Phi \times \bs V \rarrow \bs V$.
One defines a right action of $\Aut(P)$ on  $\Phi \times \bs V$ by: 
\begin{equation}
\label{Diff-action-PxV}
\begin{aligned}
\big(\Phi \times \bs V \big) \times \Aut(P) &\rarrow \Phi \times \bs V, \\
\big((\phi,  v), \psi \big) & \mapsto  \big(\psi^*\phi,\, \rho(\psi)\-  v\big)=\big(R_\psi \phi,\, \rho(\psi)\-  v\big) \rdefeq \b R_\psi (\phi,  v). 
\end{aligned}
\end{equation}
 The  bundle $\bs E$ associated to $\Phi$ via the representation $\rho$ is  the quotient of $\Phi \times \bs V$ by $\b R$:  
 \begin{equation}
\begin{aligned}
\bs E = \Phi \times_\rho \bs V \defeq   \Phi \times  \bs V /\!\sim
\end{aligned}
\end{equation}
where $(\phi',  v') \sim (\phi,  v)$ when $\exists\, \psi \in \Aut(P)$ s.t. $(\phi',  v')=\b R_\psi (\phi,  v)$. 
We write $\b \pi_{\bs E} :  \Phi \times \bs V \rarrow \bs E$. 
A point in $\bs E$ is an equivalence class $e=[\phi,  v]$. The projection of $\bs E \xrightarrow{\pi_{\bs E}} \M$ is $\pi_{\bs E}([\phi,  v]) \defeq \pi(\phi)=[\phi]$. 
It is a well-known result that there is a bijection between sections of $\bs E$ and $V$-valued $\rho$-equivariant functions on $\Phi$: 
 \begin{equation}
 \label{iso-section-equiv-fct}
\begin{aligned}
\Gamma(\bs E)\defeq \big\{ \bs s: \M \rarrow \bs E\big\}  \   \
\simeq \ \  \Omega^0_\text{eq}\big(\Phi, \rho \big)\defeq \big\{ \bs\vphi: \Phi \rarrow \bs V\, |\,  R^\star_\psi \bs\vphi =\rho(\psi)\- \bs\vphi  \big\} ,
\end{aligned}
\end{equation}
the isomorphism being $\bs s([\phi]) = [\phi, \bs \vphi(\phi)]$.  
Equivariant functions are tensorial 0-forms, so their vertical (gauge) transformations are given by \eqref{GT-tensorial}-\eqref{GT-inf-tensorial}: 
$\bs\vphi^{\bs\psi} = \rho(\bs\psi)\- \bs\vphi$ %
and 
$ \bs L_{\bs X^v} \bs\vphi = -\rho_*(\bs X) \bs\vphi$,  
for $\bs\psi \in C^\infty \big(\Phi, \Aut(P)\big) \simeq \bs\Diff_v(\Phi)$ and $\bs X \in C^\infty \big(\Phi, \aut(P)\big) \simeq \bs\diff_v(\Phi)$.
A Ehresmann connection is needed for their covariant differentiation, see section \ref{Ehresmann connections}. 
\medskip

The construction holds the same replacing representations $\rho$ by  1-cocycles 
$\ C\!: \Phi \times \Aut(P) \rarrow G$ as defined by \eqref{cocycle} in section \ref{Differential forms and their derivations}.
Given a $G$-space $\bs V$, one  defines the right action of $\Aut(P)$ on $ \Phi \times  \bs V$:
$\big(\Phi \times \bs V \big) \times \Aut(P) \rarrow \Phi \times \bs V$, 
$\big((\phi,  v), \psi \big)  \mapsto  \b R_\psi (\phi,  v)=\big(\psi^*\phi,\, C(\phi; \psi)\-  v\big)$. 
 The \emph{twisted} bundle $\widetilde{\bs E} \rarrow \M$ associated to $\Phi$ via the 1-cocycle $C$ is then:
$\widetilde{\bs E} = \Phi \times_C \bs V \defeq   \Phi \times  \bs V /\!\sim\,$,
with $\big(\psi^*\phi,\, C(\phi; \psi)\-  v\big) \sim (\phi,  v)$. 
As above, its space of sections is isomorphic to the space of \emph{twisted} equivariant function on $\Phi$:
 \begin{equation}
\begin{aligned}
\Gamma\big(\widetilde{\bs E}\big)\defeq \big\{ \t{\bs s}: \M \rarrow \widetilde{\bs E}\big\}  \   \
\simeq \ \  \Omega^0_\text{eq}\big(\Phi,  C \big)\defeq \big\{ \t{\bs\vphi}: \Phi \rarrow V\, |\,  R^\star_\psi \t{\bs\vphi} = C(\phi; \psi)\- \t{\bs\vphi}  \big\}.
\end{aligned}
\end{equation}
The vertical transformation of such  twisted equivariant functions is given by  \eqref{GT-tensorial}-\eqref{GT-inf-tensorial}. 
A twisted connection as discussed in section \ref{Twisted connections} is needed for their covariant differentiation.

\subsubsection{Associated bundle of regions} 
\label{Associated bundle of regions} 

Let $\bs E= \b {\bs V}(P)$ be the bundle canonically associated to $\Phi$ via the \emph{defining representation} of $\Aut(P)$:
the $\sigma$-algebra of open sets of $P$, 
$\bs V(P)\defeq \big\{ V \subset P\, | \, V \text{ open set} \big\}$.\footnote{It is actually the defining representation of $\Diff(P)$, as a (Lie) pseudo-group \cite{Kobayashi1972} and restricts naturally to $\Aut(P)$.} 
The~right action of $\Aut(P)$ on the product space $ \Phi \times \bs V(P) $ is
 \begin{equation}
 \label{right-action-diff}
\begin{aligned}
\big( \Phi \times \bs V(P) \big) \times \Aut(P) &\rarrow  \Phi \times \bs V(P), \\
\big( (\phi, V), \psi \big) &\mapsto \b R_\psi (\phi,  V) \defeq \big( \psi^*\phi, \psi\-(V) \big). 
\end{aligned}
\end{equation}
The  \emph{associated bundle of regions} of $P$ is thus:
 \begin{equation}
 \label{Assoc-bundle-regions}
\begin{aligned}
\b{\bs V}(P) = \Phi \times_\text{{\tiny $\Aut(P)$}} \bs V(P) \defeq   \Phi \times  \bs V(P) /\!\sim.
\end{aligned}
\end{equation}
Its space of sections $\Gamma\big(\b{\bs V}(P) \big)\defeq \big\{ \b{\bs s}: \M \rarrow \b{\bs V}(P) \big\}$ is isomorphic to
 \begin{equation}
 \label{equiv-regions}
\begin{aligned}
\Omega^0_\text{eq}\big(\Phi,  \bs V(P)  \big)\defeq \big\{ \bs V: \Phi \rarrow \bs V(P) \, |\,  R^\star_\psi {\bs V} = \psi\-({\bs V})  \big\}.
\end{aligned}
\end{equation}
A map   $\phi \rarrow \bs V(\phi)$ may be seen as a ``field-dependent" open set of $P$, a region of $P$ defined in a ``$\phi$-\emph{relative}" and $\Aut(P)$-equivariant way.
By \eqref{GT-tensorial}-\eqref{GT-inf-tensorial}, its  transformations under $\bs\Diff_v(\Phi) \simeq C^\infty \big(\Phi, \Aut(P)\big)$ and $\bs\diff_v(\Phi) \simeq C^\infty \big(\Phi, \aut(P)\big)$ are respectively: 
\begin{align}
\bs V^{\bs \psi} = \bs\psi\-(\bs V), \quad \text{ and } \quad \bs L_{\bs X^v} \bs V = -\bs X(\bs V). 
\end{align} 
Integration on $P$ provides just such an example of equivariant function, as can be shown by framing it as a natural construction over  $\Phi \times \bs V(P)$.

 \subsubsection{Integration map} 
\label{Integration map} 

Associated bundles are defined  via the  action of the structure group $\Aut(P)$ on $\Phi \times \bs V$: $\b R_\psi (\phi,  v)\defeq \big(\psi^*\phi,\, \rho(\psi)\-  v\big)$. 
The corresponding action of $C^\infty\big(\Phi, \Aut(P)\big) \simeq \bs\Diff_v(\Phi)$ is
\begin{equation}
\label{Diff_v-action-PxV}
\begin{aligned}
\big(\Phi \times \bs V \big) \times C^\infty\big(\Phi, \Aut(P)\big) &\rarrow \Phi \times \bs V, \\
\big((\phi,  v), \bs\psi \big) & \mapsto  \big(\bs\psi^*\phi,\, \rho(\bs\psi)\-  v\big) =  \big(\Xi(\phi),\, \rho(\bs\psi)\-  v\big) \rdefeq \b \Xi (\phi,  v). 
\end{aligned}
\end{equation}
In particular, for $\bs\psi \in \bs\Aut(P)$ we have $\b \Xi \bs\circ \b R_\psi = \b R_\psi \bs\circ \b \Xi$. 
The action of $C^\infty\big(\Phi, \aut(P)\big)\simeq \bs\diff_v(\Phi)$ is the linearisation  
$\big(\Phi \times \bs V \big) \times C^\infty\big(\Phi, \aut(P)\big) \rarrow V(\Phi \times \bs V)\simeq V\Phi \oplus V\bs V \subset T(P\times \bs V)$. 

The induced actions of $\Aut(P)$ and $C^\infty\big(\Phi, \Aut(P)\big)\simeq \bs\Diff_v(\Phi)$ on $ \Omega^\bullet(\Phi) \times \bs V$ are 
\begin{equation}
\begin{aligned}
\big( \Omega^\bullet(\Phi) \times \bs V \big) \times \Aut(P) &\rarrow \Omega^\bullet(\Phi)\times \bs V, \\
\big((\bs\alpha,  v), \psi \big) & \mapsto  \big(R^\star_\psi \bs\alpha,\, \rho(\psi)\-  v\big)  \rdefeq \t R_\psi (\bs\alpha,  v)
\end{aligned}
\end{equation}
and 
\begin{equation}
\label{Diff_v-action-PxV1}
\begin{aligned}
\big(\Omega^\bullet(\Phi) \times \bs V \big) \times C^\infty\big(\Phi, \Aut(P)\big) &\rarrow \Omega^\bullet(\Phi) \times \bs V, \\
\big((\bs\alpha,  v), \bs\psi \big) & \mapsto  \big(\Xi^\star \bs\alpha ,\, \rho(\bs\psi)\-  v\big)  \rdefeq \t \Xi (\bs\alpha,  v). 
\end{aligned}
\end{equation}
Correspondingly, the induced actions of $\aut(P)$ and $C^\infty\big(\Phi, \aut(P)\big)\simeq \bs\diff_v(\Phi)$ are the linearisations:
\begin{equation}
\label{linear-versions}
\begin{aligned}
\big((\bs\alpha,  v), X \big) & \mapsto  \tfrac{d}{d\tau}\, \t R_{\psi_\tau} (\bs\alpha,  v) \,\big|_{\tau=0} = \big(\bs L_{X^v} \bs\alpha,\,  v\big) \oplus \big( \bs\alpha,\, -\rho_*(X)  v\big), \\
\big((\bs\alpha,  v), \bs X \big) & \mapsto  \tfrac{d}{d\tau}\, \t \Xi_\tau (\bs\alpha,  v)\, \big|_{\tau=0} = \big(\bs L_{\bs X^v} \bs\alpha,\,  v\big) \oplus \big( \bs\alpha,\, -\rho_*(\bs X)  v\big).
\end{aligned}
\end{equation}
Furthermore, given a representation $(\t \rho, \bs W)$ of $\Aut(P)$,
\begin{equation}
\begin{aligned}
&\text{if }\ \bs\alpha \in \Omega^\bullet_{\text{eq}}(\Phi, \bs W)  \  \text{ then } \   
 \t R_\psi (\bs\alpha,  v) =  \big(R^\star_\psi \bs\alpha,\, \rho(\psi)\-  v\big)=  \big( \t\rho(\psi)\- \bs\alpha,\, \rho(\psi)\-  v\big), \\
&\text{if }\ \bs\alpha \in \Omega^\bullet_{\text{tens}}(\Phi, \bs W)  \  \text{ then } \   
 \t \Xi (\bs\alpha,  v) =  \big(\Xi^\star \bs\alpha,\, \rho(\bs \psi)\-  v\big)=  \big( \t\rho(\bs\psi)\- \bs\alpha,\, \rho(\bs\psi)\-  v\big),
\end{aligned}
\end{equation}
with linearisations read from \eqref{linear-versions}.
The exterior derivative $\bs d$ on $\Phi$  extends to $\Phi \times \bs V$ as ${\bs d} \rarrow  \bs d \times  \id$. 
Yet, after the action of  $C^\infty\big(\Phi, \aut(P)\big)\simeq \bs\diff_v(\Phi)$ and due to the $\phi$-dependence of $\bs\psi$, it will also act on the second factor $\rho(\bs\psi)\-  v$. 

Let $(\b \rho, \bs V^*)$ be a representation of $\Aut(P)$ \emph{dual} to $(\rho, \bs V)$ w.r.t. a non-degenerate $\Aut(P)$-invariant \emph{pairing} 
\begin{equation}
\begin{aligned}
\langle\ , \ \rangle: \bs V^* \times \bs V &\rarrow \RR, \\
                                                       (w, v) &\mapsto \langle w , v \rangle, \quad \text{s.t.} \quad \langle\, \b \rho(\psi) w ,   \rho(\psi)v \rangle =  \langle w , v \rangle. 
\end{aligned}
\end{equation}
Under the action of $\aut(P)$, with induced representation $\b \rho_*$ and $\rho_*$, it holds that
\begin{align}
\label{infinitesimal-equiv-pairing}
\langle\, \b \rho_*(X) w ,   v \rangle + \langle\,  w ,   \rho_*(X)v \rangle =0.
\end{align}
For $\bs\alpha \in \Omega^\bullet(\Phi, \bs V^*)$, let us define the operation $\mathcal I$ 
on $\Omega^\bullet(\Phi, \bs V^*) \times \bs V$ by,
\begin{equation}
\begin{aligned}
\mathcal I: \Omega^\bullet(\Phi, \bs V^*)  \times \bs V &\rarrow \Omega^\bullet(\Phi),\\
					(\bs\alpha, v) &\mapsto \mathcal I(\bs\alpha, v) \defeq  \langle \bs\alpha , v \rangle. 
\end{aligned}
\end{equation}
This can be seen as an object on $\Phi \times \bs V$:
\begin{equation}
\begin{aligned}
\mathcal I(\bs\alpha, \ ): \Phi \times \bs V &\rarrow \Lambda^{\!\bullet}(\Phi),\\
					(\phi, v) &\mapsto \mathcal I(\bs\alpha_{|\phi}, v) \defeq  \langle \bs\alpha_{|\phi} , v \rangle. 
\end{aligned}
\end{equation}
We thus have:
\begin{align}
\label{der-eval-integr-obj}
\bs d \mathcal I(\bs\alpha, \ ) = \mathcal I(\bs{d\alpha}, \ ), \quad \text{ and } \quad \iota_{\mathfrak X}  \mathcal I(\bs\alpha, \ ) = \mathcal I( \iota_{\mathfrak X} \bs\alpha, \ ) \quad \text{ for } \mathfrak X \in \Gamma(T\Phi).
\end{align}
The induced actions of $\Aut(P)$ and $C^\infty\big(\Phi, \Aut(P) \big) \simeq \bs\Diff_v(\Phi)$ on such objects are:
\begin{equation}
\begin{aligned}
 &\t R^\star_\psi \mathcal I(\bs\alpha,\   )_{|(\psi^*\phi,\  \rho(\psi)\- v)} \defeq  \langle\ , \ \rangle \circ \t R_\psi (\bs \alpha, v) = \langle R^\star_\psi \bs\alpha_{|\psi^*\phi},\, \rho(\psi)\-  v\rangle, \\
& \t \Xi^\star \mathcal I(\bs\alpha,\   )_{|(\Xi (\phi),\  \rho(\bs\psi)\- v)} \defeq  \langle\ , \ \rangle \circ \t \Xi (\bs \alpha, v) = \langle \Xi^\star \bs\alpha_{|\Xi(\phi)},\, \rho(\bs\psi)\-  v\rangle. 
 \end{aligned}
\end{equation}
The  actions of $\aut(P)$ and $C^\infty\big(\Phi, \aut(P)\big)\simeq \bs\diff_v(\Phi)$ are thus:
\begin{equation}
\label{linear-versions-pairing}
\begin{aligned}
&\tfrac{d}{d\tau}\, \t R_{\psi_\tau}^\star \mathcal I (\bs\alpha,  v) \,\big|_{\tau=0} = \langle \bs L_{X^v} \bs\alpha,\,  v \rangle 
																			   + \langle \bs\alpha,\, -\rho_*(X)  v \rangle, \\
&\tfrac{d}{d\tau}\, \t \Xi_\tau^\star \mathcal I  (\bs\alpha,  v)\, \big|_{\tau=0} = \langle \bs L_{\bs X^v} \bs\alpha,\,  v \rangle 
																			 + \langle \bs\alpha,\, -\rho_*(\bs X)  v \rangle.  
\end{aligned}
\end{equation}

Observe that for  $\bs\alpha \in \Omega^\bullet_{\text{eq}}(\Phi, \bs V^*)$:
\begin{equation}
\label{Int-eq}
\begin{aligned}
 \t R^\star_\psi \mathcal I(\bs\alpha,\   )_{|(\psi^*\phi,\  \rho(\psi)\- v)} \defeq 
 &\, \langle R^\star_\psi \bs\alpha_{|\psi^*\phi},\, \rho(\psi)\-  v\rangle \\
 =&\,  \langle \b\rho(\psi)\- \bs\alpha_{|\phi},\, \rho(\psi)\-  v\rangle = \langle  \bs\alpha_{|\phi},  v\rangle \rdefeq \mathcal I(\bs\alpha,\ )_{|(\phi, v)}. 
 \end{aligned}
\end{equation}
From this follows, by \eqref{linear-versions-pairing}:
\begin{equation}
\label{General-cont-eq1}
\begin{aligned}
 \langle \bs L_{X^v} \bs\alpha,\,  v \rangle + \langle \bs\alpha,\, -\rho_*(X)  v \rangle &= 0, \qquad X \in \aut(P).\\
  \langle  -\b \rho_*(X)\, \bs\alpha,\,  v \rangle + \langle \bs\alpha,\, -\rho_*(X)  v \rangle &= 0.
\end{aligned}
\end{equation}
If $\bs\alpha \in \Omega^\bullet_{\text{tens}}(\Phi, \bs V^*)$:
 \begin{equation}
  \label{Int-tens}
\begin{aligned}
 \t \Xi^\star \mathcal I(\bs\alpha,\   )_{|(\Xi (\phi),\  \rho(\psi)\- v)} \defeq  
 &\, \langle \Xi^\star \bs\alpha_{|\Xi(\phi)},\, \rho(\bs\psi)\-  v\rangle \\
 =&\,  \langle \b\rho(\bs\psi)\- \bs\alpha_{|\phi},\, \rho(\bs\psi)\-  v\rangle = \langle  \bs\alpha_{|\phi},  v\rangle \rdefeq \mathcal I(\bs\alpha,\ )_{|(\phi, v)}. 
 \end{aligned}
\end{equation}
And it follows by \eqref{linear-versions-pairing}:
\begin{equation}
\label{General-cont-eq2}
\begin{aligned}
 \langle \bs L_{\bs X^v} \bs\alpha,\,  v \rangle + \langle \bs\alpha,\, -\rho_*(\bs X)  v \rangle &= 0, \qquad \bs X \in C^\infty\big(\Phi, \aut(P)\big).\\
  \langle  -\b \rho_*(\bs X)\, \alpha,\,  v \rangle + \langle \bs\alpha,\, -\rho_*(\bs X)  v \rangle &= 0.
\end{aligned}
\end{equation} 
When $\bs\alpha$ is tensorial, $\mathcal I(\bs\alpha,\ )$ is ``basic" on $\Phi\times \bs V$, meaning it is well-defined on $\bs E=\Phi \times \bs V/\sim$. 
Since $\mathcal I(\bs \alpha,\ )$ is constant along an $\Aut(P)$-orbit in $\Phi \times \bs V$,  it  allows to define 
$\bs\vphi_{\mathcal I(\bs \alpha)}\! \in \Omega^0_\text{eq}(\Phi, \rho)$ 
 via: 
 \begin{equation}
 \label{Induced-equiv-fct}
\begin{aligned}
\bs\vphi_{\mathcal I(\bs \alpha)}(\phi)&\defeq \pi_{\bs V}(\phi,  v)_{|\mathcal I(\bs\alpha_{|\phi}, v) =\text{cst}} \equiv v, \\
\bs\vphi_{\mathcal I(\bs \alpha)}(\psi^*\phi)& \defeq \pi_{\bs V}(\psi^*\phi, \rho(\psi)\- v)_{|\mathcal I(\bs\alpha_{|\phi}, v) =\text{cst}} \equiv \rho(\psi)\- v.
 \end{aligned}
\end{equation}
By  \eqref{iso-section-equiv-fct}, the latter is equivalent to a section $\bs s_{\mathcal I(\bs \alpha)} :\M \rarrow \bs E$.  

Observe that   \eqref{Int-tens} implies that for $\bs\alpha \in \Omega^\bullet_{\text{tens}}(\Phi, \bs V^*)$ one has 
$\bs d \, \t \Xi^\star \mathcal I(\bs\alpha,\   ) = \bs d \mathcal I(\bs\alpha,\   )= \mathcal I(\bs{d\alpha},\   )$. 
Also, we derive the following lemma:
\begin{align}
  \t\Xi^\star \langle \bs{d\alpha}, v \rangle &\defeq \langle  \Xi^\star \bs d \bs\alpha, \, \rho(\bs\psi)\-v \rangle  = \langle  \bs d \,  \Xi^\star  \bs\alpha, \, \rho(\bs\psi)\-v \rangle \notag \\
					&= \langle  \bs d \, \b\rho(\bs\psi)\-  \bs\alpha, \, \rho(\bs\psi)\-v \rangle 
								 \notag \\
					&= \langle  \b\rho(\bs\psi)\- \big(  \bs d \bs \alpha -\b\rho_*(\bs{d\psi}\circ \bs\psi\-)\, \bs\alpha \big),\, \rho(\bs\psi)\-v\rangle 
								 \notag \\
					&= \langle    \bs d \bs \alpha -\b\rho_*(\bs{d\psi}\circ \bs\psi\-) \bs\alpha,\, v\rangle,
								  \notag\\[1mm]
\hookrightarrow\quad 	 \t\Xi^\star \langle \bs{d\alpha}, v \rangle 	&= \langle    \bs d \bs \alpha, v \rangle+  \langle -\b\rho_*(\bs{d\psi}\circ \bs\psi\-) \bs\alpha,\, v\rangle.
			\label{Vert-trsf-pairing-dalpha}	 
\end{align}
\medskip

We now specialise  the above construction to the fundamental representation $\bs V=\bs V(P)$ 
and the representation $\bs V^* = \Omega^\text{top}(V)$ of volume forms on $V \in \bs V(P)$ on which $\Aut(P)$ acts by pullback. These are dual under the invariant \emph{integration pairing}:
\begin{equation}
\begin{aligned}
\langle\ , \ \rangle: \bs \Omega^\text{top}(V) \times \bs V(P) &\rarrow \RR, \\[-2mm]
                                                       (\omega, V) &\mapsto \langle \omega , V \rangle \defeq \int_V \omega. 
\end{aligned}
\end{equation}
 The invariance property is the familiar identity
 \begin{align}
 \label{invariance-action}
\langle\, \psi^*\omega ,   \psi\-(V) \rangle =  \langle \omega , V \rangle \quad   \rarrow  \quad  \int_{\psi\-(V)} \psi^*\omega = \int_V \omega. 
 \end{align}
 This, as a special case of \eqref{infinitesimal-equiv-pairing} with $-\b \rho_*(X)=\mathfrak L_X$ and $-\rho_*(X)=-X$, gives:
  \begin{align}
 \label{inf-invariance-action}
\langle\, \mathfrak L_X \omega ,   V \rangle + \langle\,  \omega ,   -X(V) \rangle =0  \quad
\rarrow \quad
  \int_V   \mathfrak L_X \omega  + \int_{-X(V)} \hspace{-2mm}\omega =0.
 \end{align}
 This can be read as a  \emph{continuity equation} for the action of $\aut(P)$. 
 By Stokes theorem,
the de Rham derivative $d$ on $\Omega^\bullet(V)$ and the boundary operator $\d$ on $\bs V(P)$ are  adjoint operators w.r.t. to the integration pairing:
 \begin{align}
 \langle\, d \omega ,  V \rangle = \langle\,  \omega ,   \d V \rangle \quad
 \rarrow \quad
  \int_V   d \omega  = \int_{\d V} \hspace{-2mm}\omega. 
\end{align} 
 
Considering  $\bs \alpha \in \Omega^\bullet\big( \Phi,  \Omega^\text{top}(V) \big)$, the field-dependent volume forms, we define the integration map on $\Phi \times \bs V(P)$:
\begin{align}
\mathcal I(\bs\alpha_{|\phi}, V) = \langle \bs\alpha_{|\phi}, V \rangle \defeq \int_V \bs \alpha_{|\phi}.
\end{align}
The simplified notation $\bs\alpha_V$ may be used when more convenient.  As a special case of \eqref{der-eval-integr-obj},
we have  $\bs d \mathcal I(\bs\alpha, V) = \mathcal I(\bs{d\alpha}, V)$ and $\iota_{\mathfrak X}  \mathcal I(\bs\alpha, V ) = \mathcal I( \iota_{\mathfrak X} \bs\alpha, V)$ for $\mathfrak X \in \Gamma(T\Phi)$. 
The induced actions of $\Aut(P)$ and $C^\infty\big(\Phi, \Aut(P) \big) \simeq \bs\Diff_v(\Phi)$ on integrals are:
\begin{equation}
\begin{aligned}
 &\t R^\star_\psi \mathcal I(\bs\alpha,\   )_{|\big(\psi^*\phi,\ \psi\-(V) \big)} \defeq  \langle R^\star_\psi \bs\alpha_{|\psi^*\phi},\, \psi\-(V) \rangle = \int_{ \psi\-(V)} R^\star_\psi \bs\alpha_{|\psi^*\phi},  \\
& \t \Xi^\star \mathcal I(\bs\alpha,\   )_{|\big(\Xi (\phi),\  \bs\psi\-(V) \big)} \defeq  \langle \Xi^\star \bs\alpha_{|\Xi(\phi)},\, \bs\psi\-(V)  \rangle = \int_{ \bs\psi\-(V)}  \Xi^\star \bs\alpha_{|\Xi(\phi)}. \label{Vert-trsf-int-generic}
 \end{aligned}
\end{equation}
In the latter case,  $\bs d$ acts also on the transformed region $\bs\psi\-(V)$ due to the $\phi$-dependence of $\bs\psi$. 
We may write the above as ${\bs\alpha_V}^\psi$ and ${\bs\alpha_V}^{\bs\psi}$, respectively. 
The actions of $\aut(P)$ and $C^\infty\big(\Phi, \aut(P)\big)\simeq \bs\diff_v(\Phi)$ on integrals are 
\begin{equation}
\label{linear-versions-integral}
\begin{aligned}
&\tfrac{d}{d\tau}\, \t R_{\psi_\tau}^\star \mathcal I (\bs\alpha,  V) \,\big|_{\tau=0} = \langle \bs L_{X^v} \bs\alpha,\,  V \rangle 
																			   + \langle \bs\alpha,\, -X(V) \rangle
														           =\int_V  \bs L_{X^v} \bs\alpha + \int_{-X(V)} \bs\alpha, \\
&\tfrac{d}{d\tau}\, \t \Xi_\tau^\star \mathcal I  (\bs\alpha,  V)\, \big|_{\tau=0} = \langle \bs L_{\bs X^v} \bs\alpha,\,  V \rangle 
																			 + \langle \bs\alpha,\, -\bs X(V)   \rangle
															=\int_V  \bs L_{\bs X^v} \bs\alpha + \int_{-\bs X(V)} \bs\alpha.  
\end{aligned}
\end{equation}
When convenient, we may write the above as $\delta_X\ \!{\bs\alpha_V}$ and $\delta_{\bs X}\ \!{\bs\alpha_V}$ respectively. 

When $\bs\alpha$ is  tensorial 
on $\Phi$, as a special case of the results \eqref{Int-tens} 
above, we have that  $\bs\alpha_V=\mathcal I(\bs\alpha_{|\phi}, V)$ is $C^\infty\big(\Phi, \Aut(P) \big)$-invariant:  ${\bs\alpha_V}^{\bs \psi} =\bs\alpha_V$. 
From which follows
\begin{equation}
\label{Cont-eq-general-case}
\begin{aligned}
\delta_{\bs X} \bs\alpha_V =0 \quad \Rightarrow \quad \langle \bs L_{\bs X^v} \bs\alpha,\,  V \rangle + \langle \bs\alpha,\, -\bs X(V) \rangle &= 0, \qquad \bs X \in C^\infty\big(\Phi, \aut(P)\big).\\
  \langle  \mathfrak L_{\bs X} \alpha,\,  V \rangle + \langle \bs\alpha,\, -\bs X(V)   \rangle &= 0 \quad \rarrow \quad \int_V \mathfrak L_{\bs X} \alpha + \int_{-\bs X(V)} \bs\alpha =0,
\end{aligned}
\end{equation}
as a special case of  \eqref{General-cont-eq2} and \eqref{linear-versions-integral}.
This can be interpreted as a continuity equation.
 For $\bs\alpha$ equivariant, we have $\Aut(P)$-invariance of its integral: ${\bs\alpha_V}^{\psi} =\bs\alpha_V$, and \eqref{Cont-eq-general-case} holds mutatis mutandis ($\bs X \rarrow X$). 
For $\bs\alpha$ tensorial, $\bs\alpha_V=\mathcal I(\bs\alpha, V)$ is thus well-defined on the bundle of regions $\b{\bs V}(P)=\Phi \times \bs V(P)/\sim$, and one may  define an equivariant $\bs V(P)$-valued function on $\Phi$ as in  \eqref{Induced-equiv-fct}.
It also means that $\bs d({\bs\alpha_V}^{\bs\psi}) = \bs d \bs\alpha_V$, i.e. 
\begin{align}
\label{Invariance-tens-int}
\bs d \, \langle \bs\psi^* \bs\alpha, \bs\psi\-(V)\rangle =   \langle \bs d \bs\alpha, V\rangle = \bs d  \langle  \bs\alpha, V\rangle \quad \rarrow \quad
\bs d \int_{\bs\psi\-(V)}  \bs\psi^* \bs\alpha = \int_V \bs{d\alpha} = \bs d \int_V \bs \alpha.
\end{align}
Also, specialising \eqref{Vert-trsf-pairing-dalpha}, we get the identity
\begin{equation}
\label{Vert-trsf-int-dalpha}
\begin{aligned}
\bs (\bs{d\alpha}_V)^{\bs \psi}	 &= \bs{d\alpha}_V + \langle \mathfrak L_{\bs{d\psi}\circ \bs\psi\-} \bs\alpha,  V \rangle,  \\
 \langle  \bs{d\alpha}, V \rangle ^{\bs \psi} &= \langle\bs{d\alpha}, V \rangle + \langle \mathfrak L_{\bs{d\psi}\circ \bs\psi\-} \bs\alpha,  V \rangle. 
\end{aligned}
\end{equation}
As we will see in \ref{Associated bundle of regions and integration theory for local field theory}, the \emph{local} version of the above are relevant to the standard  formulation of  field theory, i.e. its formulation on ($U\subset$) $M$. 
The results as presented in this section are relevant to a programmatic \emph{global} formulation of gRGFT, on $P$. 
As indicated in conclusion \ref{Conclusion},
this part of our program is to be pursued in future work.

\section{The dressing field method}  
\label{The dressing field method}  

The dressing field method (DFM) is a formal but systematic,  algorithm to build basic forms on a bundle. 
In the context of gauge field theory, it allows to construct gauge-invariants, generalising \emph{Dirac variables}.
First developed for  Yang-Mills theories and gauge gravity theories formulated via Cartan geometry, i.e. for internal gauge groups \cite{GaugeInvCompFields, Francois2014, Attard_et_al2017,  Attard-Francois2016_II, Attard-Francois2016_I, Francois2018, Francois2021, Francois-et-al2021, Berghofer-et-al2023}, it is then extended to general-relativistic theories, i.e. for diffeomorphism symmetry in  \cite{Francois2023-a}. 
Its first (field-theoretic) application to supersymmetric field theory (and supergravity) is given in \cite{JTF-Ravera2024susy}.

It can be understood as the geometric framework underlying various notions encountered in recent literature on gauge theories and gravity: notably  that of \emph{edge modes} as introduced in  \cite{DonnellyFreidel2016} and further developed e.g. in \cite{Speranza2018, Chandrasekaran_Speranza2021, Speranza2022, Chandrasekaran-et-al2022}, and that of \emph{gravitational dressings} \cite{Giddings-Donnelly2016, Giddings-Donnelly2016-bis, Giddings-Kinsella2018, Giddings2019, Giddings-Weinberg2020, Witten-et-al2023, Speranza-et-al2023} -- see also  \cite{Kabel-Wieland2022}  -- as well as ``dynamical reference frames"  \cite{carrozza-et-al2022, Hoehn-et-al2022}, or  ``embedding fields"   \cite{Speranza2019, Ciambelli-et-al2022,  Speranza2022, Ciambelli2023}. 
  
In  \cite{Francois2023-a}, dealing with $\Diff(M)$, it was argued that in the most natural (favourable) situations the DFM  renders manifest the \emph{relational} character of general-relativistic physics:
It does so by systematically implementing a notion of \emph{relational observables} for $\Diff(M)$-theories \cite{Tamborino2012}. 
In the following, we unify the internal and $\Diff(M)$ versions of the DFM, treating them both at the same time by considering the extension of the method to the full automorphism group $\Aut(P)$ of a principal bundle.

\subsection{Building basic forms via dressing}  
\label{Building basic forms via dressing}  

Let $Q \rarrow N$ be a reference fiber bundle with base $N$.
The space of $\Aut(P)$-dressing fields is defined as the set of bundle morphisms   s.t.
\begin{align}
\label{Dressing-fields-space}
\D r[Q, P] \defeq \left\{\ u: Q \rarrow P\  |\ u^\psi \defeq \psi\- \circ u, \ \text{ with }\  \psi \in \Aut(P) \right\} \ \subset\  \text{Hom}(Q, P).
\end{align}
Which means that dressing fields are special morphisms in the category of fiber bundles. 
The linearisation of their defining property is then $\delta_X u \defeq  -X \circ u$ for $X \in \aut(P)$.
The dressing map is defined as: 
\begin{equation}
\begin{aligned}
\label{Dressed-field}
|^u : \Phi \ &\rarrow\ \Phi^u,  \\
\phi \ &\mapsto \ \phi^u\defeq  \ u^*\phi.
\end{aligned}
\end{equation}
The object $\phi^u$, called the \emph{dressing} of $\phi$,  is  $\Aut(P)$-invariant: explicitly, $(u^*\phi)^\psi \defeq (u^\psi)^*(\phi^\psi) = (\psi\- \circ u )^*(\psi^*\phi)= u^*\phi$. The space $\Phi^u$ of dressed fields is a subset of fields living on $Q$.

We  define  \emph{field-dependent $\Aut(P)$-dressing fields} as
\begin{equation}
\begin{aligned}
\label{Field-dep-dressing}
\bs u : \Phi \ &\rarrow\ \D r[Q, P] ,  \\
\phi \ &\mapsto \  \bs u(\phi), \qquad \text{ s.t. }\quad  R^\star_\psi \bs u = \psi\- \!\circ \bs u \quad\text{ i.e. } \quad \bs u(\phi^\psi)=\psi\-\! \circ \bs u(\phi). 
\end{aligned}
\end{equation}
Said otherwise, $\bs u$ is an equivariant 0-form on $\Phi$ with value in the representation $\D r[Q, P]$ of $\Aut(P)$. 
Its infinitesimal equivariance, i.e. $\aut(P)$-transformation, is then $\bs L_{X^v} \bs u = -X \circ \bs u$.
Therefore, its $\bs\Diff_v(\Phi)\simeq C^\infty \big(\Phi, \Aut(P) \big)$ and  $\bs\diff_v(\Phi)\simeq C^\infty \big(\Phi, \aut(P) \big)$ transformations are, respectively:
\begin{align}
\label{Vert-trsf-dressing}
\bs {u^\psi}\defeq \Xi^\star \bs u = \bs\psi\- \circ \bs u, \qquad \text{ and }\qquad \bs L_{\bs X^v} \bs u= -\bs X \circ \bs u.
\end{align}
A $\phi$-dependent dressing field induces a map  
\begin{equation}
\begin{aligned}
\label{F-map}
F_{\bs u} : \Phi \ &\rarrow\ \M   , \\
\phi \ &\mapsto \  F_{\bs u}(\phi) \defeq \bs u(\phi)^* \phi \sim [\phi],  \qquad \text{ s.t. }\quad  F_{\bs u} \circ R_\psi = F_{\bs u}. 
\end{aligned}
\end{equation}
It is constant along $\Aut(P)$-orbits: $F_{\bs u}(\phi^\psi) = \bs u(\phi^\psi)^*(\phi^\psi)=\big( \psi\-\! \circ \bs u(\phi) \big)^*\psi^*\phi= \bs u(\phi)^* \phi \rdefeq F_{\bs u}(\phi)$.  
As  $\Aut(P)$-orbits $[\phi] \in \M$ of $\phi \in \Phi$ are represented by fields $\bs u(\phi)^* \phi$ on $Q$, not on $P$, the image of $F_{\bs u}$ can be understood as a ``coordinatisation" of $\M$.
As an $\Aut(P)$-invariant, the dressed field $\phi^{\bs u}$ represents physical d.o.f. in a manifestly \emph{relational} way: the  expression $\bs u(\phi)^* \phi$ is an explicit field-dependent coordinatisation of the physical d.o.f., meaning that physical d.o.f. are defined w.r.t. to each other. 
Dressed fields $\phi^{\bs u}$ are relational Dirac variables, also called  ``complete observables" in \cite{Rovelli2002, Tamborino2012}. 
Applying the DFM then amounts to reformulate a field theory in a manifestly $\Aut(P)$-invariant and relational way.

The map \eqref{F-map} realises the bundle projection, $F_{\bs u} \sim \pi$. 
It therefore allows to build basic forms on $\Phi$, since 
$\Omega^\bullet_{\text{basic}}(\Phi) =$ Im$\,\pi^\star \simeq$ Im$\,F_{\bs u}^\star$.  
To build the  basic counterpart of a form $\bs\alpha=\alpha\big(\!\w^\bullet\! \bs d\phi; \phi \big) \in \Omega^\bullet(\Phi)$, one must first consider its formal analogue on the base space $\b{\bs \alpha}=\alpha\big( \!\w^\bullet\! \bs d[\phi]; [\phi] \big) \in  \Omega^\bullet(\M)$,  then define
\begin{align}
\label{Dressing-form}
\bs \alpha^{\bs u}\defeq F_{\bs u}^\star  \b{\bs \alpha} = \alpha \big(  \!\w^\bullet\! F_{\bs u}^\star \bs d[\phi]; F_{\bs u}(\phi)  \big) \quad \in \ \Omega^\bullet_{\text{basic}}(\Phi).
\end{align}
We call this object the \emph{dressing} of $\bs\alpha$. It is basic by construction, so it is invariant under $\bs\Diff_v(\Phi)\simeq C^\infty \big(\Phi, \Aut(P) \big)$ -- and under $\bs\Aut_v(\Phi)\simeq \bs\Aut(P)$: $ (\bs{\alpha^u})^{\bs \psi} = \bs{\alpha^u}$.

To get a final expression for $\bs \alpha^{\bs u}$,
one needs $F_{\bs u}^\star \bs d[\phi]$, a basis for basic forms, expressed in terms of $\bs d \phi$ and $\bs u$. 
We may find it via the pushforward ${F_{\bs u}}_\star : T_{\phi}\Phi \rarrow T_{F_{\bs u}(\phi)}\M$, $\mathfrak X_{|\phi} \mapsto {F_{\bs u}}_\star \mathfrak X_{|\phi}$. 
For a generic $\mathfrak X \in \Gamma(T\Phi)$ with flow $\vphi_\tau : \Phi \rarrow \Phi$, s.t. $\mathfrak X_{|\phi}  = \tfrac{d}{d\tau} \vphi_\tau(\phi) \big|_{\tau =0} = \mathfrak X(\phi)\tfrac{\delta}{\delta \phi}$, one has $F_{\bs u}^\star \bs d[\phi]_{|F_{\bs u}(\phi)}\big(\mathfrak X_{|\phi} \big)= \bs d[\phi]_{|F_{\bs u}(\phi)} \big( {F_{\bs u}}_\star \mathfrak X_{|\phi} \big)$. So, 
\begin{align*}
 {F_{\bs u}}_\star \mathfrak X_{|\phi}  \defeq  {F_{\bs u}}_\star \tfrac{d}{d\tau}\ \vphi_\tau(\phi)\ \big|_{\tau =0}  =  \tfrac{d}{d\tau}\ {F_{\bs u}} \big( \vphi_\tau(\phi)   \big)\ \big|_{\tau =0}
 							&=  \tfrac{d}{d\tau} \ \bs u\big( \vphi_\tau(\phi)\big)^* \big( \vphi_\tau(\phi)   \big) \ \big|_{\tau =0} \\
							&=  \tfrac{d}{d\tau} \bs u\big( \vphi_\tau(\phi)\big)^* \phi \ \big|_{\tau =0} +  \tfrac{d}{d\tau}\ \bs u(\phi)^* \big( \vphi_\tau(\phi)   \big)\  \big|_{\tau =0}.
 \end{align*}
Inserting $\id_Q =  \bs u(\phi)\- \!\circ \bs u(\phi)$ in the first term: $
 \tfrac{d}{d\tau}\ \bs u(\phi)^* {\bs u(\phi)\-}^{_{\text{\scriptsize	$*$} }}   \bs u\big( \vphi_\tau(\phi)\big)^* \phi \ \big|_{\tau =0} = 
 \bs u(\phi)^*  \tfrac{d}{d\tau}\ \big(    \bs u\big( \vphi_\tau(\phi)\big)   \circ   \bs u(\phi)\-   \big)^*\phi \ \big|_{\tau =0}$.
The term $\bs u\big( \vphi_\tau(\phi)\big)   \circ   \bs u(\phi)\-$ is a curve in $P$, so $ \tfrac{d}{d\tau}\   \bs u\big( \vphi_\tau(\phi)\big)   \circ   \bs u(\phi)\-   \ \big|_{\tau =0} = \bs{du}_{|\phi}(\mathfrak X_{|\phi}) \circ \bs u(\phi)\-  \ \in \Gamma(TP)$. Therefore, 
\begin{align}
\label{1st-term}
\tfrac{d}{d\tau} \bs u\big( \vphi_\tau(\phi)\big)^* \phi \ \big|_{\tau =0} = 
 \bs u(\phi)^*  \tfrac{d}{d\tau}\ \big(    \bs u\big( \vphi_\tau(\phi)\big)   \circ   \bs u(\phi)\-   \big)^*\phi \ \big|_{\tau =0} =   \bs u(\phi)^*  \mathfrak L_{\bs{du}_{|\phi}(\mathfrak X_{|\phi}) \circ \bs u(\phi)\-} \phi. 
\end{align}
The second term is 
$\tfrac{d}{d\tau}\ \bs u(\phi)^* \big( \vphi_\tau(\phi)   \big)\  \big|_{\tau =0} \rdefeq  \tfrac{d}{d\tau}\ F_{\bs u(\phi)} \big( \vphi_\tau(\phi)   \big)\  \big|_{\tau =0}
														 = F_{\bs u(\phi)\, \star} \mathfrak X_{|\phi}$.
As a vector on $\M$, we find its expression as a derivation by applying it to  $\bs g \in C^\infty(\M)$:
 \begin{align*}
\big[F_{\bs u(\phi)\, \star} \mathfrak X\,  \big( \bs g \big)\big]([\phi]) &=  \tfrac{d}{d\tau}\ \bs g \left( F_{\bs u(\phi)} \big( \vphi_\tau(\phi) \big) \right)\  \big|_{\tau =0} \\
			  &=  \left( \tfrac{\delta}{\delta [\phi]} \bs g \right) \big(F_{\bs u(\phi)}(\phi)\big)    \ \underbrace{\tfrac{d}{d\tau}\, F_{\bs u(\phi)}\big( \vphi_\tau(\phi) \big) \, \big|_{\tau=0}}_{\big[\mathfrak X(F_{\bs u(\phi)})\big] (\phi)  }
			  =  \left( \tfrac{\delta}{\delta [\phi]} \bs g \right) \big(\underbrace{\bs u(\phi)^*\phi}_{\sim [\phi]} \big)    \  \underbrace{\big( \tfrac{\delta}{\delta \phi} F_{\bs u(\phi)}\big)(\phi)}_{\tfrac{\delta}{\delta \phi} \, \bs u(\phi)^* \phi\, =\, \bs u(\phi)^*}  \ \mathfrak X(\phi) \\
			  &= \left( \tfrac{\delta}{\delta [\phi]} \bs g \right) \big([\phi]\big) \ \bs u(\phi)^* \mathfrak X(\phi)
			   = \big[ \bs u(\phi)^* \mathfrak X(\phi)  \tfrac{\delta}{\delta [\phi]} (\bs g) \big]([\phi]).
\end{align*}
We have finally
\vspace{-2mm}
\begin{align}
\label{Pushforward-X-dress}
 {F_{\bs u}}_\star \mathfrak X_{|\phi} = \bs u(\phi)^* \left( \mathfrak X(\phi) +  \mathfrak L_{\bs{du}_{|\phi}(\mathfrak X_{|\phi}) \circ \bs u(\phi)\-} \phi  \right) \,  \frac{\delta}{\delta [\phi]}_{|F_{\bs u(\phi)(\phi)}}.\\[-8mm]
 ~ \notag
\end{align}
From which we find, for any $\mathfrak X \in \Gamma(T\Phi)$:
\begin{align*}
F_{\bs u}^\star \bs d[\phi]_{|F_{\bs u(\phi)(\phi)}}\, \big(\mathfrak X_{|\phi} \big) &= \bs d[\phi]_{|F_{\bs u(\phi)(\phi)}} \, \big( {F_{\bs u}}_\star \mathfrak X_{|\phi} \big)
						= \bs d[\phi]_{|F_{\bs u(\phi)(\phi)}}\,  \left(  \bs u(\phi)^* \left( \mathfrak X(\phi) +  \mathfrak L_{\bs{du}_{|\phi}(\mathfrak X_{|\phi}) \circ \bs u(\phi)\-} \phi  \right) \,  \tfrac{\delta}{\delta [\phi]}_{|F_{\bs u(\phi)(\phi)}} \right) \\
						&=  \bs u(\phi)^* \left( \mathfrak X(\phi) +  \mathfrak L_{\bs{du}_{|\phi}(\mathfrak X_{|\phi}) \circ \bs u(\phi)\-} \phi  \right) 
						=  \bs u(\phi)^* \left( \bs d\phi_{|\phi}\, \big(\mathfrak X_{|\phi} \big) +  \mathfrak L_{\bs{du}_{|\phi}(\mathfrak X_{|\phi}) \circ \bs u(\phi)\-} \phi  \right) \\
						&=\left(   \bs u(\phi)^* \big( \bs d\phi_{|\phi}  +  \mathfrak L_{\bs{du}_{|\phi} \circ \bs u(\phi)\-} \phi  \big)   \right) (\mathfrak X_{|\phi}).
\end{align*}
We get the dressing of $\bs d\phi$, the basic 1-form basis:
\begin{align}
\label{Dressing-basis-1-form}
\bs d\phi^{\bs u}\defeq F_{\bs u}^\star \bs d [\phi] =  \bs u^* \big( \bs d\phi  +  \mathfrak L_{\bs{du} \circ \bs u\-} \phi \big) \quad \in \ \Omega^1_{\text{basic}}(\Phi).
\end{align}
Inserted into \eqref{Dressing-form}, this yields the dressing of $\bs\alpha$:
\begin{align}
\label{Dressing-form-bis}
\bs \alpha^{\bs u} =  \alpha \big(  \!\w^\bullet\! \bs d\phi^{\bs u};\, \phi^{\bs u}  \big) \ \  \in \ \Omega^\bullet_{\text{basic}}(\Phi).
\end{align}

On account of the formal similarity between $\Xi(\phi)=\bs\psi(\phi)^*\phi$ and $F_{\bs u}(\phi)=\bs u(\phi)^*\phi$,  and $\bs d\phi^{\bs \psi}$  \eqref{GT-basis-1-form} and $\bs d\phi^{\bs u}$ \eqref{Dressing-basis-1-form},  resulting into the close formal expressions of $\bs\alpha^{\bs \psi}$  \eqref{GT-general} and $\bs\alpha^{\bs u}$ \eqref{Dressing-form}-\eqref{Dressing-form-bis}, the following rule of thumb to obtain the dressing of any form $\bs\alpha$ holds: 
First compute the  $\bs\Diff_v(\Phi)\simeq C^\infty\big(\Phi, \Aut(P) \big)$ transformation $\bs\alpha^{\bs \psi}$, then substitute $\bs\psi \rarrow \bs u$ in the resulting expression to obtain $\bs\alpha^{\bs u}$. 
This rule may be used be systematically. 
\medskip

For example, for $\bs\alpha \in \Omega^\bullet_\text{tens}(\Phi, \rho)$ the rule gives us  $\bs\alpha^{\bs u} =\rho(\bs u)\-\bs\alpha$. 
 For an Ehresmann connection, in view of  \eqref{Vert-trsf-connection}, the rule ensures that $\bs\omega^{\bs u} = \bs u\-_* \bs \omega \circ \bs u + \bs u\-_* \bs{du} \in \Omega^1_\text{basic}(\Phi)$. These two results allows to write a lemma analogous to \eqref{Formula-general-rep}-\eqref{Formula-pullback-rep}: 
For $\bs\alpha, \bs{D\alpha} \in \Omega^\bullet_\text{tens}(\Phi, \rho)$, 
we have  $\bs d\, F_{\bs u}^\star \b{\bs\alpha} = \bs d\big( \rho(\bs u)\- \bs\alpha \big)$. On the other hand, since $F_{\bs u}^\star \b{\bs D} \b{\bs\alpha} = \bs d \bs\alpha^{\bs u}+ \rho_*(\bs\omega^{\bs u})\bs\alpha^{\bs u} = \rho(\bs u)\- \bs{D\alpha}$, we have 
\begin{align*}
F_{\bs u}^\star \bs d \b{\bs\alpha} &= \rho(\bs u)\- \bs{D\alpha}  -  \rho_*(\bs\omega^{\bs u})\bs\alpha^{\bs u}
				   =  \rho(\bs u)\- \bs{d\alpha} +  \rho(\bs u)\-\,  \rho_*(\bs\omega) \bs\alpha  -  \rho_*(\bs\omega^{\bs u})\bs\alpha^{\bs u}
				   = \rho(\bs u)\- \bs{d\alpha}  -  \rho_*(\bs u\-_* \bs d\bs u) \rho(\bs u)\-\bs\alpha\\
				 &= \rho(\bs u)\- \left(  \bs{d\alpha} -  \rho_*( \bs d\bs u \circ \bs u\-) \bs\alpha \right). 
\end{align*}
By $[F_{\bs u}^\star , \bs d]  =0$ (remark that the exterior derivatives belong to different spaces) we  obtain:
\begin{align}
\label{Formula-general-rep-dress}
\bs d\big( \rho(\bs u)\- \bs\alpha \big) = \rho(\bs u)\- \left(  \bs{d\alpha} -  \rho_*( \bs d\bs u \circ \bs u\-)\, \bs\alpha \right).
\end{align}
For the pullback representation $\rho(u)\-\!= u^*$, this specialises to
  \begin{align}
\label{Formula-pullback-rep-dress}
\bs d\big( \bs u^* \bs\alpha \big) = \bs u^* \left(  \bs{d\alpha} + \mathfrak L_{\bs d\bs u \circ \bs u\-} \bs\alpha \right). 
\end{align}
This identity, generalising the one appearing e.g. in \cite{DonnellyFreidel2016, Speranza2018, Kabel-Wieland2022}, must not be conflated  with  \eqref{Dressing-basis-1-form}. Despite their formal similarity, these two results have distinct geometric origin.

\subsubsection{Dressing field and flat  connections}  
\label{Dressing field and flat  connections}  

$\bullet$ A field-dependent dressing field $\bs u$, induces a \emph{flat Ehresmann connection} $\bs\omega_{\text{\tiny$0$}} \defeq -\bs{du} \circ\bs u\- \in \Omega^1(\Phi)$.  
By \eqref{Vert-trsf-dressing} we have 
\begin{align}
\label{Dress-connection-vert-prop}
{\omegaf}_{|\phi}(X^v_{|\phi})=  -\bs{du}_{|\phi}(X^v_{|\phi}) \circ \bs u(\phi)\- = -\bs L_{X^v} \bs u \circ  \bs u(\phi)\- = X \circ \bs u(\phi) \circ  \bs u(\phi)\- = X \ \in \aut(P),
\end{align}
and by \eqref{Field-dep-dressing}, using the naturality of $\bs d$, we get
\begin{equation}
\begin{aligned}
\label{Dress-connection-equiv-prop}
R^\star_\psi \omegaf &=  -R^\star_\psi \bs{du}  \circ (R^\star_\psi \bs u)\- =  - \bs d (R^\star_\psi \bs u) \circ  (R^\star_\psi \bs u)\- = - \bs d (\psi\- \circ \bs u) \circ (\psi\- \circ \bs u)\-
				= \psi\-_* \bs{du} \circ \bs u\- \circ \psi   \\
				&=  \psi\-_*\,\omegaf\!\circ \psi. 
\end{aligned}
\end{equation}
These are indeed the defining properties \eqref{Variational-connection} of an Ehresmann connection, and  
$\Omega_{\text{\tiny$0$}}=\bs d\omegaf + \tfrac{1}{2} [\omegaf, \omegaf]_{\text{{\tiny $\aut(P)$}}} \equiv0$ is immediate.
It follows that, as a special case of \eqref{Vert-trsf-connection} and  \eqref{Inf-Vert-trsf-connection},
\begin{align}
\label{Vert-trsf-flat-connection}
\omegaf^{\bs \psi}= \bs\psi\-_*\, \omegaf \circ \bs\psi + \bs\psi\-_*\bs{d\psi}, \qquad \text{ and } \qquad
\bs L_{\bs X^v} \omegaf =\bs{dX}+ [\omegaf, \bs X]_{\text{{\tiny $\aut(P)$}}\,}.
\end{align}
Thus, \eqref{Dressing-basis-1-form} can also be written as $
\bs d\phi^{\bs u}=  \bs u^* \big( \bs d\phi  -  \mathfrak L_{  \omegaf} \phi \big) \  \in \ \Omega^1_{\text{basic}}(\Phi)$.

The existence of a flat connection is a strong topological constraint on a bundle, then so is that of a global dressing field. A global dressing field may indeed give a global trivialisation of the bundle $\Phi$. However the field space may not be trivial   in general, and Gribov-like obstructions may exclude the existence  of  global dressing fields.  Local dressing fields are always compatible with the local triviality of $\Phi$.
\medskip

\noindent
$\bullet$ A dressing field may also induce a \emph{flat twisted connection} $\bs\varpi_{\text{\tiny$0$}} \defeq -\bs d C(\bs{u}) \cdot C(\bs u)\-$.
To see this, one only has to assume that the 1-cocycle \eqref{cocycle} $C: \Phi \times \Aut(P) \rarrow G$ controlling the equivariance \eqref{C-eq-form} of twisted forms has a functional expression that can be  extended to $C: \Phi \times \text{Hom}(Q, P) \rarrow G$. Then, we have the well-defined  map 
\begin{equation}
\begin{aligned}
\label{twisted-dressing}
C(\bs u): \Phi &\rarrow G,  \\
		\phi &\mapsto [C(\bs u)](\phi)\defeq C\big(\phi; \bs u(\phi) \big).
\end{aligned}
\end{equation}
By the  cocycle property \eqref{cocycle}, it is a twisted equivariant 0-form:\footnote{Remark that it is then a dressing field for twisted forms.}
\begin{equation}
\begin{aligned}
\label{eq-twisted-dressing}
\big[ R^\star_\psi C(\bs u) \big](\phi) &= \big[ C(\bs u) \circ R_\psi \big] (\phi) \defeq C\big(\phi^\psi; \bs u(\phi^\psi) \big) = C\big(\phi^\psi;  \psi\- \circ \bs u(\phi) \big)
						       = C\big(\phi^\psi;  \psi\-  \big) \cdot C\big(\phi;  \bs u(\phi) \big)  \\
						     &= C\big(\phi; \psi \big)\-\! \cdot C\big(\phi; \bs u (\phi)\big) \\
						     & \rdefeq \big[C\big( \ \,;  \psi\big)\-\!\cdot C(\bs u) \big] (\phi).
\end{aligned}
\end{equation}
Its infinitesimal equivariance is  $\bs L_{X^v} C(\bs u) = \iota_{X^v} \bs d C(\bs u) = -a(X; \ \,) \cdot C(\bs u)$. 
From these, one easily checks that  $\bs\varpi_{\text{\tiny$0$}}$ satisfies the defining properties \eqref{Variational-twisted-connection} of a twisted connection.
As a special case of \eqref{Vert-trsf-twisted-connection}-\eqref{Inf-Vert-trsf-twisted-connection}, we have
\begin{align}
\label{Vert-trsf-flat-connection1}
\bs\varpi_{\text{\tiny$0$}}^{\bs\psi } 
=\Ad\big( C(\bs\psi)\- \big) \, \bs\varpi_{\text{\tiny$0$}} + C(\bs\psi)\-\bs{d} C(\bs\psi), 
 \qquad \text{ and } \qquad
\bs L_{\bs X^v}\bs\varpi_{\text{\tiny$0$}} =\bs{d} a(\bs X)+ [\bs\varpi_{\text{\tiny$0$}}, a(\bs X)]_{\text{{\tiny $\LieG$}}}. 
\end{align}
As we will see in the next section, twisted dressings $C(\bs u)$ and associated $\bs\varpi_{\text{\tiny$0$}}$ underly the construction of Wess-Zumino counterterms in field theory.

\subsection{Composition of dressing operations}  
\label{Composition of dressing operations}  

 As stated in introduction, the dressing field method has been first developed for ``internal symmetry groups" -- for either Yang-Mills type theory, or for gauge theories of gravity based on Cartan geometry -- i.e. for $\Aut_v(P)$.
It has been developed for general-relativistic theories, i.e. for $\Diff(M)$, in \cite{Francois2023-a}.
To indicate how to recover both cases from the above construction, 
we may rely on the decomposition \eqref{Morph-graph-2} in Appendix \ref{Semi-direct product structure of Aut(P)}  of bundle morphisms: an $\Aut(P)$-dressing can be written as the composition 
$\bs u  = f_\u \circ \b{\bs \upupsilon}$, as in the following diagram
\begin{equation}
\label{Dressing-comp-diag}
\begin{tikzcd}
Q 
\arrow[bend left,"\bs u"]{rr} \arrow[r,swap,"\b{\bs \upupsilon}"] \arrow[d] & 
 P \arrow[r,swap,"f_\u"] \arrow[d] &
P \arrow[d]
\\
N \arrow[r,swap,"\bs \upupsilon"] & M \arrow[equal]{r} & M 
\end{tikzcd}
\end{equation}
where 
$\b{\bs \upupsilon} : \Phi \rarrow   \D r[Q, P]$ is  covering $\bs\upupsilon:  \Phi \rarrow   \D r[N, M]$, $\phi \mapsto \big(\, \bs\upupsilon[\phi]:  N \rarrow M \big)$ which is a $\Diff(M)$-dressing as defined in \cite{Francois2023-a}. 
The  map 
$f_\u : \Phi \rarrow   \D r[P, P]$, $\phi \mapsto \big( \, f_{\u[\phi]}: P\rarrow P \, \big)$, covers  $\id_M$.
The latter is thus, omitting the $\phi$-dependence for notational simplicity, of the form $f_{\u}(p)\defeq p \u(p)$, with $\u : P\rarrow H$. 
By definition of an $\Aut_v(P)$-dressing field, it satisfies  $f_\u^{\,\psi} = \psi\- \circ f_\u$, for $\psi\in \Aut_v(P)$, so that
\begin{align}
f^{\,\psi}_\u(p)\defeq\psi\- \circ f_\u(p)= \psi\-\big(p \u(p) \big)= \psi\-(p)\u (p) = p\gamma(p)\-\u (p),  
\end{align}
where $\gamma\- \in \H$ is the gauge group element corresponding to $\psi\-\in \Aut_v(P)$. 
One has thus that $f^{\,\psi}_\u=f_{\u^\gamma} = f_{\gamma\-\u}$, i.e. one finds that the generating element $\u$ of the $\Aut_v(P)$-dressing $f_\u$ is  
\begin{align}
    \u: P\rarrow H, \quad \text{s.t.}\quad \u^\gamma =\gamma\- \u.  
\end{align}
This is the bundle version of the original field-theoretic definition of an $\H$-dressing field as given e.g. in \cite{GaugeInvCompFields, Francois2018}.\footnote{ 
One may define a dressing for a subgroup of $\Aut_v(P)$, leading to a partial symmetry reduction, to which corresponds a $\K$-dressing field $\u: P\rarrow K\subset H$ s.t. $\u^\kappa=\kappa^{-1} \u$, for $\kappa \in\K \subset \H$. See e.g. \cite{Francois2021, Berghofer-et-al2023} for details.}

The decomposition $\bs u  = f_\u \circ \b{\bs \upupsilon}$ is suggested by the semi-direct product structure of $\Aut(P)=\overline{\Diff}\ltimes \Aut_v(P)$, discussed in Appendix \ref{Semi-direct product structure of Aut(P)}. 
As illustrated by the diagram \eqref{Dressing-comp-diag}, it shows that one may perform dressings in a stepwise manner: first dressing for $\Aut_v(P)$ via $f_{\bs\u}$ and then for $\overline{\Diff}$ via $\b{\bs\upupsilon}$. 
This  works if, again as suggested by the semi-direct structure of $\Aut(P)$, the two dressing maps  $f_{\bs \u}$ and $\b{\bs\upupsilon}$ satisfy, in addition to their  defining relations (left column below), the compatibility conditions (right column) under 
$\psi=\big(\b\uppsi, \eta \big) \in \Aut(P)=\overline{\Diff}\ltimes \Aut_v(P)$:
\begin{multicols}{2}
\noindent
\begin{align}
f_{\bs \u}^{\,\eta}& \defeq \eta\- \circ f_{\bs \u} , \label{Aut_v-dress} \\
\b{\bs\upupsilon}^{\,\b \uppsi}& \defeq \b\uppsi\- \circ \b{\bs \upupsilon} , \label{bDiff-dress}
\end{align}
\columnbreak
\begin{align}
f_{\bs \u}^{\,\b\uppsi}& \defeq \b\uppsi\- \circ f_{\bs \u} \circ \b\uppsi ,  \label{CompCond1} \\
\b{\bs\upupsilon}^{\,\eta}&\defeq \b{\bs\upupsilon} . \label{CompCond2}
\end{align}
\end{multicols}
\vspace{-8mm}
\noindent
The condition \eqref{CompCond1} ensures that the first dressed fields $f^*_{\bs \u}\phi$ retains well-defined $\overline{\Diff}$-transformations, that can be dressed for via \eqref{bDiff-dress}.
While condition \eqref{CompCond2} ensures that upon dressing by $\b{\bs\upupsilon}$ the $\Aut_v(P)$-invariance is preserved.
From these, one checks explicitly the invariance of the dressed fields $\phi^{\,\bs u}\defeq {\bs u}^* \phi = \b{\bs\upupsilon}^* f^*_{\bs\u} \phi$:
\begin{equation}
\begin{aligned}
( \phi^{\,\bs u})^{\,\eta} & = \big( \b{\bs\upupsilon}^* f^*_{\bs\u} \phi \big)^{\,\eta}\hspace{-4mm} &&=  ({\b{\bs\upupsilon}^{\,\eta}})^* (f^*_{\bs \u}\phi )^{\,\eta} = \b{\bs\upupsilon} f^*_{\bs \u}\phi = \phi^{\, \bs u} , \\
( \phi^{\,\bs u})^{\,\b\uppsi} & = ({\b{\bs\upupsilon}}^* f^*_{\bs \u}\phi )^{\,\b\uppsi} \hspace{-4mm}&& = ({\b{\bs\upupsilon}}^{\,\b\uppsi})^* (f_{\bs \u}^{\,\b\uppsi})^* \phi^{\,\b\uppsi} \\
& &&= (\b\uppsi\- \circ \b{\bs\upupsilon})^* (\b\uppsi\- \circ f_{\bs \u} \circ \b\uppsi)^* \b\uppsi^* \phi =  {\b{\bs\upupsilon}}^* ({\b\uppsi\-})^* \b\uppsi^* f^*_{\bs \u} ({\b\uppsi\-})^* \b\uppsi^* \phi = \phi^{\, \bs u} ,
\end{aligned}
\end{equation}
by \eqref{Aut_v-dress} and \eqref{CompCond2} in the first line, and by \eqref{bDiff-dress} and \eqref{CompCond1} in the second line. From this immediately follows that $[(\phi^{\,\bs u})^{\,\eta}]^{\,\b\uppsi}=[(\phi^{\,\bs u})^{\,\b\uppsi}]^{\,\eta}$.
The consecutive dressing operations can be summarised via the following sequence:
\begin{equation}
\label{stepwise-reduction}
\Big(\Aut(P)=\overline{\Diff}\ltimes \Aut_v(P)\ ;\ \phi \ \Big) \quad \xlongrightarrow{f_{\bs \u}} \quad \Big(\ \overline{\Diff}\ ;\ f^*_{\bs \u}\phi \ \Big) \quad \xlongrightarrow{{\b{\bs \upupsilon}}} \quad \Big( \ \emptyset\ ;\ {\b{\bs \upupsilon}}^* f^*_{\bs \u} \phi = (f_{\bs \u} \circ {\b{\bs \upupsilon}})^* \phi =: \u^* \phi = \phi^{\,\u} \ \Big).
\end{equation}
A similar pattern of stepwise dressings may also appear for $\Aut_v(P)$ if it has a semi-direct structure. Then dressing compatibility conditions such as \eqref{CompCond1}-\eqref{CompCond2} must hold, for the same reasons.
Such is notably the case in conformal Cartan geometry, where the DFM can be used to build conformal tractors and twistors \cite{Attard-Francois2016_I, Attard-Francois2016_II}, and applied to conformal gravity \cite{AFL2016_I}.
One may then specialise the results of section \ref{Building basic forms via dressing} to the distinct cases of $\Aut_v(P)$ and $\overline{\Diff}$ dressings.
As we shall see in section \ref{Local field theory}, this is especially relevant when dealing with (local) field theory, where the semi-direct structure of the local version of  $\Aut(P)$ is   unavoidable.

\subsection{Residual symmetries}  
\label{Residual symmetries}  

We ought to discuss two kinds of \emph{residual} symmetries in the DFM. First, the genuine residual symmetry coming from the elimination of a subgroup of the original symmetry group. Secondly, a new symmetry replacing  the eliminated one, parametrising the choice of dressing fields. Both may  be present  simultaneously, yet in the following we discuss them in turn for the sake of clarity.  

\subsubsection{Residual symmetries of the first kind}  
\label{Residual symmetries of the first kind}  

Consider a dressing field $ \ \bs u : \Phi \ \rarrow\ \D r[Q, P]\ $ whose defining equivariance is controlled by a subgroup  $\Aut_{\text{\tiny$0$}}(P)$:
\begin{align}
\label{equiv-partial-dressing}
 R^\star_\vphi \bs u = \vphi\- \circ \bs u, \quad \text{for }\  \vphi \in \Aut_{\text{\tiny$0$}}(P)\subset \Aut(P).
\end{align}   
Applying the DFM is then equivalent to building a bundle projection $\Phi \rarrow \Phi/ \Aut_{\text{\tiny$0$}}(P)\rdefeq \Phi^{\bs u}$. 
For $\Phi^{\bs u}$ to be principal (sub)bundle in its own right, the quotient  $\Aut(P)/ \Aut_{\text{\tiny$0$}}(P)$ must be a group, so $\Aut_{\text{\tiny$0$}}(P)$ has to be a \emph{normal} subgroup of $\Aut(P)$: $\Aut_{\text{\tiny$0$}}(P) \triangleleft \Aut(P)$. 
We assume  that it is and denote $\Aut_{\text{\scriptsize r}}(P) \defeq \Aut(P)/ \Aut_{\text{\tiny$0$}}(P)$ the residual structure group of $\Phi^{\bs u}$. 

For instance, one may consider the subgroup of automorphisms $\Aut_{\text{\tiny$0$}}(P)= \Aut_{\text{\tiny c}}(P)$ whose domains $P_{|D} \subset P$ have $\pi$-projections $D\subset M$ that are compact subsets of $M$.\footnote{In case the structure group $H$ of $P$ is compact, their domains are compact and these are indeed compactly supported automorphims.}
Another noteworthy example, as seen in the previous section, is the subgroup of vertical automorphisms $\Aut_{\text{\tiny$0$}}(P)= \Aut_v(P)$, in which case $\Aut_{\text{\scriptsize r}}(P) =\overline{\Diff} \simeq \Diff(M)$. These are relevant examples for local field theory applications, see section \ref{Local field theory}.
\medskip

Dressed objects $\bs\alpha^{\bs u}$ are then $\Aut_{\text{\tiny$0$}}(P)$-basic on $\Phi$, i.e. by construction invariant under  $C^\infty\big(\Phi, \Aut_{\text{\tiny 0 }}(P)\big)\subset C^\infty\big(\Phi, \Aut(P)\big) \simeq \bs\Diff_v(\Phi)$, so  in particular invariant under $\bs\Aut_{\text{\tiny$0$}}(P)\subset \bs\Aut(P)$. 
One expects them to exhibit \emph{residual transformations} under $C^\infty\big(\Phi, \Aut_{\text{\scriptsize r}}(P)\big)\subset C^\infty\big(\Phi, \Aut(P)\big)$. 
The transformation of $\bs\alpha$ under $C^\infty\big(\Phi, \Aut_{\text{\scriptsize r}}(P)\big)$ is known, so determining that of $\bs\alpha^{\bs u}$ boils down to finding the residual transformation of the dressing field $\bs u$ under $C^\infty\big(\Phi, \Aut_{\text{\scriptsize r}}(P)\big)$, which is given by its equivariance $R^\star_\psi \bs u$  for $\psi \in \Aut_{\text{\scriptsize r}}(P)$.  

We illustrate this in the following two propositions dealing with two interesting cases that can be systematically treated. 
Let us consider $\bs\alpha \in \Omega^\bullet_{\text{tens}}\big(\Phi, \rho \big)$ and $\bs \omega \in \C$, whose dressing by $\bs u$ as in \eqref{equiv-partial-dressing} above are $\bs\alpha^{\bs u}$ and $\bs\omega^{\bs u}$, both $C^\infty\big(\Phi, \Aut_{\text{\tiny 0 }}(P)\big)$-invariant.

\begin{prop}
\label{prop1}
If $ \ \bs u : \Phi \ \rarrow\ \D r[Q, P]$, with $Q\subseteq P$, is s.t. 
\begin{align}
\label{Resid-eq1}
R^\star_\psi \bs u = \psi\-\! \circ \bs u \circ \psi, \quad \text{ for } \ \psi \in \Aut_{\text{\scriptsize r}}(P),
\end{align}
then $\bs\alpha^{\bs u} \in \Omega_{\text{tens}}\big(\Phi, \rho \big)$ and $\bs \omega^{\bs u} \in \C$. 
Then, their residual $C^\infty\big(\Phi; \Aut_{\text{\scriptsize r}}(P)\big)$-transformations are
\begin{align}
\label{Residual-GT1}
(\bs\alpha^{\bs u})^{\bs\psi} =\rho(\bs\psi\-) \bs\alpha^{\bs u}, \qquad \text{ and } \qquad (\bs\omega^{\bs u})^{\bs\psi} =\bs\psi\-_* \, \bs\omega^{\bs u}  \circ \bs\psi + \bs\psi\-_*\bs{d\psi}. 
\end{align}
\end{prop}
\noindent Indeed, when $\bs\alpha$ is horizontal so is $\bs\alpha^{\bs u}=\rho(\bs u)\-\bs\alpha$, and  $R^\star_\psi\bs\alpha^{\bs u} = \rho\big(R^\star_\psi \bs u \big)\- R^\star_\psi \bs\alpha = \rho\big(\psi\-\! \circ \bs u \circ \psi \big) \- \rho(\psi)\-\bs\alpha = \rho(\psi\-) \bs\alpha^{\bs u}$. 
Also, given the linearisation of \eqref{Resid-eq1},
\begin{align}
\label{Resid-eq1-linear}
\bs L_{X^v} \bs u = \iota_{X^v}\bs{du}= \tfrac{d}{d\tau}\, R^\star_{\psi_\tau} \bs u\,\big|_{\tau=0} = -X \circ \bs u + \bs u_* X, \quad \text{ for } \ X\defeq \tfrac{d}{d\tau} \psi_\tau\,\big|_{\tau=0} \in \aut_{\text{\scriptsize r}}(P),
\end{align}
 one finds that $\bs\omega^{\bs u}(X^v)= \bs u\-_* \bs\omega(X^v) \circ \bs u + \bs u\-_*\bs d \bs u (X^v)= \bs u\-_* X \circ \bs u +  \bs u\-_*\big(-X \circ \bs u + \bs u_* X \big) =X$.
Then, by \eqref{Resid-eq1} and  \eqref{Variational-connection}, we have
$R^\star_\psi \bs\omega^{\bs u} = \psi\-_*\, \bs\omega^{\bs u} \circ \psi$, for $\psi \in \Aut_{\text{\scriptsize r}}(P)$. 
The dressed connection $\bs\omega^{\bs u}$ satisfies the defining properties \eqref{Variational-connection} of an $\Aut_{\text{\scriptsize r}}(P)$-principal connection. 
Thus, \eqref{Residual-GT1}  follows from \eqref{GT-tensorial} and \eqref{Vert-trsf-connection}. 
Remark that $\bs\omega^{\bs u}$, being $\aut(P)$-valued, splits as an $\aut_{\text{\scriptsize r}}(P)$-connection and an 
$\aut_{\text{\tiny 0}}(P)$-valued tensorial 1-form. 
In particular, from either results in \eqref{Residual-GT1}, one derives that $\bs\Omega^{\bs u} \in \Omega^2_{\text{tens}}\big(\Phi, \aut(P) \big)$, so $(\bs\Omega^{\bs u})^{\bs\psi} =\bs\psi\-_*\, \bs\Omega^{\bs u} \circ \bs\psi$. 

\begin{prop}
\label{prop2}
If  $\, \bs u : \Phi \ \rarrow\ \D r[Q, P]\, $ is s.t. 
\begin{align}
\label{Resid-eq2}
R^\star_\psi \bs u = \psi\-\! \circ \bs u \circ C(\ \, ;\psi), \quad \text{ for } \ \psi \in \Aut_{\text{\scriptsize r}}(P), 
\end{align}
where $C:\Phi \times \Aut_{\text{\scriptsize r}}(P) \rarrow \Aut(Q)$ is a special case of  1-cocycle  \eqref{cocycle},  satisfying 
$C(\phi; \psi'\circ \psi) =C(\phi; \psi') \circ C(\phi^{\psi'}; \psi)$,
then $\bs\alpha^{\bs u} \in \Omega^\bullet_{\text{tens}}\big(\Phi, C \big)$ and $\bs \omega^{\bs u} \in \b\C$. 
Their residual $C^\infty\big(\Phi, \Aut_{\text{\scriptsize r}}(P)\big)$-transformations are then
\begin{align}
\label{Residual-GT2}
(\bs\alpha^{\bs u})^{\bs\psi} =\rho\big( C(\bs\psi)\big)\- \bs\alpha^{\bs u}, \qquad \text{ and } \qquad (\bs\omega^{\bs u})^{\bs\psi} =C(\bs\psi)\-_* \bs\omega^{\bs u}  \circ C(\bs\psi) + C(\bs\psi)\-_*\bs d C(\bs\psi). 
\end{align}
\end{prop}
As above, if $\bs\alpha$ is horizontal, so is $\bs\alpha^{\bs u}=\rho(\bs u)\-\bs\alpha$, and $R^\star_\psi\bs\alpha^{\bs u} = \rho\big(R^\star_\psi \bs u \big)\- R^\star_\psi \bs\alpha = \rho\big(\psi\-\! \circ \bs u \circ C(\ \,;\psi) \big) \- \rho(\psi)\-\bs\alpha = \rho\big(C(\ \,;\psi) \big)\- \bs\alpha^{\bs u}$. For $\rho$ the pullback action, $\rho(\psi)\-=\psi^*$, we have  
$R^\star_\psi \bs\alpha^{\bs u}  =C(\ \,;\psi)^* \bs\alpha^{\bs u}$ and 
$(\bs\alpha^{\bs u})^{\bs\psi} = C(\bs\psi)\big)^* \bs\alpha^{\bs u}$.
The~linear version of \eqref{Resid-eq2} reads
\begin{align}
\label{Resid-eq2-linear}
\bs L_{X^v} \bs u= \iota_{X^v}\bs{du}= \tfrac{d}{d\tau}\, R^\star_{\psi_\tau} \bs u\,\big|_{\tau=0} = -X \circ \bs u + \bs u_*\, \tfrac{d}{d\tau} C(\ \,;\psi_\tau)\,\big|_{\tau=0}= -X \circ \bs u + \bs u_* \, a(X;\ \,), \quad \text{ for } \ X  \in \aut_{\text{\scriptsize r}}(P).
\end{align}
 So one has 
 $\bs\omega^{\bs u}(X^v)= \bs u\-_* \bs\omega(X^v) \circ \bs u + \bs u\-_*\bs d \bs u (X^v)= \bs u\-_* X \circ \bs u +  \bs u\-_*\big(-X \circ \bs u + \bs u_* \, a(X;\ \,) \big) =a(X;\ \, )\ \in \aut(Q)$.
Also, given \eqref{Resid-eq2} and  \eqref{Variational-connection}, it is easily shown that: 
$R^\star_\psi \bs\omega^{\bs u} = C(\ \, ;\psi)\-_*\, \bs\omega^{\bs u} \circ C(\ \, ;\psi) + C(\ \, ;\psi)\-_*\bs d C(\ \, ;\psi)$, for $\psi \in \Aut_{\text{\scriptsize r}}(P)$. 
The dressed connection $\bs\omega^{\bs u}$ satisfies the defining properties \eqref{Variational-twisted-connection}-\eqref{Special-eq-twisted} of a $\Aut_{\text{\scriptsize r}}(P)$-twisted connection. 
The~transformations \eqref{Residual-GT2} follow from \eqref{GT-tensorial} and \eqref{Vert-trsf-twisted-connection}. 
From either follows that $\bs\Omega^{\bs u} \in \Omega^2_{\text{tens}}\big(\Phi, C \big)$, so $(\bs\Omega^{\bs u})^{\bs\psi} =C(\bs\psi)\-_*\, \bs\Omega^{\bs u} \circ C(\bs\psi)$.

\subsubsection{Residual symmetries of the second kind}  
\label{Residual symmetries of the second kind}  

The defining equivariance $R^\star_\psi \bs u=\psi\- \circ \bs u$ of a dressing field $\, \bs u : \Phi \ \rarrow\ \D r[Q, P]\, $ means that, as a  bundle map $\bs u(\phi): Q \rarrow P$,  $\Aut(P)$ acts on its target space. But then a priori there is a natural right action of $\Aut(Q)$ on its source space, which we write: 
\begin{equation}
\begin{aligned}
\D r[Q, P] \times \Aut(Q) &\rarrow \D r[Q, P] \\
(\bs u, \vphi) &\mapsto \vphi^*\bs u= \bs u \circ \vphi \rdefeq \bs u^\vphi.  \label{dressing-ambiguity}
\end{aligned}
\end{equation}
One way to interpret this, is that two candidates dressing fields $\bs u'$ and $\bs u$ may a priori be related by an element $\vphi \in \Aut(Q)$: $\bs u '=\bs u \circ \vphi = \bs u^\vphi$. 
The group $\Aut(Q)$ does not act on $\Phi$, so we write $\phi^\vphi=\phi$, but it acts on $\phi^{\bs u} = F_{\bs u}(\phi)\defeq \bs u^* \phi$ as a field on $Q$:
\begin{align}
\phi^{\bs u} \mapsto   \phi^{\bs u^\vphi} \defeq (\bs u^\vphi)^*\phi = ( \bs u \circ \vphi)^*\phi =  \vphi^* ({\bs u}^*\phi) =  \vphi^* \phi^{\bs u}. 
\end{align}
The space $\Phi^{\bs u}$ of dressed fields is thus fibered by the right action of $\Aut(Q)$:
\begin{equation}
\label{dressing-ambiguity-bis}
\begin{aligned}
  \Phi^{\bs u} \times \Aut(Q) &\rarrow \Phi^{\bs u},  \\
               (\phi^{\bs u}, \vphi) &\mapsto R_\vphi \phi^{\bs u}:= \vphi ^*\phi^{\bs u}.
\end{aligned}
\end{equation}
So, in analogy with the original field space $\Phi$, the space of dressed fields 
is a principal bundle $\Phi^{\bs u} \xrightarrow{\pi}   \Phi^{\bs u}/\Aut(Q) \rdefeq \M^{\bs u}$.\footnote{Provided that the action of $\Aut(Q)$ is free and transitive, which requires in general some restriction on either $\Aut(Q)$ or $\Phi^{\bs u}$. See \cite{Singer1978, Singer1981} in the case of the action of a gauge group $\H \simeq \Aut_v(P)$ on the connection space $\Phi=\C$ of $P$. } 
We have the SES of groups  
 \begin{align}
 \label{SESgroup2}
\bs\Aut(Q) \simeq \bs{\Aut}_v(\Phi^{\bs u})  \rarrow   \bs{\Aut}(\Phi^{\bs u}) \rarrow \bs{\Diff}(\M^{\bs u}),
 \end{align}
where $\bs\Aut(Q) := \big\{ \bs\vphi : \Phi^{\bs u} \rarrow   \Aut(Q) \, |\,  R^\star_\vphi \bs\vphi  =\vphi^{-1} \circ \bs\vphi \circ \vphi   \big\} $
is the gauge group of $\Phi^{\bs u}$. The latter is a subgroup of $C^\infty\big( \Phi^{\bs u}, \Aut(Q)\big)  \simeq \bs\Diff_v(\Phi^{\bs u})$, acting on $\Gamma(T\Phi^{\bs u})$ and $\Omega^\bullet(\Phi^{\bs u})$ as previously described. 
In particular,  in exact analogy with \eqref{GT-general} and \eqref{GT-basis-1-form}, 
 a dressed form $\bs\alpha^{\bs u}=\alpha\big( \! \w^\bullet \! \bs d\phi^{\bs u}; \phi^{\bs u}\big) \in \Omega^\bullet(\Phi^{\bs u})$ transforms under  $\bs\vphi \in C^\infty\big( \Phi^{\bs u}, \Aut(Q)\big)$ as:
\begin{align}
\label{GT-general-bis} 
(\bs\alpha^{\bs u})^{\bs\vphi}  &= \alpha\big(\! \w^\bullet \! (\bs d\phi^{\bs u})^{\bs \vphi};\,  (\phi^{\bs u})^{\bs\vphi} \big),  \\[.5mm]
 \text{with}\qquad ( \bs d\phi^{\bs u})^{\bs\vphi} &= \bs\vphi^*   \big(  \bs d\phi^{\bs u}   +   \mathfrak L_{ \{\bs d \bs\vphi  \circ \bs\vphi\-\}} \phi^{\bs u}  \big),     \label{GT-basis-1-form-bis}
 \end{align}
 from which follows that
$(\bs\alpha^{\bs u})^{\bs\vphi}  =R^\star_{\bs \vphi} \bs{\alpha^u} + \iota_{\{\bs d \bs\vphi  \circ \bs\vphi\-\}^v} \bs{\alpha^u}$.
\medskip

We notice that $(\phi^{\bs u})^{\vphi}$ is $\Aut(P)$-invariant for all $\vphi\in \Aut(Q)$, 
meaning that all representatives in the $\Aut(Q)$-orbit $\O_{\Aut(Q)}[\phi^{\bs u}]$ of $\phi^{\bs u}$  are valid coordinatisations of $[\phi]\in \M$. 
So, a priori $\O_{\Aut(Q)}[\phi^{\bs u}] \simeq  \O_{\Aut(P)}[\phi]$, and $\M^{\bs u} \simeq \M$. 
By~\eqref{SESgroup2}, it is clear that $\bs\Aut(Q)\simeq \bs\Aut_v(\Phi^{\bs u})$  is isomorphic to  the original gauge group $\bs\Aut(P)\simeq \bs\Aut_v(\Phi)$, and  that $\bs\Diff(\M^{\bs u}) \simeq \bs\Diff(\M)$, which are ``physical symmetries".  
A  priori, the new symmetry $ \bs\Diff_v(\Phi^{\bs u}) \supset \bs\Aut_v(\Phi^{\bs u})$ does not enjoy a more direct physical interpretation than the original symmetry it replaces. 

The group $C^\infty\big( \Phi^{\bs u}, \Aut(Q)\big)  \simeq \bs\Diff_v(\Phi^{\bs u})$, and transformations like \eqref{dressing-ambiguity}, \eqref{dressing-ambiguity-bis} and \eqref{GT-basis-1-form-bis},  encompass what is known, in the covariant phase space literature on edge modes, as ``surface symmetries" or ``corner symmetries" -- or more rarely as ``physical symmetries", which is misleading, per our previous remark. 

We observe that, in concrete situations, the process by which one constructs a dressing field $\bs u(\phi)$ out of the field content of a theory may be such that the choice reflected in the relation \eqref{dressing-ambiguity} is parametrized by only a subgroup of
 $\Aut(Q)$, possibly even a discrete one. 
A possibility forfeited for dressing fields introduced by hand in a theory.

 \subsection{Dressed regions and integrals} 
 \label{Dressed regions and integrals} 
  
  In section \ref{Integration map}, we observed that for $V \in \bs V(P)$ and $\bs \alpha \in \Omega^\bullet \big(\Phi, \Omega^{\text{top}}(V) \big)$,
    integrals $\bs\alpha_V=\langle \bs \alpha, V \rangle =\int_V \bs \alpha$ are objects on $\Phi \times \bs V(P)$, with values in $\Omega^\bullet(\Phi)$. 
    Integrals of tensorial integrand are invariant under the action of $C^\infty\big(\Phi, \Aut(P) \big)\simeq \bs\Diff_v(\Phi)$, defined by \eqref{Vert-trsf-int-generic}. 
  Their projection along $\b\pi : \Phi \times \bs V(P) \rarrow \b{\bs V}(P)$, $(\phi, V) \mapsto [\phi, V]=[\psi^*\phi, \psi\-(V)] $ is well-defined, and they can be said ``basic" w.r.t. $\b \pi$. 
So, they descend  on the associated bundle of regions $\b{\bs V}(P)\defeq \Phi \times \bs V(P)/\sim$, which is the quotient of the product space by the action of $\Aut(P)$  \eqref{right-action-diff}. 
    
   In section \ref{Building basic forms via dressing},  dressed objects were defined as being in Im$\,\pi^\star$ (i.e. basic on $\Phi$) with the projection realised via a dressing field, $F_{\bs u} \sim \pi$.
Relying on the formal similarity of the actions of $ F_{\bs u}$ and $ \Xi \in \bs\Diff_v(\Phi) \simeq C^\infty\big(\Phi, \Aut(P) \big)$, the rule of thumb to obtain the dressing $\bs\alpha^{\bs u}$ 
of a form $\bs\alpha$ is to replace the field-dependent parameter $\bs\psi$ in $\bs\alpha^{\bs\psi}\defeq \Xi^\star \bs \alpha$ 
by the dressing field $\bs u$.  
    In the same way, we define dressed integrals as being basic on $\Phi \times \bs V(P)$, i.e. in Im$\,\b\pi^\star$, with the projection realised as:
\begin{equation}   
\begin{split} 
    \b F_{\bs u}: \Phi \times \bs V(P) &\rarrow \b{\bs V}(P) \simeq \Phi^{\bs u} \times \bs V(Q), \\
    				(\phi, V) &\mapsto    \b F_{\bs u}(\phi, V)\defeq \big(F_{\bs u}(\phi), \bs u\-(V) \big) =  \big( \phi^{\bs u}, \bs u\-(V) \big).
\end{split}    
\end{equation}    
We highlight the fact that the region $V^{\bs u} \defeq  \bs u\-(V) \in \bs V(Q)$ is a map $V^{\bs u}: \Phi \times \bs V(P) \rarrow \bs V(Q)$ s.t. for $\psi \in \Aut(P)$:
\begin{align}
(V^{\bs u})^\psi \defeq \b R^\star_\psi \, V^{\bs u} = (R^\star_\psi \bs u)\- \circ \psi\-(V) = (\psi\- \circ \bs u)\-\circ \psi\-(V) = \bs u\-(V) \rdefeq V^{\bs u}. 
\end{align}
The same then holds for $\bs \psi \in C^\infty\big(\Phi, \Aut(P) \big)\simeq  \bs\Diff_v(\Phi)$: $(V^{\bs u})^{\bs\psi} \defeq \b \Xi^\star  V^{\bs u} =V^{\bs u}$. 
Therefore, $V^{\bs u}$ is a $\phi$-dependent $\Aut(P)$-invariant region of what we may call, as we did in section \ref{Relationality in  general-relativistic gauge field theory}, the  physical \emph{relationally defined  enriched spacetime}. 
As we have seen there, the generalised hole argument and the generalised point-coincidence argument establish that the physical enriched spacetime is defined  \emph{relationally}, in an $\Aut(P)$-invariant way, by its physical gauge field content. 
A fact that is tacitly encoded by the $\Aut(P)$-covariance/invariance of general-relativistic gauge field theories, and made manifest via the DFM:
$V^{\bs u}$ are manifesly $\Aut(P)$-invariant and manifestly $\phi$-relationally defined  regions, faithfully representing regions of the physical enriched spacetime, on which relationally defined and $\Aut(P)$-invariant fields $\phi^{\bs u}\defeq \bs u^*\phi$ live -- and might  be integrated over.
From this follows that the physical, \emph{relational boundary} $\d V^{\bs u}$ of a true enriched spacetime region $V^{\bs u}$ is necessarily $\Aut(P)$-invariant. This observation is key to dissolve the so-called ``boundary problems" often discussed in field theory, as we will see in section \ref{Dressed integrals and integration on spacetime}. 

So, on the space $\Omega^\bullet \big(\Phi, \Omega^{\text{top}}(V) \big) \times \bs V(P)$ we define:
\begin{equation}   
\begin{split} 
    \t F_{\bs u}: \Omega^\bullet \big(\Phi, \Omega^{\text{top}}(V) \big) \times \bs V(P) &\rarrow \Omega^\bullet_\text{\text{basic}}\big(\Phi, \Omega^{\text{top}}(Q) \big) \times \bs V(Q) \\
    				(\bs \alpha, V) &\mapsto    \t F_{\bs u}^\star (\bs \alpha, V)\defeq \big(\bs \alpha^{\bs u}, \bs u\-(V) \big). 
\end{split}    
\end{equation}  
With indeed $ \Omega^\bullet_\text{\text{basic}}\big(\Phi, \Omega^{\text{top}}(Q) \big) \simeq \Omega^\bullet \big(\Phi^{\bs u}, \Omega^{\text{top}}(Q) \big)$, as  dressed fields $\phi^{\bs u}$ live on $Q$. 
Then, in formal analogy with \eqref{Vert-trsf-int-generic}, the dressing of an integral $\bs\alpha_V \defeq \langle \bs\alpha, V \rangle=\int_V \bs\alpha$ is
\begin{equation}   
\label{Dressed-integral}
\begin{split} 
({\bs\alpha_V})^{\bs u} = \langle\ ,\ \rangle \circ  \t F_{\bs u}(\bs \alpha, V) = \langle \bs\alpha^{\bs u} , \bs u\-(V)   \rangle 
= \int_{\bs u\-(V)} \hspace{-4mm} \bs\alpha^{\bs u}. 
\end{split}    
\end{equation}  
We remark that the residual symmetry $C^\infty\big(\Phi^{\bs u}, \Aut(Q) \big) \simeq \bs\Diff_v(\Phi^{\bs u})$ may  act on dressed integrals, analogously to \eqref{Vert-trsf-int-generic}, as
\begin{align}
\label{residual-trsf-dressed-int}
\left(({\bs\alpha_V})^{\bs u} \right)^{\bs \vphi} = \int_{\bs \vphi \-(V^{\bs u})}  (\bs\alpha^{\bs u})^{\bs \vphi}. 
\end{align}
Now, for $\bs \alpha \in \Omega^\bullet_\text{tens} \big(\Phi, \Omega^{\text{top}}(V) \big)$ one has 
\begin{equation}   
\begin{split} 
{\bs\alpha_V}^{\bs u} = \langle \bs\alpha^{\bs u} , \bs u\-(V)   \rangle = \langle \bs u^*\bs\alpha, \bs u\-(V)   \rangle = \langle \bs\alpha, V   \rangle = \bs\alpha_V,
\end{split}    
\end{equation}  
by the invariance property \eqref{invariance-action} of the integration pairing. 
Therefore, ${\bs d \bs\alpha}_V = \langle \bs d \bs\alpha, V \rangle = \bs d  \langle \bs u^*\bs\alpha, \bs u\-(V)   \rangle = \bs d ({\bs\alpha_V}^{\bs u})$. 
The latter result can also be proven 
using the lemma  \eqref{Formula-general-rep-dress}-\eqref{Formula-pullback-rep-dress} and concluding by the invariance property  \eqref{inf-invariance-action}. This~calculation also supplies the analogue of  \eqref{Vert-trsf-pairing-dalpha}/\eqref{Vert-trsf-int-dalpha}: 
\begin{equation}
\label{Dressing-int-dalpha}
\begin{aligned}
\bs (\bs{d\alpha}_V)^{\bs u}	 &= \bs{d\alpha}_V + \langle \mathfrak L_{\bs{du}\circ \bs u\-} \bs\alpha,  V \rangle,  \\
 \langle  \bs{d\alpha}, V \rangle ^{\bs u} &= \langle\bs{d\alpha}, V \rangle + \langle \mathfrak L_{\bs{du}\circ \bs u\-} \bs\alpha,  V \rangle.
\end{aligned}
\end{equation}
The local version is key to the interplay between the DFM and the variational principle, as  shown in   section \ref{Dressed integrals and integration on spacetime}.

\section{Local field theory} 
\label{Local field theory} 

Local field theory is usually expressed not directly via fields on a principal bundle $P$, but via local representatives of these global objects on (open sets of) the base space $M$. 
In this section, we thus consider the local counterpart of the above formalism to obtain applications to field theory in its standard formulation. 
As we shall see, the local case features several subtleties. 
We thus define the elementary notions 
and provide the most important relations necessary for field theory.

\subsection{Local field space}  
\label{Local field space} 

For local field theory, the field space $\Phi$ is now the space of local representatives $\phi=\{A, F,  \upphi, D\upphi, \ldots\}$ on $M$ of global object defined on $P$: e.g. $A$ is a gauge potential over $U\subset M$ representing the connection $\omega$ over $P_{|U} \subset P$, while $F$ is the field strength representing the curvature $\Omega$. 

Locally, over $U\subset M$, the group $\Aut(P)$ of automorphisms -- the structure group of $\Phi$ seen as a principal bundle -- is represented as the semi-direct product group $\Diff(M)\ltimes \Hl$, where $\Hl$ is the local representative of the gauge group $\H$ of $P$ (isomorphic to $\Aut_v(P)$):
\begin{align}
\label{Hloc-def}
    \Hl = \left\{\, \upgamma, \upeta :U \rightarrow H\ |\  \upeta^\upgamma\defeq \upgamma^{-1} \upeta\upgamma\, \right\}.
\end{align}
This field-theoretic characterisation of the local (active) gauge group flows from the geometric definition of the gauge group of $P$.
Similarly, the action of $\Hl$ on a (local) field $\phi$ is \emph{defined} as the local version of the action of $\H$ on the corresponding global object: 
For example, the $\H$-gauge transformation of $\omega$ is $\omega^\gamma \defeq \psi^* \omega = \gamma\- \omega \gamma +\gamma\-d\gamma$, for $\psi \in \Aut_v(P) \sim \gamma \in \H$ -- remind the isomorphism $\psi(p)=p\gamma(p)$ --  which locally gives $A^\upgamma \!\defeq  \upgamma\- A \upgamma +\upgamma\-d\upgamma$, for $\upgamma \in \Hl$.
Likewise, for $\beta \in \Omega^\bullet_\text{tens}(P, \rho)$, one has $\beta^\gamma \defeq \psi^* \beta = \rho(\gamma)\- \beta$, which gives locally $b^\upgamma \!\defeq \rho(\upgamma)\- b$.  
Take~$b=\{ F, \upphi, D\upphi, \ldots\}$ with their respective representation $\rho$. 
The right action of the structure group on $\Phi$, the~local version of \eqref{Right-action-on-Phi}, is then
\begin{equation}
\label{group-law-loc-aut}
\begin{aligned}
\Phi \times \big(\!\Diff(M)\ltimes \Hl \big) &\rarrow \Phi, \\
\left(\phi, \big(\uppsi, \upgamma \big) \right) &\mapsto R_{(\uppsi, \upgamma)}\,\phi 
= (\uppsi, \upgamma)^*\phi
\defeq 
\uppsi^*(\phi^\upgamma). 
\end{aligned}
\end{equation}
Remark that it does not matter which of $\Diff(M)$ or $\Hl$ transformation applies first. This follows from the fact that $(\uppsi, \upgamma)$ is the local representative of a given automorphism $\psi=(\b\uppsi, \eta) \in \Aut(P)=\overline{\Diff}\ltimes \Aut_v(P)$ which, by \eqref{Aut-graph}, is written as  $\psi(p)=\big(\eta \circ \b\uppsi\big)(p)=\b\uppsi(p) \gamma\big(\b\uppsi(p)\big)=\b\uppsi(p)\big(\b\uppsi^*\gamma\big)(p)$.
The action  \eqref{group-law-loc-aut} is the local version of the pullback~action~$\psi^*$.
The semi-direct structure \eqref{semi-dir-DiffM-H} of the local group $\Diff(M)\ltimes \Hl$ is inherited from the semi-direct structure of $\Aut(P)$ as detailed in Appendix \ref{Semi-direct product structure of Aut(P)}.
It can also be derived from field-theoretic considerations, by iterating twice the transformation \eqref{group-law-loc-aut}:
\begin{equation}
\begin{aligned}
        \phi \mapsto & \ {\uppsi'}^*(\phi^{\upgamma'}) = ({\uppsi'}^* \phi)^{{\uppsi'}^*\upgamma'} \\
     \mapsto & \ \uppsi^*\Big(\big[ ({\uppsi'}^* \phi)^{{\uppsi'}^*\upgamma'} \big]^{\upgamma} \Big)
    = \Big[(\uppsi^* {\uppsi'}^*\phi)^{\uppsi^*{\uppsi'}^*\upgamma'} \Big]^{{\uppsi}^*\upgamma} 
    =\Big[(\uppsi^* {\uppsi'}^*\phi)^{\uppsi^*{\uppsi'}^*\upgamma'} \Big]^{{\uppsi}^*{\uppsi'}^*{{\uppsi'}\-}^* \upgamma} \\
    & \phantom{ \ \uppsi^*\Big(\big[ ({\uppsi'}^* \phi)^{{\uppsi'}^*\upgamma'} \big]^{\upgamma} \Big)} = \uppsi^* \uppsi'{}^* \Big(\big[\phi^{\upgamma'}\big]^{\uppsi'{}\-{}^* \upgamma}\Big)
    =(\uppsi'\circ \uppsi)^* \Big(\phi^{\,\gamma' \cdot\, (\uppsi'{}\-{}^* \gamma)} \Big). 
\end{aligned}
\end{equation}
One may test this more explicitly for the simple case $\phi=b$.
Given the right action of $\Diff(M)\ltimes \Hl$  on $\Phi$, $R_{(\uppsi, \upgamma)}\, R_{(\uppsi', \upgamma')}\,\phi = R_{ (\uppsi', \upgamma') \cdot (\uppsi, \upgamma)}\, \phi$, its \emph{semi-direct product} is found to be
\begin{align}
\label{semi-dir-prod-DiffM-H-2}
 \big( \uppsi' , \upgamma' \big) \cdot \big( \uppsi, \upgamma \big) = \left( \uppsi' \circ \uppsi , \, \upgamma' \!\cdot\big({\uppsi'}{}\-{}^* \upgamma \big)  \right).  
\end{align}
The corresponding Lie algebra 
$\diff(M) \oplus  \text{Lie}\Hl$ is a semi-direct sum with Lie bracket 
\begin{align}
\label{Liebracket-local-aut}
\big[(X', \lambda'), (X, \lambda)\big]_\text{\tiny{Lie}}
=\Big(\,  [X', X]_{\text{\tiny{$\diff(M)$}}}\,,\  [\lambda', \lambda]_\text{\tiny{Lie$H$}} - X'(\lambda) + X(\lambda')\, \Big).
\end{align} 
Remark that $[X',X]_{\text{\tiny{$\diff(M)$}}} \defeq- [X',X]_\text{\tiny{$\Gamma(TM)$}}$.
The bracket \eqref{Liebracket-local-aut} can also be understood as the bracket on sections of the (locally trivialised) Atiyah Lie algebroid associated to $P$ -- see e.g. \cite{Masson-Lazz} -- described by the SES
\begin{align}
    0 \rarrow \text{Lie}\Hl \rarrow \diff(M) \oplus  \text{Lie}\Hl \rarrow  \diff(M) \rarrow 0.
\end{align}

The vertical subbundle of field space $\Phi \xrightarrow{\pi} \Phi/ \big(\!\Diff(M)\ltimes \Hl\big)\rdefeq \M$, with $\M$ the moduli space of gauge orbits, is $V\Phi\defeq \ker \pi_*$. The fundamental vertical vector field are 
\begin{align}
\label{fund-vect-field-loc}
(X, \lambda)^v_{|\phi}\defeq \tfrac{d^2}{d\tau ds} \, R_{(\uppsi_\tau, \upgamma_s)} \, \phi \, \big|_{\tau=0,\, s=0}
=  \tfrac{d^2}{d\tau ds} \,  \uppsi^*_\tau \big(\phi^{\upgamma_s}\big) \, \big|_{\tau=0,\, s=0}
=\mathfrak{L}_X \phi + \delta_{\!\lambda} \phi = X^v_{\,|\phi} + \lambda^v_{\,|\phi},
\end{align}
with $\mathfrak{L}_X \phi$ is the Lie derivative of $\phi$ along the vector field $X\in \Gamma(TM)$ generating the diffeomorphism $\uppsi_\tau \in \Diff(M)$, and $\delta_\lambda \phi$ is the infinitesimal (active) gauge transformation of $\phi$ by the element $\lambda: U\rarrow$ Lie$H$ $\in$ Lie$\Hl$. For example, for $\phi=A$ this is $\delta_{\!\lambda} A = D\lambda$,
for $\phi=b$ this is $\delta_{\!\lambda} b = -\rho_*(\lambda) b$.
The pushforward by the $\big(\!\Diff(M)\ltimes \Hl\big)$-right action of a fundamental vector field is 
\begin{align}
\label{pushf-fund-vec-loc}
R_{(\uppsi, \upgamma)\star} \ (X, \lambda)^v_{|\phi} 
\defeq \Big(  \Ad_{(\uppsi, \upgamma)\-} (X, \lambda) \,\Big)^v_{|\,R_{(\uppsi, \upgamma)} \phi 
}
\end{align}
where $\big(\uppsi, \upgamma \big)\-=\big( \uppsi\-, \uppsi^*\upgamma\- \big)$ and  the adjoint action is given by
\begin{align}
    \Ad_{(\uppsi, \upgamma)\-} (X, \lambda)
= \Big( \, \uppsi\-_* \, X \circ \uppsi , \ \uppsi^* \big( \Ad_{\upgamma\-} \lambda - \upgamma\- \mathfrak{L}_X \, \upgamma \big) \, \Big) ,  \label{SD-loc-adj-action2}
\end{align}
as proven in \eqref{SD-loc-adj-action-inverse} of Appendix \ref{Semi-direct product structure of Aut(P)}.

The maximal group of transformations of the \emph{local} field space $\Phi$ is again 
\begin{align}
    \bs\Aut(\Phi) \defeq \Big\{\, \Xi \in \bs\Diff(\Phi) \, |\, \Xi \circ R_{(\uppsi, \upgamma)} = R_{(\uppsi, \upgamma)} \circ  \Xi\,\Big\},
\end{align}
whose elements cover those of $\bs\Diff(\M)$.   
The vertical diffeomorphisms 
\begin{align}
    \bs\Diff_v(\Phi)\defeq \Big\{\, \Xi \in \bs\Diff(\Phi)\, |\, \pi \circ \Xi =\pi \,\Big\}
\end{align}
cover the identity $\id_\M$. 
Therefore, a vertical diffeomorphism is given by $\Xi(\phi)=R_{\big(\bs\uppsi(\phi),\bs\upgamma(\phi)\big)}\,\phi:=\big(\bs\uppsi(\phi),\bs\upgamma(\phi) \big)^*\phi$,
with the maps $\big(\bs\uppsi, \bs\upgamma \big): \Phi \rarrow \Diff(M)\ltimes \Hl$, i.e. one has $\bs\Diff(\Phi)\simeq C^\infty\big(\Phi, \Diff(M)\ltimes \Hl \big)$. The former is the action Lie groupoid of $\Phi$, the latter its group of sections. 
We have that, as the local version of \eqref{twisted-comp-law},
\begin{align}
\label{twisted-comp-law2}
 \Xi' \circ \Xi \ \in \bs\Diff(\Phi)\ \  \text{ corresponds to } \ \ 
 \big(\bs\uppsi, \bs\upgamma\big) \cdot \Big( \big( \bs\uppsi', \bs\upgamma'\big) \circ R_{(\bs\uppsi, \bs\upgamma)}\Big)\ \in \ C^\infty\big(\Phi, \Diff(M)\ltimes \Hl \big). 
\end{align}
This extends the normal subgroup of vertical automorphisms, 
\begin{align}
  \bs\Aut_v(\Phi)\defeq  \big\{\,  \Xi \in \bs \Aut(\Phi)\,|\, \pi\circ\Xi =\pi \, \big\} \triangleleft \bs\Aut(\Phi),
\end{align}
which is isomorphic, still via 
$\Xi(\phi)=R_{\big(\bs\uppsi(\phi),\bs\upgamma(\phi)\big)}\,\phi:=\big(\bs\uppsi(\phi),\bs\upgamma(\phi) \big)^*\phi$, to the gauge group
\begin{align}
\bs\Diff(M)\ltimes \bs\Hl \defeq \Big\{\, (\bs\uppsi, \bs\upgamma): \Phi \rarrow \Diff(M)\ltimes \Hl\, |\ R^\star_{(\uppsi, \upgamma)}\, \big(\bs\uppsi, \bs\upgamma \big)= \text{Conj}(\uppsi, \upgamma)\- \big( \bs\uppsi, \bs\upgamma\big) \, \Big\}.
\end{align}
By the equivariance of gauge group elements, one has as a special case of \eqref{twisted-comp-law2} that to $\Xi'\circ\Xi \in \Aut_v(\Phi)$ corresponds the gauge group element $\big(\bs\uppsi', \bs\upgamma'\big) \cdot \big(\bs\uppsi, \bs\upgamma\big) \in \bs\Diff(M)\ltimes \bs\Hl$. 
The SES associated to the local field space is 
\begin{align}
 \label{SES-grp-loc-field-space}   
 \id_\Phi \rarrow \bs\Diff(M)\ltimes \bs\Hl \simeq \bs\Aut_v(\Phi) \xlongrightarrow{\triangleleft} \bs\Aut(\Phi) \longrightarrow \bs\Diff(\M) \rarrow \id_\M.
\end{align}

We have the  Lie algebra morphism 
$\diff_v(\Phi) \simeq \C^\infty\big(\Phi, \diff(M)\oplus\text{Lie}\Hl \big)$ induced by the ``verticality map" \eqref{fund-vect-field-loc} $|^v: \C^\infty\big(\Phi, \diff(M)\oplus\text{Lie}\Hl \big) \rarrow \bs\diff_v(\Phi)$, $(\bs X, \bs \lambda)\mapsto(\bs X, \bs \lambda)^v$.
The bracket is
\begin{align}
\label{FN-bracket-loc}
\big[(\bs X, \bs\lambda)^v, (\bs X', \bs\lambda')^v\big]_\text{\tiny $\Gamma(T\Phi)$} = \left\{ \big[(\bs X, \bs\lambda), (\bs X', \bs\lambda')\big]_\text{\tiny{Lie}} 
+ (\bs X, \bs\lambda)^v \big((\bs X', \bs\lambda')\big)
- (\bs X', \bs\lambda')^v \big((\bs X, \bs\lambda)\big) \right\}^v
\rdefeq \big\{(\bs X, \bs\lambda), (\bs X', \bs\lambda') \big\}^v.
\end{align}
It is the local   version of \eqref{extended-bracket1}.
The \emph{extended bracket} $\{\ , \,\}$ on $C^\infty\big(\Phi, \diff(M)\oplus\text{Lie}\Hl\big)$  and can be understood both as the action Lie algebroid bracket $A=\Phi \rtimes \big(\diff(M)\oplus\text{Lie}\Hl \big) \rarrow \Phi$ bracket, and as the Frölicher-Nijenhuis bracket on $\Omega^0(\Phi, V\Phi)$. 
Observe that, as already mentioned, $[\ ,\,]_\text{\tiny{Lie}}$ is the trivial Atiyah Lie algebroid bracket \eqref{Liebracket-local-aut}. 
So the  (action Lie algebroid, FN) bracket $\{\ ,\, \}$ extends the trivial Atiyah Lie algebroid bracket $[\ ,\,]_\text{\tiny{Lie}}$ to field-dependent parameters  \emph{with unspecified equivariance}.

The  right-invariant vector fields $\Gamma_\text{\tiny inv}(T\Phi)$ constitute the Lie algebra $\bs\aut(\Phi)$ of $\bs\Aut(\Phi)$, with Lie ideal $\bs\aut_v(\Phi)$. 
As a special case of the above, we have the  isomorphism 
$\bs\aut_v(\Phi) \simeq \bs\diff(M)\oplus\text{\bf Lie}\bs\Hl$,
with the gauge Lie algebra 
\begin{align}
\bs\diff(M)\oplus\text{\bf Lie}\bs\Hl \defeq \Big\{\, (\bs X, \bs\lambda): \Phi \rarrow \diff(M)\oplus\text{Lie}\Hl\, |\ R^*_{(\uppsi, \upgamma)}\, \big(\bs X, \bs\lambda \big)= \text{Conj}(\uppsi, \upgamma)\- \big( \bs X, \bs\lambda\big) \, \Big\}.
\end{align}
So, by the  equivariance of gauge Lie algebra elements, linearly given by the adjoint action $\ad_{(X, \lambda)}= [(X, \lambda),\,\,\, ]_\text{\tiny{Lie}}$,
\eqref{FN-bracket-loc} specialises to
\begin{align}
\big[(\bs X, \bs\lambda)^v, (\bs X', \bs\lambda')^v\big]_\text{\tiny $\Gamma(T\Phi)$} = \left\{ -\big[(\bs X, \bs\lambda), (\bs X', \bs\lambda')\big]_\text{\tiny{Lie}}
\right\}^v.
\end{align}
We have the SES of Lie algebras
\begin{align}
 \label{Atiyah-Algebroid-loc}
0\rarrow \bs\diff(M)\oplus \text{\bf{Lie}}\bs\Hl  \simeq \bs\aut_v(\Phi)  \xlongrightarrow{|^v}   \bs{\mathfrak{aut}}(\Phi) \xlongrightarrow{\pi_\star} \bs{\diff}(\M) \rarrow 0,
\end{align}
describing the Atiyah Lie algebroid associated to the local field space. 

The pushforward of $\mathfrak X \in \Gamma(T\Phi)$ by $\Xi \in \bs\Diff_v(\Phi)\sim (\bs\uppsi, \bs\upgamma) \in C^\infty\big(\Phi, \Diff(M)\ltimes \Hl \big)$ -- a local version of \eqref{pushforward-X} --  is
\begin{equation}
\label{pushforward-X-loc}
\begin{aligned}
\Xi_\star \mathfrak X_{|\phi}
& = R_{(\bs\uppsi, \bs\upgamma ) \star}\mathfrak X_{|\phi} 
+ \left\{(\bs\uppsi, \bs\upgamma)^{-1}\! \cdot  \big(\bs d \bs\uppsi , \bs d\bs \upgamma\, \big)_{|\phi} (\mathfrak X_{|\phi}) \right\}^v_{|\Xi(\phi)} \\
& = R_{(\bs\uppsi, \bs\upgamma ) \star}\mathfrak X_{|\phi} 
+ \left\{
\big(
\bs\uppsi\-_* \bs d\bs \uppsi, 
\bs\uppsi^*\, (\bs\upgamma\- \bs d\bs \upgamma)
\big)_{|\phi} (\mathfrak X_{|\phi})
\right\}^v_{|\Xi(\phi)} \\[2mm]
\Xi_\star \mathfrak X_{|\phi}
&=  R_{(\bs\uppsi, \bs\upgamma ) \star}\left(
\mathfrak X_{|\phi} + 
\left\{
 \Big(\bs d \bs\uppsi , \bs d\bs \upgamma\, \Big)_{|\phi} (\mathfrak X_{|\phi})
 \cdot \big(\bs\uppsi, \bs\upgamma \big)\-
\right\}^v_{|\phi} 
\right) \\
&=  R_{(\bs\uppsi, \bs\upgamma ) \star}\left(
\mathfrak X_{|\phi} + 
\left\{\Big( 
\bs{d\uppsi} \circ \bs\uppsi\-,\ 
\bs{d\upgamma} \, \bs\upgamma\-  - \bs\upgamma\, \mathfrak L_{\bs{d\uppsi} \circ \bs\uppsi\-} \bs\upgamma\-
\Big)_{|\phi}(\mathfrak X_{|\phi})
\right\}^v_{|\phi} 
\right)
\end{aligned}
\end{equation}
as proven in \eqref{Pushforward-X-loc-1}-\eqref{Pushforward-X-loc-2}  in Appendix \ref{Lie algebra (anti-)isomorphisms}.
It is necessary to compute geometrically vertical and gauge transformations on the local field space $\Phi$.

\subsubsection{Forms on local field space and their transformations}
\label{Forms on local field space and their transformations}

Forms $\bs\alpha = \alpha\big(\wedge^\bullet\!\bs d\phi; \phi \big) \in\Omega^\bullet(\Phi)$ on the local field space, together with the basic operations $\bs d$, $\iota_\mathfrak X$, $\bs L_\mathfrak X$, for $\mathfrak X \in \Gamma(T\Phi)$ -- as well as the extensions via the Nijenhuis-Richardson and Frölicher-Nijenhuis brackets -- are defined as in section~\ref{Differential forms and their derivations}.
Their~equivariance is defined by (the result of) $R^*_{(\uppsi, \upgamma)} \bs\alpha$, with infinitesimal version given by $\bs L_{(X, \lambda)^v} \bs\alpha$. 
Of special interest are tensorial forms: 
For $(\rho, \bs V)$ a representation for $\Diff(M)\ltimes \Hl$, one defines
\begin{align}
 \label{tens-forms-loc}
\Omega_\text{tens}^\bullet(\Phi, \rho)\defeq \left\{ \, \bs\alpha \in \Omega^\bullet(\Phi, \bs V)\,|\, R^\star_{(\uppsi, \upgamma)}\bs\alpha=\rho\big(\uppsi, \upgamma\big)\-\bs\alpha,\, \text{ \& }\ \iota_{(X, \lambda)^v}\bs\alpha=0\, \right\}. 
\end{align}
The infinitesimal equivariance is then 
$\bs L_{(X, \lambda)^v} \bs\alpha = -\rho_*\big( X, \lambda\big)\, \bs\alpha$. 
Given a 1-cocycle 
\begin{equation}
\begin{aligned} 
 C: \Phi \times \big(\!\Diff(M)\ltimes \Hl\big) &\rarrow G, \\
 \big(\phi; (\uppsi, \upgamma) \big) &\mapsto C\big(\phi; (\uppsi, \upgamma) \big), \\[1mm]
  \text{s.t. } \quad
 C\big(\phi; (\uppsi', \upgamma') \cdot (\uppsi, \upgamma) \big) &= C\big(\phi; (\uppsi', \upgamma') \big)\, C\big(R_{(\uppsi', \upgamma')}\phi; (\uppsi, \upgamma) \big),
\end{aligned}
\end{equation}
and $\bs V$ a $G$-space, one extends the above to  \emph{twisted tensorial} forms 
 \begin{align}
 \label{twisted-tens-forms-loc}
\Omega_\text{tens}^\bullet(\Phi, C)\defeq \left\{ \, \bs\alpha \in \Omega^\bullet(\Phi, \bs V)\,|\, R^\star_{(\uppsi, \upgamma)}\bs\alpha=C\big(\phi; (\uppsi, \upgamma)\big)\-\bs\alpha,\, \text{ \& }\ \iota_{(X, \lambda)^v}\bs\alpha=0\, \right\}. 
\end{align}
The infinitesimal equivariance is then 
$\bs L_{(X, \lambda)^v} \bs\alpha = -a\big(  ( X, \lambda); \phi \big)\, \bs\alpha$, where 
$a: \Phi \times \big(\diff(M)\oplus \text{Lie}\Hl\big) \rarrow \LieG$ is the infinitesimal 1-cocycle associated to $C$.
Acting via $[\bs L_{(X, \lambda)^v}, \bs L_{(X', \lambda')^v}]=\bs L_{[(X, \lambda)^v,(X', \lambda')^v]}=\bs L_{\big(\left[(X, \lambda),(X', \lambda')\right]_\text{\tiny{Lie}}\big)^v}$  reproduces the defining $\big(\diff(M)\oplus \text{Lie}\Hl\big)$-1-cocycle property
\begin{align}
\label{cc-gen-anom-loc0}
\big(X, \lambda\big)^v a\big((X', \lambda'); \phi \big)
-
\big(X', \lambda'\big)^v a\big((X, \lambda); \phi \big)
+\big[ 
a\big((X, \lambda); \phi \big), a\big((X', \lambda'); \phi \big)
\big]_\text{\tiny{$\LieG$}}
=a\Big(\big[(X, \lambda),\, (X', \lambda') \big]_\text{\tiny{Lie}};\phi \Big).
\end{align}
This, as we will see, generalises the Wess-Zumino consistency condition for anomalies. 
Fundamental to our discussions are the \emph{basic forms}, 
\begin{equation}
\begin{aligned}
 \label{basic-forms-loc}
\Omega_\text{basic}^\bullet(\Phi, C)\defeq& \left\{ \, \bs\alpha \in \Omega^\bullet(\Phi)\,|\, R^\star_{(\uppsi, \upgamma)}\bs\alpha= \bs\alpha,\, \text{ \& }\ \iota_{(X, \lambda)^v}\bs\alpha=0\, \right\}\\
=&\left\{ \, \bs\alpha \in \Omega^\bullet(\Phi)\,|\, \bs\alpha=\pi^\star \bs \beta, \text{ for } \bs\beta \in\Omega^\bullet(\M) \right\}.
\end{aligned}
\end{equation}
Our aim in the next section will be to write basic counterparts of forms on $\Phi$ via the DFM. 

\paragraph{Vertical and gauge transformations}
The finite vertical and gauge transformations of forms are defined by the pullback actions of $\bs\Diff_v(\Phi)$ and $\bs\Aut_v(\Phi)$ respectively: 
$\bs\alpha \mapsto \bs\alpha^{(\bs\uppsi, \bs\upgamma)}\defeq \Xi^\star \bs\alpha$, with $\Xi$ generated by $(\bs\uppsi, \bs\upgamma)$, element of $\C^\infty\big(\Phi, \Diff(M)\ltimes\Hl \big)$ or $\bs\Diff(M)\ltimes\bs\Hl$,
and computed geometrically via \eqref{pushforward-X-loc}:
\begin{align}
\label{Vertic-trsf-loc}
{\bs\alpha^{(\bs\uppsi, \bs\upgamma)}}_{|\phi} \big(\mathfrak X_{|\phi}, \ldots \big) \defeq \Xi^\star \bs\alpha_{|\, \Xi(\phi)} \big(\mathfrak X_{|\phi}, \ldots \big)
=R_{(\bs\uppsi, \bs\upgamma)}^\star\, \bs\alpha_{|\, \Xi(\phi)} \left(  \mathfrak X_{|\phi} + \left\{ \Big(\bs d \bs\uppsi , \bs d\bs \upgamma\, \Big)_{|\phi} (\mathfrak X_{|\phi})
 \cdot \big(\bs\uppsi, \bs\upgamma \big)\-
\right\}^v_{|\phi} , \ldots \right).
\end{align}
Clearly, the transformation of a form is controlled by its equivariance and verticality properties.
The infinitesimal transformation is given
 \begin{align}
\label{Inf-GT-general-loc}
\bs L_{(\bs X, \bs\lambda)^v} \bs\alpha = \tfrac{d}{d\tau} \, R_{(\bs\uppsi_\tau, \bs\upgamma_\tau)}^\star \,\bs\alpha\, \big|_{\tau=0} 
+ \iota_{\big( \bs{dX}, \bs{d\lambda}\big)^v}\, \bs\alpha,
\end{align}
with  $(\bs X, \bs\lambda)\defeq \tfrac{d}{d\tau}\, \big(\bs\uppsi_\tau, \bs\upgamma_\tau \big)\, \big|_{\tau=0}$ in $\C^\infty\big(\Phi, \diff(M)\oplus\text{Lie}\Hl \big)$ or $\bs\diff(M)\oplus\text{\bf{Lie}}\bs\Hl$.
One may observe that $\big( \bs{dX}, \bs{d\lambda}\big)^v$ can be seen as an element of $\Omega^1(\Phi, V\Phi)$, so $\iota_{\big( \bs{dX}, \bs{d\lambda}\big)^v}$ is a degree 0 algebraic derivation, as discussed in section \ref{Differential forms and their derivations}. 
For standard and twisted equivariant forms respectively, the above gives
\begin{align}
\label{Inf-GT-eq-loc}
\bs L_{(\bs X, \bs\lambda)^v} \bs\alpha = \left\{   \begin{matrix}   -\rho_*\big(\bs X, \bs\lambda\big)\, \bs\alpha + \iota_{\big( \bs{dX},\, \bs{d\lambda}\big)^v}\, \bs\alpha, \\[3mm]          
-a\big(\bs X, \bs\lambda\big)\, \bs \alpha + \iota_{\big( \bs{dX},\, \bs{d\lambda}\big)^v}\, \bs\alpha,
					  \end{matrix} \right.
\end{align}
where we  introduce the notation $\big[a(\bs X, \bs\lambda)\big](\phi)\defeq a\Big(\bs X(\phi), \bs\lambda(\phi); \phi\Big)$ for the linearised 1-cocycle. 
In particular,  
for $\Omega^\bullet(M)$-valued forms or tensor-valued forms, i.e. for the natural representation
$\rho(\uppsi, \upgamma)\-= (\uppsi^*,\{-\}^\upgamma )$ the above gives
\begin{align}
\label{Inf-GT-diff-rep-loc}
\bs L_{(\bs X, \bs\lambda)^v} \bs\alpha =     \big( \mathfrak L_{\bs X}, \delta_{\bs\lambda} \big)\, \bs\alpha + \iota_{\big( \bs{dX},\, \bs{d\lambda}\big)^v}\, \bs\alpha.
\end{align}
This both generalises and  clarifies the geometrical meaning of the ``anomaly operators", $\Delta_{\bs X}$ and $\Delta_{\bs \lambda}$, featuring in \cite{Hopfmuller-Freidel2018, Chandrasekaran_Speranza2021, Freidel-et-al2021, Freidel-et-al2021bis, Speziale-et-al2023}: in our notation  
 $\big(\Delta_{\bs X}, \Delta_{\bs\lambda} \big)\defeq \bs L_{(\bs X, \bs\lambda)^v} -  \big( \mathfrak L_{\bs X}, \delta_{\bs\lambda} \big) - \iota_{\big( \bs{dX},\, \bs{d\lambda}\big)^v}$. 
 As we signaled in the global case,
this operator is  non-zero only in theories admitting background non-dynamical structures or fields ``breaking" $\big(\!\Diff(M)\ltimes \Hl\big)$-covariance.
Such theories thus fail to comply with the core physical principles of gRGFT. 
 
Observe that the field-dependent gauge algebra is given by the commutator of Lie derivatives and involves the FN brackets \eqref{FN-bracket-loc} of the field-dependent parameters
\begin{align}
\label{comm-Lie-der-FN-loc}
    \big[\bs L_{(\bs\uppsi, \bs\upgamma)^v}, \bs L_{(\bs\uppsi', \bs\upgamma')^v} \big]
    = \bs L_{\big[(\bs\uppsi, \bs\upgamma)^v, (\bs\uppsi' , \bs\upgamma')^v\big]_\text{\tiny{FN}}} 
    = \bs L_{\big\{(\bs\uppsi, \bs\upgamma), (\bs\uppsi', \bs\upgamma')\big\}^v}.
\end{align}

As a special case of \eqref{Vertic-trsf-loc} and \eqref{Inf-GT-eq-loc}, a tensorial form $\bs\alpha \in \Omega^\bullet_{\text{tens}} (\Phi,\rho) $ transforms as
\begin{align}
\label{vert-trsf-tens-form-loc}
\bs\alpha^{(\bs\uppsi, \bs\upgamma)} = \rho\big(\bs\uppsi, \bs\upgamma\big)\- \bs \alpha, \quad \text{so} \quad 
\bs L_{(\bs X,\bs \lambda)^v} \bs\alpha = -\rho_*\big(\bs X, \bs\lambda\big)\, \bs\alpha.
\end{align}
For a twisted tensorial form $\bs\alpha \in \Omega^\bullet_{\text{tens}} (\Phi,C)$, instead, the transformation reads
\begin{align}
\label{vert-trsf-twisted-tens-form-loc}
\bs\alpha^{(\bs\uppsi, \bs\upgamma)} = C\big(\bs\uppsi, \bs\upgamma\big)\- \bs \alpha,  \quad \text{so} \quad 
\bs L_{(\bs X,\bs \lambda)^v} \bs\alpha = -a\big(\bs X,\bs \lambda\big)\, \bs\alpha,
\end{align}
where we introduce the notation $[C(\bs\uppsi,\bs\upgamma)](\phi)\defeq C\Big(\phi; \big(\bs\uppsi(\phi),\bs\upgamma(\phi)\big) \Big)$ and $[a(\bs X,\bs \lambda)](\phi):=a\big( ( \bs X(\phi), \bs \lambda(\phi) ); \phi\big)$.
Acting via \eqref{comm-Lie-der-FN-loc}  reproduces the  $(\bs\diff(M)\oplus \text{\bf{Lie}}\Hl)$-1-cocycle property
\begin{align}
\label{cc-gen-anom-loc}
\big(\bs X, \bs\lambda\big)^v a\big((\bs X', \bs\lambda'); \phi \big)
-
\big(\bs X', \bs\lambda'\big)^v a\big((\bs X, \bs\lambda); \phi \big)
+\big[ 
a\big((\bs X, \bs\lambda); \phi \big), a\big((\bs X', \bs\lambda'); \phi \big)
\big]_\text{\tiny{$\LieG$}}
=a\Big(\big\{(\bs X, \bs\lambda),\, (\bs X',\bs\lambda') \big\};\phi \Big).
\end{align}
This actually reduces to \eqref{cc-gen-anom-loc0}, all computation done.
As we know, basic forms $\bs\alpha \in \Omega^\bullet_{\text{basic}} (\Phi)$ are invariant:
\begin{align}
    \bs\alpha^{(\bs\uppsi, \bs\upgamma)} = \bs \alpha , \quad \text{so} \quad L_{(\bs X,\bs \lambda)^v} \bs \alpha = 0 .
\end{align}

A crucial example is the basis 1-form $\bs d \phi \in \Omega^1(\Phi)$, whose equivariance and verticality properties are defined~as
\begin{align}
    R^\star_{(\uppsi,\upgamma)} \bs d \phi =(\uppsi,\upgamma)^* \bs d \phi \defeq \uppsi^*\big(\bs d \phi ^\upgamma \big), \qquad \text{and} \qquad \bs \iota_{(X,\lambda)^v} \bs d \phi = \bs d \phi_{|\munderline{blue}{\phi}} \big[ (X,\lambda)^v_{|\munderline{blue}{\phi}} \big] := \mathfrak{L}_{X} \munderline{blue}{\phi} + \delta_\lambda \munderline{blue}{\phi}. \label{basis-1-form-prop-loc}
\end{align}
Then, by \eqref{pushforward-X-loc} and \eqref{basis-1-form-prop-loc}, the $\bs\Diff_v(\Phi)\simeq C^\infty\big(\Phi, \Diff(M)\ltimes \Hl \big)$-transformation of the basis 1-form is
\begin{equation}
\begin{aligned}
{\bs d\phi^{(\bs\uppsi, \bs\upgamma)}}_{|\phi} \big( \mathfrak X_{|\phi} \big)
\defeq&\, \Xi^\star \bs d \phi_{|\,\Xi(\phi)} \big(\mathfrak{X}_{|\phi} \big) \\
=&\, \bs d \phi_{|\,\Xi(\phi)} \ \big( \Xi_\star \mathfrak{X}_{|\phi} \big) \\
=&\, R^\star_{(\bs \uppsi,\bs \upgamma)}\bs d \phi_{|\,\Xi(\phi)} \left( \mathfrak{X}_{|\phi} + \Big\{ \big[ \big( \bs d \bs \uppsi, \bs d \bs \upgamma \big) \cdot \big( \bs \uppsi , \bs \upgamma \big)\- \big]_{|\phi} \big(\mathfrak{X}_{|\phi} \big)_{|\phi} \Big\}^v_{|\phi} \right) \\
=&\, \big( \bs \uppsi , \bs \upgamma \big)^* \bs d \phi_{|\phi} \left( \mathfrak{X}_{|\phi} +  \Big\{ \big[ \big(\bs d \bs \uppsi \circ \bs \uppsi\- , \bs d \bs \upgamma \, \bs \upgamma\- - \bs \upgamma \, \mathfrak{L}_{\bs d \bs \uppsi \circ \bs \uppsi\-} \bs \upgamma\-  \big) \big]_{|\phi} \big(\mathfrak{X}_{|\phi} \big)_{|\phi} \Big\}^v_{|\phi} \right) \\
=&\, \big( \bs \uppsi , \bs \upgamma \big)^* \left[ \bs d \phi_{|\phi} + \mathfrak{L}_{\bs d \bs \uppsi \circ \bs \uppsi\-} \phi + \delta_{\bs d \bs \upgamma \, \bs \upgamma\- - \, \bs \upgamma \, \mathfrak{L}_{\bs d \bs \uppsi \circ \bs \uppsi\-} \bs \upgamma\-} \phi \right] \big( \mathfrak{X}_{|\phi} \big) \\
=&\,  \big( \bs \uppsi , \bs \upgamma \big)^* \left[ \bs d \phi_{|\phi} + \mathfrak{L}_{\bs d \bs \uppsi \circ \bs \uppsi\-} \phi + \delta_{\bs d \bs \upgamma \, \bs \upgamma\-} \phi + \delta_{(\mathfrak{L}_{\bs d \bs \uppsi \circ \bs \uppsi\-} \bs \upgamma)\, \bs \upgamma\-} \phi \right] \big( \mathfrak{X}_{|\phi} \big).
\end{aligned}
\end{equation}
Hence the final result, 
\begin{align}
\label{dphi-vert-trsf-loc}
    {\bs d\phi^{(\bs\uppsi, \bs\upgamma)}} 
    =
    \bs \uppsi^* 
    \Big[
    \Big( \bs d \phi + \mathfrak{L}_{\bs d \bs \uppsi \circ \bs \uppsi\-} \, \phi + \delta_{(\mathfrak{L}_{\bs d \bs \uppsi \circ \bs \uppsi\-} \bs \upgamma)\,  \bs \upgamma\-} \, \phi + \delta_{\bs d \bs \upgamma  \bs \upgamma\-} \, \phi \Big)^{\bs \upgamma}
    \Big],
\end{align}
To crosscheck this result one may compute it in another, less straightforward but still geometrical, way.
From \eqref{basis-1-form-prop-loc}  one can  derive
\begin{equation}
\begin{aligned}
    & R^\star_{(\uppsi,\upgamma)} \bs d \phi_{|R_{(\uppsi,\upgamma)}\phi} \big[ (X,\lambda)^v_{|\phi} \big] = \uppsi^* \Big( \mathfrak{L}_X \big( \phi^\upgamma \big) + \delta_{(\Ad_{\upgamma\-}\lambda \, - \, \upgamma\- \mathfrak{L}_X \, \upgamma)} \phi^\upgamma \Big) \\[1.5mm]
    & \hookrightarrow \big( \uppsi,\upgamma \big)^* \bs d \phi_{|\phi} \big[ \big( X, \lambda \big)^v_{|\phi} \big] = \uppsi^* \Big( \big( \mathfrak{L}_X \phi \big)^\upgamma \Big) + \uppsi^* \Big( \big( \delta_\lambda \phi \big)^\upgamma \Big)  = \uppsi^* \Big( \big( \mathfrak{L}_X \phi \big)^\upgamma \Big) + \uppsi^* \Big( \delta_{\Ad_{\upgamma\-}\lambda} \, \phi^\upgamma \Big),
\end{aligned}
\end{equation}
where, in the second line, we used the well-known identity
\begin{align}
\label{wellknown-id}
    \big( \delta_\lambda \phi \big)^\upgamma = \delta_{\Ad_{\upgamma\-}\lambda} \, \phi^\upgamma.
\end{align}
From this follows the useful identity
\begin{align}
\label{useful-Lie-id-loc}
    \mathfrak{L}_X \big( \phi^\upgamma \big) = \big( \mathfrak{L}_X \phi + \delta_{\mathfrak{L}_X \, \upgamma \, \cdot \, \upgamma\-} \phi \big)^\upgamma.
\end{align}
Then, the alternative way to compute  the $\bs\Diff_v(\Phi)\simeq C^\infty\big(\Phi, \Diff(M)\ltimes \Hl \big)$-transformation of the basis 1-form is
\begin{equation}
\begin{aligned}
{\bs d\phi^{(\bs\uppsi, \bs\upgamma)}}_{|\phi} \big( \mathfrak X_{|\phi} \big)
\defeq&\, \Xi^\star \bs d \phi_{|\,\Xi(\phi)} \big(\mathfrak{X}_{|\phi} \big) \\
=&\, \bs d \phi_{|\,\Xi(\phi)} \ \big( \Xi_\star \mathfrak{X}_{|\phi} \big) \\
=&\, \bs d \phi_{|\,\Xi(\phi)} \ \left( R_{(\bs \uppsi,\bs \upgamma)\star} \mathfrak{X}_{|\phi} + \left[\Big( \bs \uppsi\-_* \bs d \bs \uppsi , \bs \uppsi^* (\bs \upgamma\- \bs d \bs \upgamma ) \Big)_{|\phi} \big( \mathfrak{X}_{|\phi} \big) \right]^v_{|\,\Xi(\phi)} \right) \\
=&\, R^\star_{(\bs \uppsi,\bs \upgamma)}\, \bs d \phi_{|\,\Xi(\phi)} \big( \mathfrak{X}_{|\phi} \big) 
+ \bs d \phi_{|\,\Xi(\phi)} \left( \left[ \Big( \bs \uppsi\-_* \bs d \bs \uppsi, \bs \uppsi^* (\bs \upgamma\- \bs d \bs \upgamma) \Big)_{|\phi} \big(\mathfrak{X}_{|\phi} \big) \right]^v_{|\,\Xi(\phi)} \right) \\
=&\, \big( \bs \uppsi , \bs \upgamma \big)^* \bs d \phi_{|\phi}\, \big( \mathfrak{X}_{|\phi} \big) + \mathfrak{L}_{\bs \uppsi\-_* \bs d \bs \uppsi_{|\phi}(\mathfrak{X}_{|\phi})} \, \Xi(\phi) + \delta_{\left[\bs \uppsi^*(\bs \upgamma\- \bs d \upgamma)\right]_{|\phi} (\mathfrak{X}_{|\phi})} \, \Xi (\phi) \\
=&\, \bs \uppsi^* \Big( \big[\bs d \phi_{|\phi}\big]^{\bs \upgamma} \big( \mathfrak{X}_{|\phi}\big) \Big) + \mathfrak{L}_{\bs \uppsi\-_* \bs d \bs \uppsi_{|\phi}(\mathfrak{X}_{|\phi})} \, \bs \uppsi^* \big( \phi^{\bs \upgamma} \big) + \delta_{\left[\bs \uppsi^*(\bs \upgamma\- \bs d \upgamma)\right]_{|\phi} (\mathfrak{X}_{|\phi})} \, \bs \uppsi^* \big( \phi^{\bs \upgamma} \big)  \\
=&\, \bs \uppsi^* \Big( \big[\bs d \phi_{|\phi}\big]^{\bs \upgamma} \big( \mathfrak{X}_{|\phi}\big) \Big) + \bs \uppsi^* \Big( \mathfrak{L}_{\bs d \bs \uppsi_{|\phi} (\mathfrak{X}_{|\phi})\circ \bs \uppsi\-} \phi^{\bs \upgamma} \Big) + \uppsi^* \Big( \delta_{(\bs \upgamma\- \bs d \upgamma)_{|\phi} (\mathfrak{X}_{|\phi})} \, \phi^{\bs \upgamma}  \Big),
\end{aligned}    
\end{equation}
where \eqref{lemma1} is used for the second term of the last line. 
Using \eqref{wellknown-id} for the third term, we get the  result
\begin{align}
    {\bs d\phi^{(\bs\uppsi, \bs\upgamma)}} = \bs \uppsi^* \left( \bs d \phi^{\bs \upgamma} + \mathfrak{L}_{\bs d \bs \uppsi \circ \bs \uppsi\-} \, \phi^{\bs \upgamma} + \big( \delta_{\bs d \bs \upgamma  \bs \upgamma\-} \, \phi \big)^{\bs \upgamma} \right).
\end{align}
Using the identity \eqref{useful-Lie-id-loc} to rewrite the second term, we get the alternative form
\begin{align}
    {\bs d\phi^{(\bs\uppsi, \bs\upgamma)}} 
    =
    \bs \uppsi^* 
    \Big[
    \Big( \bs d \phi + \mathfrak{L}_{\bs d \bs \uppsi \circ \bs \uppsi\-} \, \phi + \delta_{(\mathfrak{L}_{\bs d \bs \uppsi \circ \bs \uppsi\-} \bs \upgamma) \,  \bs \upgamma\-} \, \phi + \delta_{\bs d \bs \upgamma  \bs \upgamma\-} \, \phi \Big)^{\bs \upgamma}
    \Big],
\end{align}
which matches \eqref{dphi-vert-trsf-loc}. 
It is interesting to see the semi-direct structure of $\Diff(M)\ltimes\Hl$ manifest.
These results encompass both the pure 
$\Diff(M)$ case -- see \cite{Francois2023-a} equation (66) -- and the pure $\Hl$ case (i.e.
``internal", Yang-Mills and/or Cartan geometric) -- see \cite{Francois-et-al2021} equation (7), also \cite{Francois2021}.\footnote{
In the last case, in \cite{Francois-et-al2021}, the notation $R^\star_{\gamma}\bs d\phi = \uprho(\gamma)\-\bs d \phi$ is used, where $\uprho =\big(\Ad, \rho \big)$ on $\bs d\phi=(\bs d A,\bs d\vphi)$, where $\vphi$ is a matter field transforming via the representation $\rho$. Indeed, $\bs dA$ transforms $\Ad$-tensorially under $\Hl$ because the space of gauge potentials (connections) is affine modeled on $Ad$-tensorial forms. 
}

\paragraph{Connections on local field space}
We now briefly cover the two notions of connections adapted to the covariant derivation of tensorial and twisted tensorial forms on the local field space $\Phi$. 
\medskip

\noindent $\bullet$ \emph{Ehresmann connection} 1-forms on $\Phi$ are $\bs\omega=\big(\bs \omega_\text{\tiny{$\Diff$}}, \bs\omega_\text{\tiny{$\Hl$}}\big) \in \Omega^1_\text{eq}\big(\Phi, \diff(M)\oplus\text{Lie}\Hl\big)$ s.t. 
\begin{equation}
\begin{aligned}
 \label{connection-loc}  
 \bs\omega_{|\phi}\,\big((X, \lambda)^v_{|\phi} \big) &= (X, \lambda) \quad \in  \diff(M)\oplus\text{Lie}\Hl, \\
 R^\star_{(\uppsi, \upgamma)} \bs\omega_{\,|R_{((\uppsi, \upgamma))}\phi} &= \Ad_{(\uppsi, \upgamma)\-}\,\bs\omega_{|\phi}.
\end{aligned}
\end{equation}
The equivariance property can be written more explicitly as 
\begin{align}
R^\star_{(\uppsi, \upgamma)} \big(\bs \omega_\text{\tiny{$\Diff$}}, \ \bs\omega_\text{\tiny{$\Hl$}}\big) = \left( \, \uppsi\-_* \, \bs \omega_\text{\tiny{$\Diff$}} \circ \uppsi , \ \, \uppsi^* \left( \Ad_{\upgamma\-}\, \bs\omega_\text{\tiny{$\Hl$}} - \upgamma\- \mathfrak{L}_{\bs \omega_\text{\tiny{$\Diff$}}} \, \upgamma \right) \, \right).
\end{align}
The infinitesimal equivariance is thus given by
\begin{equation}
\begin{aligned}
\bs L_{(X, \lambda)^v}\, \bs\omega & = \big[\bs\omega , (X, \lambda) \big]_\text{\tiny{Lie}},\\[1mm]
\bs L_{(X, \lambda)^v}\big(\bs \omega_\text{\tiny{$\Diff$}}, \ \bs\omega_\text{\tiny{$\Hl$}}\big) &= 
\left[\big(\bs \omega_\text{\tiny{$\Diff$}}, \bs\omega_\text{\tiny{$\Hl$}}\big), (X, \lambda) \right]_\text{\tiny{Lie}}\\
&= \Big(\,  [\omega_\text{\tiny{$\Diff$}},X]_{\text{\tiny{$\diff(M)$}}}\,,\  [\bs\omega_\text{\tiny{$\Hl$}}, \lambda]_\text{\tiny{Lie$H$}} -\bs\omega_\text{\tiny{$\Diff$}}(\lambda) + X\big(\bs\omega_\text{\tiny{$\Hl$}}\big)\, \Big),
\end{aligned}
\end{equation}
as is also checked via the formula  \eqref{Liebracket-local-aut} for the Lie   bracket.
The  associated curvature 2-form  $\bs\Omega=\big(\bs \Omega_\text{\tiny{$\Diff$}}, \bs\Omega_\text{\tiny{$\Hl$}}\big) \in \Omega^2_\text{tens}\big(\Phi, \Ad \big)$ is given by the Cartan structure equation
\begin{equation}
\begin{aligned}
\bs \Omega &= \bs d \bs \omega + \tfrac{1}{2} [\bs \omega,\bs \omega]_\text{\tiny{Lie}},\\[1mm]
\hookrightarrow\quad 
\Big( \bs \Omega_\text{\tiny{$\Diff$}} , \bs\Omega_\text{\tiny{$\Hl$}} \Big)
&=\left(
\bs d\bs\omega_\text{\tiny{$\Diff$}} + \tfrac{1}{2}\big[\bs \omega_\text{\tiny{$\Diff$}},\ \bs \omega_\text{\tiny{$\Diff$}}\big]_\text{\tiny{$\diff$(M)}}\, ,
\ \bs d \bs\omega_\text{\tiny{$\Hl$}} \!+ \tfrac{1}{2}\big[\bs \omega_\text{\tiny{$\Hl$}}, \bs \omega_\text{\tiny{$\Hl$}}\big]_\text{\tiny{Lie}}
- \bs\omega_\text{\tiny{$\Diff$}}(\bs\omega_\text{\tiny{$\Hl$}} )
\right).
\end{aligned}
\end{equation}
The connection allows to define a covariant derivative on tensorial forms
$\bs D: \Omega^\bullet_\text{tens}(\Phi, \rho) \rarrow \Omega^{\bullet+1}_\text{tens}(\Phi, \rho)$, $\alpha \mapsto \bs D\bs \alpha= \bs{d\alpha}+ \rho_*(\bs \omega)\,\bs\alpha=
\bs{d\alpha}+ \rho_*\big(\bs \omega_\text{\tiny{$\Diff$}}, \bs\omega_\text{\tiny{$\Hl$}}\big)\,\bs\alpha$.
In particular, $\bs\Omega$ satisfies the Bianchi identity  $\bs D \bs \Omega= \bs d \bs \Omega + [\bs \omega,\bs \Omega]_\text{\tiny{Lie}}\equiv 0$. 

The $\bs\Diff_v(\Phi)\simeq C^\infty\big(\Phi, \Diff(M)\ltimes \Hl \big)$-transformation of a connection are easily found via
by \eqref{connection-loc} and \eqref{pushforward-X-loc}:
\begin{equation}
\begin{aligned}
{\bs\omega^{(\bs\uppsi, \bs\upgamma)}}_{|\phi} \big( \mathfrak X_{|\phi} \big)
\defeq&\, \Xi^\star \bs\omega_{|\,\Xi(\phi)} \big(\mathfrak{X}_{|\phi} \big) \\
=&\, \bs\omega_{|\,\Xi(\phi)} \ \big( \Xi_\star \mathfrak{X}_{|\phi} \big) \\
=&\, \bs\omega_{|\,\Xi(\phi)} \ \left( R_{(\bs \uppsi,\bs \upgamma)\star} \mathfrak{X}_{|\phi} + \left[\Big( \bs \uppsi\-_* \bs d \bs \uppsi\, ,\ \bs \uppsi^* (\bs \upgamma\- \bs d \bs \upgamma ) \Big)_{|\phi} \big( \mathfrak{X}_{|\phi} \big) \right]^v_{|\,\Xi(\phi)} \right)\\
=&\, \Ad_{(\bs\uppsi, \bs\upgamma)\-}\,\bs\omega_{|\phi} \, \big(\mathfrak{X}_{|\phi} \big) 
+ \Big( \bs \uppsi\-_* \bs d \bs \uppsi \, ,\ \bs \uppsi^* (\bs \upgamma\- \bs d \bs \upgamma ) \Big)_{|\phi} \big( \mathfrak{X}_{|\phi} \big).
\end{aligned}
\end{equation}
Which, by the explicit form of the adjoint action,  gives
\begin{equation}
\begin{aligned}
 \bs \omega^{(\bs\uppsi,\bs\upgamma)} \defeq \Xi^\star \bs \omega 
&= \Ad_{(\bs\uppsi, \bs\upgamma)\-} \bs\omega + 
\Big(
\bs\uppsi\-_*\bs{d\uppsi}\, ,\ \bs\uppsi^*(\bs\upgamma\-\bs{d\upgamma} )
\Big), \\[2mm]
\hookrightarrow\quad \Big(\bs \omega_\text{\tiny{$\Diff$}}^{(\bs\uppsi,\bs\upgamma)} , \ \bs\omega_\text{\tiny{$\Hl$}}^{(\bs\uppsi,\bs\upgamma)}\Big) 
&= \left( \, \bs \uppsi\-_* \, \bs \omega_\text{\tiny{$\Diff$}} \circ \bs \uppsi + \bs\uppsi\-_*\bs{d\uppsi}\, , \ \, 
\bs\uppsi^* \left( \Ad_{\bs\upgamma\-} \bs\omega_\text{\tiny{$\Hl$}}\!+\bs\upgamma\-\bs{d\upgamma} \,  - \bs\upgamma\- \mathfrak{L}_{\bs\omega_\text{\tiny{$\Diff$}}} \, \bs\upgamma \right) \, \right).
\end{aligned}
\end{equation}
The transformations under $C^\infty \big(\Phi, \diff(M)\oplus\text{Lie}\Hl \big)$ --  and $\bs\diff(M)\oplus\text{\bf{Lie}}\Hl$ -- are, by \eqref{Inf-GT-eq-loc},
\begin{equation}
\begin{aligned}
\bs L_{(\bs X, \bs\lambda)^v} \bs\omega   
&= -\ad_{(\bs X, \bs\lambda)} \bs\omega+ \big(\bs{dX},\, \bs{d\lambda} \big) \\
&=\bs d(\bs X,\, \bs\lambda) + \big[\bs \omega, (\bs X, \bs \lambda) \big]_\text{\tiny{Lie}}, \\[2mm] 
\hookrightarrow \quad
\bs L_{(\bs X, \bs\lambda)^v} 
\big(
\bs \omega_\text{\tiny{$\Diff$}}, \ \bs\omega_\text{\tiny{$\Hl$}}
\big)
&= 
\Big(\, \bs{dX}+ [\omega_\text{\tiny{$\Diff$}},\bs X]_{\text{\tiny{$\diff(M)$}}}\,,\  
\bs{d\lambda} + [\bs\omega_\text{\tiny{$\Hl$}}, \bs\lambda]_\text{\tiny{Lie$H$}} -\bs\omega_\text{\tiny{$\Diff$}}(\bs\lambda) + \bs X\big(\bs\omega_\text{\tiny{$\Hl$}}\big)\, 
\Big).
\end{aligned}
\end{equation}
For $(\bs X, \bs \lambda) \in \bs\diff(M)\oplus\text{\bf{Lie}}\Hl$, i.e. tensorial 0-forms, one may legitimately write the result as their (geometric) covariant derivative: $\bs L_{(\bs X, \bs\lambda)^v} \bs\omega = D(\bs X, \bs \lambda)$. 

Similarly, the $\bs\Diff_v(\Phi)\simeq C^\infty\big(\Phi, \Diff(M)\ltimes \Hl \big)$-transformation of the curvature $\bs\Omega\in \Omega_\text{tens}^\bullet\big(\Phi, \Ad\big)$ is given by
\begin{equation}
\begin{aligned}
 \bs \Omega^{(\bs\uppsi,\bs\upgamma)} \defeq \Xi^\star \bs \Omega 
&= \Ad_{(\bs\uppsi, \bs\upgamma)\-} \bs\Omega , \\[2mm]
\hookrightarrow\quad \Big(\bs \Omega_\text{\tiny{$\Diff$}}^{(\bs\uppsi,\bs\upgamma)} , \ \bs\Omega_\text{\tiny{$\Hl$}}^{(\bs\uppsi,\bs\upgamma)}\Big) 
&= \left( \, \bs \uppsi\-_* \, \bs \Omega_\text{\tiny{$\Diff$}} \circ \bs \uppsi \, , \ \, 
\bs\uppsi^* \big(\Ad_{\bs\upgamma\-} \bs\Omega_\text{\tiny{$\Hl$}} - \bs\upgamma\- \mathfrak L_{\bs\Omega_\text{\tiny{$\Diff$}}} \bs\upgamma\big) \, \right),
\end{aligned}
\end{equation}
with linear version
\begin{equation}
\begin{aligned}
\bs L_{(\bs X, \bs\lambda)^v} \bs\Omega   
&= -\ad_{(\bs X, \bs\lambda)} \bs\Omega  
= \big[\bs \Omega, (\bs X, \bs \lambda) \big]_\text{\tiny{Lie}}, \\[1mm] 
\hookrightarrow \quad
\bs L_{(\bs X, \bs\lambda)^v} 
\big(
\bs \Omega_\text{\tiny{$\Diff$}}, \ \bs\Omega_\text{\tiny{$\Hl$}}
\big)
&= 
\Big(\, [\Omega_\text{\tiny{$\Diff$}},\bs X]_{\text{\tiny{$\diff(M)$}}}\,,\  
 [\bs\Omega_\text{\tiny{$\Hl$}}, \bs\lambda]_\text{\tiny{Lie$H$}} -\bs\Omega_\text{\tiny{$\Diff$}}(\bs\lambda) + \bs X\big(\bs\Omega_\text{\tiny{$\Hl$}}\big)\, 
\Big).
\end{aligned}
\end{equation}
More could be said, but we refrain from pursuing too much tangent observations.
\medskip

\noindent $\bullet$ \emph{Twisted connection} 1-forms on $\Phi$ are $\bs\varpi \in \Omega^1_\text{eq}\big(\Phi, \LieG\big)$ s.t. 
\begin{equation}
\begin{aligned}
 \label{twisted-connection-loc}  
 \bs\varpi_{|\phi}\,\big((X, \lambda)^v_{|\phi} \big) &= \tfrac{d}{d\tau}\, C\big(\phi; (\uppsi_\tau, \upgamma_\tau) \big)\, \big|_{\tau=0}= a\big( (X, \lambda); \phi \big) \ \  \in  \LieG, \\
 R^\star_{(\uppsi, \upgamma)} \bs\varpi_{\,|R_{((\uppsi, \upgamma))}\phi} &= \Ad_{C\left(\phi^{\phantom{|}}\!;\, (\uppsi, \upgamma)\right)\-}\,\bs\varpi_{|\phi} 
 + C\big(\phi; (\uppsi, \upgamma)\big)\- 
 \bs d C\big(\ \,; (\uppsi, \upgamma)\big)_{|\phi}.
\end{aligned}
\end{equation}
The infinitesimal equivariance is
\begin{equation}
\begin{aligned}
\bs L_{(X, \lambda)^v}\, \bs\varpi  
= \bs d a\big((X, \lambda);\ \ \big)
+
\big[\bs\varpi , a\big(\phi; (X, \lambda) \big)\big]_{\text{{\tiny $\LieG$}}}.
\end{aligned}
\end{equation}
The  twisted curvature 2-form  $\b{\bs\Omega}\in \Omega^2_\text{tens}\big(\Phi, C \big)$ is \emph{defined} by the Cartan structure equation
\begin{align}
\b{\bs\Omega} \defeq \bs d \bs\varpi + \tfrac{1}{2} [\bs\varpi,\bs\varpi]_{\text{{\tiny $\LieG$}}}.
\end{align}
The twisted connection allows to define a  covariant derivative on twisted forms
$\b{\bs D}: \Omega^\bullet_\text{eq}(\Phi, C) \rarrow \Omega^{\bullet+1}_\text{tens}(\Phi, C)$, $\alpha \mapsto \b{\bs D}\bs \alpha= \bs{d\alpha}+ \bs\varpi\,\bs\alpha$.
In particular, $\b{\bs\Omega}$ satisfies the Bianchi identity  $\b{\bs D} \b{\bs\Omega}= \bs d \b{\bs\Omega} + [\bs\varpi,\bs \Omega]_{\text{{\tiny $\LieG$}}}= 0$.

The vertical transformation under $\bs \Diff_v(\Phi)\simeq C^\infty\big(\phi,\Diff(M)\ltimes \Hl \big)$ of a twisted connection is
\begin{align}
\label{Vert-trsf-twisted-connection-loc}
    \bs \varpi^{(\bs \uppsi,\bs \upgamma)} \defeq \Xi^\star \bs \varpi = \Ad_{C(\bs \uppsi,\bs \upgamma)\-} \, \bs \varpi + C(\bs \uppsi,\bs \upgamma)\- \bs d C(\bs \uppsi,\bs \upgamma).
\end{align}
The infinitesimal version, under $\bs \diff_v(\Phi)\simeq C^\infty(\Phi, \diff(M)\oplus \text{Lie}\Hl)$, is given by
\begin{align}
\label{Inf-Vert-trsf-twisted-connection-loc}
\bs L_{(\bs X,\bs \lambda)^v} \bs\varpi =\bs{d} a(\bs X,\bs \lambda)+ [\bs\varpi, a(\bs X,\bs \lambda)]_{\text{{\tiny $\LieG$}}}. 
\end{align}
Finite and infinitesimal general vertical transformations of the curvature are given by, 
\begin{align}
\label{GT-twisted-curvature-loc}
\b{\bs \Omega}^{(\bs\uppsi, \bs\upgamma) }\defeq \Xi^\star \b{\bs\Omega} =\Ad_{C(\bs\uppsi, \bs\upgamma)\-} \, \b{\bs\Omega}, \qquad \text{so} \qquad \bs L_{(\bs X,\bs \lambda)^v}\b{\bs\Omega} =[\b{\bs\Omega}, a(\bs X,\bs \lambda)]_{\text{{\tiny $\LieG$}}}. 
\end{align}
Twisted geometry features in an essential way into gRGFT, as it controls the phenomenon of $\Hl$ and $\Diff(M)$ \emph{anomalies} \cite{Bertlmann, Bonora2023}.

\subsubsection{Associated bundle of regions and integration theory for local field theory}
\label{Associated bundle of regions and integration theory for local field theory}

\paragraph{Associated bundle of regions for local field space}

Given a representation space $(\rho, \bs V)$ for $\Diff(M) \ltimes \Hl$, and defining the right action
\begin{equation}
\label{action-dir-prod-phi-V}
\begin{aligned}
    (\Phi \times \bs V ) \times \big(\!\Diff(M) \ltimes \Hl\big)  & \rarrow  \Phi \times \bs V, \\
    \Big((\phi, v),\, (\uppsi, \upgamma) \Big) & \mapsto \left(R_{(\uppsi, \upgamma)}\phi, \, \rho(\uppsi, \upgamma)\- v \right) \rdefeq \b R_{(\uppsi, \upgamma)} (\phi, v),
\end{aligned}
\end{equation}
one defines the associated bundle
$\bs E \defeq \Phi \times_\rho \bs V \defeq \Phi \times \bs V / \sim$,
with $\sim$ the equivalence relation under  the right action $\b R$.
As usual one has the isomorphism of spaces
\begin{equation}
\label{iso-section-equiv-fct-loc}
\begin{aligned}
\Gamma(\bs E)\defeq \big\{ \bs s: \M \rarrow \bs E\big\}  \   \
\simeq \ \  \Omega^0_\text{eq}\big(\Phi, \rho \big)\defeq \left\{\, \bs\vphi: \Phi \rarrow \bs V\, |\,  R^\star_{(\uppsi, \upgamma)}\, \bs\vphi =\rho(\uppsi, \upgamma)\- \bs\vphi \, \right\} ,
\end{aligned}
\end{equation}
The same construction holds when $\bs V$ is a G-space and $\rho$ is replaced  by a 1-cocycle $C: \Phi \times \big(\!\Diff(M) \ltimes \Hl\big) \rarrow G$, $\big(\phi, (\uppsi, \upgamma)\big) \mapsto C\big(\phi;  (\uppsi, \upgamma) \big)$,  
so that one may build twisted associated bundles $\t E \defeq \Phi \times_C \bs V$ 
whose sections are twisted tensorial 0-forms.

One considers $\bs V=\bs U(M) \defeq \{ U \subset M \,|\, U \text{ open set} \}$, the $\sigma$-algebra of open sets of $M$, and the corresponding \emph{associated bundle of regions} of $M$,
\begin{align}
\label{Assoc-bundle-regions-loc}
\b{\bs U}(M) = \Phi \times_\text{\tiny{$\Diff(M) \ltimes \Hl$}} \bs U(M) \defeq \Phi \times \bs U(M)/ \sim 
\end{align}
where one the equivalence relation is under the right action
\begin{equation}
\label{right-action-bundle-region}
\begin{aligned}
    (\Phi \times \bs U(M) ) \times \big(\!\Diff(M) \ltimes \Hl\big)  & \rarrow  \Phi \times \bs U (M), \\
    \Big((\phi, U),\, (\uppsi, \upgamma) \Big) & \mapsto \left(R_{(\uppsi, \upgamma)} \phi,\, \uppsi\- (U) \right)
    =\left( \uppsi^*(\phi^\upgamma), \, \uppsi\- (U) \right)\rdefeq \b R_{(\uppsi, \upgamma)} (\phi, U).
\end{aligned}
\end{equation}
As a special case of the above, we have
\begin{equation}
\begin{aligned}
\Gamma(\b{\bs U}(M))\defeq \big\{ \bs s: \M \rarrow \b{\bs U}(M)\big\}  \   \
\simeq \ \  \Omega^0_\text{eq}\big(\Phi, \bs U(M) \big)\defeq \left\{\, \bs U: \Phi \rarrow  \bs U(M)\, |\,  R^\star_{(\uppsi, \upgamma)}\, \bs U =\uppsi\- (\bs U) \, \right\}.
\end{aligned}
\end{equation}
More explicitly, the equivariance can be written as 
\begin{align}
 \bs U\big(\uppsi^*(\phi^\upgamma)\big)  
 =\uppsi\-\big( \bs U(\phi) \big).
\end{align}
Such $\bs U$'s can be understood as field-dependent, or $\phi$-relative, regions of $M$.
Notice that, naturally, $\Hl$ has trivial action on $\bs U(M)$ since $M$ is defined as the space of fibers of $P$. 
So, $\bs U$'s are $\Hl$-basic 0-forms on $\Phi$, and project onto the $\Diff(M)$-subbundle $\Phi'\subset \Phi$. 

The transformations of these equivariant 0-forms under $\bs\Diff_v(\Phi) \simeq C^\infty\big( \Phi, \Diff(M)\ltimes \Hl\big)$ and $\bs\diff_v(\Phi) \simeq C^\infty\big( \Phi, \diff(M)\oplus\text{Lie}\Hl\big)$ are 
\begin{align}
    \bs U^{(\bs \uppsi,\bs \upgamma)} = \bs\uppsi\-(\bs U), \quad \text{ and } \quad \bs L_{(\bs X, \bs \lambda)^v} = - \bs X(\bs U).
\end{align}
In the following, we formulate integration on $M$ as a natural operation involving the 
above objects.

\paragraph{Integration in local field theory}
For pedagogical benefit, let us start from general considerations before specialising to our case of interest.
The action  \eqref{action-dir-prod-phi-V} induces the action  
\begin{equation}
\label{Diff_v-action-PxV-loc}
\begin{aligned}
\big(\Phi \times \bs V \big) \times C^\infty\big(\Phi, \Diff(M)\ltimes \Hl\big) &\rarrow \Phi \times \bs V, \\
\big((\phi,  v), (\bs\uppsi, \bs\upgamma) \big) & \mapsto  \Big(R_{(\bs\uppsi, \bs\upgamma)}\phi,\, \rho(\bs\uppsi, \bs\upgamma)\-  v\Big) =  \Big(\Xi(\phi),\, \rho(\bs\uppsi, \bs\upgamma)\-  v\Big) \rdefeq \b \Xi (\phi,  v). 
\end{aligned}
\end{equation}
The corresponding linearisation  is:
$\big(\Phi \times \bs V \big) \times C^\infty\big(\Phi, \diff(M)\oplus\text{Lie}\Hl\big) \rarrow V(\Phi \times \bs V)\simeq V\Phi \oplus V\bs V \subset T(P\times \bs V)$. 
The induced actions of $\Diff(M)\ltimes \Hl$ and $C^\infty\big(\Phi, \Diff(M)\ltimes \Hl\big)\simeq \bs\Diff_v(\Phi)$ on $ \Omega^\bullet(\Phi) \times \bs V$ are 
\begin{equation}
\begin{aligned}
\big( \Omega^\bullet(\Phi) \times \bs V \big) \times \big(\!\Diff(M)\ltimes \Hl\big) &\rarrow \Omega^\bullet(\Phi)\times \bs V, \\
\big((\bs\alpha,  v), (\uppsi, \upgamma) \big) & \mapsto  \Big(R^\star_{(\uppsi, \upgamma)} \bs\alpha,\, \rho(\uppsi, \upgamma)\-  v\Big)  \rdefeq \t R_\psi (\bs\alpha,  v)
\end{aligned}
\end{equation}
and 
\begin{equation}
\label{Diff_v-action-PxV1-loc}
\begin{aligned}
\big(\Omega^\bullet(\Phi) \times \bs V \big) \times C^\infty\big(\Phi, \Diff(M)\ltimes \Hl\big) &\rarrow \Omega^\bullet(\Phi) \times \bs V, \\
\big((\bs\alpha,  v), (\bs\uppsi, \bs\upgamma) \big) & \mapsto  \big(\Xi^\star \bs\alpha ,\, \rho(\bs\uppsi, \bs\upgamma)\-  v\big)  \rdefeq \t \Xi (\bs\alpha,  v). 
\end{aligned}
\end{equation}
The induced actions of $\diff(M)\oplus\text{Lie}\Hl$ and $C^\infty\big(\Phi, \diff(M)\oplus\text{Lie}\Hl\big)\simeq \bs\diff_v(\Phi)$ are the linearisations:
\begin{equation}
\label{linear-versions-loc}
\begin{aligned}
\big((\bs\alpha,  v), (X, \lambda \big) & \mapsto  \tfrac{d}{d\tau}\, \t R_{(\uppsi_\tau, \upgamma_\tau)} (\bs\alpha,  v) \,\big|_{\tau=0} = \big(\bs L_{(X, \lambda)^v} \bs\alpha,\,  v\big) \oplus \big( \bs\alpha,\, -\rho_*(X, \lambda)  v\big), \\
\big((\bs\alpha,  v), (\bs X, \bs\lambda) \big) & \mapsto  \tfrac{d}{d\tau}\, \t \Xi_\tau (\bs\alpha,  v)\, \big|_{\tau=0} = \big(\bs L_{(\bs X, \bs\lambda)^v} \bs\alpha,\,  v\big) \oplus \big( \bs\alpha,\, -\rho_*(\bs X, \bs\lambda)  v\big).
\end{aligned}
\end{equation}
Given a representation space $(\t \rho, \bs W)$ of $\Diff(M)\ltimes \Hl$,
\begin{equation}
\begin{aligned}
&\text{if }\ \bs\alpha \in \Omega^\bullet_{\text{eq}}(\Phi, \bs W)  \  \text{ then } \   
 \t R_\psi (\bs\alpha,  v) =  \big(R^\star_{(\uppsi,\upgamma)} \bs\alpha,\, \rho(\uppsi,\upgamma)\-  v\big)=  \big( \t\rho(\uppsi,\upgamma)\- \bs\alpha,\, \rho(\uppsi,\upgamma)\-  v\big), \\
&\text{if }\ \bs\alpha \in \Omega^\bullet_{\text{tens}}(\Phi, \bs W)  \  \text{ then } \   
 \t \Xi (\bs\alpha,  v) =  \big(\Xi^\star \bs\alpha,\, \,\rho(\bs\uppsi, \bs\upgamma)\-  v\big)=  \big( \t\rho(\bs\uppsi, \bs\upgamma)\- \bs\alpha,\,\, \rho(\bs\uppsi, \bs\upgamma)\-  v\big),
\end{aligned}
\end{equation}
with linearisations being read from \eqref{linear-versions-loc}.
The exterior derivative $\bs d$ on $\Phi$  extends to $\Phi \times \bs V$ as ${\bs d} \rarrow  \bs d \times  \id$. 
But applying after the action of  $C^\infty\big(\Phi, \diff(M)\oplus \text{Lie}\Hl\big)\simeq \bs\diff_v(\Phi)$, due to the $\phi$-dependence of $(\bs\uppsi, \bs\upgamma)$, it will  act on the  factor $\rho(\bs\uppsi, \bs\upgamma)\-  v$. 

Consider $(\b \rho, \bs V^*)$ a representation of $\Diff(M)\ltimes \Hl$ \emph{dual} to $(\rho, \bs V)$ w.r.t. a non-degenerate 
\emph{pairing} 
\begin{equation}
\begin{aligned}
\langle\ , \ \rangle: \bs V^* \times \bs V &\rarrow \RR, \\
 (w, v) &\mapsto \langle w , v \rangle. 
\end{aligned}
\end{equation}
If it is $\big(\!\Diff(M)\ltimes \Hl\big)$-invariant, it satisfies
\begin{equation}
\begin{aligned}
\label{inv-pairing-loc}
&\langle\, \b \rho(\uppsi,\upgamma) w, \rho(\uppsi,\upgamma)v \rangle 
 =  \langle w , v \rangle, \\[1mm]
\hookrightarrow \quad &\langle\, \b \rho_*(X, \lambda) w ,   v \rangle + \langle\,  w ,   \rho_*(X, \lambda)v \rangle =0,
\end{aligned}
\end{equation}
 where the second line features the induced representation $\b \rho_*$ and $\rho_*$ for the action of $\diff(M)\oplus\text{Lie}\Hl$. 
For $\bs\alpha \in \Omega^\bullet(\Phi, \bs V^*)$, we define an operation $\mathcal I$ 
on $\Omega^\bullet(\Phi, \bs V^*) \times \bs V$ by
\begin{equation}
\begin{aligned}
\mathcal I: \Omega^\bullet(\Phi, \bs V^*)  \times \bs V &\rarrow \Omega^\bullet(\Phi),\\
					(\bs\alpha, v) &\mapsto \mathcal I(\bs\alpha, v) \defeq  \langle \bs\alpha , v \rangle,
\end{aligned}
\end{equation}
which can be understood as an object on $\Phi \times \bs V$ by
\begin{equation}
\begin{aligned}
\mathcal I(\bs\alpha, \ ): \Phi \times \bs V &\rarrow \Lambda^{\!\bullet}(\Phi),\\
					(\phi, v) &\mapsto \mathcal I(\bs\alpha_{|\phi}, v) \defeq  \langle \bs\alpha_{|\phi} , v \rangle. 
\end{aligned}
\end{equation}
It is therefore the case that
\begin{align}
\label{der-eval-integr-obj-loc}
\bs d \mathcal I(\bs\alpha, \ ) = \mathcal I(\bs{d\alpha}, \ ) \quad \text{ and } \quad \iota_{\mathfrak X}  \mathcal I(\bs\alpha, \ ) = \mathcal I( \iota_{\mathfrak X} \bs\alpha, \ ), \quad \text{ for } \mathfrak X \in \Gamma(T\Phi).
\end{align}
The induced actions of $\Diff(M)\ltimes \Hl$ and $C^\infty\big(\Phi, \Diff(M)\ltimes \Hl \big) \simeq \bs\Diff_v(\Phi)$ on such objects are:
\begin{equation}
\label{Grps-action-loc}
\begin{aligned}
 &\t R^\star_\psi \mathcal I(\bs\alpha,\   )_{|(R_{(\uppsi, \upgamma)}\phi,\  \rho(\uppsi, \upgamma)\- v)} \defeq  \langle\ , \ \rangle \circ \t R_{(\uppsi, \upgamma)} (\bs \alpha, v) = \langle R^\star_{(\uppsi, \upgamma)} \bs\alpha_{|\, R_{(\uppsi, \upgamma)}\phi},\, \rho(\uppsi, \upgamma)\-  v\rangle, \\
& \t \Xi^\star \mathcal I(\bs\alpha,\   )_{|(\Xi (\phi),\  \rho(\bs\uppsi, \bs\upgamma)\- v)} \defeq  \langle\ , \ \rangle \circ \t \Xi (\bs \alpha, v) = \langle \Xi^\star \bs\alpha_{|\Xi(\phi)},\, \rho(\bs\uppsi, \bs\upgamma)\-  v\rangle. 
 \end{aligned}
\end{equation}
The actions of $\diff(M)\oplus\text{Lie}\Hl$ and $C^\infty\big(\Phi, \diff(M)\oplus\text{Lie}\Hl\big)\simeq \bs\diff_v(\Phi)$ are
\begin{equation}
\label{linear-versions-pairing-loc}
\begin{aligned}
&\tfrac{d}{d\tau}\, \t R_{(\bs\psi_\tau, \bs\upgamma_\tau)}^\star \mathcal I (\bs\alpha,  v) \,\big|_{\tau=0} 
= \langle \bs L_{(X, \lambda)^v} \bs\alpha,\,  v \rangle 
 + \langle \bs\alpha,\, -\rho_*(X, \lambda)  v \rangle, \\
&\tfrac{d}{d\tau}\, \t \Xi_\tau^\star \mathcal I  (\bs\alpha,  v)\, \big|_{\tau=0} = \langle \bs L_{(\bs X, \bs\lambda)^v} \bs\alpha,\,  v \rangle 
 + \langle \bs\alpha,\, -\rho_*(\bs X, \bs\lambda)  v \rangle.
\end{aligned}
\end{equation}
For $\bs\alpha \in \Omega^\bullet_{\text{eq}}(\Phi, \bs V^*)$ we have
\begin{equation}
\label{Int-eq-loc}
\begin{aligned}
 &\t R^\star_{(\uppsi, \upgamma)} \mathcal I(\bs\alpha,\ \,  )_{|\,(R_{(\uppsi, \upgamma)}\phi,\  \rho(\uppsi, \upgamma)\- v)} \defeq 
 \, \langle R^\star_{(\uppsi, \upgamma)} \bs\alpha_{|R_{(\uppsi, \upgamma)}\phi},\ \, \rho(\uppsi, \upgamma)\-  v\rangle 
 =\,  \langle \b\rho(\uppsi, \upgamma)\- \bs\alpha_{|\phi},\ \,\rho(\uppsi, \upgamma)\-  v\rangle,
 \\[1mm]
 &\hookrightarrow\quad \langle \bs L_{(X, \lambda)^v} \bs\alpha,\,  v \rangle 
 + \langle \bs\alpha,\, -\rho_*(X, \lambda)  v \rangle = \langle  -\b \rho_*(X,\lambda)\, \bs\alpha,\,  v \rangle + \langle \bs\alpha,\, -\rho_*(X,\lambda)  v \rangle.
 \end{aligned}
\end{equation}
If the pairing is invariant, by \eqref{inv-pairing-loc} we then get 
\begin{equation}
\begin{aligned}
\t R^\star_{(\uppsi, \upgamma)} \mathcal I(\bs\alpha,\ \,  )_{|\,(R_{(\uppsi, \upgamma)}\phi,\  \rho(\uppsi, \upgamma)\- v)} = \langle  \bs\alpha_{|\phi},  v\rangle &\rdefeq \mathcal I(\bs\alpha,\ )_{|(\phi, v)}, \\[1mm] 
 \hookrightarrow\quad \langle  -\b \rho_*(X,\lambda)\, \bs\alpha,\,  v \rangle + \langle \bs\alpha,\, -\rho_*(X,\lambda)  v \rangle &= 0.
\end{aligned}
\end{equation}
If $\bs\alpha \in \Omega^\bullet_{\text{tens}}(\Phi, \bs V^*)$:
 \begin{equation}
  \label{Int-tens-loc}
\begin{aligned}
 &\t \Xi^\star \mathcal I(\bs\alpha,\   )_{|(\Xi (\phi),\  \rho(\uppsi,\upgamma)\- v)} \defeq 
  \langle \Xi^\star \bs\alpha_{|\Xi(\phi)},\, \rho(\bs\uppsi,\bs\upgamma)\-  v\rangle 
 =  \langle \b\rho(\bs\uppsi,\bs\upgamma)\- \bs\alpha_{|\phi},\, \rho(\bs\uppsi,\bs\upgamma)\-  v\rangle,\\[1mm]
 &\hookrightarrow\quad \langle \bs L_{(\bs X, \bs\lambda)^v} \bs\alpha,\,  v \rangle 
 + \langle \bs\alpha,\, -\rho_*(\bs X, \bs\lambda)  v \rangle = \langle  -\b \rho_*(\bs X, \bs\lambda)\, \bs\alpha,\,  v \rangle + \langle \bs\alpha,\, -\rho_*(\bs X,\bs\lambda)  v \rangle.
 \end{aligned}
\end{equation}
If the pairing is invariant, by \eqref{inv-pairing-loc} we get
\begin{equation}
\label{Field-dep-trsf-inv-pair}
\begin{aligned}
&\t \Xi^\star \mathcal I(\bs\alpha,\   )_{|(\Xi (\phi),\  \rho(\uppsi,\upgamma)\- v)}= \langle  \bs\alpha_{|\phi},  v\rangle \rdefeq \mathcal I(\bs\alpha,\ )_{|(\phi, v)} , \\[1mm]
& \hookrightarrow\quad  \langle  -\b \rho_*(\bs X,\bs \lambda)\, \alpha,\,  v \rangle + \langle \bs\alpha,\, -\rho_*(\bs X,\bs \lambda)  v \rangle = 0.
\end{aligned}
\end{equation} 
Whenever $\bs\alpha \in \Omega^\bullet_{\text{tens}}(\Phi, \bs V^*)$  and the pairing is invariant, $\mathcal I(\bs\alpha,\ )$ is then ``basic" on $\Phi\times \bs V$, inducing a well-defined object on $\bs E=\Phi \times \bs V/\sim$. 
Being  constant along a $\big(\!\Diff(M)\ltimes \Hl\big)$-orbit in $\Phi \times \bs V$,  $\mathcal I(\bs \alpha,\ )$ allows to define 
$\bs\vphi_{\mathcal I(\bs \alpha)}\! \in \Omega^0_\text{eq}(\Phi, \rho)$ 
 via
 \begin{equation}
 \label{Induced-equiv-fct-loc}
\begin{aligned}
\bs\vphi_{\mathcal I(\bs \alpha)}(\phi)&\defeq \pi_{\bs V}(\phi,  v)_{|\mathcal I(\bs\alpha_{|\phi}, v) =\text{cst}} \equiv v, \\
\bs\vphi_{\mathcal I(\bs \alpha)}\big(R_{(\uppsi, \upgamma)}\phi\big)& \defeq \pi_{\bs V}\big(R_{(\uppsi, \upgamma)}\phi, \rho(\psi)\- v\big)_{|\mathcal I(\bs\alpha_{|\phi}, v) =\text{cst}} \equiv \rho(\uppsi, \upgamma)\- v.
 \end{aligned}
\end{equation}
As expressed by \eqref{iso-section-equiv-fct-loc}, the latter is equivalent to a section $\bs s_{\mathcal I(\bs \alpha)} :\M \rarrow \bs E$. 
By equation \eqref{Field-dep-trsf-inv-pair} one can write
$\bs d \, \t \Xi^\star \mathcal I(\bs\alpha,\   ) = \bs d \mathcal I(\bs\alpha,\   )= \mathcal I(\bs{d\alpha},\   )$. 
In that case, one also finds the following lemma to hold:
\begin{align}
  \t\Xi^\star \langle \bs{d\alpha}, v \rangle &\defeq \langle  \Xi^\star \bs d \bs\alpha, \, \rho(\bs\uppsi, \bs\upgamma)\-v \rangle 
  = \langle  \bs d \,  \Xi^\star  \bs\alpha, \, \rho(\bs\uppsi, \bs\upgamma)\-v \rangle \notag \\
  &= \langle  \bs d \, \big(\b\rho(\bs\uppsi, \bs\upgamma)\-  \bs\alpha\big), \, \rho(\bs\uppsi, \bs\upgamma)\-v \rangle \notag \\
&= \left\langle  \b\rho(\bs\uppsi, \bs\upgamma)\- \Big(  \bs d \bs \alpha -\b\rho_*\big(\bs d(\bs\uppsi, \bs\upgamma)\cdot (\bs\uppsi, \bs\upgamma)\-\big)\, \bs\alpha \Big),\, \rho(\bs\uppsi, \bs\upgamma)\-v\right\rangle \notag \\ 
&= \langle    \bs d \bs \alpha -\b\rho_*\big((\bs{d\uppsi}, \bs{d\upgamma})\cdot (\bs\uppsi, \bs\upgamma)\-\big)\, \bs\alpha,\, v\rangle,\notag\\[2mm]
\hookrightarrow\quad 	 \t\Xi^\star \langle \bs{d\alpha}, v \rangle 	
&= \langle    \bs d \bs \alpha, v \rangle+  \langle -\b\rho_*\big((\bs{d\uppsi}, \bs{d\upgamma})\cdot (\bs\uppsi, \bs\upgamma)\-\big)\, \bs\alpha,\, v\rangle  
\notag\\
&=\langle    \bs d \bs \alpha, v \rangle 
 +  
 \left\langle 
 -\b\rho_*\Big( \bs{d\uppsi} \circ \bs\uppsi\-,\ 
\bs{d\upgamma} \, \bs\upgamma\-  - \bs\upgamma\, \mathfrak L_{\bs{d\uppsi} \circ \bs\uppsi\-} \bs\upgamma\- \Big)\, \bs\alpha, 
 \, v\,
 \right\rangle, 
\label{Vert-trsf-pairing-dalpha-loc} 
\end{align}
using the result \eqref{linear-right-action-loc} in the last line.
We will see shortly a useful application of this result. 
\medskip

The above construction specialises to the fundamental representation spaces  $\bs V=\bs U(M)$ 
and  $\bs V^* = \Omega^\text{top}(U)$ (top forms on $U \in \bs U(M)$) for the action of $\Diff(M)\ltimes \Hl$. 
They are dual under the \emph{integration pairing}:
\begin{equation}
\begin{aligned}
\langle\ \,, \ \rangle: \bs \Omega^\text{top}(U) \times \bs U(M) &\rarrow \RR, \\[-2mm]
(\omega, U) &\mapsto \langle \omega , U \rangle \defeq \int_U \omega.
\end{aligned}
\end{equation}
This  pairing is invariant if and only if $\omega$ is $\Hl$-invariant, $\omega^\upgamma=\omega$, since $\Hl$ does not act on $U\in \bs U(M)$. 
We have then the special case of \eqref{inv-pairing-loc}, for $\b\rho(\uppsi, \upgamma)\-=\uppsi^*$ and $\rho(\uppsi, \upgamma)\-(U)=\uppsi\-(U)$, yielding the  identity:
 \begin{align}
 \label{invariance-int-loc}
\langle\, \uppsi^*\omega ,   \uppsi\-(U) \rangle =  \langle \omega , U \rangle \quad   \rarrow  \quad  \int_{\uppsi\-(U)} \uppsi^*\omega = \int_U \omega, 
 \end{align}
 reproducing the well-known $\Diff(M)$-invariance of  integration as an intrinsic operation on $M$.
 This implies, for $-\b \rho_*(X,\lambda)=\mathfrak L_X$ and $-\rho_*(X,\lambda)(U)=-X(U)$:
  \begin{align}
 \label{inf-invariance-action-loc}
\langle\, \mathfrak L_X \omega ,   U \rangle + \langle\,  \omega ,   -X(U) \rangle =0  \quad
\rarrow \quad
  \int_U   \mathfrak L_X \omega  + \int_{-X(U)} \hspace{-2mm}\omega =0,
 \end{align}
which can be understood as a sort of   \emph{continuity equation} for the action of $\diff(M)$.  
 By Stokes theorem,
the de Rham derivative $d$ on $\Omega^\bullet(U)$ and the boundary operator $\d$ on $\bs U(M)$ are  adjoint operators w.r.t. to the integration pairing:
 \begin{align}
 \label{Stokes}
 \langle\, d \omega ,  U \rangle = \langle\,  \omega ,   \d U \rangle \quad
 \rarrow \quad
  \int_U   d \omega  = \int_{\d U} \hspace{-2mm}\omega. 
\end{align} 
 
Considering  $\bs \alpha \in \Omega^\bullet\big( \Phi,  \Omega^\text{top}(U) \big)$, the field-dependent top forms, we define the integration map on $\Phi \times \bs U(M)$:
\begin{align}
\mathcal I(\bs\alpha_{|\phi}, U) = \langle \bs\alpha_{|\phi}, U \rangle \defeq \int_U \bs \alpha_{|\phi}.
\end{align}
We may use  the notation $\bs\alpha_U$   when convenient.  
Equation \eqref{der-eval-integr-obj-loc} holds here as a special case.
The induced actions of $\Diff(M)\ltimes \Hl$ and $C^\infty\big(\Phi, \Diff(M)\ltimes \Hl \big) \simeq \bs\Diff_v(\Phi)$ on integrals are, respectively,
\begin{equation}
\label{Vert-trsf-int-generic-loc}
\begin{aligned}
 &\t R^\star_{(\uppsi,\upgamma)} \mathcal I(\bs\alpha,\   )_{|\,(R_{(\uppsi,\upgamma)} \phi,\ \uppsi\-(U) )} \defeq  \langle R^\star_{(\uppsi,\upgamma)} \bs\alpha_{|\,R_{(\uppsi,\upgamma)}\phi},\, \uppsi\-(U) \rangle = \int_{ \uppsi\-(U)} R^\star_{(\uppsi,\upgamma)} \bs\alpha_{|\,\uppsi^*(\phi^{\upgamma})} ,  \\
& \t \Xi^\star \mathcal I(\bs\alpha,\   )_{|\,(\Xi (\phi),\  \bs\uppsi\-(U) )} \defeq  \langle \Xi^\star \bs\alpha_{|\,\Xi(\phi)},\, \bs\uppsi\-(U)  \rangle = \int_{ \bs\uppsi\-(U)}  \Xi^\star \bs\alpha_{|\,\bs\uppsi^*(\phi^{\bs\upgamma})}. 
 \end{aligned}
\end{equation}
We may write the above as ${\bs\alpha_U}^{(\uppsi,\upgamma)}$ and ${\bs\alpha_U}^{(\bs\uppsi,\bs\upgamma)}$, respectively. 
Applying $\bs d$ on the second line, it will  also act on the transformed region $\bs\uppsi\-(U)$ due to the $\phi$-dependence of~$\bs\uppsi$. 
One shows indeed that
\begin{align}
\label{commut-bXi-d-int}
\bs d \left( \b\Xi^\star \langle \bs\alpha, U \rangle \right)
&=
\langle \bs d \Xi^\star \bs\alpha, \bs\uppsi\-(U) \rangle 
+
\langle \Xi^\star \bs\alpha, \bs d\bs\uppsi\-(U)\rangle \notag\\
&=
\langle \Xi^\star \bs d\bs\alpha, \bs\uppsi\-(U) \rangle 
+
\langle \Xi^\star \bs\alpha, - \bs\uppsi
\-_* \bs d\bs\uppsi\circ \bs\uppsi\-
(U) \rangle \notag \\
&=
\b \Xi^\star \langle  \bs d\bs\alpha, U \rangle 
-
\langle \mathfrak L_{ \bs\uppsi
\-_* \bs d\bs\uppsi}\Xi^\star \bs\alpha, 
\bs\uppsi\-(U) \rangle, \notag\\[.5mm]
\text{or}\quad
\bs d\big({\bs\alpha_U}^{(\bs\uppsi,\bs\upgamma)} \big)
&=
\left( \bs d\bs\alpha_U\right)^{(\bs\uppsi,\bs\upgamma)}
- \langle \mathfrak L_{ \bs\uppsi
\-_* \bs d\bs\uppsi}\Xi^\star \bs\alpha, 
\bs\uppsi\-(U)  \rangle.
\end{align}
The identity \eqref{inf-invariance-action-loc} has been used to conclude.
This means that on $\Phi \times \bs U(M)$ we have $[\b{\bs \Xi}^\star, \bs d]\neq 0$ and the commutator is a boundary term, $\langle \iota_{ \bs\uppsi
\-_* \bs d\bs\uppsi}\Xi^\star \bs\alpha , \d\big( \bs\uppsi\-(U)\big) \rangle$, since $\bs\alpha$ is a top form  on $U$ and by \eqref{Stokes}. This should be contrasted to the standard relation $[{\bs \Xi}^\star, \bs d] =  0$ holding on $\Phi$.

From \eqref{Vert-trsf-int-generic-loc} the induced actions of $\diff(M) \oplus\text{Lie}\Hl$ and $C^\infty\big(\Phi, \diff(M) \oplus\text{Lie}\Hl\big)\simeq \bs\diff_v(\Phi)$ on integrals are 
\begin{equation}
\label{linear-versions-integral-loc}
\begin{aligned}
&\tfrac{d}{d\tau}\, \t R_{(\uppsi_\tau, \upgamma_\tau)}^\star \mathcal I (\bs\alpha,  U) \,\big|_{\tau=0} = \langle \bs L_{(X, \lambda)^v} \bs\alpha,\,  U \rangle 
+ \langle \bs\alpha,\, -X(U) \rangle
 =\int_U  \bs L_{(X, \lambda)^v} \bs\alpha + \int_{-X(U)} \bs\alpha, \\
&\tfrac{d}{d\tau}\, \t \Xi_\tau^\star \mathcal I  (\bs\alpha,  U)\, \big|_{\tau=0} = \langle \bs L_{(\bs X, \bs\lambda)^v} \bs\alpha,\,  U \rangle 
 + \langle \bs\alpha,\, -\bs X(U)   \rangle
=\int_U  \bs L_{(\bs X, \bs\lambda)^v} \bs\alpha + \int_{-\bs X(U)} \bs\alpha.  
\end{aligned}
\end{equation}
If convenient, we may use the notation $\delta_{(X, \lambda)}\ \!{\bs\alpha_U}$ and $\delta_{(\bs X, \bs\lambda)}\ \!{\bs\alpha_U}$ for the above results. 

We shall consider in particular equivariant  and tensorial forms for the natural action of $\Diff(M)\ltimes \Hl$, satisfying respectively
\begin{equation}
\label{equi-tens-forms-loc}
\begin{aligned}
R^\star_{(\uppsi, \upgamma)} \bs\alpha &= 
\bs\alpha^{(\uppsi, \upgamma)}
=\b\rho(\uppsi, \upgamma)\-\bs\alpha
=\uppsi^*(\bs\alpha^\upgamma), \qquad 
\bs L_{(X, \lambda)^v}\, \bs\alpha 
=-\b\rho_*(X, \lambda)\,\bs\alpha
= \mathfrak L_X \bs\alpha + \delta_{\!\lambda}\, \bs\alpha,
\\
\Xi^\star_{(\uppsi, \upgamma)} \bs\alpha &= \bs\alpha^{(\bs\uppsi, \bs\upgamma)}
=\b\rho(\bs\uppsi, \bs\upgamma)\-\bs\alpha
=\bs\uppsi^*(\bs\alpha^{\bs\upgamma}), \qquad 
\bs L_{(\bs X, \bs\lambda)^v}\, \bs\alpha 
=-\b\rho_*(\bs X, \bs\lambda)\,\bs\alpha
= \mathfrak L_{\bs X} \bs\alpha + \delta_{\!\bs\lambda}\, \bs\alpha.
\end{aligned}
\end{equation}
Then we have that 
\begin{align}
\label{d-trsf-int-special}
d\big({\bs\alpha_U}^{(\bs\uppsi,\bs\upgamma)} \big) 
=
\bs d \langle \bs\uppsi^*(\bs\alpha^{\bs\upgamma}), \bs \uppsi\-(U)\rangle 
=
\bs d \langle \bs\alpha^{\bs\upgamma}, U\rangle 
=
\langle \bs d  \big(\bs\alpha^{\bs\upgamma} \big), U\rangle.
\end{align}
And the relation \eqref{commut-bXi-d-int} specialises to 
\begin{equation}
\label{commut-bXi-d-special}
\begin{aligned}
\bs d\big({\bs\alpha_U}^{(\bs\uppsi,\bs\upgamma)} \big)
&=
\left( \bs d\bs\alpha_U\right)^{(\bs\uppsi,\bs\upgamma)}
- \langle \mathfrak L_{ \bs\uppsi
\-_* \bs d\bs\uppsi}\, \bs\uppsi^*\big(\bs\alpha^{\bs\upgamma} \big), 
 \bs\uppsi\-(U)  \rangle \\
&=
\left( \bs d\bs\alpha_U\right)^{(\bs\uppsi,\bs\upgamma)}
-
\langle \bs\uppsi^* \mathfrak L_{\bs d\bs\uppsi \circ \bs \uppsi\-}\big(\bs\alpha^{\bs\upgamma} \big), 
\bs\uppsi\-(U)  \rangle \\
&=
\left( \bs d\bs\alpha_U\right)^{(\bs\uppsi,\bs\upgamma)}
-
\langle  \mathfrak L_{\bs d\bs\uppsi \circ \bs \uppsi\-}\big(\bs\alpha^{\bs\upgamma} \big), 
U  \rangle.
\end{aligned}
\end{equation}
Taken together the last two results imply to derive the following lemma:
\begin{align}
\bs d \big( \bs\uppsi^* \bs\alpha \big)
=
\bs\uppsi^*\big( \bs d\bs\alpha +  \mathfrak L_{\bs d\bs\uppsi \circ \bs \uppsi\-} \bs\alpha \big).
\end{align}

When $\bs\alpha$ is  $\Diff(M)$-tensorial \emph{and} $\Hl$-basic on $\Phi$ -- i.e. $\bs\alpha^{\bs\gamma}=\bs\alpha$, and $\delta_{\bs\lambda}\bs\alpha=0$ --  equation \eqref{Field-dep-trsf-inv-pair} holds
and we have that  $\bs\alpha_U=\mathcal I(\bs\alpha_{|\phi}, U)$ is $C^\infty\big(\Phi, \Diff(M)\ltimes \Hl \big)$-invariant:  ${\bs\alpha_U}^{(\bs \uppsi, \bs\upgamma)} =\bs\alpha_U$, and 
\begin{equation}
\label{Cont-eq-general-case-loc}
\begin{aligned}
\delta_{(\bs X, \bs\lambda)} \bs\alpha_U =0 \quad \Rightarrow \quad \langle \bs L_{(\bs X, \bs\lambda)^v} \bs\alpha,\,  U \rangle + \langle \bs\alpha,\, -\bs X(U) \rangle &= 0, \\
  \langle  \mathfrak L_{\bs X} \alpha,\,  U \rangle + \langle \bs\alpha,\, -\bs X(U)   \rangle &= 0 \quad \rarrow \quad \int_U \mathfrak L_{\bs X} \alpha + \int_{-\bs X(U)} \bs\alpha =0.
\end{aligned}
\end{equation}
 For $\bs\alpha$ $\Diff(M)$-equivariant only, and $\Hl$-basic, we have invariance of its integral: ${\bs\alpha_U}^{(\uppsi, \upgamma)} =\bs\alpha_U$, and \eqref{Cont-eq-general-case-loc} holds substituting $(\bs X, \bs \lambda) \rarrow (X, \lambda)$. 
 
For $\bs\alpha$ $\Diff(M)$-tensorial \emph{and} $\Hl$-basic, we thus have that $\bs\alpha_U=\mathcal I(\bs\alpha, U)$ induces a  well-defined object on the bundle of regions $\b{\bs U}(M)=\Phi \times \bs U(M)/\sim$ (so  one may  define an associated equivariant $\bs U(M)$-valued function on $\Phi$). 
 It also holds that $\bs d\left({\bs\alpha_U}^{(\bs\uppsi,\bs\upgamma)}\right) = \bs d \bs\alpha_U$, i.e. 
\begin{align}
\label{Invariance-tens-int-loc}
\bs d \, \langle \bs\alpha^{(\bs\uppsi,\bs\upgamma)}, \bs\uppsi\-(U)\rangle =   \langle \bs d \bs\alpha, U\rangle = \bs d  \langle  \bs\alpha, U\rangle \quad \rarrow \quad
\bs d \int_{\bs\uppsi\-(U)}  \bs\uppsi^* \bs\alpha = \int_U \bs{d\alpha} = \bs d \int_U \bs \alpha.
\end{align}
Besides, specialising \eqref{Vert-trsf-pairing-dalpha-loc}, we get the identities
\begin{equation}
\label{Vert-trsf-int-dalpha-loc}
\begin{aligned}
\bs (\bs{d\alpha}_U)^{(\bs\uppsi, \bs\upgamma)}	 
&= \bs{d\alpha}_U + \langle \mathfrak L_{\bs{d\uppsi}\circ \bs\uppsi\-} \bs\alpha,  U \rangle,  \\
 \langle  \bs{d\alpha}, U \rangle ^{(\bs \uppsi, \bs\upgamma)} 
 &= \langle\bs{d\alpha}, U \rangle + \langle \mathfrak L_{\bs{d\uppsi}\circ \bs\uppsi\-} \bs\alpha,  U \rangle.
\end{aligned}
\end{equation}
As we now show, all this is relevant to local gRGFT and the variational principle. When $\bs \alpha$ is a $\Hl$-invariant Lagrangian \mbox{0-form} on $\Phi$ and top form on ($U\subset$) $M$, equations \eqref{Invariance-tens-int-loc}-\eqref{Vert-trsf-int-dalpha-loc} yield a \emph{well-defined variational principle}, meaning that the space of solutions is preserved under $C^\infty\big(\Phi,\Diff(M)\ltimes \Hl\big)\simeq \bs\Diff_v(\Phi)$ transformations.

\subsubsection{Lagrangian, action and geometric prescription for the variational principle in field theory}
\label{Geometric prescription for the variational principle in field theory}

\paragraph{Lagrangian, action and anomalies}
A general-relativistic gauge field theory over a region $U\subset M$ is usually given by a choice of Lagrangian form 
$L:\Phi \rarrow \Omega^n(U)$, $\phi\mapsto L(\phi)$, with $n=$ dim$M$.\footnote{Or an element of the $d$-cohomology class of $L$, as they give rise to the same field equations, as we are about to see. Only the presymplectic potential of the theory differs between members of the same class, which may have relevant consequences regarding the symplectic structures of the theories, and their quantization.
We may write $L=L'+d\ell$, with $\ell$ the so-called ``boundary Lagrangian". }
A priori, it may support a non-trivial action of $\Diff(M)\ltimes \Hl$, so that 
\begin{align}
\label{equiv-Lagrangian}
    R^\star_{(\uppsi, \upgamma)}L = \uppsi^*(L^\upgamma)  
    =  \uppsi^* \big(L + c(\ ;\upgamma) \big),
\end{align}
 where $c: \Phi \times \Hl \rarrow \Omega^n(U)$ is a $\Hl$-1-cocyle (with value in an additive Abelian group) satisfying indeed $c(\phi; \upgamma\,\upgamma')= c(\phi; \gamma) + c(\phi^\gamma; \gamma')$. It is automatic, following from the definition $c(\ ;\upgamma)\defeq R^\star_\upgamma L -L = L^\gamma -L$. 
 Defining similarly the $\Diff(M)$-1-cocycle $c: \Phi \times \Diff(M) \rarrow \Omega^n(U)$, $(\phi, \uppsi)\mapsto c(\phi; \uppsi)\defeq R^\star_\uppsi L - L= \uppsi^* L-L$, one finds the total $\big(\!\Diff(M)\ltimes \Hl\big)$-1-cocycle to be
 \begin{align}
 c\big(\phi; (\uppsi, \upgamma) \big) 
\defeq R^\star_{(\uppsi,\upgamma)} L - L
= c(\phi; \uppsi) + \psi^* c(\phi; \upgamma). 
 \end{align}
Linearising this 1-cocycle we get the  $\big(\diff(M)\oplus\text{Lie}\H\big)$-1-cocycle $a: \Phi \times \big(\diff(M)\oplus\text{Lie}\H\big) \rarrow \Omega^n(U)$:
\begin{align}
    a\big((X,\lambda);\ \, \big) 
    \defeq 
    \tfrac{d}{d\tau}\, c\big(\ \, ;(\uppsi_\tau,\upgamma_\tau)\big)\,\big|_{\tau=0} 
    = \tfrac{d}{d\tau} \left(R^\star_{(\uppsi_\tau,\upgamma_\tau)} L - L \right)\,\Big|_{\tau=0} 
    = \bs L_{(X,\lambda)^v} L .
\end{align}
In other words, a Lagrangian is a priori a twisted tensorial form, $L\in \Omega^0_\text{tens}\big(\Phi, \Omega^n(U)\big)$. 
One then finds that
\begin{equation}
\label{lin-anom-locfs}
\begin{aligned}
 \bs L_{(X,\lambda)^v} L
 = \tfrac{d}{d\tau}\, c\big(\ \ ;(\uppsi_\tau,\upgamma_\tau)\big)\,\big|_{\tau=0} &=  \tfrac{d}{d\tau}
 \left( c(\ \, ;\uppsi_\tau) + \tfrac{d}{d\tau} \uppsi^*_\tau \, c(\ \, ;\upgamma_\tau)\right)\,\Big|_{\tau=0} \\
 &= a(X;\ \, ) + \mathfrak{L}_X\, c(\ \, ;\upgamma_{\tau=0}) + \uppsi^*_{\tau=0} \, a(\lambda;\ \, ) \\
 &= a(X;\ \, )+a(\lambda;\ \, ) \\
 &= \mathfrak{L}_X L + \delta_\lambda L
 = a\big((X,\lambda); \ \, \big).
\end{aligned}
\end{equation}
We used the fact that $\upgamma_{\tau=0}=\id_{\Hl}$, $\uppsi_{\tau=0}=\id_M$, so $c(\ ; \id_{\Hl})=0$. 
The term $a\big((X,\lambda); \ \, \big)$ is what one may call the \emph{classical} $\big(\!\Diff(M)\ltimes \Hl\big)$-anomaly. As a special case of \eqref{cc-gen-anom-loc0}, it satisfies the identity
\begin{align}
\label{CC-class-anom-loc}
\big(X, \lambda\big)^v a\big((X', \lambda'); \phi \big)
-
\big(X', \lambda'\big)^v a\big((X, \lambda); \phi \big)
=a\Big(\big[(X, \lambda),\, (X', \lambda') \big]_\text{\tiny{Lie}};\phi \Big),
\end{align}
which is the classical analogue of the ``Wess-Zumino consistency condition", 
and contains the combined consistency conditions for both the $\Diff(M)$-anomaly $a(X;\phi)$ and the  $\Hl$-anomaly $a(\lambda;\phi) $.
Indeed, we remind that $(X, \lambda)^v=X^v + \lambda^v$ and that  the Lie bracket in $\diff(M)\oplus\text{Lie}\Hl$ is \eqref{Liebracket-local-aut}, so \eqref{CC-class-anom-loc} splits as:
\begin{align}
&X^v\, a(X';\phi) - {X'}^v\, a(X;\phi) &&\hspace{-4.5cm}= a\big([X, X']_\text{\tiny{$\diff(M)$}}; \phi \big) \label{class-Diff-anomaly}\\
&+  &&\hspace{-4.5cm}\phantom{=\ }\ +\notag\\
&\lambda^v\, a(\lambda';\phi) - {\lambda'}^v\, a(\lambda;\phi) &&\hspace{-4.5cm}= a\big([\lambda, \lambda']_\text{\tiny{Lie$H$}}; \phi \big)\label{class-Hl-anomaly}\\
&+ &&\hspace{-4.5cm}\phantom{=\ }\ +\notag\\
&X^v\, a(\lambda';\phi) + \lambda^v\, a(X';\phi) 
&&\hspace{-4.5cm}= a\big( -X'(\lambda) + X(\lambda');\phi \big). \label{crossed-terms-anomaly}\\
&\phantom{=} - {X'}^v\, a(\lambda;\phi) - {\lambda'}^v\, a(X;\phi)&& \notag
\end{align}
The line \eqref{class-Diff-anomaly} is the  
$\Diff(M)$-anomaly consistency condition,   \eqref{class-Hl-anomaly} is that of the $\Hl$-anomaly, while \eqref{crossed-terms-anomaly} is the mutual consistency condition between the two, reflecting (again) the semi-direct structure of $\Diff(M)\ltimes \Hl$. 

Remembering that $L$ has to be a top form on $M$ (so $dL\equiv0$), and requiring that the field equations be preserved under $\Hl$-transformations (i.e. imposing 
$\delta_{\!\lambda} L=d\beta(\lambda; \ \, )$), one may write
\begin{align}
\label{inf-eq-L-loc}
\bs L_{(X,\lambda)^v} L=
a\big((X,\lambda); \ \, \big)
=
a(X;\ \, )+a(\lambda;\ \, )
=\mathfrak{L}_X L + \delta_\lambda L=
d\,\big (\,\iota_X L + \beta(\lambda; \ \, ) \big) 
\rdefeq d\,\beta\big((X, \lambda); \ \, \big). 
\end{align}
This requirement is essential for a consistent treatment via covariant phase space methods, as we shall see elsewhere.

The Lagrangian being tensorial, its $\bs\Diff_v(\Phi)\simeq C^\infty\big(\Phi, \Diff(M)\ltimes \Hl \big)$-transformation is thus easily found to be
\begin{align}
\label{L-vert-trsf}
L^{(\bs\uppsi, \bs\upgamma)} \defeq \Xi^\star L = \bs\uppsi^*(L^{\bs\upgamma}),
\end{align}
while its $\bs\diff_v(\Phi)\simeq C^\infty\big(\Phi, \diff(M)\oplus\text{Lie}\Hl \big)$-transformation is
\begin{align}
\label{L-vert-trsf-lin}
\bs L_{(\bs X, \bs\lambda)^v}L  = a\big((\bs X, \bs\lambda); \phi \big)= \mathfrak L_{\bs X}L + \delta_{\!\bs\lambda}L. 
\end{align}
The anomaly for $\phi$-dependent parameters satisfies, as a special case of \eqref{cc-gen-anom-loc}, the consistency condition
\begin{align}
\label{CC-comb-anomaly-bis}
\big(\bs X, \bs\lambda\big)^v a\big((\bs X', \bs\lambda'); \phi \big)
-
\big(\bs X', \bs\lambda'\big)^v a\big((\bs X, \bs\lambda); \phi \big)
=a\Big(\big\{(\bs X, \bs\lambda),\, (\bs X',\bs\lambda') \big\};\phi \Big),
\end{align}
\medskip
which,  using the bracket of $C^\infty\big(\Phi, \diff(M)\oplus\text{Lie}\Hl \big)$ featuring in \eqref{FN-bracket-loc}, can be shown to reduce  to 
\eqref{CC-class-anom-loc}.
\medskip

The action functional over $U$ is the map 
\begin{equation}
\begin{aligned}
S: \Omega^0\big(\Phi, \Omega^n(U) \big) \times \b{\bs U}(M) &\rarrow \Omega^0(\Phi),\\
(L, U) &\mapsto  S = \langle L, U \rangle \defeq \int_U L.
\end{aligned}
\end{equation}
It is always $\Diff(M)$-invariant, by \eqref{invariance-int-loc}, so there can be no $\Diff(M)$-anomaly for the action. 
A priori, there may be a $\Hl$-anomaly: 
By \eqref{Vert-trsf-int-generic-loc}, the action of $\Diff(M)\ltimes \Hl$ on the action is
\begin{align}
\label{equiv-action-loc}
S^{(\uppsi, \upgamma)}\defeq \t R^\star_{(\uppsi, \upgamma)} S 
= \int_{\uppsi\-(U)} R^\star_{(\uppsi, \upgamma)} L 
= \int_{\uppsi\-(U)} \uppsi^*(L^\upgamma
) 
= \int_U L + c(\ \, ;\upgamma) \rdefeq S + \mathsf c(\ \, ;\upgamma). 
\end{align}
The term $\mathsf c: \Phi \times \Hl \rarrow \RR$, $(\phi, \upgamma)\mapsto \mathsf c(\phi; \upgamma)$, is a $\Hl$-1-cocycle. 
It is the classical counterpart of the so-called Wess-Zumino term, or  yet ``integrated anomaly", as its linearisation is 
\begin{align}
 \mathsf a(\lambda;\phi)
 \defeq \tfrac{d}{d\tau}\,   \mathsf c(\phi; \gamma_\tau)\, \big|_{\tau=0}
 = \int_U   \tfrac{d}{d\tau}\,  c(\phi; \gamma_\tau)\, \big|_{\tau=0} 
 \rdefeq \int_U a(\lambda; \phi),
\end{align}
featuring the $\Hl$-anomaly. 
Naturally, it satisfies the consistency condition
\begin{align}
\label{WZ-CC-Hl}
  \lambda^v\, \mathsf a(\lambda';\phi) 
- {\lambda'}^v\, \mathsf a(\lambda;\phi) 
= \mathsf a\big([\lambda, \lambda']_\text{\tiny{Lie$H$}}; \phi \big)
\end{align}
analogue to the non-integrated one \eqref{class-Hl-anomaly}.
One may write the linearisation of \eqref{equiv-action-loc}, by \eqref{linear-versions-integral-loc} and \eqref{inf-invariance-action-loc}: 
\begin{align}
 \label{linear-trsf-action-loc}   
 \delta_{(X, \lambda)} S 
 =\int_{U}  \bs L_{(X, \lambda)^v} L
 + \int_{-X(U)}   \hspace{-3mm} L \
 =\int_{U}  \delta_{\!\lambda}L + \mathfrak L_X L\ \ 
 + \int_{-X(U)}  \hspace{-3mm} L\ 
 =\int_U \delta_{\!\lambda} L 
 = \int_U a(\lambda; \phi) 
 =\mathsf a(\lambda; \phi),
\end{align}
which shows indeed that there is no classical $\Diff(M)$-anomaly.
Observe that, since $\Hl$ does not act on $\b{\bs U}(M)$, by \eqref{der-eval-integr-obj-loc} it holds that:  $\delta_{(X, \lambda)} S=\bs L_{\lambda^v} S = \iota_{\lambda^v}\bs d S$. 

The action can thus be seen as a twisted tensorial object. 
The action of $C^\infty\big(\Phi, \Diff(M)\ltimes \Hl \big)$ is,  by \eqref{Vert-trsf-int-generic-loc}, 
\begin{align}
\label{S-vert-trsf}
S^{(\bs\uppsi, \bs\upgamma)} \defeq \t\Xi^\star S 
= \int_{\bs\uppsi\-(U)} \Xi^* L
= \int_U L + c(\ \, ;\bs\upgamma) \rdefeq S + \mathsf c(\ \, ;\bs\upgamma), 
\end{align}
while the action of $C^\infty\big(\Phi, \diff(M)\oplus\text{Lie}\Hl \big)$ is, by \eqref{linear-versions-integral-loc}
\begin{align}
\label{S-vert-trsf-lin}
\delta_{(\bs X, \bs\lambda)} S  
=\int_{U}  \bs L_{(\bs X, \bs\lambda)^v} L
 + \int_{-\bs X(U)}   \hspace{-3mm} L \
= \int_U \delta_{\!\bs\lambda}L
= \mathsf a(\bs\lambda; \phi ). 
\end{align}
Again, one has that $\delta_{(\bs X, \bs\lambda)} S=\bs L_{\bs\lambda^v} S = \iota_{\bs\lambda^v}\bs d S$.
One may thus define the twisted tensorial form $Z \in \Omega^0_\text{tens}(\Phi, C)$, 
\begin{align}
\label{Z-action}
Z\defeq \exp{i\, S}, 
\quad \text{s.t.} \quad 
Z^{(\uppsi, \upgamma)} = C\big(\ \, ; \upgamma \big)\- Z, 
\quad \text{with} \quad 
C\big(\ \, ; \upgamma \big) \defeq \exp{\{-i\, \mathsf c(\ \,;\upgamma)\}},
\end{align}
where 
$C: \Phi \times \Hl \rarrow U(1)$ is a $\Hl$-1-cocycle, with phase the Wess-Zumino term. As a twisted object, it must be acted upon via a twisted covariant derivative, $\b {\bs D}\defeq \bs d\ \, + \bs\varpi$, where, by  \eqref{twisted-connection-loc}, the twisted connection satisfies: 
\begin{equation}
\begin{aligned}  
 \bs\varpi\,\big((X, \lambda)^v \big) &= \tfrac{d}{d\tau}\, C\big(\phi;  \upgamma_\tau \big)\, \big|_{\tau=0}= -i\, \mathsf a\big( \lambda; \phi \big) \qquad  \in  \mathfrak u(1)=i \, \RR, \\
 R^\star_{(\uppsi, \upgamma)} \bs\varpi 
 &= \bs\varpi 
 + C\big(\ \, ;  \upgamma\big)\- 
 \bs d C\big(\ \,;  \upgamma\big) 
 = \bs\varpi - i\bs d \mathsf c(\ \, ;\upgamma).
\end{aligned}
\end{equation}
The horizontality of the curvature $\b{\bs\Omega} =\bs{d\varpi} \in \Omega^2_\text{tens}(\Phi, C)$, i.e. 
 $\b{\bs\Omega}(\lambda^v,   {\lambda'}^v)\equiv 0$, encodes/reproduces the consistency condition \eqref{WZ-CC-Hl}.
It is by definition the case that $\b{\bs D}Z \in \Omega^1_\text{tens}(\Phi, C)$, but then since
\begin{align}
&\b{\bs D}Z
= \bs d Z + \bs\varpi Z
=\big(i\bs d S + \bs\varpi \big) Z, \notag\\
&\text{it follows that} \quad i\bs d S + \bs\varpi \in \Omega^1_\text{basic}(\Phi),
\label{Gen-WZ-action}
\end{align}
because $Z$ carries the (twisted) equivariance. The object \eqref{Gen-WZ-action}, that one may write $\bs d S - i\bs\varpi$, is what one may call a ``generalised Wess-Zumino (WZ) improved action":
Indeed, the usual WZ improved action is recovered as a special case for flat twisted connections $\bs\varpi_{\text{\tiny$0$}}$, which are necessarily written in terms of dressing fields, as we will see in the next section. The WS trick is then explained as coming from twisted covariant derivation. 
\medskip

When there is no $\Hl$-anomaly, i.e. when $S$ is $\big(\!\Diff(M)\ltimes \Hl\big)$-invariant, it induces a well-defined section on the bundle of regions $\b{\bs U}(M)=\Phi \times \bs U(M)/\sim$. 
It thus equivalently defines a $\Diff(M)(\ltimes \Hl)$-equivariant (tensorial) $\b{\bs U}(M)$-valued 0-form,
\begin{equation}
\begin{aligned}
\bs U_ S(\phi)&\defeq \pi_{\bs U}(\phi,  U)_{|\, S =\text{cst}} \equiv U, \\
\bs U_ S\big(\uppsi^*(\phi^\upgamma)\big)&
\defeq \pi_{\bs U}\big(\uppsi^*(\phi^\upgamma),  \uppsi\-(U) \big)_{|\, S =\text{cst}} 
\equiv \uppsi\-(U).
\end{aligned}
\end{equation}
Its $C^\infty\big(\Phi, \Diff(M)\ltimes \Hl\big)$-transformation is then
\begin{align}
 \bs U_S^{(\bs\uppsi, \bs\upgamma)}  = \bs\uppsi\-(\bs U_S), \quad \text{so} \quad
 \bs L_{(\bs X, \bs\lambda)^v} \bs U_S= -\bs X(\bs U_S).
\end{align}
The map $\bs U_S$ we may interpret as defining   $\Diff(M)$-equivariant ``$\phi$-relative" regions of $M$. 
This may be understood as a first formal step towards be the concrete  translation of the conceptual insight, exposed in section \ref{Relationality in general-relativistic gauge field theory}, that theories implementing the symmetries of gRGFT (of GR in particular) -- i.e. enforcing principles of epistemic democracy -- 
have a relational definition of spacetime: that is, define spacetime regions relative to the d.o.f. of their fields content.\footnote{It is only a first step because, as we have shown, $M$ is not spacetime. 
Still, $\bs U_S$  formally expresses, through its $\Diff(M)$-equivariance, the  covariant relation between field d.o.f. and physical points/regions of spacetime.
We may call manifolds defined via the set of values (open sets) of objects like $\bs U_S$,  ``\emph{manifields}".}
It is of course only half the full picture, which requires to see how fields d.o.f. relationally co-define each other. 
A more complete formal relational picture is provided in section \ref{Relational formulation via dressing}.
\medskip

We remark that a quantum theory is also a  priori a twisted tensorial object, even for an invariant action $S$:
\begin{align}
{\sf {Z}}\defeq \int \delta \phi\, \exp{\tfrac{i}{\hbar} S}, \quad \text{s.t.} \quad 
{\sf {Z}}^{(\uppsi, \upgamma)} = C\big(\ \, ; (\uppsi,\upgamma) \big)\- {\sf {Z}}, 
\end{align}
with $\delta \phi$ is an integration measure on $\Phi$, 
and where $C:\Phi \times \big(\!\Diff(M)\ltimes \Hl\big) \rarrow U(1)$ is a 1-cocycle, 
whose linearisation is the the combined $\big(\!\Diff(M)\ltimes \Hl\big)$-anomaly. 
It is the transformation of the measure $\delta \phi$ that may generate a non-zero cocycle.
Therefore, a $\Diff(M)$-anomaly potentially occurs only in quantum field theories. 
The~twisted covariant derivative adapted to ${\mathsf Z} \in\Omega^0_\text{tens}(\Phi, C) $  is $\b{\bs D}{\sf Z}=\bs d{\sf Z} + \bs\varpi {\sf Z} \in \Omega^1_\text{tens}(\Phi, C)$, with twisted connection satisfying
\begin{equation}
\begin{aligned}  
 \bs\varpi\,\big((X, \lambda)^v \big) &= \tfrac{d}{d\tau}\, C\big(\phi;  (\uppsi_\tau,\upgamma_\tau) \big)\, \big|_{\tau=0}= -\tfrac{i}{\hbar}\, \mathsf a\big( (X, \lambda); \phi \big) \qquad  \in  \mathfrak u(1)=i \, \RR, \\[1mm]
 R^\star_{(\uppsi, \upgamma)} \bs\varpi 
 &= \bs\varpi 
 + C\big(\ \, ; (X, \lambda)\big)\- 
 \bs d C\big(\ \,; (X, \lambda)\big). 
\end{aligned}
\end{equation}
The horizontality of the curvature $\b{\bs\Omega} =\bs{d\varpi} \in \Omega^2_\text{tens}(\Phi, C)$, i.e. 
 $\b{\bs\Omega}\big((X,\lambda)^v,   (X',{\lambda'}^v)\big)\equiv 0$, encodes/reproduces the WZ consistency condition of the combined $\Diff(M)\ltimes \Hl$ quantum anomaly, directly analogous to \eqref{CC-class-anom-loc}.
 We may observe that acting only with $\bs d$ on ${\sf Z}$ is a well-defined geometric operation only if ${\sf Z}\in \Omega^0_\text{basic}(\Phi)$ -- i.e. in the absence of quantum anomaly -- as then $\bs d$ is indeed a covariant derivative on basic forms, so that $\bs d {\sf Z}\in \Omega^1_\text{basic}(\Phi)$.

\paragraph{Variational principle in local field theory} 

We want to write the variational principle for a general-relativistic gauge field theory, and examine how the objects involved, notably the field equations, transform under the action of $C^\infty\big(\Phi, \Diff(M)\ltimes \Hl \big)$.
We write the variational principle as
\begin{equation}
\label{dS-dL}
\begin{aligned}
\bs dS_{|\phi}=0 \quad
\text{ with } \quad
\bs d L_{|\phi} &= \bs E_{|\phi} + d\bs \theta_{|\phi} \\
&= E(\bs d\phi; \phi) + d \theta(\bs d\phi; \phi),
\end{aligned}
\end{equation}
with $\bs E \in \Omega^1\big(\Phi, \Omega^n(U) \big)$ the field equations 1-form, and $\bs \theta \in \Omega^1\big(\Phi, \Omega^{n-1}(U) \big)$ the so-called ``presymplectic potential" of the theory. 
Remember that in $\bs dS$, the derivative $\bs d$ is extended from $\Phi$  to $\Phi \times \bs U(M)$ by
\eqref{der-eval-integr-obj-loc}.
The space of solutions is defined as $\S\defeq \big\{\phi \in \Phi \ |\ \bs E_{|\phi}=0 \big\}$.
Our goal is  to assess its stability under the action of $\bs\Diff_v(\Phi)$. 
\medskip

The equivariance and verticality properties of $\bs dL$ are  easily found. 
By, \eqref{equiv-Lagrangian} we have
\begin{align}
\label{equiv-dL}
 R^\star_{(\uppsi, \upgamma)} \bs d L 
 = \bs d     R^\star_{(\uppsi, \upgamma)} L 
 = \bs d \, \uppsi^*\big(L + c(\ \, ; \upgamma) \big) 
 = \uppsi^* \big( \bs dL +  \bs d c(\ \, ; \upgamma)  \big).
\end{align}
The verticality property of $\bs d L$ is just another way to see the infinitesimal equivariance of $L$, given by \eqref{inf-eq-L-loc}, so:
\begin{equation}
\label{vert-dL}
\begin{aligned}
\iota_{(X, \lambda)^v} \bs d L 
&=
a\big((X,\lambda); \ \, \big)
=
a(X;\ \, )+a(\lambda;\ \, )\\
&=
d\,\beta\big((X, \lambda); \ \, \big)
=
d\,\big (\,\iota_X L + \beta(\lambda; \ \, ) \big).
\end{aligned}
\end{equation}
This allows to compute geometrically the  $\bs\Diff_v(\Phi)\simeq C^\infty\big(\Phi, \Diff(M)\ltimes \Hl \big)$-transformation of $\bs d L$, 
using \eqref{pushforward-X-loc}:
\begin{equation}
\begin{aligned}
\bs dL^{(\bs\uppsi, \bs\upgamma)}_{|\phi}(\mathfrak X_{|\phi})
\defeq&\,
\Xi^\star \bs dL_{|\phi} (\mathfrak X_{|\phi}) 
=
\bs dL_{|\,\Xi(\phi)}\big(\Xi_\star \mathfrak X_{|\phi} \big)\\
=&\,
\bs dL_{|\,\Xi(\phi)}
\left(
R_{(\bs\uppsi, \bs\upgamma ) \star}\mathfrak X_{|\phi} 
+ \left\{
\big(
\bs\uppsi\-_* \bs d\bs \uppsi,\, 
\bs\uppsi^*\,(\bs\upgamma\- \bs d\bs \upgamma)
\big)_{|\phi} (\mathfrak X_{|\phi})
\right\}^v_{|\,\Xi(\phi)}
\right)\\
=&\,
R_{(\bs\uppsi, \bs\upgamma)}^\star \bs dL_{|\,\Xi(\phi)} (\mathfrak X_{|\phi})
+ \bs dL_{\Xi(\phi)} \left(
\left\{
\big(
\bs\uppsi\-_* \bs d\bs \uppsi,\, 
\bs\uppsi^*\,(\bs\upgamma\- \bs d\bs \upgamma)
\big)_{|\phi} (\mathfrak X_{|\phi})
\right\}^v_{|\,\Xi(\phi)}
\right)\\
=&\,
\bs\uppsi(\phi)^*\Big( \bs dL + \bs d c\big(\ \ ; \bs\upgamma(\phi)\big)\,\Big)_{|\phi}(\mathfrak X_{|\phi})
+ 
\left[a\left(\bs\uppsi\-_* \bs d\bs \uppsi; \,\Xi(\phi)  \right)
+a\left(\bs\uppsi^* (\bs\upgamma\- \bs d \bs \upgamma); \,\Xi(\phi)  \right)
\right]_{|\phi}(\mathfrak X_{|\phi}).
\end{aligned}
\end{equation}
The second term is readily found to be
\begin{equation}
\begin{aligned}
a\left(\bs\uppsi\-_* \bs d\bs \uppsi; \,\Xi(\phi)  \right)
&=
\mathfrak L_{\bs\uppsi\-_* \bs d\bs \uppsi} L\big( \bs\uppsi^*(\phi^{\bs\upgamma})\big)
=
\mathfrak L_{\bs\uppsi\-_* \bs d\bs \uppsi} \bs\uppsi^*L\big( \phi^{\bs\upgamma}\big)
=
\bs\uppsi^* \mathfrak L_{\bs d\bs \uppsi \circ \bs\uppsi\-} L (\phi^{\bs\upgamma}) \\
&=
\bs\uppsi^* \left( \mathfrak L_{\bs d\bs \uppsi \circ \bs\uppsi\-} \big(\, L + c(\ \ ;\bs\upgamma)\,\big) \, \right).
\end{aligned}
\end{equation}
The last term is a bit more subtle: using the definition of the infinitesimal cocycle $a$ and the defining property of $c$, 
\begin{equation}
\begin{aligned}
a\left(\bs\uppsi^* (\bs\upgamma\- \bs d \bs \upgamma)_{|\phi}(\mathfrak X_{|\phi}); \,\Xi(\phi)  \right) 
&=
\bs\uppsi^*\, a\big( \bs\upgamma\- \bs d \bs \upgamma_{|\phi}(\mathfrak X_{|\phi}); \, \phi^{\bs\upgamma}  \big) \\
&=
\bs\uppsi^*\,\tfrac{d}{d\tau}\, 
c\left(\phi^{\bs\upgamma} ;\, \exp{\{ \tau\, (\bs\upgamma\- \bs d \bs \upgamma)_{|\phi}(\mathfrak X_{|\phi})\} }\ \right)\,\big|_{\tau=0}\\
&=
\bs\uppsi^*\,\tfrac{d}{d\tau}\, 
c\left(\phi^{\bs\upgamma} ;\, \bs\upgamma(\phi)\-\exp{\{ \tau\, ( \bs d \bs \upgamma \bs\upgamma\-)_{|\phi}(\mathfrak X_{|\phi})\} \bs\upgamma(\phi)}\ \right)\,\big|_{\tau=0}\\
&=
\bs\uppsi^*\,\tfrac{d}{d\tau}\,
c\left(\phi ;\, \exp{\{ \tau\, ( \bs d \bs \upgamma \bs\upgamma\-)_{|\phi}(\mathfrak X_{|\phi})\} \bs\upgamma(\phi)}\ \right) 
-
 c\big(\phi;  \bs\upgamma(\phi) \big)
\ \big|_{\tau=0}\\
&=
\bs\uppsi^*\,\bs d
c\left(\phi ;\ \  \right)_{|\bs\upgamma(\phi)} \circ \bs d\bs\upgamma_{|\phi}(\mathfrak X_{|\phi}) \\
&=
\bs\uppsi^*\,\bs d c\big(\phi; \bs\upgamma \big)_{|\phi}(\mathfrak X_{|\phi}).
\end{aligned}    
\end{equation}
Gathering all results, we get
\begin{equation}
\begin{aligned}
\bs dL^{(\bs\uppsi, \bs\upgamma)}_{|\phi}(\mathfrak X_{|\phi})
&=
\bs\uppsi(\phi)^*\left(
\bs dL_{|\phi} + \bs d c\big(\ \ ; \bs\upgamma(\phi)\big)_{|\phi}\,
+
\bs d c\big(\phi; \bs\upgamma \big)_{|\phi}
+ 
\mathfrak L_{\bs d\bs \uppsi \circ \bs\uppsi\-} \big(\, L + c(\ \ ;\bs\upgamma)\,\big)_{|\phi} 
\right)(\mathfrak X_{|\phi})\\
&=
\bs\uppsi(\phi)^*\left(
\bs dL_{|\phi} + \bs d c\big(\ \ ; \bs\upgamma\big)_{|\phi}\,
+ 
\mathfrak L_{\bs d\bs \uppsi \circ \bs\uppsi\-} \big(\, L + c(\ \ ;\bs\upgamma)\,\big)_{|\phi} 
\right)(\mathfrak X_{|\phi}).
\end{aligned}
\end{equation}
Stating the result as valid $\forall \phi \in \Phi$ and $\forall \mathfrak X \in \Gamma(T\Phi)$, this is
\begin{equation}
\label{GT-dL}
\begin{aligned}
\bs dL^{(\bs\uppsi, \bs\upgamma)}
=
\bs\uppsi^*\left(
\bs dL + \bs d c\big(\ \ ; \bs\upgamma\big)
+ 
\mathfrak L_{\bs d\bs \uppsi \circ \bs\uppsi\-} \big(\, L + c(\ \ ;\bs\upgamma)\,\big) 
\right).
\end{aligned}
\end{equation}  
This can be cross-checked by using $[\Xi^\star, \bs d]=0$ and computing $\bs d\, \Xi^\star L = \bs d(L^{(\bs\uppsi, \bs\upgamma)})$. 
This generalises eq.(233)  in \cite{Francois2023-a}.

We observe that for the $\big(\!\Diff(M)\ltimes\Hl\big)$-transformed Lagrangian  $L^{(\uppsi, \upgamma)}=R^\star_{(\uppsi, \upgamma)} L$ to have the same field equations as $L$, 
it must be the case that 
\begin{align}
\label{Hyp0}
  \bs d c(\phi ;\, \upgamma) 
  =d\bs b(\phi ;\,\upgamma),
\end{align}
where $\bs b(\phi ;\upgamma)=b\big(\bs d\phi; \phi, \upgamma\big)$ is linear in $\bs d\phi$.
This is a separate hypothesis, that must be stressed as such. It is realised for example in the case of 3D Chern-Simons theory, where 
$c(A; \upgamma) 
= \Tr\big( d(\upgamma d \upgamma\- A) -\tfrac{1}{3}(\upgamma\-d \upgamma)^3
\big)$
so that 
$\bs b(A ;\upgamma) =
\Tr\big( \upgamma d \upgamma\- \bs d A \big)$.
Under hypothesis \eqref{Hyp0},  the result 
\eqref{GT-dL} gives
\begin{equation}
\label{GT-dL-bis}
\begin{aligned}
\bs dL^{(\bs\uppsi, \bs\upgamma)}
&=
\bs\uppsi^*\left(
\bs dL +  d \bs b\big(\ \ ; \bs\upgamma\big)
+ 
\mathfrak L_{\bs d\bs \uppsi \circ \bs\uppsi\-} \big(\, L + c(\ \ ;\bs\upgamma)\,\big) 
\right) \\
&=
\bs\uppsi^*\left(
\bs dL +  d  \left[ \bs b\big(\ \ ; \bs\upgamma\big)
+ 
\iota_{\bs d\bs \uppsi \circ \bs\uppsi\-}  L 
+ 
\iota_{\bs d\bs \uppsi \circ \bs\uppsi\-}\, c(\ \ ;\bs\upgamma)\,\right] 
\right),
\end{aligned}
\end{equation}
where we  use the fact that $L$ and $c(\ \ ;\upgamma)$ are top forms on $U\subset M$.
Hence, $\bs dL$ transforms under 
$\bs\Diff_v(\Phi)\simeq C^\infty\big(\Phi, \Diff(M)\ltimes \Hl \big)$
by a boundary term. 
\medskip

By \eqref{Vert-trsf-int-generic-loc}, we find the $C^\infty\big(\Phi, \Diff(M)\ltimes \Hl \big)$-transformation of $\bs d S$ to be
\begin{equation}
\begin{aligned}
 \bs dS^{(\bs\uppsi, \bs\upgamma)}
&=
 \int_{\bs\uppsi\-(U)} \bs d L^{(\bs\uppsi, \bs\upgamma)}
&= \int_U 
\bs dL + \bs d c\big(\ \ ; \bs\upgamma\big)
+ 
\mathfrak L_{\bs d\bs \uppsi \circ \bs\uppsi\-} \big(\, L + c(\ \ ;\bs\upgamma)\,\big).
 \end{aligned}
 \end{equation}
Under the hypotesis \eqref{Hyp0}, this gives
\begin{equation}
\begin{aligned}
\bs dS^{(\bs\uppsi, \bs\upgamma)}
&=
 \int_{U} \bs dL + d  \left[ \bs b\big(\ \ ; \bs\upgamma\big)
+ 
\iota_{\bs d\bs \uppsi \circ \bs\uppsi\-}  L 
+ 
\iota_{\bs d\bs \uppsi \circ \bs\uppsi\-}\, c(\ \ ;\bs\upgamma)\,\right] \\
&=
\bs dS+ \int_{\d U} 
\bs b\big(\ \ ; \bs\upgamma\big)
+ 
\iota_{\bs d\bs \uppsi \circ \bs\uppsi\-}  L 
+ 
\iota_{\bs d\bs \uppsi \circ \bs\uppsi\-}\, c(\ \ ;\bs\upgamma).
\end{aligned}
\end{equation}
We see that if $L$ is $\Hl$-invariant, so that $c=0=\bs b$, this result is indeed a special case of
\eqref{Vert-trsf-int-dalpha-loc}.
The fact that $\bs dL$, and $\bs dS$, transform with a boundary term hint at the fact that the variational principle remains well-defined under field-dependent  transformations, and that the space of solutions is preserved. 
\medskip

The fact of the latter matter can be  assessed by directly computing the transformation of the field equations $\bs E$.
For this, we need both the equivariance and verticality properties of $\bs E$. 
By 
\eqref{dS-dL}-\eqref{equiv-dL}, we get 
\begin{align}
 R^\star_{(\uppsi, \upgamma)} \bs dL 
 =
 R^\star_{(\uppsi, \upgamma)} \bs E + R^\star_{(\uppsi, \upgamma)} d \bs \theta 
 =
 \uppsi^*\big( 
 \bs E + d\bs \theta + \bs d  c(\ \ ; \upgamma)\,
 \big).
\end{align}
As things stand, we would not know how to split the $\Hl$-equivariance of $\bs dL$ as contributions coming from that of $\bs E$ and $d\bs \theta$. 
There are only two options to consider.
The first would be that $\bs E$ contributes non-trivially, alongside~$\bs\theta$: In~such cases, the action of $\Diff(M)\ltimes \Hl$ does not preserve the space of solutions $\S$ (a $\Hl$-transformation moves us ``off-shell"). 
As this assumption explicitly contravenes the foundational principles of gRGFT as reminded in section \ref{Relationality in  general-relativistic gauge field theory}, we reject it.\footnote{We remark that the extreme case where only $\bs E$ contributes, and $\bs \theta$ is invariant, is realised e.g. in the case of Massive Yang-Mills theory. See appendix E in \cite{Frankel2011}.}
The only option compatible with gRGFT is that $\bs E$ is $\Hl$-invariant. 
This then implies that \eqref{Hyp0} holds, a condition thus revealed to be essential to gRGFT. 
We therefore have the equivariance
\begin{equation}
\label{equiv-E-theta}
\begin{aligned}
R^\star_{(\uppsi, \upgamma)} \bs E = \uppsi^* \bs E \qquad \text{and} \qquad
R^\star_{(\uppsi, \upgamma)} \bs \theta = \uppsi^*\big( \bs\theta + \bs b(\ \; \upgamma)\big).  
\end{aligned}
\end{equation}

We need now only find the verticality property of $\bs E$. 
First, from \eqref{inf-eq-L-loc} and \eqref{dS-dL}  we get:
\begin{equation}
\label{vert-E-1}
\begin{aligned}
\iota_{(X, \lambda)^v}\bs E 
&=
d\left( \beta\big((X, \lambda); \phi \big)- \iota_{(X, \lambda)^v}\bs\theta \right)\\
\iota_{X^v} \bs E \ +\ \iota_{\lambda^v} \bs E &= d\big( \iota_X L - \iota_{ X^v}\bs\theta \big)\ + \ d\big( \beta(\lambda; \phi )-  \iota_{ \lambda^v}\bs\theta \big). 
 \end{aligned}
\end{equation}
We may use these expressions to compute $\bs E^{(\bs\uppsi, \bs\upgamma)}$, but the result would not be conceptually enlightening. 
It turns out to be possible to write each of the two contributions on the r.h.s. above in terms of the functional expression of $\bs E$ itself. 
To see this requires of us to have a closer look at how \eqref{dS-dL} is obtained.

\bigskip
Since we are considering  general-relativistic gauge field theories, we remind that our space of fields is made of $\phi=\{A, b\}$, where $A=A_\text{{\tiny YM}}+ A_\text{{\tiny Cartan}}$ with $A_\text{{\tiny YM}}$  the local representative of a Ehresmann connection (i.e. a Yang-Mills gauge potential) and 
$A_\text{{\tiny Cartan}}$ the local representative of a Cartan connection (i.e. a gravitational gauge potential), and $b=\{\vphi, F, \ldots\}$ are $\Hl$-tensorial fields, with $\vphi$ a matter field and $F=F_\text{{\tiny YM}}+ F_\text{{\tiny Cartan}}$ the Yang-Mills and gravitational field strengths. 
We have $DA=F$ and $Db=db+\rho_*(A)b$. 
Since $DF=0$, the Bianchi identity, and $DDb=\rho_*(F)b$, one shows that $D^{2p} b= \rho_*(F^p)b$ and $D^{2p+1} b = \rho_*(F^p)Db$. Thus, $\{\phi, D\phi\}$ is an algebraically closed set of variables under the action of $D$; we may write this $D\{\phi\} \subset \{\phi\}$. 
One easily shows that 
$\mathcal L_X A = D(\iota_XA) + \iota_X F$ 
and 
$\mathcal L_Xb= [\iota_X, D] b -\rho_*(\iota_XA)b$,
so we may write the generic formula
\begin{align}
\label{identity1}
\iota_{X^v} \bs d \phi =\mathcal L_X \phi= \iota_X D\phi + D(\iota_X \phi) - \rho_*(\iota_X A) \phi. 
\end{align}

The Lagrangian $L$ in gRGFT must be (or in most relevant cases, is) built from an invariant polynomial $P$ on Lie$H$ and representation $V$ of $H$ -- in the latter case $P$ is usually simply a $H$-invariant bilinear form on $V$: i.e. it is s.t. 
$P\big(\rho(h) v_1,\ldots, \rho(h) v_i, \ldots \big)=P\big(v_1, \ldots, v_i, \ldots \big)$ for $h \in H$ and $v_i$ Lie$H$- and/or $V$-valued variables ($\,\rho=\Ad$ in the former case). 
More symbolically, 
$P \circ \rho(H) =P$.\footnote{
This means that a Lagrangian $L=P(b)$ would be $\Hl$-invariant: a prototypical example is  YM theory,  $L_\text{{\tiny YM}}(\phi)=P(F,*F)=\Tr(F *\!F)$.  Such is not always the case, e.g.  3D Chern-simons theory is $L_\text{{\tiny CS}}(\phi)=P(A, F)=\Tr(AF-\sfrac{1}{3}A^3)$. Remark that in both cases $P=\Tr$.}
By linearizing, this implies $\Sigma_i P(\ldots,  \rho_*(\text{Lie}H)v_i,\ldots) = 0 $,
and the identity
$\Sigma_i (-)^{p(k_1+ \ldots + k_{i-1})}P\big( \ldots, \rho_*(\eta) v_i ,\ldots \big) =0$, 
for $\eta$ a Lie$H$-valued $p$-form and the variables $v_i$ are now $k_i$-forms.

Now, we may assume the Lagrangian to have the form $L(\phi)=\t L(\phi; D\phi)$ -- meaning it is defined on  $J^1\Phi$, the $1^\text{st}$ jet bundle of field space. Then we have, 
\begin{align}
\bs d L_{|\phi}= \t L_0(\bs d\phi; \{\phi\}) + \t L_1(\bs d D\phi; \{\phi\}),
\end{align} 
where $\{\phi\}$ means the collection of remaining $\phi$ and $D\phi$ in the respective functional expressions $\t L_{0/1}$, which are linear in their first argument.
All functional expressions $L$, $\t L$ and $\t L_{0/1}$ are built from the $H$-invariant polynomial $P$.  
 Using that $\bs d D\phi = D(\bs d \phi) + \rho_*(\bs d A)\phi$
 and integrating by parts, we obtain the formal expressions of the field equation and presymplectic potential:
 \begin{equation}
 \label{variational-principle}
 \begin{aligned}
\bs d L_{|\phi}
&=
\t L_0\big(\bs d\phi; \{\phi\}\big) 
+ \t L_1\big( D(\bs d \phi) 
+ \rho_*(\bs d A)\phi; \{\phi\}\big),  \\
&= 
\t L_0\big(\bs d\phi; \{\phi\}\big)
+
d \t L_1\big( \bs d \phi; \{\phi\}\big)
-(-)^{|\phi|} \t L_1\big(\bs d \phi; d\{\phi\}\big) 
+ 
\t L_1\big( \rho_*(A) \bs d\phi; \{\phi\} \big)
+
\t L_1\big(\rho_*(\bs dA)\phi; \{\phi\}\big) 
\\
&=
\t L_0\big(\bs d\phi; \{\phi\}\big)
+
d \t L_1\big( \bs d \phi; \{\phi\}\big)
-(-)^{|\phi|} \t L_1\big(\bs d \phi; d\{\phi\}\big) 
- 
(-)^{|\phi|} \t L_1\big( \bs d\phi; \rho_*(A) \{\phi\} \big) 
+
\t L_1\big(\rho_*(\bs dA)\phi; \{\phi\}\big) 
\\
&=
\t L_0\big(\bs d\phi; \{\phi\}\big)
- 
(-)^{|\phi|} \t L_1\big( \bs d\phi; D\{\phi\} \big) 
+
\t L_1\big(\rho_*(\bs dA)\phi; \{\phi\}\big) 
\ + \
d \t L_1\big( \bs d \phi; \{\phi\}\big)
\\
&\rdefeq 
E(\bs d\phi; \phi)+ d\theta (\bs d\phi; \phi)  \\
&=  \bs E_{|\phi} + d\bs \theta_{|\phi},
\end{aligned}
 \end{equation}
where we denote the form degree $|\bs d \phi|=|\phi|$, and we used the identity 
$\t L_1\big(\rho_*(A) \bs d\phi; \{\phi\} \big) +(-)^{|\phi|} \t L_1\big( \bs d\phi; \rho_*(A) \{\phi\} \big) =0$ 
stemming from the fact that $\t L_1$ is built from $P$.

By a similar formal computation, and  using \eqref{identity1}, we get the expression of the evaluation of the Lagrangian $n$-form on a vector field $X\in \Gamma(TM)\simeq \diff(M)$, acting as a derivation (like $\bs d$):
\begin{equation}
\label{iota-X-L-id}
\begin{aligned}
 \iota_{X} L(\phi) 
 &=
 \iota_{X} \t L(\phi; D\phi) \\
 &= 
 \t L_0\big( \iota_{X} \phi; \{\phi\} \big) 
 +
 \t L_1\big( \iota_{X} D\phi; \{\phi\}\big), \\
 &=
 \t L_0\big( \iota_{X} \phi; \{\phi\}\big) 
 +
 \t L_1\big( \iota_{X^v} \bs d \phi ; \{\phi\} \big)
-
\t L_1\big( D(\iota_X \phi) ; \{\phi\}\big)
+ 
\t L_1\big(\rho_*(\iota_X A) \phi; \{\phi\} \big), \\
&=
\t L_0\big( \iota_{X} \phi; \{\phi\}\big)
+
\t L_1( \iota_{X^v} \bs d \phi ; \{\phi\}) 
- 
d\t L_1(  \iota_{X} \phi ; \{\phi\}) 
+(-)^{|\iota_X\phi|}
\t L_1( \iota_{X} \phi ; D\{\phi\})
+
\t L_1\big( \rho_*(\iota_X A) \phi; \{\phi\} \big)  \\
&=
\t L_0\big( \iota_{X} \phi; \{\phi\}\big) 
-(-)^{|\phi|}
\t L_1( \iota_{X} \phi ; D\{\phi\})
+
\t L_1\big( \rho_*(\iota_X A) \phi; \{\phi\} \big) 
\ + \
\t L_1( \iota_{X^v} \bs d \phi ; \{\phi\})
\ - \
d\t L_1(  \iota_{X} \phi ; \{\phi\})\\
&= 
E( \iota_X\phi; \phi)
+
\iota_{X^v} \bs \theta 
- d\theta( \iota_X\phi; \phi).
 \end{aligned}
 \end{equation}
From this we obtain 
$\iota_X L - \iota_{X^v} \bs \theta = E( \iota_X\phi; \phi)
- d\theta( \iota_X\phi; \phi)$, so that we get the first contribution in  \eqref{vert-E-1} 
\begin{align}
\label{iota-X-E}
\iota_{X^v} \bs E = d E( \iota_X\phi; \phi).
\end{align}

There remains only to find $\iota_{\lambda^v}\bs E$. The most straightforward way to do so is to attempt to compute it formally from the expressiond derived in \eqref{variational-principle}. 
Using the relevant part in \eqref{basis-1-form-prop-loc}, i.e. $\iota_{\Lambda^v}\bs d\phi = \delta_\lambda \phi$, considering further that $\delta_\lambda \phi= (\delta_\lambda A, \delta_\lambda b)=(D\lambda, -\rho_*(\lambda) b)$, and integrating by part as above, one may be easily convinced that
\begin{equation}
\begin{aligned}
\iota_{\lambda^v}\bs E
&=
d \left(
\t L_0\big( \lambda; \{\phi\}\big) 
-(-)^{|\phi|}
\t L_1( \lambda ; D\{\phi\})
+
\t L_1\big( \rho_*(\lambda) \phi; \{\phi\} \big) 
\right)
\ +\ \t L_2\big( \lambda; \{\phi\} \big) \\
&= dE(\lambda; \phi) \ +\  \t L_2\big( \lambda; \{\phi\} \big),
\end{aligned}
\end{equation}
where $\t L_2$ is a functional expression (depending on $\t L_0$ and $\t L_1$) linear in its first argument, an \emph{underived} $\lambda$. 
But since from more general considerations we have that $\iota_{\lambda^v}\bs E = d\big( \beta(\lambda; \phi) - \iota_{\lambda^v} \bs \theta\big)$, i.e. $\iota_{\lambda^v}\bs E$ is $d$-exact, it must be that $\t L_2\equiv 0$. 
So we get the result 
\begin{align}
\label{iota-lambda-E}
\iota_{\lambda^v} \bs E = d E( \lambda; \phi).
\end{align}
Finally, we thus get to write the verticality property of $\bs E$:
\begin{equation}
\label{vert-E}
\begin{aligned}
\iota_{(X, \lambda)^v}\bs E 
&= 
dE\big(\iota_X \phi ;\phi\big)
+ dE\big(\lambda ;\phi\big) \rdefeq
 dE\big((\iota_X \phi, \lambda) ;\phi\big).
 \end{aligned}
\end{equation}
If one fails to be convinced by the above argumentation, one may take \eqref{vert-E}, especially \eqref{iota-lambda-E}, as an hypothesis or working assumption that needs to be checked in specific theories whenever one wishes to apply/use the results presented hereafter. 

The $\bs\Diff_v(\Phi)\simeq C^\infty\big(\Phi, \Diff(M)\ltimes \Hl \big)$-transformation of $\bs E$ is then easily found to be,
using \eqref{pushforward-X-loc}:
\begin{equation}
\begin{aligned}
\bs E^{(\bs\uppsi, \bs\upgamma)}_{|\phi}(\mathfrak X_{|\phi})
\defeq&\,
\Xi^\star \bs E_{|\phi} (\mathfrak X_{|\phi}) 
=
\bs E_{|\,\Xi(\phi)}\big(\Xi_\star \mathfrak X_{|\phi} \big)\\
=&\,
\bs E_{|\,\Xi(\phi)}
\left(
R_{(\bs\uppsi, \bs\upgamma ) \star}
\left[
\mathfrak X_{|\phi} + 
\left\{\Big( 
\bs{d\uppsi} \circ \bs\uppsi\-,\ 
\bs{d\upgamma} \, \bs\upgamma\-  - \bs\upgamma\, \mathfrak L_{\bs{d\uppsi} \circ \bs\uppsi\-} \bs\upgamma\-
\Big)_{|\phi}(\mathfrak X_{|\phi})
\right\}^v_{|\phi} 
\right]
\right)\\
=&\,
R_{(\bs\uppsi, \bs\upgamma )}^\star \bs E_{|\,\Xi(\phi)} 
\left(
\mathfrak X_{|\phi} + 
\left\{\Big( 
\bs{d\uppsi} \circ \bs\uppsi\-,\ 
\bs{d\upgamma} \, \bs\upgamma\-  - \bs\upgamma\, \mathfrak L_{\bs{d\uppsi} \circ \bs\uppsi\-} \bs\upgamma\-
\Big)_{|\phi}(\mathfrak X_{|\phi})
\right\}^v_{|\phi} 
\right)\\
=&\,
\bs\uppsi^* \bs E_{|\phi}
\left(
\mathfrak X_{|\phi} + 
\left\{\Big( 
\bs{d\uppsi} \circ \bs\uppsi\-,\ 
\bs{d\upgamma} \, \bs\upgamma\-  - \bs\upgamma\, \mathfrak L_{\bs{d\uppsi} \circ \bs\uppsi\-} \bs\upgamma\-
\Big)_{|\phi}(\mathfrak X_{|\phi})
\right\}^v_{|\phi} 
\right)\\
=&\,
\bs\uppsi^* \bs E_{|\phi}
\left(
\mathfrak X_{|\phi}\right)
+ 
\bs\uppsi^* dE \left( 
\iota_{\bs{d\uppsi}\circ\bs \uppsi\-_{|\phi}(\mathfrak X_{|\phi})} \phi; \phi
\right)
+
\bs\uppsi^* dE \left( \Big(\bs{d\upgamma} \, \bs\upgamma\-  - \bs\upgamma\, \mathfrak L_{\bs{d\uppsi} \circ \bs\uppsi\-} \bs\upgamma\-
\Big)_{|\phi}(\mathfrak X_{|\phi}) ; \phi
\right). 
\end{aligned}
\end{equation}
Hence, written as valid at all points $\phi \in \Phi$ and $\forall \mathfrak X \in \Gamma(T\Phi)$, this is
\begin{align}
\label{GT-E}
\bs E^{(\bs\uppsi, \bs\upgamma)} 
=
\bs\uppsi^* \left( 
\bs E 
+ 
dE \big( 
\iota_{\bs{d\uppsi}\circ\bs \uppsi\-} \phi; \phi
\big)
+
 dE \big(\bs{d\upgamma} \, \bs\upgamma\-  - \bs\upgamma\, \mathfrak L_{\bs{d\uppsi} \circ \bs\uppsi\-} \bs\upgamma\-
 ; \phi \big)\,
\right).
\end{align}
This is one of the most important equations of this paper. 
It shows that the space of solutions $\S$ of a general-relativistic gauge theory
is stable under field-dependent transformations: $\S$ is fibered by the action of $\bs\Diff_v(\Phi)\simeq C^\infty\big(\Phi, \Diff(M)\ltimes \Hl \big)$, and hence a principal bundle in its own right. 
Which means that  gRGFT admits a much bigger set of symmetries than initially assumed ($\Diff(M)\ltimes \Hl$) and encoded in its foundational principles. 
To the best of our knowledge, this fact was first noticed in the case of GR by \cite{Bergmann1961, Bergmann-Komar1972}, and later by \cite{Salisbury-Sundermeyer1983}, where metric-dependent diffeomorphisms were considered.\footnote{And where instances of the Frölicher-Nijenhuis bracket for metric-dependent vector fields was computed heuristically. 
See eqs.(3.1)-(3.2) in \cite{Bergmann-Komar1972} and eq.(2.1) in \cite{Salisbury-Sundermeyer1983}.} 
The result \eqref{GT-E} specialises to
\begin{equation}
\label{GT-E-special}
\begin{aligned}
\bs E^{\bs\uppsi} 
&=
\bs\uppsi^* \left( 
\bs E 
+ 
dE \big( 
\iota_{\bs{d\uppsi}\circ\bs \uppsi\-} \phi; \phi
\big)
\right), \\
\bs E^{ \bs\upgamma} 
&=
\bs E 
+
 dE \big(\bs{d\upgamma} \, \bs\upgamma\-  
 ; \phi \big),
\end{aligned}
\end{equation}
which reproduce and/or generalise respectively e.g. eq.(233) in \cite{Francois2023-a}, and eq.(103) in \cite{Francois2021} and eq.(74) in \cite{Francois-et-al2021}. 

By the same logic, from \eqref{equiv-E-theta} and  using \eqref{iota-X-L-id} and \eqref{iota-lambda-E} to find the verticality property if $\bs\theta$, one may obtain its 
$\bs\Diff_v(\Phi)\simeq C^\infty\big(\Phi, \Diff(M)\ltimes \Hl \big)$-transformation --
showing explicitly that $\bs dL^{(\bs\uppsi, \bs\upgamma)} =\bs E^{(\bs\uppsi, \bs\upgamma)} + d\bs\theta^{(\bs\uppsi, \bs\upgamma)}$. This result, as well as the transformation of the symplectic 2-form $\bs\Theta \defeq \bs{d\theta}$, are relevant to the covariant phase space analysis of gRGFT. A study we postpone to a forthcoming part of this series, which will recover and expand on previous works such as \cite{Prabhu2017, Geiller2018, Francois2023-a, Francois2021, Francois-et-al2021, Francois2023-a}.  
For now, our focus is on the relational reformulation of local gRGFT via dressing. 

\subsection{Relational formulation via dressing}  
\label{Relational formulation via dressing} 

The dressing field method for local theory parallels the treatment of the global case detailed in section \ref{The dressing field method}. 
As the local case has many interesting specific features, we again provide as much details as needed to get the clearest conceptual picture. 

\subsubsection{Building basic forms on local field space}
\label{Building basic forms on local field space}

We remind that $Q \rarrow N$ is the reference bundle with base $N$ featuring in the global DFM.
We define the space of $\Diff(M)\ltimes \Hl$-dressing fields as
\begin{equation}
\label{Dressing-field-full}
\begin{aligned}
\D r[N; M, H] \defeq \Big\{ 
\ &\big( \upupsilon, \text u  \big) \ \text{ with }\ \upupsilon: N \rarrow M \ \text{ and } \ \text u: M \rarrow H\ \ 
| \\[1mm] 
\ & \big( \upupsilon, \text u  \big)^{(\uppsi, \upgamma)}
\defeq
\Big(\uppsi\-\circ \upupsilon ,\ \uppsi^*(\text u^\upgamma)\,\Big)
\defeq
\Big(\uppsi\-\circ \upupsilon ,\ \uppsi^*(\upgamma\- \text u)\,\Big)
=
(\uppsi, \upgamma)\- \cdot 
\big( \upupsilon, \text u  \big)
\ \Big\},
\end{aligned}
\end{equation}
where $(\uppsi, \upgamma)\-=(\uppsi\-, \uppsi^* \upgamma\-)$, and the group product law \eqref{semi-dir-prod-DiffM-H-2} (
/\eqref{semi-dir-DiffM-H}) is used (and extended) to write the last equality in the defining $\Diff(M)\ltimes \Hl$-transformation of the dressing field $\big( \upupsilon, \text u  \big)$. 
The linearisation of which is 
\begin{align}
\label{inf-dress-trsf}
\delta_{(X, \lambda)} \big( \upupsilon, \text u  \big)
=
\big( -X \circ \upupsilon,\ \mathfrak L_X \text u  - \lambda \text u \big).
\end{align}
We will usually refer to $\big( \upupsilon, \text u  \big)$ as \emph{a} dressing field, despite it being a pair with $\upupsilon$ a $\Diff(M)$-dressing field as defined in \cite{Francois2023-a} and $\text u$ a $\Hl$-dressing field as originally defined in \cite{GaugeInvCompFields, Francois2014, Francois2021}.

We  define the space of dressed fields $\Phi^{( \upupsilon, \text u  )}$ by
\begin{equation}
\label{dressing-map-loc}
\begin{aligned}
|^u : \Phi \ &\rarrow\ \Phi^{( \upupsilon, \text u  )},  \\
\phi \ &\mapsto \ \phi^{( \upupsilon, \text u  )}\defeq  \ \upupsilon^*(\phi^{\text u}).
\end{aligned}
\end{equation}
We refer to $ \phi^{( \upupsilon, \text u  )}$ as a dressed field, or the dressing of $\phi$.
The notation $\phi^{\text u}$ means the same functional expression as the $\Hl$-transformation of $\phi$, i.e. $\phi^\upgamma$, but with $\upgamma \in \Hl$ replaced by $\text u$. For example, the $\Hl$-dressing of a gauge potential $A$ is $A^{\text u}\defeq \text u\- A \text u + \text u\- d\text u$. 
The $\Hl$-dressing of a tensorial field $b$ is $b^{\text u}\defeq\rho(\text u\-)\, b$. In particular the $\Hl$-dressing of a field strength $F$ is $F^{\text u}\defeq \text u\- F \text u $, while that of a matter fields $\varphi$ is $\varphi^{\text u}\defeq\rho(\text u\-)\, \varphi$.
As is clear from these examples, we have that $(\phi^{\text u})^\upgamma = (\phi^\upgamma)^{\text u^\upgamma} = (\phi^\upgamma)^{\upgamma\-\text u} =\phi^{\text u}$. 
By construction, dressed fields are $\Diff(M)\ltimes \Hl$-invariant:
\begin{align}
\big( \phi^{( \upupsilon, \text u  )} \big)^{(\uppsi, \upgamma)}
=
\big( \phi^{(\uppsi, \upgamma)} \big)^{( \upupsilon, \text u  )^{(\uppsi, \upgamma)}}
=
\big( \phi^{(\uppsi, \upgamma)} \big)^{(\uppsi, \upgamma)\- \cdot ( \upupsilon, \text u )}
=
 \phi^{(\uppsi, \upgamma) \cdot (\uppsi, \upgamma)\- \cdot ( \upupsilon, \text u )}
=
\phi^{( \upupsilon, \text u  )}.
\end{align}
Or, more explicitly, 
\begin{equation}
\label{dressed-field-inv-loc-bis}
\begin{aligned}
\big( \phi^{( \upupsilon, \text u  )} \big)^{(\uppsi, \upgamma)}
&=
\big( \phi^{(\uppsi, \upgamma)} \big)^{( \upupsilon, \text u  )^{(\uppsi, \upgamma)}}
=
\big( \phi^{(\uppsi, \upgamma)} \big)^{\big(\uppsi\- \circ \upupsilon, \ \uppsi^*(\upgamma\- \text u) \big)}\\
&=
(\uppsi \circ \upupsilon)^*
\Big([\phi^{(\uppsi, \upgamma)}]^{\uppsi^*(\upgamma\-\text u)} \Big) 
=
(\uppsi \circ \upupsilon)^*
\Big( \uppsi^*[\phi^ \upgamma]^{\uppsi^*(\upgamma\-\text u)} \Big) \\
&= 
\upupsilon^*\circ \uppsi\- \Big( \uppsi^* [(\phi^\upgamma)^{\upgamma\-\text u}] \Big)\\
&=
\upupsilon^*(\phi^{\text u}) 
=
 \phi^{( \upupsilon, \text u  )}.
\end{aligned}
\end{equation}
Observe that we have automatic compatibility conditions for $\upupsilon$ and $\text u$, analogous to those displayed in section \ref{Composition of dressing operations}, ensuring the stepwise  reduction
\eqref{stepwise-reduction}:
Indeed, from \eqref{Dressing-field-full} we have on the one hand that $\text u^\uppsi = \uppsi^* \text u$ which ensures that the $\Hl$-invariant dressed field $\phi^{\text u}$ still has a well-defined $\Diff(M)$-transformation, that one can dress for via $\upupsilon$.
On the other hand,
 we have $\upupsilon^\upgamma=\upupsilon$, which secures the fact that the $\Hl$-invariance achieved via $\text u$ is not spoiled after further dressing via $\upupsilon$. 
So, whenever dressed fields are defined $\phi^{( \upupsilon, \text u  )}\in \Phi^{( \upupsilon, \text u  )} \sim [\phi] \in \M$. 

We now define \emph{field-dependent} $\Diff(M)\ltimes \Hl$-dressing fieds as
\begin{equation}
\label{phi-dep-dressing-loc}
\begin{aligned}
\big(\bs\upupsilon, \bs\u\big): \Phi &\rarrow \D r[N; M, H], &&\\
\phi &\mapsto \big(\bs\upupsilon(\phi), \u(\phi)\big), 
\quad &&\text{ s.t. } \quad 
R^\star_{(\uppsi, \upgamma)}\big(\bs\upupsilon, \u\big) 
= (\uppsi, \upgamma)\- \cdot \big(\bs\upupsilon, \u \big) 
=
\Big(\uppsi\-\circ \bs\upupsilon ,\ \uppsi^*(\upgamma\- \u)\,\Big),\\
&\ && \text{ i.e. } \quad
\Big(\,\bs\upupsilon\big(\uppsi^*(\phi^\upgamma)\big),\ \u \big(\uppsi^*(\phi^\upgamma) \big)\,\Big) 
= (\uppsi, \upgamma)\- \cdot \big(\bs\upupsilon(\phi),\ \u (\phi )\big).
\end{aligned}
\end{equation}
The linear version of this defining equivariance is 
\begin{align}
\label{linear-equi-phi-dep-dressing}
\bs L_{(X, \lambda)^v} \big(\bs\upupsilon, \u \big)
=
\big( 
-X \circ \bs\upupsilon,\, \mathfrak L_X \u - \lambda\u
\big). 
\end{align} 
Such a dressing field can be understood as an equivariant 0-form, i.e. tensorial, with values in $\D r[N; M, H]$ seen as a representation space for (a right action of) $\Diff(M)\ltimes \Hl$. 
Therefore, its $\bs\Diff_v(\Phi)\simeq C^\infty\big(\Phi, \Diff(M)\ltimes \Hl \big)$ and $\bs\diff_v(\Phi)\simeq C^\infty\big(\Phi, \diff(M)\oplus \text{Lie}\Hl \big)$ transformations are  
\begin{align}
\big(\bs\upupsilon, \u\big)^{(\bs\uppsi, \bs\upgamma)} 
=
(\bs\uppsi, \bs\upgamma)\-\cdot  \big(\bs\upupsilon, \u\big), 
\quad \text{ and } \quad 
\bs L_{(\bs X, \bs\lambda)^v}\big(\bs\upupsilon, \u\big)
=
\big( - \bs X \circ \bs\upupsilon,\ \mathfrak L_{\bs X} \text \u  - \bs\lambda \u \big).
\end{align}
A $\phi$-dependent dressing field induces a map
\begin{equation}
\label{F-map-loc}
\begin{aligned}
F_{(\bs\upupsilon, \u)} : \Phi \ &\rarrow\ \M   , \\
\phi \ &\mapsto \  F_{(\bs\upupsilon, \u)}(\phi) \defeq \bs\upupsilon(\phi)^*\big( \phi^{\bs\u(\phi)} \big)=\phi^{(\bs\upupsilon, \u)} \sim [\phi],  \qquad \text{ s.t. }\quad  F_{(\bs\upupsilon, \u)} \circ R_{(\uppsi, \upgamma)} = F_{(\bs\upupsilon, \u)}. 
\end{aligned}    
\end{equation}
It tells us that a $\Diff(M)\ltimes \Hl$-orbit, a point $[\phi]\in \M$ is represented by a dressed field $\phi^{(\bs\upupsilon, \u)}$,
so the image of $F_{(\bs\upupsilon, \u)}$ is a \emph{relational coordinatisation} of $\M$:
Indeed, the expression of the $\Diff(M)\ltimes \Hl$-invariant variable $\phi^{(\bs\upupsilon, \u)}=\bs\upupsilon(\phi)^*\big( \phi^{\bs\u(\phi)} \big)$ is an explicit field-dependent coordinatisation of  the physical d.o.f. with respect to each other. 

Let us unpack this further.
The field $\phi^{\u(\phi)}$ can be understood to coordinatise the internal d.o.f. of $\phi$ w.r.t each other, exhibiting how the true \emph{physical internal} d.o.f.  encoded in $\phi$ are co-defined -- Remind that $\phi$ is a priori a collection of fields.
Thus, $\phi^{\u(\phi)}$ is a $\Hl$-invariant assignation of internal d.o.f. over points of $M$. 
It is the formal implementation, locally on $M$, of the core conceptual gauge theoretic insight stemming from the internal hole argument and the internal point-coincidence argument as detailed in section \ref{Relationality in  general-relativistic gauge field theory}. 

The field $\bs\upupsilon(\phi)^*\big( \phi^{\bs\u(\phi)} \big)$ can be understood to coordinatise the external d.o.f. of $\phi^{\u(\phi)}$ w.r.t each other. 
It exhibits that the true \emph{physical spatio-temporal} d.o.f. encoded in $\phi^{\u(\phi)}$ are co-defined. 
It is the formal implementation of the core general-relativistic insight stemming from  articulating of the hole argument and the point-coincidence argument... Or at least half of it:
The fact that $\bs\upupsilon(\phi)^*\big( \phi^{\bs\u(\phi)} \big)$ is a field on $N$ further stresses that $M$ was never the physical spacetime. 
We will indicate shortly, when considering dressed integrals below,  how the DFM makes manifest how the physical spacetime is defined via the field content. 

In short, the dressed field $\phi^{(\bs\upupsilon, \u)}$ represents the physical spatio-temporal and internal d.o.f. in a manifestly $\Diff(M)\ltimes \Hl$-invariant and \emph{manifestly relational} way.

\medskip 
It is clear that the map \eqref{F-map-loc} realises the bundle projection, $F_{(\bs\upupsilon, \u)} \sim \pi$, 
and thus allows to build basic forms on the local field space $\Phi$ via
$\Omega^\bullet_{\text{basic}}(\Phi) =$ Im$\,\pi^\star \simeq$ Im$\,F_{(\bs\upupsilon, \u)}^\star$.  
Our aim will be in general to build the basic counterpart of a form $\bs\alpha=\alpha\big(\!\w^\bullet\! \bs d\phi; \phi \big) \in \Omega^\bullet(\Phi)$. For this we  must first consider its formal analogue on the base space $\M$ of $\Phi$, $\b{\bs \alpha}=\alpha\big( \!\w^\bullet\! \bs d[\phi]; [\phi] \big) \in  \Omega^\bullet(\M)$, and then define the \emph{dressing} of $\bs\alpha$ as
\begin{align}
\label{Dressing-form-loc}
\bs \alpha^{(\bs\upupsilon, \u)}\defeq F_{(\bs\upupsilon, \u)}^\star  \b{\bs \alpha} = \alpha \big(  \!\w^\bullet\! F_{(\bs\upupsilon, \u)}^\star \bs d[\phi];\ F_{(\bs\upupsilon, \u)}(\phi)  \big) \quad \in \ \Omega^\bullet_{\text{basic}}(\Phi).
\end{align}
Since it is basic by construction, it is automatically invariant under $\bs\Diff_v(\Phi)\simeq C^\infty \big(\Phi, \Diff(M)\ltimes \Hl \big)$.

To obtain a concrete expression of $\bs \alpha^{(\bs\upupsilon, \u)}$, one may use the usual rule of thumb of the DFM: replace in the expression for the transformation $\bs \alpha^{(\bs\uppsi, \bs\upgamma)}$ the parameters  $(\bs\uppsi, \bs\upgamma)$ by the dressing field $(\bs\upupsilon, \u)$. 
But for the sake of completness, let us see how we may confirm the validity of this rule. 

From \eqref{Dressing-form-loc}, it is clear that all we need is $F_{(\bs\upupsilon, \u)}^\star \bs d[\phi]$ expressed in terms of $\bs d \phi$ and $(\bs\upupsilon, \u)$. 
The obvious way to try and find it is to write 
$F_{(\bs\upupsilon, \u)}^\star \bs d[\phi]_{|F_{(\bs\upupsilon, \u)}(\phi)}\big(\mathfrak X_{|\phi} \big)= \bs d[\phi]_{|F_{(\bs\upupsilon, \u)}(\phi)} \big( {F_{(\bs\upupsilon, \u)}}_\star \mathfrak X_{|\phi} \big)$, 
for a generic $\mathfrak X \in \Gamma(T\Phi)$ with flow $\vphi_\tau : \Phi \rarrow \Phi$, and s.t. $\mathfrak X_{|\phi}  = \tfrac{d}{d\tau} \vphi_\tau(\phi) \big|_{\tau =0} = \mathfrak X(\phi)\tfrac{\delta}{\delta \phi}$. 
So, we need the pushforward $F_{(\bs\upupsilon, \u)\star} : T_{\phi}\Phi \rarrow T_{F_{(\bs\upupsilon, \u)}(\phi)}\M$, $\mathfrak X_{|\phi} \mapsto F_{(\bs\upupsilon, \u)\star} \mathfrak X_{|\phi}$.    
Using the expression of the map \eqref{F-map-loc} we have
\begin{align*}
 F_{(\bs\upupsilon, \u)\star} \mathfrak X_{|\phi}  
 \defeq&\,
 F_{(\bs\upupsilon, \u)\star} \tfrac{d}{d\tau}\ \vphi_\tau(\phi)\ \big|_{\tau =0} 
 =
 \tfrac{d}{d\tau}\ F_{(\bs\upupsilon, \u)} \big( \vphi_\tau(\phi)   \big)\ \big|_{\tau =0}
 =
 \tfrac{d}{d\tau} \ 
\bs\upupsilon\big( \vphi_\tau(\phi)\big)^*\Big( \phi^{\,\bs\u\left( \vphi_\tau(\phi)\right)} \Big)   
 \ \big|_{\tau =0} \\[1mm]
 =&\,
  \tfrac{d}{d\tau} \ 
\bs\upupsilon\big( \vphi_\tau(\phi)\big)^* \left( \phi^{\u(\phi)}\right)
  \ \big|_{\tau =0}
  +
  \tfrac{d}{d\tau} \ 
\bs\upupsilon(\phi)^* \left(\vphi_\tau(\phi)^{\u(\phi)}\right)
  \ \big|_{\tau =0}
  +
  \tfrac{d}{d\tau} \ 
\bs\upupsilon(\phi)^* \left(\phi^{\u \left(\vphi_\tau(\phi)\right)}\right)
  \ \big|_{\tau =0}.
 \end{align*}
The result may be written as $\t{\mathfrak X}([\phi])\tfrac{\delta}{\delta[\phi]}$ as a derivation of $C^\infty(\M)$. 
Let us work out each term in turn. 
Inserting $\bs\upupsilon(\phi)\-\circ \bs\upupsilon(\phi) =\id_M$, the 1st term is
$
 \tfrac{d}{d\tau} \ 
 \bs\upupsilon(\phi)^* \bs\upupsilon(\phi)^{-1*}
\bs\upupsilon\big( \vphi_\tau(\phi)\big)^* \left( \phi^{\u(\phi)}\right)
  \ \big|_{\tau =0}
=
\bs\upupsilon(\phi)^*
\tfrac{d}{d\tau} \ 
\left( \bs\upupsilon\big( \vphi_\tau(\phi)\big) \circ \bs\upupsilon(\phi)\-\right)^* \left( \phi^{\u(\phi)}\right)
  \ \big|_{\tau =0}
$.
The term $ \bs\upupsilon\big( \vphi_\tau(\phi)\big) \circ \bs\upupsilon(\phi)\-$ is a curve in $M$, 
so $\tfrac{d}{d\tau}\ \bs\upupsilon\big( \vphi_\tau(\phi)\big) \circ \bs\upupsilon(\phi)\-\,  \ \big|_{\tau =0} = \bs d \bs\upupsilon \circ {\bs \upupsilon\-}_{|\phi}(\mathfrak X_{|\phi}) \, \in \Gamma(TM)$. 
Thus, 
\begin{equation}
\begin{aligned}
 \tfrac{d}{d\tau} \ 
\bs\upupsilon\big( \vphi_\tau(\phi)\big)^* \left( \phi^{\u(\phi)}\right)
  \ \big|_{\tau =0}
&=\bs\upupsilon(\phi)^*
\tfrac{d}{d\tau} \ 
\big( \bs\upupsilon\big( \vphi_\tau(\phi)\big) \circ \bs\upupsilon(\phi)\-\big)^* \left( \phi^{\u(\phi)}\right)
  \ \big|_{\tau =0}\\
&=
 \bs\upupsilon(\phi)^* \left(
 \mathfrak L_{\bs d \bs\upupsilon \circ {\bs \upupsilon\-}_{|\phi}(\mathfrak X_{|\phi})} \ \phi^{\u(\phi)}
 \right)\\
&=
 \bs\upupsilon(\phi)^*
 \left[
 \Big( 
 \mathfrak L_{\bs d \bs\upupsilon \circ {\bs \upupsilon\-}_{|\phi}(\mathfrak X_{|\phi})} \ \phi
 + 
 \delta_{ \big(\mathfrak L_{ \{ \bs d \bs\upupsilon \circ {\bs \upupsilon\-}_{|\phi}(\mathfrak X_{|\phi})\}  } \u(\phi)\big)\, \u(\phi)\-} \phi
 \Big)^{\u(\phi)}
 \right],
\end{aligned}
\end{equation}
 where we used a direct analogue of the identity \eqref{useful-Lie-id-loc} in the last step.
Inserting $\u(\phi)\, \u(\phi)\-=\id_{\Hl}$ in the 3rd term we get
$
  \bs\upupsilon(\phi)^*  \tfrac{d}{d\tau} \ 
 \left(\phi^{\u(\phi)\, \u(\phi)\-\u \left(\vphi_\tau(\phi)\right)}\right)
  \ \big|_{\tau =0}
=
 \bs\upupsilon(\phi)^*  \tfrac{d}{d\tau} \ 
 \big(\phi^{\u(\phi)} \big)^{\u(\phi)\-\u \left(\vphi_\tau(\phi)\right)}
  \ \big|_{\tau =0}
$.
The term $\u(\phi)\-\u \left(\vphi_\tau(\phi)\right)$ is a curve in $\Hl$ through the identity, so 
$\tfrac{d}{d\tau}\ \u(\phi)\-\u \left(\vphi_\tau(\phi)\right)\,  \ \big|_{\tau =0} = \u\-\bs d \u_{|\phi}(\mathfrak X_{|\phi}) \, \in \text{Lie}\Hl$. 
Thus,
\begin{align}
 \tfrac{d}{d\tau} \ 
\bs\upupsilon(\phi)^* \left(\phi^{\u \left(\vphi_\tau(\phi)\right)}\right)
  \ \big|_{\tau =0}
=
\bs\upupsilon(\phi)^* \delta_{\{ \u\-\bs d \u_{|\phi}(\mathfrak X_{|\phi}) \}} \big(\phi^{\u(\phi)}\big)
=
\bs\upupsilon(\phi)^*
\left[
\left(
\delta_{\{ \bs d \u{\u\-}_{|\phi}(\mathfrak X_{|\phi}) \}} \phi 
\right)^{\u(\phi)}
\right],
\end{align}
where we used an exact analogue of \eqref{wellknown-id} in the last step. 
Finally, the 2nd term is 
$\tfrac{d}{d\tau} \ 
\bs\upupsilon(\phi)^* \left(\vphi_\tau(\phi)^{\u(\phi)}\right)
  \ \big|_{\tau =0} 
\rdefeq  
\tfrac{d}{d\tau}\ F_{\left(\bs\upupsilon(\phi), \u(\phi) \right)} \left( \vphi_\tau(\phi)  \right)\  \big|_{\tau =0} 
=
F_{\left(\bs\upupsilon(\phi), \u(\phi) \right)\star} \mathfrak X_{|\phi}$.
As a vector on $\M$, we may find its expression as a derivation by applying it to  $\bs g \in C^\infty(\M)$:
  \begin{align*}
\big[F_{\left(\bs\upupsilon(\phi), \u(\phi) \right)\, \star} \mathfrak X\,  \big( \bs g \big)\big]([\phi]) 
&=
\tfrac{d}{d\tau}\ \bs g \left( F_{\left(\bs\upupsilon(\phi), \u(\phi) \right)} \big( \vphi_\tau(\phi) \big) \right)\  \big|_{\tau =0} \\
&=
 \ \underbrace{\tfrac{d}{d\tau}\, F_{\left(\bs\upupsilon(\phi), \u(\phi) \right)}\big( \vphi_\tau(\phi) \big) \, \big|_{\tau=0}}_{\big[\mathfrak X(F_{\left(\bs\upupsilon(\phi), \u(\phi) \right)})\big] (\phi)  }\, 
\left( \tfrac{\delta}{\delta [\phi]} \bs g \right) \big(F_{\left(\bs\upupsilon(\phi), \u(\phi) \right)}(\phi)\big)   \\
&=
 \underbrace{\big( \tfrac{\delta}{\delta \phi} F_{\left(\bs\upupsilon(\phi), \u(\phi) \right)}\big)(\phi)}_{\tfrac{\delta}{\delta \phi} \, \bs\upupsilon(\phi)^*\big( \phi^{\u(\phi)} \big)\, 
=\,
\bs\upupsilon(\phi)^*\big( (\ \ ) ^{\u(\phi)} \big)}  \hspace{-5mm}\mathfrak X(\phi) \ \ 
\left( \tfrac{\delta}{\delta [\phi]} \bs g \right) \big(\underbrace{\bs\upupsilon(\phi)^*\big( \phi^{\u(\phi)} \big)}_{\sim [\phi]} \big)  
\\
&=
\left[ \bs\upupsilon(\phi)^* \left( \mathfrak X(\phi)^{\u(\phi)}  \right)\tfrac{\delta}{\delta [\phi]} (\bs g) \right]
([\phi]).
\end{align*}
Gathering all three terms we have finally
\begin{align*}
\label{Pushforward-X-dress-loc}
 F_{\left(\bs\upupsilon(\phi), \u(\phi) \right)\star} \mathfrak X_{|\phi} 
 =
\bs\upupsilon(\phi)^* 
\left[
\Big(\ 
\bs d\phi_{|\phi} \big(\mathfrak X_{|\phi}  \big)
+
\mathfrak L_{\bs d \bs\upupsilon \circ {\bs \upupsilon\-}_{|\phi}(\mathfrak X_{|\phi})} \ \phi
+ 
\delta_{ \big(\mathfrak L_{ \{ \bs d \bs\upupsilon\circ {\bs \upupsilon\-}_{|\phi}(\mathfrak X_{|\phi})\}  } \u(\phi)\big)\, \u(\phi)\-} \phi
+
\delta_{\{ \bs d \u{\u\-}_{|\phi}(\mathfrak X_{|\phi}) \}} \phi 
\ \Big)^{\u(\phi)} 
\right]
\frac{\delta}{\delta [\phi]}_{|\, F_{(\bs\upupsilon, \u)}(\phi)}
\end{align*}
from which we find 
\begin{equation}
\begin{aligned}
{\bs d\phi^{ (\bs\upupsilon, \u ) }}_{|\phi}
\big(\mathfrak X_{|\phi} \big)
\defeq&\,
F_{(\bs\upupsilon, \u)}^\star \bs d[\phi]_{|F_{(\bs\upupsilon, \u)}(\phi)}\big(\mathfrak X_{|\phi} \big)= \bs d[\phi]_{|F_{(\bs\upupsilon, \u)}(\phi)} \big( {F_{(\bs\upupsilon, \u)}}_\star \mathfrak X_{|\phi} \big)    \\
=&\,
\bs\upupsilon(\phi)^* 
\Big[
\Big(\
\bs d\phi_{|\phi} \big( \mathfrak X_{|\phi} \big)
+
\mathfrak L_{\bs d \bs\upupsilon \circ {\bs \upupsilon\-}_{|\phi}(\mathfrak X_{|\phi})} \ \phi
+ 
\delta_{ \big(\mathfrak L_{ \{ \bs d \bs\upupsilon\circ {\bs \upupsilon\-}_{|\phi}(\mathfrak X_{|\phi})\}  } \u(\phi)\big)\, \u(\phi)\-} \phi
+
\delta_{\{ \bs d \u{\u\-}_{|\phi}(\mathfrak X_{|\phi}) \}} \phi 
\ \Big)^{\u(\phi)}
\Big].
\end{aligned}
\end{equation}
Which defines the dressing of $\bs d \phi$, the basis 1-form of basic forms on $\Phi$:
\begin{align}
\label{basic-basis-1-form-loc}
\bs d\phi^{(\bs\upupsilon, \u )}
=
\bs\upupsilon^* 
\Big[
\Big(\
\bs d\phi
+
\mathfrak L_{\bs d \bs\upupsilon \circ {\bs \upupsilon\-}} \phi
+ 
\delta_{ \big(\mathfrak L_{ \{ \bs d \bs\upupsilon\circ {\bs \upupsilon\-}\} } \u \big)\, \u\-} \phi
+
\delta_{ \bs d \u{\u\-}  } \phi 
\ \Big)^{\u}
 \Big]
 \quad \in \ \Omega^1_{\text{basic}}(\Phi).
\end{align}
This specialises to 
\begin{equation}
\label{special-case-dphi-dress}
\begin{aligned}
\bs d\phi^{\bs\upupsilon} 
&=
\bs\upupsilon^* \Big(\,
\bs d\phi
+
\mathfrak L_{\bs d \bs\upupsilon \circ {\bs \upupsilon\-}} \phi\,  \Big),\\
\bs d\phi^{\u} 
&= 
\Big(\,
\bs d\phi
+
\delta_{\bs d \u{\u\-} } \phi 
\, \Big)^{\u},
\end{aligned}
\end{equation}
which are respectively the $\Diff(M)$-dressing and $\Hl$-dressing of $\bs d\phi$, showing that 
\eqref{basic-basis-1-form-loc} generalises 
eq.(149) in \cite{Francois2023-a}, 
as well as eqs.(37)-(38) in \cite{Francois2021} and eq.(49) in \cite{Francois-et-al2021}. 
Such formulas feature in the sub-literature of covariant phase space dealing with ``edge modes" \cite{DonnellyFreidel2016}. 

Inserting \eqref{basic-basis-1-form-loc} into \eqref{Dressing-form-loc}, this yields the dressing of $\bs\alpha$:
\begin{align}
\label{Dressing-form-loc-bis}
\bs \alpha^{(\bs\upupsilon, \u)} =  \alpha \Big(  \!\w^\bullet\! \bs d\phi^{(\bs\upupsilon, \u)};\, \phi^{(\bs\upupsilon, \u)}  \Big) \ \  \in \ \Omega^\bullet_{\text{basic}}(\Phi).
\end{align}
On account of the formal similarity between $\Xi(\phi)=\bs\uppsi^*\big (\phi^{\bs\upgamma} \big)$ and  $F_{(\bs\upupsilon, \u)}(\phi)=\bs\upupsilon^*\big(\phi^\u \big)$,  
and between $\bs d\phi^{(\bs\uppsi, \bs\upgamma)}$ \eqref{dphi-vert-trsf-loc} 
and $\bs d\phi^{(\bs\upupsilon, \u)}$ \eqref{basic-basis-1-form-loc},  resulting into the close formal expressions of $\bs\alpha^{(\bs\uppsi, \bs\upgamma)}$  
and $\bs\alpha^{(\bs\upupsilon, \u)}$ \eqref{Dressing-form-loc}-\eqref{Dressing-form-loc-bis}, 
we confirm the rule of thumb to obtain the dressing of any form $\bs\alpha$ on $\Phi$: 
First compute the  $\bs\Diff_v(\Phi)\simeq C^\infty\big(\Phi, \Diff(M)\ltimes \Hl \big)$ transformation $\bs\alpha^{(\bs\uppsi, \bs\upgamma)}$, then substitute $(\bs\uppsi, \bs\upgamma) \rarrow (\bs\upupsilon, \u)$ in the resulting expression to obtain $\bs\alpha^{(\bs\upupsilon, \u)}$. 
This rule will be used systematically in what follows. 

\subsubsection{Dressing and flat connections}
\label{Dressing and flat connections}

As we already observed in section \ref{Dressing field and flat  connections}, dressing fields induce flat connections, and flat connections are expressible in terms of dressing fields. 
\medskip

\noindent
$\bullet$ A $\phi$-dependent dressing field induces the \emph{flat Ehresmann connection} 
\begin{equation}
\begin{aligned}
\bs\omegaf\defeq&\, 
-\bs d (\bs\upupsilon, \u)\cdot (\bs\upupsilon, \u)\- \quad \in \Omega^1_\text{eq}\big( \Phi, \diff(M)\oplus\text{Lie}\Hl \big) ,\\[.5mm]
\big({\bs \omegaf}_\text{\tiny{$\,|\,\Diff$}} \, ,\ {\bs\omegaf}_\text{\tiny{$\,|\,\Hl$}} \big) 
=&\, 
-\left(
\bs{d \upupsilon}\circ \bs\upupsilon\- , \,
\bs d\u\, \u\- 
- \u \mathfrak L_{ \bs{d \upupsilon}\circ \bs\upupsilon\- } \u\-
\right),
\end{aligned}
\end{equation}
using \eqref{linear-right-action-loc}-\eqref{right-action-tangent} in the second equality.
Its $\Ad$-equivariance is easily proven, 
\begin{equation}
\begin{aligned}
 R^\star_{(\uppsi, \upgamma)}  \bs\omegaf
 &=
 -\bs d \left( R^\star_{(\uppsi, \upgamma)} (\bs\upupsilon, \u)\right) \cdot
 \left( R^\star_{(\uppsi, \upgamma)} (\bs\upupsilon, \u) \right)\-
 =
 -(\uppsi, \upgamma)\- \cdot \bs d(\bs\upupsilon, \u) \cdot (\bs\upupsilon, \u)\-\cdot (\uppsi, \upgamma) \\
 &=
 \Ad_{(\uppsi, \upgamma)\-} \bs\omegaf.
\end{aligned}
\end{equation}
The verticality is also easily found, using \eqref{linear-equi-phi-dep-dressing} and \eqref{right-action-tangent},
\begin{equation}
\begin{aligned}
\iota_{(X, \lambda)^v} \bs\omegaf
&=
-\iota_{(X, \lambda)^v} \bs d\big( \bs\upupsilon, \u\big)  \cdot \big( \bs\upupsilon, \u\big)\-
=
-\big( -X \circ \bs\upupsilon\,,\ \mathfrak L_X \u -\lambda \u \big)\cdot \big( \bs\upupsilon, \u\big)\-\\
&=
-\left( -X\circ \bs\upupsilon\circ\upupsilon \-\,,\ 
(\mathfrak L_X\u -\lambda \u) \u\- - \u \mathfrak L_{\{-X\circ \bs\upupsilon\circ\upupsilon\- \}} \u\-
\right)
=(X, \lambda).
\end{aligned}
\end{equation}
Or, done another way, 
\begin{equation}
\begin{aligned}
\iota_{(X, \lambda)^v} \bs\omegaf
&=
-\iota_{(X, \lambda)^v}
\left(
\bs{d \upupsilon}\circ \bs\upupsilon\- , \,
\bs d\u\, \u\- 
- \u \mathfrak L_{ \bs{d \upupsilon}\circ \bs\upupsilon\- } \u\-
\right) \\
&=
-\left(
-X  \circ \bs\upupsilon\circ\upupsilon\-\,,\
(\mathfrak L_X \u \, \u\- - \lambda \u\, \u\- )
- \u \mathfrak L_{\{-X\circ \bs\upupsilon\circ\upupsilon\-\}} \u\-
\right)
=
(X, \lambda).
\end{aligned}
\end{equation}
\medskip
This shows that $\bs\omegaf$ satisfies the defining properties of a Ehresmann connection \eqref{connection-loc}. 
Its flatness is manifest from $\bs\omegaf\defeq 
-\bs d (\bs\upupsilon, \u)\cdot (\bs\upupsilon, \u)\-$, but less trivial to check from the explicit form $\bs\omegaf = -\left(
\bs{d \upupsilon}\circ \bs\upupsilon\- , \,
\bs d\u\, \u\- 
- \u \mathfrak L_{ \bs{d \upupsilon}\circ \bs\upupsilon\- } \u\-
\right)$ and the bracket \eqref{Liebracket-local-aut}.
\begin{align}
\bs \Omega_\text{\tiny{0}} &= \bs d \bs \omegaf + \tfrac{1}{2} [\bs \omegaf,\bs \omegaf]_\text{\tiny{Lie}} =0,\\[1mm]
\hookrightarrow\quad 
\Big( \bs \Omega_\text{\tiny{$0\,|\Diff$}} , \bs\Omega_\text{\tiny{$0\,|\Hl$}} \Big)
&=\left(
\bs d\bs\omega_\text{\tiny{$0\,|\Diff$}} + \tfrac{1}{2}\big[\bs \omega_\text{\tiny{$0\,|\Diff$}},\ \bs \omega_\text{\tiny{$0\,|\Diff$}}\big]_\text{\tiny{$\diff$(M)}}\, ,
\ \bs d \bs\omega_\text{\tiny{$0\,|\Hl$}} \!+ \tfrac{1}{2}\big[\bs \omega_\text{\tiny{$0\,|\Hl$}}, \bs \omega_\text{\tiny{$0\,|\Hl$}}\big]_\text{\tiny{Lie}}
- 
\omega_\text{\tiny{$0\,|\Diff$}} \big(\omega_\text{\tiny{$0\,|\Hl$}}\big)
\right) =(0, 0). \notag
\end{align}
The non-trivial part to check is  $\bs\Omega_\text{\tiny{$0\,|\Hl$}} =0$. 
Let us show it explicitly, for the benefit of the reader: 
\begin{align*}
\bs\Omega_\text{\tiny{$0\,|\Hl$}} 
&=
\bs d \big( -\bs{d\u}\u\- + \u \mathfrak L_{\bs{d\upupsilon}\circ\bs\upupsilon\-} \u\-\big)
\ &&\hspace{-13mm}+\ 
\tfrac{1}{2}\big[
-\bs{d\u}\u\- + \u \mathfrak L_{\bs{d\upupsilon}\circ\bs\upupsilon\-} \u\-, 
-\bs{d\u}\u\- + \u \mathfrak L_{\bs{d\upupsilon}\circ\bs\upupsilon\-} \u\-
\big]\\
& \ &&\hspace{-13mm}-\ 
(-\bs{d\upupsilon}\circ\bs \upupsilon)
\left(
-\bs{d\u}\u\- + \u \mathfrak L_{\bs{d\upupsilon}\circ\upupsilon\-} \u\-
\right)  \\[1mm]
&=
\cancel{\bs d(-\bs d\u \u\-) }
+
\bs d \big(  \u \mathfrak L_{\bs{d\upupsilon}\circ\bs\upupsilon\-} \u\- \big)
&&\hspace{-13mm}+ 
\tfrac{1}{2}\cancel{\big[
\bs{d\u}\u\-, \bs{d\u}\u\- \Big]}
- 
\big[ 
\u \mathfrak L_{\bs{d\upupsilon}\circ\upupsilon\-} \u\-, \bs d\u\u\-
\big] \\
& \ &&\hspace{-13mm}+
\tfrac{1}{2} \big[ 
\u \mathfrak L_{\bs{d\upupsilon}\circ\upupsilon\-} \u\-, \u \mathfrak L_{\bs{d\upupsilon}\circ\upupsilon\-} \u\-
\big] \\
& \ &&\hspace{-13mm}-
\mathfrak L_{\bs{d\upupsilon}\circ\upupsilon\-} (\bs d\u\u\-) 
+
\mathfrak L_{\bs{d\upupsilon}\circ\upupsilon\-}
\big( \u \mathfrak L_{\bs{d\upupsilon}\circ\bs\upupsilon\-} \u\-   \big) \\[1mm]
&= \cancel{\bs d\u \big( \mathfrak L_{\bs{d\upupsilon}\circ\upupsilon\-} \u\- \big)}
+ \u \bs d \big(\mathfrak L_{\bs{d\upupsilon}\circ\upupsilon\-} \u\- \big)
&&\hspace{-13mm}-
 \u  \mathfrak L_{\bs{d\upupsilon}\circ\upupsilon\-} \u\- \cdot \bs d\u\u\- 
 - 
 \cancel{\bs d\u\u\- \u  \mathfrak L_{\bs{d\upupsilon}\circ\upupsilon\-} \u\-} \\
 &\ &&\hspace{-13mm}+  
 \bcancel{\u  \mathfrak L_{\bs{d\upupsilon}\circ\upupsilon\-} \u\- \cdot  \u  \mathfrak L_{\bs{d\upupsilon}\circ\upupsilon\-} \u\- }\\
 & \ &&\hspace{-13mm}- 
 \u \u\- \mathfrak L_{\bs{d\upupsilon}\circ\upupsilon\-} \big( \bs d\u\u\-\big)
 +
 \bcancel{
 \mathfrak L_{\bs{d\upupsilon}\circ\upupsilon\-} \u \mathfrak L_{\bs{d\upupsilon}\circ\upupsilon\-} \u\-}
 + 
 \u \mathfrak L_{\bs{d\upupsilon}\circ\upupsilon\-}\mathfrak L_{\bs{d\upupsilon}\circ\upupsilon\-} \u\- \\[1mm]
 &= 
 \u \, \mathfrak L_{\bs d(\bs{d\upupsilon}\circ \bs\upupsilon\-)} \u\- 
 -
 \cancel{
 \u \mathfrak L_{\bs{d\upupsilon}\circ \bs\upupsilon\-} \bs d \u\-}
 &&\hspace{-13mm}-
 \cancel{\u \mathfrak L_{\bs{d\upupsilon}\circ \bs\upupsilon\-} \big( \u\-\bs d\u \u\- \big)}
 + 
 \u\  \tfrac{1}{2}\big[
 \mathfrak L_{\bs{d\upupsilon}\circ\upupsilon\-}, \mathfrak L_{\bs{d\upupsilon}\circ\upupsilon\-}
 \big]_{\text{\tiny{$\Gamma(TM)$}}} \u\- \\[1mm]
 &=
 \u \, \mathfrak L_{\bs{d\upupsilon}\circ \bs\upupsilon\-} \u\-
 + 
 \u \, \tfrac{1}{2}\mathfrak L_{ \big[\bs{d\upupsilon}\circ \bs\upupsilon\-,\, \bs{d\upupsilon}\circ \bs\upupsilon\- \big]_{\text{\tiny{$\Gamma(TM)$}}}}&&
 \hspace{-8mm} \u\- \\[1mm]
 &= 
 -\u \, \mathfrak L_{-\bs{d\upupsilon}\circ \bs\upupsilon\-} \u\-
 - 
 \u \, \tfrac{1}{2}\mathfrak L_{ \big[\bs{d\upupsilon}\circ \bs\upupsilon\-,\, \bs{d\upupsilon}\circ \bs\upupsilon\- \big]_{\text{\tiny{$\diff(M)$}}}}&&
 \hspace{-8mm} \u\- \\[1mm]
 &= - \u \, \tfrac{1}{2}\mathfrak L_{\{ \Omega_\text{\tiny{$0\,|\Diff$}} \}} \u\- =0. 
\end{align*}
As usual, $\bs\omegaf$ induces a covariant derivative $\bs D_\text{\tiny{$0$}} = \bs d\  + \rho_*(\bs\omegaf)$ on $\Omega^\bullet_\text{tens}(\Phi, \rho)$. 
In this case is satisfies $\bs D_\text{\tiny{$0$}}^2=0$. 
\medskip

\noindent
$\bullet$ Consider a $G$-valued 1-cocycle $C$ for the action of $\Diff(M)\ltimes \Hl$ on $\Phi$, satisfying by definition 
\begin{align}
C\big(\phi; (\uppsi, \upgamma)\cdot(\uppsi', \upgamma') \big) 
=
C\big(\phi; (\uppsi, \upgamma)\big)\cdot 
C\big(\phi^{(\uppsi, \upgamma)}; (\uppsi', \upgamma') \big),
\end{align}
where the product ``$\cdot$" on the r.h.s. is in the group $G$. From this we have that the inverse is 
\begin{align}
C\big(\phi; (\uppsi, \upgamma) \big)\-  
=
C\big(\phi^{(\uppsi, \upgamma)}; (\uppsi, \upgamma)\- \big).   
\end{align}
Its functional form allows to define a \emph{twisted dressing field} defined by $C\big(\ \ ; (\bs\upupsilon, \u) \big)$ whose equivariancce is indeed, 
\begin{align}
\label{twisted-dressing-loc}
\big[R^\star_{(\uppsi, \upgamma)} C\big(\ \ ; (\bs\upupsilon, \u) \big)\big] (\phi)
&=
C\big( \phi^{(\uppsi, \upgamma)}; (\uppsi, \upgamma)\- (\bs\upupsilon, \u)\big)
=
C\big( \phi^{(\uppsi, \upgamma)}; (\uppsi, \upgamma)\- \big) \cdot 
C\big(\phi; (\bs\upupsilon, \u) \big) \notag\\
&=
C\big(\phi; (\uppsi, \upgamma) \big)\-
 \cdot 
C\big(\phi; (\bs\upupsilon, \u) \big), \notag\\[1mm]
\text{so } \quad 
R^\star_{(\uppsi, \upgamma)} C\big(\ \ ; (\bs\upupsilon, \u) \big)
&=
C\big(\ \ ; (\uppsi, \upgamma) \big)\-
 \cdot 
C\big(\ \ ; (\bs\upupsilon, \u) \big).
\end{align}
The linear version of which is
\begin{align}
 \label{inf-equiv-twisted-dressing-loc}  
 \bs L_{(X, \lambda)^v}  C\big(\ \ ; (\bs\upupsilon, \u) \big)
 =
 -a\big((X, \lambda); \ \ \big)\cdot  C\big(\ \ ; (\bs\upupsilon, \u) \big).
\end{align}
Such a twisted dressing field is a tensorial 0-form, so its $\Diff_v(\Phi)\simeq C^\infty \big(\Phi, \Diff(M)\ltimes \Hl \big)$ and 
$\diff_v(\Phi)\simeq C^\infty \big(\Phi, \diff(M)\oplus \text{Lie}\Hl \big)$ transformations are, as a special case of \eqref{vert-trsf-twisted-tens-form-loc}
\begin{align}
C\big(\ \ ; (\bs\upupsilon, \u) \big)^{(\bs\uppsi, \bs\upgamma)}
=
C (\bs\uppsi, \bs\upgamma)\-
 \cdot 
C\big(\ \ ; (\bs\upupsilon, \u) \big)
\quad \text{ and }\quad 
\bs L_{(\bs X, \bs\lambda)^v} C\big(\ \ ; (\bs\upupsilon, \u) \big)
= - a(\bs X, \bs\lambda) \cdot C\big(\ \ ; (\bs\upupsilon, \u) \big),
\end{align}
and we remind the  notation $[C(\bs\uppsi,\bs\upgamma)](\phi)\defeq C\Big(\phi; \big(\bs\uppsi(\phi),\bs\upgamma(\phi)\big) \Big)$ and $[a(\bs X,\bs \lambda)](\phi):=a\big( ( \bs X(\phi), \bs \lambda(\phi) ); \phi\big)$, 
so that we may likewise write the twisted dressing field as $C(\bs\upupsilon, \u)$.

The latter induces the \emph{flat twisted connection}
\begin{align}
\bs\varpi_\text{\tiny{$0$}}
 \defeq
-\bs d C(\bs\upupsilon, \u) \cdot C(\bs\upupsilon, \u)\- \quad \in \ \Omega^1_{\text{eq}}(\Phi, C),
\end{align}
whose equivariance is easily found to be
\begin{equation}
\begin{aligned}
 R^\star_{(\uppsi, \upgamma)}  \bs\varpi_\text{\tiny{$0$}}
 &=
 -\bs d \left( R^\star_{(\uppsi, \upgamma)} C(\bs\upupsilon, \u)\right) \cdot
 \left( R^\star_{(\uppsi, \upgamma)} C(\bs\upupsilon, \u) \right)\- \\
 &=
 -C \big(\ \ ;(\bs\uppsi, \bs\upgamma) \big)\- \big[\bs dC(\bs\upupsilon, \u) \cdot C(\bs\upupsilon, \u)\-\big] C \big(\ \ ;(\bs\uppsi, \bs\upgamma) \big)
 -
\bs d C \big(\ \ ;(\bs\uppsi, \bs\upgamma) \big)\- \cdot C \big(\ \ ;(\bs\uppsi, \bs\upgamma)
 \big)\\
 &=\Ad_{C \big(\ \ ;(\bs\uppsi, \bs\upgamma) \big)\-} \bs\varpi_\text{\tiny{$0$}}
 + 
 C \big(\ \ ;(\bs\uppsi, \bs\upgamma) \big)\- \bs d C \big(\ \ ;(\bs\uppsi, \bs\upgamma).
\end{aligned}
\end{equation}
Its verticality property is derived from \eqref{inf-equiv-twisted-dressing-loc},
\begin{align}
\iota_{(X, \lambda)^v)} \bs\varpi_\text{\tiny{$0$}}
=
-\iota_{(X, \lambda)^v)} \bs d C(\bs\upupsilon, \u) \cdot C(\bs\upupsilon, \u)v\-
= a\big((X, \lambda); \ \ \big) \quad \in \LieG. 
\end{align}
These are indeed the defining properties \eqref{twisted-connection-loc} of a twisted connection. 
Its flatness,  $\b{\bs\Omega}_\text{\tiny{$0$}} =\bs d \bs\varpi_\text{\tiny{$0$}} + \tfrac{1}{2}[\bs\varpi_\text{\tiny{$0$}}, \bs\varpi_\text{\tiny{$0$}}]_\text{\tiny $\LieG$} =0$,
is obvious. 
As usual, $\bs\varpi_\text{\tiny{$0$}}$ induces a twisted covariant derivative $\b{\bs D}_\text{\tiny{$0$}} \defeq \bs d \ \ +\bs\varpi_\text{\tiny{$0$}}$ on $\Omega^\bullet_\text{tens}(\Phi, C)$, s.t. $\b{\bs D}_\text{\tiny{$0$}}^2 =0$.

In case $G=U(1)$, we have 
$C(\bs\upupsilon, \u)= \exp{\{-i\, \mathsf c (\bs\upupsilon, \u ) \}}$, with the \emph{Abelian twisted dressing field} 
\begin{equation}
\label{abelian-twisted-dressing}
\begin{aligned}
\mathsf c (\bs\upupsilon, \u )  \defeq&\, \mathsf c \big(\ \ ; (\bs\upupsilon, \u ) \big) ,\\
\text{s.t. }\quad 
R^\star_{(\uppsi, \upgamma)} \mathsf c (\bs\upupsilon, \u )
=&\, \mathsf c (\bs\upupsilon, \u )
 - \mathsf c \big(\ \ ; (\uppsi
 , \upgamma)\big). 
\end{aligned}
\end{equation}
The associated twisted connection is 
$\bs\varpi_\text{\tiny{$0$}}
 \defeq
-\bs d C(\bs\upupsilon, \u) \cdot C(\bs\upupsilon, \u)\- = i \bs d \mathsf c (\bs\upupsilon, \u )$, i.e. it is $\bs d$-exact as one would expect, so that its flatness is manifest: $\b{\bs\Omega}_\text{\tiny{$0$}} = \bs d \bs\varpi_\text{\tiny{$0$}} =0$.
For a twisted tensorial form $\bs\alpha=\exp{ \{i\bs\theta\} } \in \Omega^\bullet_\text{tens}(\Phi, C)$, the twisted covariant derivative gives 
\begin{equation}
\label{Flat-twisted-cov-deriv}
\begin{aligned}
&\b{\bs D}_\text{\tiny{$0$}} \alpha 
=
\bs d \bs\alpha + \bs\varpi_\text{\tiny{$0$}} \bs \alpha 
=
i\bs d \big(\bs\theta + \mathsf c (\bs\upupsilon, \u )\big)\, \bs\alpha
\quad \in \Omega^{\bullet+1}_\text{tens}(\Phi, C),\\
&\text{where }\quad 
\bs\theta + \mathsf c (\bs\upupsilon, \u ) \quad \in \Omega^\bullet_\text{basic}(\Phi).
\end{aligned}
\end{equation}
The invariant object $\bs\theta + \mathsf c (\bs\upupsilon, \u ) $ is also seen to be the phase of the dressing of $\bs\alpha$:
\begin{align}
\label{dressing-twisted-form}
\bs\alpha^{(\bs\upupsilon, \u)} 
=
C( \bs\upupsilon
, \u)\-\bs\alpha
=
\exp{ i\{ \bs\theta +\mathsf c (\bs\upupsilon, \u ) \} }
\rdefeq \exp{i\, \bs\theta^{(\bs\upupsilon, \u)}}
\quad \in \Omega^\bullet_\text{basic}(\Phi).
\end{align}
This is the geometry underlying the notion of  action/Lagrangian ``improved" via  Wess-Zumino counterterms implementing anomaly cancellation -- 
see e.g. section 12.3 in \cite{GockSchuck}, Chap.15 in \cite{Bonora2023}, or the end of Chap.4 in \cite{Bertlmann}.

\subsubsection{Residual transformations, and composition of dressing operations}
\label{Residual transformations, and composition of dressing operations}

We here details  the topic of possible residual transformations within the local DFM, counterpart of section \ref{Residual symmetries}.

\paragraph{Residual transformations of the first kind}
These are expected when one has a dressing field by which to reduce only a subgroup of  $\Diff(M)\ltimes \Hl$. 
For residual transformations to be well-defined as a group $\J$, the equivariance group of the dressing field, i.e. the group under which it has its defining transformation -- which may of may not be identical the group in which it takes value, called its target group -- must be a \emph{normal subgroup}~$\K$. 
Then,~the residual transformation group is $\J =\Diff(M)\ltimes \Hl)/ \K$.
The residual $\J$-transformations of the dressed fields $\phi^{(\bs\upupsilon, \u)}$ is then determined by the $\J$-transformation of $\phi$, which is known immediately as a special case of their $\big(\!\Diff(M)\ltimes \Hl\big)$-transformation, and of the $\K$-dressing field $(\bs\upupsilon, \u)$, which should naturally arise from its $\phi$-dependence.
From~this will follow  the residual $\J$-equivariance and verticality property of 
any $\K$-dressed/basic quantity $\bs\alpha^{(\bs\upupsilon, \u)}$ -- which is  $\bs \K= C^\infty(\Phi, \K)$-invariant -- and thus its transformation under $\bs \J \defeq C^\infty\big(\Phi, \J \big) \subset \bs\Diff_v(\Phi)\simeq C^\infty\big(\Phi, \Diff(M)\ltimes \Hl \big)$.

In~favorable circumstances, the $\J$-transformation of the dressing field $(\bs\upupsilon, \u)$ is s.t. $\phi^{(\bs\upupsilon, \u)}$ have \emph{standard} transformations under $\J$ -- i.e. identical to the $\J$-transformations of the bare fields $\phi$.
Thus, the residual $\J$-equivariance and verticality property of 
a dressed quantity $\bs\alpha^{(\bs\upupsilon, \u)}$
 will be the same as that of $\bs\alpha$, and therefore will share the same $\bs \J \defeq C^\infty\big(\Phi, \J \big)$-transformation.
This leaves open the possibility that a second $\J$-dressing field may be found and used to produce completely invariant dressed variables, provided that this second dressing field is also $\K$-invariant. 
These are the compatibility conditions analogous to \eqref{Aut_v-dress}-\eqref{CompCond2} discussed in the global case, section \ref{Composition of dressing operations}.
\medskip

 As observed below \eqref{dressed-field-inv-loc-bis}, this occurs naturally when considering the $\Hl$-dressing $\u$, in which  case we have $\K=\Hl  \triangleleft \big(\!\Diff(M)\ltimes \Hl\big)$ and $\J=\Diff(M)$. As shown there, the dressed field $\phi^\u$ is $\Hl$-invariant. 
 Since furthermore, by \eqref{phi-dep-dressing-loc}, $R^\star_\uppsi \u = \uppsi^*\u$,  the residual $\J=\Diff(M)$-transformation of $\phi^\u$ is the standard form $R_\uppsi \phi^\u = \uppsi^*(\phi^\u)$, as one would expect. 
Therefore any dressed form $\bs\alpha^\u=\alpha\big(\!\w^\bullet\!\bs d\phi^\u ;\,\phi^\u \big)$, for which one may use \eqref{special-case-dphi-dress} or a special case of \eqref{Dressing-form-loc-bis}, will have the same  $\J=\Diff(M)$-equivariance and verticality property as $\bs\alpha$, and thus the same $\bs \J =C^\infty\big(\Phi, \Diff(M)\big)$-transformation.
 
 This allows for the possibility of using a second $\J=\Diff(M)$-dressing field $\bs\upupsilon$, which again by 
\eqref{phi-dep-dressing-loc}
 satisfies the compatiblity condition $R^\star_\upgamma \bs\upupsilon = \upupsilon$, i.e. it is $\K=\Hl$-invariant. 
 Hence the possibility to write the completely invariant dressed field $(\phi^\u)^{\bs\upupsilon}=\phi^{(\bs\upupsilon, \u)}$, and to build basic objects $(\bs\alpha^\u)^{\bs\upupsilon}
 =
 \bs\alpha^{(\bs\upupsilon, \u)}$. 
 \medskip

 Another such case  presents itself when reducing only part of the gauge group $\Hl$ via a $\K$-dressing field $\u$, with $\K \triangleleft \Hl$ so that there is a residual gauge group $\Hl^{\text{{\tiny res}}}\defeq \Hl/\K$, and 
 the residual transformation group is 
 $\J=\Diff(M)\ltimes \Hl^{\text{{\tiny res}}}$. 
We may then have an  analogue of Prop.\ref{prop1}   in section \ref{Residual symmetries of the first kind}. 

\begin{prop}
\label{prop1-loc}
Given a $\K$-dressing field $\u$, if its  $\J$-residual equivariance is 
\begin{align}
\label{comp-cond-1st-dressing}
  R^\star_{(\uppsi, \upeta)} \u = \uppsi^*(\u^\upeta)= \uppsi^*\big(\upeta\- \u\, \upeta\big) 
 \qquad \text{ for }\ (\uppsi,\upeta) \in \J=\Diff(M)\ltimes \Hl^{\text{{\tiny res}}},
\end{align}
then the dressed fields $\phi^\u$ are standard $\Hl^{\text{{\tiny res}}}$-gauge fields, i.e. have $\Hl^{\text{{\tiny res}}}$-transformations identical to their bare counterparts $\phi$: $R_{(\uppsi, \upeta)} \phi^\u= \uppsi^*\big( (\phi^\u)^\upeta \big)$.
Any dressed form $\bs\alpha^\u$ is $\K$-basic but has identical $\J$-equivariance and verticality, hence the same
$\bs\J =C^\infty\big(\Phi, \Diff(M)\ltimes \Hl^{\text{{\tiny res}}} \big)$-transformation, as its bare counterpart $\bs\alpha$.
For example,  for $\bs d\phi^\u$ we have
\begin{equation}
\begin{aligned}
(\bs d\phi^\u)^{\bs\upgamma} 
&=
\bs d\phi^\u \quad \text{ for } \bs\upgamma \in \bs \K, \\
(\bs d\phi^\u)^{(\uppsi, \upeta)}
&=
\bs \uppsi^* 
\Big[
\Big( \bs d \phi^\u + \mathfrak{L}_{\bs d \bs \uppsi \circ \bs \uppsi\-} \, \phi^\u + \delta_{(\mathfrak{L}_{\bs d \bs \uppsi \circ \bs \uppsi\-} \bs \upeta)\,  \bs \upeta\-} \, \phi^\u + \delta_{\bs d \bs \upeta  \bs \upeta\-} \, \phi^\u \Big)^{\bs \upeta}
\Big],
\quad
\text{ for } (\bs\uppsi, \bs\upeta) \in  \bs\J,
\end{aligned}
\end{equation}
the last line being  entirely analogous to $\bs d\phi^{(\bs\uppsi, \bs\upgamma)}$ \eqref{dphi-vert-trsf-loc}. 
\end{prop}
If one happens to identify a $\Hl^{\text{{\tiny res}}}$-dressing field $\u'$ satisfying the compatibility condition
\begin{align}
\label{comp-cond-2nd-dressing}
    R^\star_{(\uppsi, \upgamma
    )}\u' = \uppsi^*\u' 
    \qquad \textbf{i.e. } \quad 
    R^\star_\upgamma \u' = \u'
    \quad \text{ for }\ \upgamma \in \K,
\end{align}
then one may build the $\Hl$-invariant  dressed fields $(\phi^\u)^{\u'}=\phi^{\u\,\u'}$, 
where the composite $\u\, \u'$ is a $\Hl$-dressing field, 
since \eqref{comp-cond-1st-dressing} and 
 \eqref{comp-cond-2nd-dressing} together ensure that $R^\star_\upgamma \u\,\u'= \upgamma  \-  \u\,\u'$ and $R^\star_\upeta \u\,\u'= \upeta  \-  \u\,\u'$. 
 We can thus produce the $\Hl$-basic form 
 $(\bs\alpha^\u)^{\u'}  = \bs\alpha^{\u\,\u'}$.
 
 By \eqref{comp-cond-2nd-dressing} still, it is manifest that $\phi^{\u\,\u'}$ have standard $\Diff(M)$-residual transformation: $R_\uppsi\,\phi^{\u\,\u'} =\uppsi^*( \phi^{\u\,\u'})$. 
 So~any dressed form $\bs \alpha^{\u\,\u'}$ shares the same $\Diff(M)$-equivariance and verticality as its bare counterpart $\bs \alpha$, and thus the same $C^\infty\big(\Phi, \Diff(M) \big)$-transformation. 
 For example, for $\bs d\phi^{\u\,\u'}$,
 \begin{equation}
\begin{aligned}
(\bs d\phi^{\u\,\u'})^{\bs\upgamma} 
&=
\bs d\phi^\u \quad \text{ for } \bs\upgamma \in \bs \Hl, \\
(\bs d\phi^{\u\,\u'})^{\uppsi}
&=
\bs \uppsi^* 
\Big( \bs d \phi^{\u\,\u'} + \mathfrak{L}_{\bs d \bs \uppsi \circ \bs \uppsi\-} \, \phi^{\u\,\u'}  \Big)
\quad
\text{ for } \bs\uppsi \in  C^\infty\big(\Phi, \Diff(M) \big).
\end{aligned}
\end{equation}
The residual $\Diff(M)$-transformation can be reduced if one finds a $\Diff(M)$-dressing field $\bs\upupsilon$  satisfying the compatibility condition
\begin{align}
R^\star_{(\uppsi, \upgamma)} \bs\upupsilon = \uppsi\-\circ \bs\upupsilon
\qquad \text{i.e. }\quad 
R^\star_\upgamma \bs\upupsilon =\bs\upupsilon \ \text{ for }\ \upgamma \in \Hl.
\end{align}
Then the final dressed fields $(\phi^{\u\,\u'})^{\bs\upupsilon}=\phi^{(\bs\upsilon,\, \u\,\u')}$ are fully $\big(\!\Diff(M)\ltimes \Hl\big)$-invariant, and one may produce finally basic forms $(\bs\alpha^{\u\,\u'})^{\bs\upupsilon}=\bs\alpha^{(\bs\upsilon, \u\,\u')} \in \Omega^\bullet_\text{basic}(\Phi)$.\footnote{
Conceivably one may have reduced for only a (normal) subgroup of $\Diff(M)$, e.g. compactly supported diffemorphisms, but this seems of limited interest. See \cite{Francois2023-a}  for a brief discussion.}

We will provide an example illustrating this in section \ref{Examples} below, when treating Einsein-Maxwell theory with  charged matter modeled by a complex scalar field. 

\medskip

\paragraph{Residual transformations of the second kind}
This distinct  type represents a parametrization of the ambiguity, or arbitrariness, in the choice of dressing field. 
It amounts to the statement that two dressing fields $(\upupsilon', \text u')$ and $(\upupsilon, \text u)$ in $\D r[N; M, H]$
are a priori related by an element $(\upvarphi, \upzeta) \in \Diff(N)\times \G_\text{\tiny{loc}}$, with $ \G_\text{\tiny{loc}}\defeq \big\{ \upzeta: U\subset M \rarrow \mathsf G\ |\ \upzeta^\upgamma=\upzeta \big\}$ and $\mathsf G$ is the target Lie group of the dressing field $\text u$ (which may be either s.t. $\mathsf G \subset H$ or s.t. $\mathsf G \supset H$, and was initially put as $\mathsf G=H$ in \eqref{Dressing-field-full} for simplicity of exposition). 
That is, we have
\begin{align}
(\upupsilon', \text u') 
=
\big( \upupsilon \circ \upvarphi,  \,\text u \upzeta \,\big). 
\end{align}
Defining $\b\upzeta \defeq \upupsilon^* \upzeta: N \rarrow \mathsf G$, and $\b \G_\text{\tiny{loc}} \defeq \upupsilon^* \G_\text{\tiny{loc}}$ , the above can be rewritten as 
\begin{align}
(\upupsilon', \text u') 
=
\left( \upupsilon \circ \upvarphi,  \,\text u\, 
 (\upupsilon\-)^*\b \upzeta \,\right)
 =
 (\upupsilon, \text u)\cdot (\upvarphi, \upzeta),
\end{align}
extending the semi-direct product rule \eqref{semi-dir-prod-DiffM-H-2}. 
Furthermore, we write
\begin{align}
(\upupsilon'', \text u'') 
&=
\big( \upupsilon' \circ \upvarphi',  \,\text u' \upzeta' \,\big) \notag\\
&=
\big( \upupsilon' \circ \upvarphi',  \,\text u' ({\upupsilon'}\-)^* \b\upzeta' \,\big) \notag\\
&=
\left( \upupsilon \circ \upvarphi\circ \upvarphi', 
\text u\, 
 (\upupsilon\-)^*\b \upzeta\, [(\upupsilon\circ \upvarphi)\-]^* \b \upzeta'
\right) \notag\\
&=
\left( \upupsilon \circ \upvarphi\circ \upvarphi', 
\text u\, 
 (\upupsilon\-)^* \big( \b \upzeta\  (\upvarphi\-)^* \b \upzeta'\big)
\right) \notag\\
&=
 (\upupsilon, \text u)\cdot \left(\upvarphi\circ \upvarphi',  \b \upzeta\  (\upvarphi\-)^* \b \upzeta'
 \right) \notag\\
 &=
 (\upupsilon, \text u)\cdot  \left( (\upvarphi , \b\upzeta ) \cdot (\upvarphi', \b\upzeta')
 \right),
\end{align}
where in the last step we define the semi-direct product in $\Diff(N) \ltimes \b\G_\text{\tiny{loc}}$. 
The last two equations show that there is a right action of $\Diff(N) \ltimes \b\G_\text{\tiny{loc}}$ on the space of dressing fields $D r[N; M, H]$:
\begin{equation}
\label{residual-2-on-dressings-loc}
\begin{aligned}
    D r[N; M, H] \times \big(\!\Diff(N) \ltimes \b\G_\text{\tiny{loc}} \big)
    &\rarrow D r[N; M, H], \\
    \Big((\upupsilon, \text u)\,,\ (\upvarphi, \b\upzeta) \Big) &\mapsto 
     R_{(\upvarphi, \b\upzeta)} (\upupsilon, \text u) \defeq (\upupsilon, \text u)\cdot (\upvarphi, \b\upzeta).
\end{aligned}
\end{equation}
This can be understood as the local counterpart of section \ref{Residual symmetries of the second kind}, 
where $\Diff(N) \ltimes \b\G_\text{\tiny{loc}}$ is the local version of $\Aut(Q)$. This action formally extends, a priori, to $\phi$-dependent dressing fields $(\bs\upupsilon, \u)$. 

Then, since $\phi$ are local representatives of fields on $P$, they do not support the action of $\Diff(N) \ltimes \b\G_\text{\tiny{loc}}$, which we may denote $\phi^{(\upvarphi, \upzeta)}=\phi$.
This, combined to \eqref{residual-2-on-dressings-loc}, implies that there is a right action of $\Diff(N) \ltimes \b\G_\text{\tiny{loc}}$ on the space of dressed fields $\Phi^{(\upupsilon, \text u)}$
\begin{equation}
\label{residual-2-right-action}
\begin{aligned}
    \Phi^{(\upupsilon, \text u)} \times \big(\! \Diff(N) \ltimes \b\G_\text{\tiny{loc}} \big)
    &\rarrow \Phi^{(\upupsilon, \text u)},\\
    \Big(\phi^{(\upupsilon, \text u)}\,,\ (\upvarphi, \b\upzeta) \Big) &\mapsto 
     R_{(\upvarphi, \upzeta)} \,\phi^{(\upupsilon, \text u)} 
     \defeq
     \upvarphi^* \left( \left[\phi^{(\upupsilon, \text u)}\right]^{\b\upzeta} \right).
\end{aligned}
\end{equation}
It is all but analogous to the action \eqref{group-law-loc-aut} of $\Diff(M)\ltimes \Hl$ on $\Phi$. 
Therefore, the space of dressed fields $\Phi^{(\upupsilon, \text u)}$ is, a priori, a $\big(\!\Diff(N) \ltimes \b\G_\text{\tiny{loc}}\big)$-principal bundle. It is then possible to elaborate on all the stuctures -- tangent bundle, vertical bundle, spaces of forms, etc. -- that exist on 
$\Phi^{(\upupsilon, \text u)}$ as we did for $\Phi$. 
In particular, dressed forms $\bs \alpha^{(\upupsilon, \text u)}$ \eqref{Dressing-form-loc-bis}
 are clearly naturally interpretable as forms on 
$\Phi^{(\upupsilon, \text u)}$. 
We might then write down their vertical transformations under $\bs\Diff\big(\Phi^{(\upupsilon, \text u)} \big) \simeq C^\infty \left(\Phi^{(\upupsilon, \text u)}, \Diff(N)  \ltimes \b\G_\text{\tiny{loc}}\right)$. 
It is easy to see that they would be formally identical to the $\bs{\Diff}_v(\Phi)\simeq C^\infty\big(\Phi, \Diff(M)\ltimes \Hl \big)$-transformation of their bare counterpart $\bs \alpha$ on $\Phi$. 
By this rule of thumb, we immediately write for example, as an exact analogue of \eqref{dphi-vert-trsf-loc},
\begin{align}
\label{res-2-dphi-dressed}
    {\big(\bs d\phi^{(\upupsilon, \text u)}\big)^{(\bs\upvarphi, \b{\bs\upzeta})}} 
    =
    \bs \upvarphi^* 
    \Big[
    \Big( \bs d \phi^{(\upupsilon, \text u)} + \mathfrak{L}_{\bs d \bs \upvarphi \circ \bs \upvarphi\-} \, \phi^{(\upupsilon, \text u)} + \delta_{(\mathfrak{L}_{\bs d \bs \upvarphi \circ \bs \upvarphi\-} \b{\bs\upzeta} )\,  \b{\bs\upzeta}\-} \, \phi^{(\upupsilon, \text u)} + \delta_{\bs d \b{\bs\upzeta}  \b{\bs\upzeta}\-} \, \phi^{(\upupsilon, \text u)} \Big)^{\b{\bs \upzeta}}
    \Big].
\end{align}

We may observe that \eqref{residual-2-on-dressings-loc}-\eqref{res-2-dphi-dressed} reproduce and generalise formulas found in the edge mode literature, and pertaining to what is called there ``surface symmetries" or ``corner symmetries".
Compare \eqref{residual-2-on-dressings-loc} with e.g. eq.(2.43) and eq.(3.44) in \cite{DonnellyFreidel2016} or eq.(3.68) and eq.(4.64) in \cite{Geiller2017}.
The name stems from the fact that in this literature, dressing fields are introduced as ``edge modes", i.e. as d.o.f. living on a codimension 2 submanifold of $M$ (a corner), which is then also the support of a transformation that they alone support, and is none other than our  \eqref{residual-2-on-dressings-loc}. 

As we have stated in introducing the topic, the  group $\Diff(N)  \ltimes \b\G_\text{\tiny{loc}}$ parametrizes the a priori ambiguity in the choice of dressing field. 
It is essentially isomorphic to $\Diff(M)\ltimes \Hl$, and thus  encodes neither more nor less information. 
Which is expected from the global fact that the dressing is a bundle morphism $\u: Q \rarrow P$, so one indeed would expect the natural morphism $\Aut(Q) \simeq \Aut(P)$, from which descends the morphism $\Diff(N)  \ltimes \b\G_\text{\tiny{loc}} \simeq \Diff(M)\ltimes \Hl$.  

This ``residual" group  is naturally there, and hardly avoidable, when one introduces an \emph{ad hoc} dressing field in a theory. 
By which we mean that it is introduced as d.o.f. independent from the original set of fields $\phi$ and for the main purpose of either implementing a $\Diff(M)\ltimes \Hl$ symmetry in a theory which initially does not enjoy it, or, relatedly, to ``restore" a $\Diff(M)\ltimes \Hl$ symmetry broken by a fixed background structure -- in effect the latter moves amount to concealing the background structure.
In such cases,  \emph{ad hoc} dressing fields generalise Stueckelberg~fields. 

The only hope for the residual group to be reduced, perhaps even to a discrete group, is with $\phi$-dependent dressing field, i.e. those built from the original d.o.f. of the theory, when the relational interpretation of the formalism is the most natural and compelling. 
We will comment further on this in section \ref{Discussion}. 

Before, we need to complete the technical and conceptual picture of the local DFM by discussing dressed integrals and how it instantiate a notion of integration on  \emph{physical} spacetime.

\subsubsection{Dressed integrals and integration on spacetime}
\label{Dressed integrals and integration on spacetime}

We can now elaborate on the local version of section \ref{Dressed regions and integrals}. 
In section \ref{Associated bundle of regions and integration theory for local field theory}, we observed that for $U \in \bs U(M)$ and $\bs \alpha \in \Omega^\bullet \big(\Phi, \Omega^{\text{top}}(U) \big)$, 
integrals $\bs\alpha_U=\langle \bs \alpha, U \rangle =\int_U \bs \alpha$ are objects on $\Phi \times \bs U(M)$ with values in $\Omega^\bullet(\Phi)$.
Furthermore, integrals with tensorial integrand are invariant under the action of $C^\infty\big(\Phi, \Diff(M)\ltimes \Hl \big)$ as defined by \eqref{Vert-trsf-int-generic-loc}. 
Their projection along $\b\pi : \Phi \times \bs U(M) \rarrow \b{\bs U}(M)$, $(\phi, U) \mapsto [\phi,U]=[\psi^*\phi, \psi\-(U)] $ is well-defined, so that they can be said ``basic" w.r.t. $\b \pi$. 
Such integrals descend to the associated bundle of regions $\b{\bs U}(M)\defeq \Phi \times \bs U(M)/\sim$, which is the quotient of the product space by the action \eqref{right-action-bundle-region} of $\Diff(M)\ltimes \Hl$. 

In section \ref{Building basic forms on local field space}, dressed forms on $\Phi$ were defined as  basic on $\Phi$, i.e. in Im$\,\pi^\star$ with the projection realised via a dressing field,  $F_{(\bs\upupsilon,  \u)} \sim \pi$.
As argued there, we can then rely on the formal similarity of the actions of $ F_{\bs u}$ and $ \Xi \in \bs\Diff_v(\Phi) \simeq C^\infty\big(\Phi, \Diff(M)\ltimes \Hl \big)$ to read the dressing $\bs\alpha^{(\bs\uppsi, \u)}$ of a form $\bs\alpha$ from its vertical transformation 
  $\bs\alpha^{(\bs\uppsi, \bs\upgamma)}$.
  Likewise, we define \emph{dressed integrals} as being basic 
  on $\Phi \times \bs U(M)$, i.e. in Im$\,\b\pi^\star$, with the projection realised as:
 \begin{equation}   
 \label{Proj-map-complete-loc}
 \begin{split} 
     \b F_{(\bs\upupsilon,  \u)}: \Phi \times \bs U(M) &\rarrow \b{\bs U}(M) \simeq \Phi^{\bs u} \times \bs U(N), \\
  (\phi, U) &\mapsto  
  \b F_{(\bs\upupsilon,  \u)}(\phi, U)
  \defeq \big(F_{(\bs\upupsilon,  \u)}(\phi), \bs\upupsilon\-(U) \big) =  \big( \bs\upupsilon^*(\phi^{\u}),  \bs\upupsilon\-(U) \big).
 \end{split}    
 \end{equation} 
 We remind that, by definition, $U\in \bs U(M)$ is acted upon trivially by $\Hl$. 
 
 Let us highlight that \eqref{Proj-map-complete-loc} features the notion of a \emph{dressed 
 region} $U^{\bs\upupsilon}\defeq \bs\upupsilon\-(U)$, which is a map
 \begin{align}
 \label{dressed-region-loc}
 U^{\bs\upupsilon}: \Phi \times \bs U(M) \rarrow \bs U(N),
  \end{align}
s.t. for $(\uppsi, \upgamma) \in \Diff(M)\ltimes \Hl$, 
 \begin{align}
 (U^{\bs\upupsilon})^{(\uppsi, \upgamma)} 
 \defeq 
 \b R^\star_{(\uppsi, \upgamma)} \,  U^{\bs\upupsilon}
 =
 (R^\star_{(\uppsi, \upgamma)} \bs\upupsilon)\- \circ \uppsi\-(U)
 = 
 (\uppsi\- \circ \bs\upupsilon)\-\circ \psi\-(U) 
 = 
 \bs\upupsilon\-(U) \rdefeq U^{\bs\upupsilon}.
 \end{align}
From which follows that under the action of $(\bs\uppsi, \bs\upgamma) \in \C^\infty\big(\Phi, \Diff(M)\ltimes\Hl\big)$ we have, 
\begin{align}
\label{Dressed-regions-inv-loc}
(U^{\bs\upupsilon})^{(\bs\uppsi, \bs\upgamma)} \defeq \b \Xi^\star  U^{\bs\upupsilon} 
=
U^{\bs\upupsilon}.
\end{align}
In other words, $U^{\bs\upupsilon}$ is a $\phi$-dependent $\big(\!\Diff(M)\ltimes \Hl\big)$- \emph{and} $C^\infty\big(\Phi, \Diff(M)\ltimes \Hl\big)$-invariant region of the physical \emph{relationally defined} \mbox{spacetime},
which we may denote $M^{(\bs\upupsilon, \u)}$. 
As reminded in section \ref{Relationality in  general-relativistic gauge field theory}, 
the  hole argument and the  point-coincidence argument establish that  physical spacetime is defined  \emph{relationally}, in a $\Diff(M)$-invariant way, by its gauge field content. 
This fact is usually tacitly encoded by the $\Diff(M)$-covariance of general-relativistic gauge field theories, but made manifest via the DFM:
$U^{\bs\upupsilon}$~are manifesly $\Diff(M)$-invariant and manifestly $\phi$-relationally defined  regions,  representing faithfully regions of the physical  spacetime, on which relationally defined and $\big(\!\Diff(M)\ltimes \Hl \big)$-invariant fields $\phi^{(\bs\upupsilon, \u)}\defeq \bs\upupsilon^*(\phi^{\u})$ live, and can  be integrated over. 
This is the second part of the picture whose first part, unpacked below \eqref{F-map-loc}, is that $\phi^\u$ implements the internal point-coincidence argument, showing how the internal d.o.f. of the fields $\phi$ co-define each other in a $\Hl$-invariant way. 
These invariant internal d.o.f. in turn coordinatise the ``points" of the internal structure (the fiber) of the enriched physical spacetime.

With \eqref{Proj-map-complete-loc} we  thus get the complete formal implementation of the relational core of gRGFT arising from the generalised hole and point-coincidence arguments. 
But observe that we further get a relational definition of spacetime invariant under the much bigger \emph{field-dependent} transformation group $C^\infty\big(\Phi, \Diff(M)\ltimes \Hl \big)$, which is indeed the largest symmetry group of gRGFT, as observed at the end of section \ref{Geometric prescription for the variational principle in field theory}.

Observe that an immediate consequence of the above is  that the physical 
 \emph{relational boundary} $\d U^{\bs\upupsilon}$ of a physical spacetime region $U^{\bs\upupsilon}$ is necessarily $\big(\!\Diff(M)\ltimes \Hl\big)$-invariant. 
This invalidate the claims, often repeated in the gauge field theory and gravity literature, that boundaries ``break" either $\Diff(M)$-invariance or $\Hl$-invariance, or~both.  
These claims are indeed logically equivalent to the  hole argument and its generalisation, as they commit the conceptual mistake of thinking of points of $M$ as physical entities defined independently of the field content. 
Such ``boundary problems" overlook the relational resolution offered by the (generalised) point-coincidence argument.\footnote{The notion of  ``edge modes" d.o.f. -- 
i.e. \emph{ad hoc} dressing fields -- sometimes introduced, and interpreted as the Goldstone bosons associated to these $\Diff(M)$ or $\Hl$ symmetry breaking, is then  a solution to an arguably non-existing problem. Or rather to a problem artificially introduced in gRGFT, i.e. treating a boundary as a background structure, and then trying to conceal that background structure. See section \ref{Discussion} next.}
We~shall elaborate further on this precise point in a forthcoming paper \cite{JTF-Ravera2024bdy}.
\medskip

We have all we need to defined integration \emph{on the physical spacetime}.
On  $\Omega^\bullet \big(\Phi, \Omega^{\text{top}}(U) \big) \times \bs U(M)$ we define:
\begin{equation}  
\label{dressed-regions-bundle-loc}
\begin{split} 
\t F_{(\bs\upupsilon, \u)}: \Omega^\bullet \big(\Phi, \Omega^{\text{top}}(U) \big) \times \bs U(M) &\rarrow 
\Omega^\bullet_\text{\text{basic}}\big(\Phi, \Omega^{\text{top}}(N) \big) \times \bs U(N), \\
(\bs \alpha, U) &\mapsto
\t F_{(\bs\upupsilon, \u)}^\star (\bs \alpha, U)\defeq \big(\bs \alpha^{{(\bs\upupsilon, \u)}}, \bs\upupsilon\-(U) \big), 
\end{split}    
\end{equation}  
with indeed $ \Omega^\bullet_\text{\text{basic}}\big(\Phi, \Omega^{\text{top}}(N) \big) \simeq \Omega^\bullet \big(\Phi^{(\bs\upupsilon, \u)}, \Omega^{\text{top}}(N) \big)$, as  dressed fields $\phi^{(\bs\upupsilon, \u)}$ live on Im$(\bs\upupsilon\-)\subset N$. 
Then, in formal analogy with \eqref{Vert-trsf-int-generic-loc}, the dressing of an integral $\bs\alpha_U \defeq \langle \bs\omega, U \rangle=\int_U \bs\alpha$ is
\begin{equation}   
\label{Dressed-integral-loc}
\begin{split} 
{\bs\alpha_U}^{(\bs\upupsilon, \u)} 
\defeq 
\langle\ \, ,\ \rangle \circ  \t F_{(\bs\upupsilon, \u)}(\bs \alpha, U) 
= 
\langle \bs\alpha^{(\bs\upupsilon, \u)} , \bs\upupsilon\-(U)   \rangle 
= 
\int_{\bs\upupsilon\-(U)} \hspace{-4mm} \bs\alpha^{(\bs\upupsilon, \u)}
=
\int_{U^{\bs\upupsilon}} \hspace{-4mm} \bs\alpha^{(\bs\upupsilon, \u)}
\rdefeq
\big(\bs\alpha^{(\bs\upupsilon, \u)}\big)_{U^{\bs\upupsilon}}. 
\end{split}    
\end{equation}  
Such dressed integrals define integration on physical spacetime, insofar as $U^{\bs\upupsilon}$ is interpretable as an invariantly and relationally defined region of spacetime.\footnote{
We remark that, if it exists, the residual transformation of the 2nd kind $C^\infty\big(\Phi^{(\bs\upupsilon, \u)}, \Diff(N)\ltimes \b\G_\text{\tiny{loc}} \big)$ may  act on dressed integrals, analogously to \eqref{Vert-trsf-int-generic-loc}, as
\begin{align*}
\left(({\bs\alpha_U})^{(\bs\upupsilon, \u)} \right)^{(\bs\upvarphi, \b{\bs\upzeta})} 
= 
\int_{\bs\upvarphi\-(U^{\bs\upupsilon})}  \left(\bs\alpha^{(\bs\upupsilon, \u)}\right)^{(\bs\upvarphi, \b{\bs\upzeta})}. 
\end{align*}
But again, this situation may likely arise only for \emph{ad hoc} dressing fields, when the relational interpretation of the DFM is unavailable, or implausible. See next section.}
Clearly, when $\bs d$ acts on such an integral, it sees the dressed region $U^{\bs\upupsilon}$ due to its $\phi$-dependence. Thefore we have, in analogy with \eqref{commut-bXi-d-int},
\begin{align}
\label{commut-dress-d-int}
\bs d \left(
\big(\bs\alpha^{(\bs\upupsilon, \u)}\big)_{U^{\bs\upupsilon}} 
\right)
&=
\left( \bs d\bs\alpha_U\right)^{(\bs\upupsilon,\u)}
- \langle \mathfrak L_{ \bs\upupsilon
\-_* \bs d\bs\upupsilon}\, \bs\alpha^{(\bs\upupsilon,\u)}, 
U^{\bs\upupsilon}  \rangle.
\end{align}
That is, $[ \t F_{(\bs\upupsilon, \u)}^\star, \bs d]\neq 0$ and the commutator is a boundary term, $\langle\, \iota_{ \bs\uppsi
\-_* \bs d\bs\uppsi}\, \bs\alpha^{(\bs\upupsilon,\u)} , \d\big( U^{\bs\upupsilon}\big) \,\rangle$,  by \eqref{Stokes} and the fact that $\bs\alpha^{(\bs\upupsilon,\u)}$ is a top form  on $U^{\bs\upupsilon}$.
Observe that is means basicity is lost, the boundary term being manifestly non-horizontal. This impacts our discussion of the \emph{relational variational principle} in the next section \ref{Relational field theory}.

For $\bs \alpha \in \Omega^\bullet_\text{tens} \big(\Phi, \Omega^{\text{top}}(U) \big)$, i.e. $\bs\alpha^{(\bs\uppsi, \bs\upgamma)}=\bs\uppsi^*(\bs\alpha^{\bs\upgamma})$ by \eqref{equi-tens-forms-loc}, a dressed integral is
\begin{equation}   
\label{int-on-spacetime-loc}
\begin{split} 
{\bs\alpha_U}^{(\bs\upupsilon, \u)} 
=
\big(\bs\alpha^{(\bs\upupsilon, \u)}\big)_{U^{\bs\upupsilon}}
=
\langle \bs\alpha^{(\bs\upupsilon, \u)} , U^{\bs\upupsilon}  \rangle 
=
\langle  \bs\upupsilon^*(\bs\alpha^\u), \bs\upupsilon\-(U)   \rangle 
=
\langle \bs\alpha^\u, U
\rangle 
=\int_U \bs\alpha^\u
\rdefeq
(\bs\alpha^\u)_U,
\end{split}    
\end{equation} 
by the invariance property \eqref{invariance-int-loc} of the integration pairing.
Meaning that the integral of the $\Hl$-invariant dressed object $\bs\alpha^\u$ over $U$ is numerically identical to the integral of the physical quantity  $\bs\alpha^{(\bs\upupsilon, \u)}$  over the spacetime region~$U^{\bs\upupsilon}$. 
Then we have  
\begin{align}
\label{d-dress-int-special}
d\big({\bs\alpha_U}^{(\bs\upupsilon,\u)} \big) 
=
\bs d \langle \bs\uppsi^*(\bs\alpha^{\u}), \bs \upupsilon\-(U)\rangle 
=
\bs d \langle \bs\alpha^{\u}, U\rangle 
=
\langle \bs d  \big(\bs\alpha^\u \big), U\rangle.
\end{align}
And the relation \eqref{commut-dress-d-int} specialises to 
\begin{equation}
\label{commut-dress-d-special}
\begin{aligned}
\bs d\left(
\big(\bs\alpha^{(\bs\upupsilon, \u)}\big)_{U^{\bs\upupsilon}} 
\right)
&=
\left( \bs d\bs\alpha_U\right)^{(\bs\upupsilon,\u)}
- \langle \mathfrak L_{ \bs\upupsilon
\-_* \bs d\bs\upupsilon}\, \bs\upupsilon^*\big(\bs\alpha^{\u} \big), 
 U^{\bs\upupsilon}  \rangle \\
&=
\left( \bs d\bs\alpha_U\right)^{(\bs\upupsilon,\u)}
-
\langle \bs\upupsilon^* \mathfrak L_{\bs d\bs\upupsilon \circ \bs \upupsilon\-}\big(\bs\alpha^{\u} \big), 
\bs\upupsilon\-(U)  \rangle \\
&=
\left( \bs d\bs\alpha_U\right)^{(\bs\upupsilon,\u)}
-
\langle  \mathfrak L_{\bs d\bs\upupsilon \circ \bs \upupsilon\-}\big(\bs\alpha^{\u} \big), 
U  \rangle.
\end{aligned}
\end{equation}
The last two relation imply  the following lemma:
\begin{align}
\bs d \big( \bs\upupsilon^* \bs\alpha \big)
=
\bs\upupsilon^*\big( \bs d\bs\alpha +  \mathfrak L_{\bs d\bs\upupsilon \circ \bs \upupsilon\-} \bs\alpha \big).
\end{align}
In case $\bs \alpha$ is $\Diff(M)$-equivariant \emph{and} $\Hl$-invariant, it is s.t. $\bs \alpha^\u=\bs \alpha$ by the DFM rule of thumb, so   
\begin{equation}   
\label{dressed-int-eq-bare-int}
{\bs\alpha_U}^{(\bs\upupsilon, \u)} 
=
\big(\bs\alpha^{(\bs\upupsilon, \u)}\big)_{U^{\bs\upupsilon}}
=
\langle \bs\alpha^{(\bs\upupsilon, \u)} , U^{\bs\upupsilon}   \rangle 
=
\langle  \bs\upupsilon^*\bs\alpha, \bs\upupsilon\-(U)   \rangle 
=
\langle \bs\alpha, U   \rangle = \bs\alpha_U.
\end{equation}
Which means that the integral of the bare unphysical quantity $\bs\alpha$ over $U$ is numerically identical to the integral of the physical quantity  $\bs\alpha^{(\bs\upupsilon, \u)}$  over the true spacetime~region~$U^{\bs\upupsilon}$. 
It~is interesting to notice that this is an instance where a quantity computed in the ``bare" formalism  gives the correct results for the corresponding  physical quantity.    
In the latter case we have, furthermore, ${\bs d \bs\alpha}_U  = \bs d \big({\bs\alpha_U}^{(\bs\upupsilon, \u)}\big)$,
and using   \eqref{Vert-trsf-pairing-dalpha-loc}-\eqref{Vert-trsf-int-dalpha-loc} we have that: 
\begin{equation}
\label{dress-int-dalpha-loc}
\begin{aligned}
\bs (\bs{d\alpha}_U)^{(\bs\upupsilon, \u)}	 
&= 
\bs{d\alpha}_U + \langle \mathfrak L_{\bs{d\upupsilon}\circ \bs\upupsilon\-} \bs\alpha,  U \rangle,  \\
 \langle  \bs{d\alpha}, U \rangle ^{(\bs\upupsilon, \u)} 
 &=
 \langle\bs{d\alpha}, U \rangle + \langle \mathfrak L_{\bs{d\upupsilon}\circ \bs\upupsilon\-} \bs\alpha,  U \rangle.
\end{aligned}
\end{equation}
This has immediate consequences for the variational principle in field theory, to which we  turn after taking stock of a few noteworthy conceptual points.

\subsubsection{Discussion}
\label{Discussion}

As we have established, 
the DFM is a systematic and formal way to obtain objects, built from fields $\phi$ and their variations $\bs d \phi$, that are invariant under $\Diff(M)\ltimes \Hl$:  
they are \emph{basic} objects on field space $\Phi$ and on $\Phi \times \bs U(M)$, so that they descend to, or represent, objects on the moduli space $\M$ and on the bundle of regions $\b{\bs U}(M)=\Phi\times \bs U(M)/\sim$.
Together the latter contain the relevant physical quantities: physical field d.o.f. for $\M$, integral quantities over physical spacetime for $\b{\bs U}(M)$.
Dressed fields and dressed integrals can be understood as coordinatisations for them.

Here we elaborate on the constraints in the existence of dressing fields, on the distinction to be made between \emph{ad hoc} dressings and genuine ones, and on the issue of locality of observables within the DFM, as well as how ``physical integrals" can be understood as local observables.

\paragraph{Existence and globality of dressings}
The DFM is a \emph{conditional} proposition: \emph{If} one can find, or build, a dressing field, \emph{then} it gives the algorithm to produce these invariants (and analyse possible residual symmetries). 
Let us discuss some of what may hide behind the conditional ``if".

It should be stressed that the existence of global dressing fields $(\upupsilon, \text u)$ is not guaranteed, with at least two senses of the word ``global" to be distinguished.
Firstly, it is clear that $\phi$-independent $\Diff(M)$-dressing fields of the type  $\upupsilon \in \D r[\RR^n, M]$, i.e. $\upupsilon: \RR^n \rarrow M$, are nothing but \emph{global} coordinate charts for $M$, which may not exist depending on the topology of $M$. 
The same is true in the $\phi$-dependent case $\bs\upupsilon: \RR^n \rarrow M$, as  topological obstructions may imply that no field $\phi$ is globally defined, or non-vanishing, everywhere on $M$ and fit to serve as a global coordinatisation.
Likewise, a  $\phi$-independent $\Hl$-dressing field $\text u \in \D r[H, H]$, i.e. $\text u: M\rarrow H$, is the local representative of a global dressing $u: P \rarrow H$,  the latter providing a global trivialisation of $P$. 
So, if $P$ is non-trivial,  $\Hl$-dressing $\text u$ fields exist only locally over $U\subset M$. 
This holds the same for a $\phi$-dependent $\u$. In such cases, the patching of multiple dressing fields necessary to globally cover $M$ is achieved via what we called above residual symmetries of the 2nd kind. Again, only in the $\phi$-dependent case may the group $\Diff(N)\ltimes \b\G_\text{\tiny{loc}}$ controlling the patching relations  be discrete, counting the finite number or ways to built the dressings from $\phi$.

Secondly, as noted in section \ref{Dressing and flat connections},  dressing fields   
induce flat Ehresmann connections. If they are defined globally  \emph{on field space} $\Phi$, so are their associated flat connections, which implies $\Phi$ is globally trivial as a bundle. 
If $\Phi$ is non-trivial, dressing fields would only be available locally, over open subsets $\Phi_{|\U}$ with  $\U \subset \M$ (being compatible with the local triviality of $\Phi$). 
In that respect, the situation is not unlike the Gribov-Singer obstruction to global gauge-fixings, i.e. global sections $\bs \s: \M \rarrow \Phi$ \cite{Gribov, Singer1978, Singer1981, Fuchs-et-al1994}.

\paragraph{Relational view \emph{vs} \emph{ad hoc} dressing fields}
As we have argued extensively, $\phi$-dependent dressing fields $\bs u=\bs u(\phi)$, whose d.o.f. are extracted from the original field space $\Phi$, 
allow a formal implementation of a \emph{relational description} of the physics of gRGFT:
The invariant dressed fields $\phi^{(\bs\upupsilon(\phi), \u(\phi))}$ 
are readily understood to represent  physical d.o.f. relationally because the  d.o.f. of the fields $\phi$ are \emph{coordinatised relative to each other}.
As a result, only $\big(\!\Diff(M)\ltimes \Hl\big)$-invariant, physical relational d.o.f. (both spatio-temporal and internal) are manifest.
Relatedly, as observed several times, only by \emph{constructively} producing a $\phi$-dependent dressing field  is there  any chance  for the residual symmetry of the 2nd kind  $\Diff(N)\ltimes \b\G_\text{\tiny{loc}}$ to be reduced to a small, discrete, or trivial subgroup. In concrete situations, it may be reduced to a discrete choice: that of reference coordinatising field among the collection $\phi$. 

So, explicit in the DFM is the proposition that the dressing field is to be found among, or built from, the existing set of fields $\phi$ of a theory.
 This is the situation of most fundamental physical  significance for the analysis of gRGFT: it is
 when the relational interpretation of the formalism is the most compelling and transparent.
But then one must be careful when considering cases where a dressing fields is introduced in a theory \emph{independently} from the original set of d.o.f. $\phi$: i.e. when extending the field space to  $\Phi'=\Phi + (\upupsilon, \text u)$ and admitting $\bs d (\upupsilon, \text u) \neq 0$.
Naturally, in that extended context, one may  see the newly introduced dressing field as trivially field-dependent,  by writing 
\begin{equation}
\begin{aligned}
(\bs \upupsilon, \u) : \Phi' &\rarrow \D r[N; M, H], \\
\big\{\phi, (\upupsilon, \text u)\big\} &\mapsto \bs u\big(\phi, (\upupsilon, \text u)\big)\defeq (\upupsilon, \text u),
\end{aligned}
\end{equation}
and $\bs d$ being understood as the exterior derivative on $\Phi'$. 
This amounts to considering a different  theory, or model, where not only the kinematics and dynamics are altered, but also where  the \emph{physical signature} of the symmetries can be radically different. 

There is a priori no issue if the initial theory one starts with already has a native $\Diff(M)\ltimes \Hl$-symmetry, and one simply considers a variant with another dynamical field that happens to be a dressing field. 
Identifying the dressing field in the variant model, with $\Phi'$, may be considered a case of ``constructive" building of dressing, as just mentioned.
Such is the case e.g. in variants on the theme of ``scalar coordinatisations" of GR, as we will illustrate in section \ref{Examples}.
Both the initial and extended theories, with respective field spaces $\Phi$ and $\Phi'$, enjoy $\Diff(M)\ltimes \Hl$ symmetry,
which in both cases  encodes what we may  call their ``general relationality", characteristic of the gRGFT framework: i.e. the fact, illustrated in Fig.\ref{Diag3}, that the physical invariant field d.o.f. and spacetime points are relationally and dynamically defined, so that there is no non-dynamical non-relational background structures.

The situation is  quite different when one starts with a theory \emph{without} $\Diff(M)\ltimes \Hl$ symmetries, i.e. having some background (non-dynamical, non-relational) structures ``breaking" $\Diff(M)\ltimes \Hl$, and one introduces by \emph{fiat} a dressing field $(\upupsilon, \text u)$ whose sole purpose is to ``implement" or ``restore" $\big(\!\Diff(M)\ltimes \Hl\big)$-invariance. 
The latter we call \emph{ad hoc} dressing fields.
We may signal two typical instances covered by this \emph{ad hoc} case of the DFM, appearing quite frequently in the literature. 
\medskip

The first is  the  starting  premise of the literature on ``edge modes" as introduced, and elaborated on, in \cite{DonnellyFreidel2016, Speranza2022, Geiller2017, Speranza2018, Geiller2018, Chandrasekaran_Speranza2021, Freidel-et-al2020-1, Freidel-et-al2020-2, Freidel-et-al2020-3, Ciambelli2023}:
 One starts from a 
theory where
$\Diff(M)\ltimes \Hl$ is seemingly broken by the 
boundary $\d U$ of a $(n-1)$-dimensional region $U \subset M$,
and then introduces extra d.o.f. living on $\d U$, the so-called edge modes, whose transformation properties are so that they compensate for the symmetry breaking at $\d U$, and thus restore $\big(\!\Diff(M)\ltimes \Hl\big)$-invariance there.
A significant drawback of this whole approach is that, since such edge modes are exactly \emph{ad hoc} dressing fields, introduced by \emph{fiat}, there is a priori no way to restrict the ambiguity in their choice represented by the residual group $\Diff(N)\ltimes \b\G_\text{\tiny{loc}}$. 
In the edge mode literature, the latter group is (or was, initially) seen as supported, like edge modes themselves, on $\d U$ and is thus call ``surface symmetry" or ``corner symmetry" ($\d U$ being a ``corner", a codimension 2 submanifold in $M$). 
It is manifest that, contrary to what is claimed in the edge mode literature, the residual group encodes neither more nor less information than the original (physically meaningful) group $\Diff(M)\ltimes \Hl$.

It should be further stressed that the starting premise of the edge mode approach, that boundaries ``break" $\Diff(M)\ltimes \Hl$, the so-called ``boundary problem", is faulty. 
As already previewed in section \ref{Dressed integrals and integration on spacetime}, above \eqref{dressed-regions-bundle-loc}, 
it has indeed the same logical structure as the hole argument: 
It  considers  boundaries $\d U$ ``drawn" on $M$, and their constitutive points, as physical entities defined independently of the field content. 
As such it artificially introduces a background structure, a \emph{non-relational boundary}, and then introduces  an \emph{ad hoc} dressing field (edge modes) whose sole purpose is to conceal it. 
It is clear from section \ref{Relationality in  general-relativistic gauge field theory} that such ``boundary problems" are a non-starter once the point-coincidence argument is brought to bear:
As argued there, points of $M$ are not spacetime points, nor its region are spacetime regions. 
Rather, spacetime points and regions are defined \emph{relationally} by the fields $\phi$, as is formally implemented by dressed regions $U^{\bs\upsilon(\phi)}$ \eqref{dressed-region-loc}-\eqref{Dressed-regions-inv-loc}. 
So a physical spacetime  boundary $\d U^{\bs\upsilon(\phi)}$ is relational and necessarily $\big(\!\Diff(M)\ltimes \Hl\big)$-invariant. 
\medskip

 A second situation is when one starts from a theory which does not enjoy a $\Diff(M)\ltimes \Hl$ symmetry as it has non-relational/non-dynamical background structures featuring in its field equations -- e.g. a background metric or geometry (preserved by a subgroup of $\Diff(M)\ltimes \Hl$ which is the genuine symmetry of the theory). 
 But  then a $\Diff(M)\ltimes \Hl$ symmetry is enforced via the introduction of an \emph{ad hoc} dressing field: those generalise  Stueckelberg fields \cite{Ruegg-Ruiz, Stueckelberg1938-I, Stueckelberg1938-II-III}, and the move is usually seen as ``restoring" or ``implementing $\big(\!\Diff(M)\ltimes \Hl\big)$-invariance". 
 Such is typically the case in the literature on massive gravity, bi-gravity, and sometimes string theory -- see e.g. \cite{deRham2014, Green-Thorn1991, Siegel1994}. 
It should be clear that such artificially enforced $\Diff(M)\ltimes \Hl$ symmetries do not have the same physical status as  native ones, as they are  grafted onto a theory essentially to hide its background non-dynamical structures.

This is what motivates their denomination as
  ``\emph{artificial} symmetries" in the philosophy of physics literature \cite{Pitts2005, Pitts2008, Pitts2009, Pitts2012}.
  Indeed, enforcing a symmetry via \emph{ad hoc} DFM (generalising the Stueckelberg trick) is an instance of ``Kretschmannisation" of a theory, i.e. rewriting it so as to display a strictly formal symmetry devoid of physical signature or content. 
The name stems from the notorious ``Kretschmann objection" against general covariance as a fundamental feature of GR, 
according to which any theory can be made $\Diff(M)$-invariant (or generally covariant). 
A closer analysis resulted in the realisation that one may distinguish between \emph{substantive} $\Diff(M)$/general covariance as a distinctive native feature of a theory, whose physical  signature is that all physical structures and d.o.f. are dynamically and relationally defined -- so that no background structures are needed  --  from \emph{artificial} $\Diff(M)$/general covariance which is implemented by hand (often via the introduction of extra objects/fields, e.g. à la Stueckelberg) and thus hides background structures. 

A comparable distinction has been developed for internal gauge symmetries, in response to a ``generalised Kretschmann objection" to the gauge principle \cite{Pitts2009, Francois2018, Berghofer-et-al2023} -- according to which any field theory can be rewritten so as to display a gauge symmetry. 
In that case, a consensual proposition regarding the physical signature of \emph{substantive gauge symmetries} is the trade-off between gauge invariance and locality of a theory, which is absent for \emph{artificial gauge symmetries}. 
See also \cite{JTF-Ravera2024c} for a possible counterpoint, and a broader discussion on the physical meaning of local symmetries.

Theories obtained through Kretschmannisation  are only superficially and formally alike gRGFT,  
but  betray their key physical insight, i.e. the \emph{complete} (or general) relationality of physical structures implying the absence of any background structure/field, which is what a  ``substantive" $\Diff(M)\ltimes \Hl$ transformation encodes. 
So, if one can still hold on to a ``partially relational" view of the DFM  when analysing such theories, they in fact do not enjoy  the signature ``general relationality" of  genuine general-relativistic gauge field theories. 

\medskip
From the above follows that the DFM as developed here is the common geometric framework underpinning a number of notions appearing in recent years in the literature on gravity and gauge theories. 
For example, as mentioned, the \emph{ad hoc} DFM underlies  massive gravity and bi-gravity theories  \cite{deRham2014}, as well as the notion, closely related to edge modes,  of ``embedding fields" \cite{Speranza2019, Freidel:2021, Ciambelli-et-al2022, Speranza2022, Ciambelli2023}  -- see also  \cite{Kabel-Wieland2022}.
The DFM also grounds the notion of ``gravitational dressings" as proposed in  \cite{Giddings-Donnelly2016, Giddings-Donnelly2016-bis, Giddings-Kinsella2018, Giddings2019, Giddings-Weinberg2020}, as well as that  of ``dynamical reference frames” as proposed in  \cite{carrozza-et-al2022, Hoehn-et-al2022} -- which also has an explicit relational angle.

\paragraph{Locality of observables, and  relational integrals as local observables}
Let us offer some comments on the issue of  locality \emph{vs} non-locality of observables in the DFM. 
We should start by defining our terms.
Here, by ``local" we shall mean ``field theoretically local", 
which may be understood to consists in at least three desiderata: 
(1) Relativistic causality: Physical
processes (carrying energy/information) propagate within the light-cone structure of spacetime. 
(2) Field locality: Physical properties/magnitudes are  (smoothly) assigned to arbitrarily small regions of spacetime,  or, in the idealised limit, to spacetime points, and thus interact pointwisely in regions where they do not vanish. 
Relatedly one may consider  (3) Separability: The physical properties/magnitudes (fields configurations) of  regions of spacetime are recovered from
physical properties/magnitudes of its constitutive subregions.
Relativistic causality (1) remains non-negotiable, so failure of non-locality means allowing infringement of (2) and/or (3). 
Both have been suggested. It is famously the case that accounts of the Aharonov-Bohm effect, which has often been argued to require a form of non-locality meant to imply a  ``non-separability" that is  argued to be a  systemic feature of gauge theory -- see e.g. \cite{Nguyen-et-al2017, Dougherty2017}.

The DFM accommodates both the possibilities that observables be local or non-local. 
If ${\big(\bs\upupsilon(\phi), \u(\phi) \big)}$ is a local functional of the bare fields $\phi$, then the invariant dressed fields $\phi^{(\bs\upupsilon, \u)}$ are local relational variables and so is any dressed form $\bs\alpha^{(\bs\upupsilon, \u )}=\alpha\big(\!\wedge^{\!\bullet}\! {\bs d}\phi^{(\bs\upupsilon, \u )}; \phi^{(\bs\upupsilon, \u )} \big)$  built from them: these are potential local observables.
If ${\big(\bs\upupsilon(\phi), \u(\phi) \big)}$ is a non-local functional of $\phi$, involving e.g. integrals or inversion of differential operators, then $\phi^{(\bs\upupsilon, \u)}$ are invariant non-local relational variables, idem for $\bs\alpha^{(\bs\upupsilon, \u)}$.
The framework developed here is therefore  a priori neutral regarding the issue of locality of the observables.
Which  of the above cases actually obtains depends essentially on the model under consideration, i.e.  on the precise field content available. See e.g. \cite{Francois2018, Berghofer-et-al2023} for discussions of this point in the context of internal gauge theories.
\medskip

 As an important example, consider the natural class of invariants, and potential observables,  given by integrals $\langle \bs\alpha ; U\rangle = \int_U \bs \alpha$ of $\Hl$-invariant local functionals $\bs \alpha$ of the fields $\phi$  -- by which we mean that the value of $\bs\alpha$ at $x\in U\subset M$ depends only on the $r$-jet of $\phi$ at $x$ for finite~$r$.
Such integrals may be considered  non-local in the sense that a priori they invariantly assign real numbers not to points, but rather to whole regions $U\subset M$.
This appears to be somewhat in tension with  desideratum (2). 
Some of these integral quantities, representing key physical quantities, have indeed been called ``quasi-local": e.g. the quasi-local mass/energy of isolated gravitational systems~\cite{Yau-Wang2009}.
However, a relational understanding of such integral quantities -- as only \emph{indirectly} representing integration on spacetime -- allows to see that they do  actually satisfy (2).
They indeed arise from the physical properties of invariant relationally defined physical field d.o.f. $\phi^{(\bs\upupsilon, \u)}$ over relationally defined \emph{physical spacetime points} and subregions $U^{\bs\upupsilon}$. 
This is shown in a manifest way by the notion of dressed integrals \eqref{Dressed-integral-loc},
which by \eqref{dressed-int-eq-bare-int} coincide with ``bare" integrals for $\Diff(M)$-tensorial and $\Hl$-basic integrand $\bs\alpha$.

\subsubsection{Relational formulation of general-relativistic gauge field theories theory}
\label{Relational field theory}

We have developed all that is needed to obtain a relational reformulation of gRGFT. 
Let us consider a Lagrangian $L \in \Omega^0_\text{tens}\big(\Phi, \Omega^\text{top}(U) \big)$ as supporting a non-trivial action of $\Diff(M)\ltimes \Hl$ \eqref{equiv-Lagrangian}. 
Its  $C^\infty\big(\Phi,  \Diff(M)\ltimes \Hl\big)\simeq \bs\Diff_v(\Phi)$-transformation is given by \eqref{L-vert-trsf}, so by the DFM rule of thumb its dressing is immediately found to be
\begin{align}
\label{dressed-L}
L^{(\bs\upupsilon, \u)} 
=
\bs\upupsilon^*(L^{\u}) 
=
\bs\upupsilon^*\big(L + c(\ \ ; \u) \big)
\quad \in \Omega^0_\text{basic}\big(\Phi, \Omega^\text{top}(U) \big),
\end{align}
where the 1-cocycle is extended to obtain the Abelian twisted dressing field $c(\ \ ;\u)$ as in \eqref{abelian-twisted-dressing}, section \ref{Dressing and flat  connections}.
The~dressed Lagrangian, explicitly $L^{(\bs\upupsilon, \u)}(\phi) = L\big( \phi^{(\bs\upupsilon, \u)} \big)
= L\big(\bs\upupsilon^*(\phi^\u)\big)$,  
 is a manifestly relational and $\big(\!\Diff(M)\ltimes \Hl\big)$-invariant reformulation of the bare theory $L$. 
 
Being basic on $\Phi$, a dressed Lagrangian \eqref{dressed-L}  represents (or coordinatises) a physical quantity on the moduli space $\M$.
It should not be confused with a \emph{gauge-fixed} Lagrangian: 
A gauge-fixing is the datum of a (local) section $\bs\s :\U \subset \M \rarrow \Phi_{|\U}$, a gauge-fixed Lagrangian is then $\bs\s^\star L = L \circ \bs\s \in \Omega^0\big(\U,\Omega^\text{top}(U) \big)$. 
Observe furthermore that under transformation by $(\uppsi, \upgamma)\in \Diff\ltimes \Hl$, a new gauge-fixing section is $\bs\s'\defeq R_{(\uppsi, \upgamma)} \circ \bs \s$, and the gauge-fixed Lagrangian transforms to ${\bs\s'}^\star L 
=
L\circ \bs\s'
=
L\circ R_{(\uppsi, \upgamma)} \circ \bs \s 
=
\big(R^\star_{(\uppsi, \upgamma)}L \big) \circ \bs \s 
= \left(\uppsi^*\big(L + c(\ \ ;\upgamma) \big)\right) \circ \s
\neq \bs\s^\star L$. 
In~other words, a gauge-fixed Lagrangian is not invariant under the action of $\Diff(M)\ltimes \Hl$. 
These two elements should help make clear how dressed and gauge-fixed Lagrangians differ: the former is basic on $\Phi$, the latter lives on $\U\subset \M$ and is not invariant under $\Diff(M)\ltimes \Hl$. 
This indeed reflects the fundamental distinction between dressing, which realises the projection $\pi :\Phi \rarrow \M$, and gauge-fixing, which goes the exact opposite way $\bs\s : \M \rarrow \Phi$. 

Let~us also keep in mind that dressings have a natural relational interpretation that gauge-fixings do not.
See \cite{Francois-Berghofer2024} for a comprehensive exposition of the distinction between dressing and gauge fixing, which further stresses how what is often considered a case of the latter is actually a case of the former (e.g. the Lorenz ``gauge" in classical electrodynamics). 
Many standard ``gauge fixings" turn out to actually be instances of dressings. 
As shown in \cite{JTF-Ravera2024susy},
the observation extends to supersymmetric field theory and supergravity, with the transverse and divergenceless ``gauges" for the Rarita-Schwinger field and the gavitino.
\medskip

The action $S=\langle L, U \rangle=\int_U L$ transforms under $C^\infty\big(\Phi,  \Diff(M)\ltimes \Hl\big)$  by
\eqref{S-vert-trsf}, from which we read  its dressing
\begin{align}
\label{S-dressed}
S^{(\bs\upupsilon, \u)} 
\defeq
\int_{U^{\bs\upupsilon}}  L^{(\bs\upupsilon, \u)}
=
\int_U L + c(\ \, ;\u) 
\rdefeq
S + \mathsf c(\ \, ;\u), 
\end{align}
which indeed a special case of , or by 
\eqref{Dressed-integral-loc}-\eqref{int-on-spacetime-loc}.
The dressed action $S^{(\bs\upupsilon, \u)}$ is basic on $\Phi \times \bs U(M)$ and thus represents a physical quantity defined on the bundle of regions $\b{\bs U}(M)$.
It is clear that $S^{(\bs\upupsilon, \u)}$ is not a gauge-fixed action, as we see \emph{by definition} integration over the physical region of spacetime $U^{\bs\upupsilon}$, and there is certainly nothing like a ``gauge-fixed region of spacetime" in the standard bare formalism.

Remark that for an $\Hl$-invariant Lagrangian we have $L^\u=L$, and the above  specialises to  
$S^{(\bs\upupsilon, \u)}=S$, so that the computation of the action in the bare formalism gives the correct physical result.
This is a first hint as to why, in theories strictly respecting the principle of gRGFT -- in that case, the gauge principle, so that $L^\upgamma=L$ -- the bare formalism can give sensible results even without solving first the issue of extracting the physical d.o.f. of the theory. 

We may observe that, writing $Z=\exp{i\, S}$ as in \eqref{Z-action}, the dressed action \eqref{S-dressed} is also obtained from the \emph{twisted} covariant derivative induced by the flat twisted $\Hl$-connection $\bs\varpi_\text{\tiny{$0$}}=i\bs d \mathsf c(\ \ ;\u)$,
writing 
\begin{align}
\b{\bs D}_\text{\tiny{$0$}} Z 
= \bs d Z + \bs\varpi_\text{\tiny{$0$}} Z 
= i\bs d \big(S + \mathsf c( \ \ ;\u) \big)\, Z,
\end{align}
which is a special case of 
\eqref{Flat-twisted-cov-deriv}-\eqref{dressing-twisted-form}
-- itself a special case of the non-flat case \eqref{Gen-WZ-action}.

We further highlights that the dressed Lagrangian and action \eqref{dressed-L}-\eqref{S-dressed} have by definition a built-in (classical) $\Hl$-anomaly cancellation mechanism: the twisted dressing field $\mathsf c(\ \ ; \u)$ plays the role of a Wess-Zumino term whose linear transformation cancels that of the Lagrangian -- as described e.g. in section 12.3 of \cite{GockSchuck}, Chap.15 in \cite{Bonora2023}, or the end of Chap.4 in \cite{Bertlmann}.  
Indeed,  the defining  property of a WZ term is that expected from the prepotential of a flat twisted connection: 
$\bs L_{\lambda^v} \mathsf c(\ \ ;\u) =\iota_{\lambda^v}\bs d \mathsf c(\ \ ;\u)
=- \iota_{\lambda^v} i\, \bs\varpi_\text{\tiny{$0$}}
\defeq  \mathsf a(\lambda;\phi)$,
for $\lambda \in$ Lie$\Hl$ and $\mathsf a(\lambda;\phi)= \int_U a(\lambda;\phi)$ the integraded $\Hl$-anomaly. 

\medskip
We now consider the behavior under dressing of  the variational principle:
$\bs dL =\bs E + d\bs \theta$ and $\bs dS =0 \ \xrightarrow{b.c.} \ \bs E=0$,
with boundary conditions (b.c.) imposed at $\d U$.
We want to establish a \emph{relational variational principle}. 
The most immediate thing to propose would be to vary  the dressed action, $\bs d S^{(\bs\upupsilon, \u)}$. 
However, by \eqref{commut-dress-d-special}, this is
\begin{equation}
\label{failed-rel-var-princ}
\begin{aligned}
\bs d S^{(\bs\upupsilon, \u)}
&=
( \bs d S)^{(\bs\upupsilon,\u)}
- 
\langle \mathfrak L_{ \bs\upupsilon
\-_* \bs d\bs\upupsilon}\, L^{(\bs\upupsilon, \u)} \big), 
 U^{\bs\upupsilon}  \rangle, \\
 &=
( \bs d S)^{(\bs\upupsilon,\u)}
- 
\langle \iota_{ \bs\upupsilon
\-_* \bs d\bs\upupsilon}\, L^{(\bs\upupsilon, \u)}, 
 \d U^{\bs\upupsilon}  \rangle,
\end{aligned}
\end{equation}
which is not basic, since as a rule $[ \t F_{(\bs\upupsilon, \u)}^\star, \bs d]\neq 0$.
So, this first guess does not qualify as a good candidate for relational variational principle.
However,  the quantity $( \bs d S)^{(\bs\upupsilon,\u)}$ is basic by definition, and differ from $\bs d S^{(\bs\upupsilon, \u)}$ by a boundary term. 
Hence,  both will give the same field equations, which furthermore will be basic. 
Therefore, the basic object $( \bs d S)^{(\bs\upupsilon,\u)}$, representing a well-defined quantity on the bundle of regions $\b{\bs U}(M)\defeq \Phi\times \bs U(M)/\sim$, is the natural quantity implementing a relational variational principle.\footnote{The fact that the most natural quantity $\bs d S^{(\bs\upupsilon, \u)}$ fails is a hint at the fact that a genuine relational  variational principle for gRGFT should involve a variational operator on $\Phi \times \bs U(M)$ extending $\bs d$ on $\Phi$ -- which only varies fields. 
We shall develop this viewpoint in a forthcoming paper \cite{JTF-Ravera2025relvarpr}.}

It is easy to write it down: 
The 1-form $\bs dL$ transforms under $\bs\Diff_v(\Phi) \simeq C^\infty\big(\Phi,  \Diff(M)\ltimes \Hl\big)$ by \eqref{GT-dL}, from which we read its dressing
\begin{equation}
\label{dressing-dL-1}
\begin{aligned}
(\bs dL)^{(\bs\upupsilon, \u)}
=
\bs\upupsilon^*\left(
\bs dL + \bs d c\big(\ \ ; \u\big)
+ 
\mathfrak L_{\bs d\bs \upupsilon \circ \bs\upupsilon\-} \big(\, L + c(\ \ ;\u)\,\big) 
\right).
\end{aligned}
\end{equation}
Here we have indeed, $(\bs dL)^{(\bs\upupsilon, \u)}= \bs dL^{(\bs\upupsilon, \u)}$ since 
$[  F_{(\bs\upupsilon, \u)}^\star, \bs d] = 0$ on $\Phi \rarrow \M$. 
Correspondingly, 
\begin{equation}
\begin{aligned}
 (\bs dS)^{(\bs\upupsilon, \u)}
&=
 \int_{U^{\bs\upupsilon}} L^{(\bs\upupsilon, \u)}
&= \int_U 
\bs dL + \bs d c\big(\ \ ; \u\big)
+ 
\mathfrak L_{\bs d\bs \upupsilon \circ \bs\upupsilon\-} \big(\, L + c(\ \ ;\u)\,\big).
 \end{aligned}
 \end{equation}
For a Lagrangian satisfying the principles of gRGFT, i.e. satisfying $\bs d c(\phi ;\, \upgamma) 
  =d\bs b(\phi ;\,\upgamma)$ under  hypothesis
\eqref{Hyp0} , this specialises to 
\begin{equation}
\begin{aligned}
(\bs dS)^{(\bs\upupsilon, \u)}
&=
 \int_{U} \bs dL + d  \left[ \bs b\big(\ \ ; \u\big)
+ 
\iota_{\bs d\bs \upupsilon \circ \bs\upupsilon\-}  L 
+ 
\iota_{\bs d\bs \upupsilon \circ \bs\upupsilon\-}\, c(\ \ ;\u)\,\right] \\
&=
\bs dS+ \int_{\d U} 
\bs b\big(\ \ ; \u\big)
+ 
\iota_{\bs d\bs \upupsilon \circ \bs\upupsilon\-}  L 
+ 
\iota_{\bs d\bs \upupsilon \circ \bs\upupsilon\-}\, c(\ \ ;\u).
\end{aligned}
\end{equation}
The fact that $\bs dS$ and its dressing $(\bs dS)^{(\bs\upupsilon, \u)}$ differ by  boundary terms implies that the bare variational principle and relational variational principle are ``equivalent", in that  the space $\S^{\bs u}$ of relational solutions  is isomorphic to the space $S$ of  solutions of the bare theory. 
This we  may show explicitly. 

The relational (dressed) field equations are directly found via $(\bs dL)^{(\bs\upupsilon, \u)} = \bs E^{(\bs\upupsilon, \u)} + d\bs\theta^{(\bs\upupsilon, \u)}$, and  read immediately from the $\bs\Diff_v(\Phi)\simeq C^\infty\big(\Phi, \Diff(M)\ltimes \Hl \big)$-transformation \eqref{GT-E} of the bare field equations $\bs E$,
\begin{equation}
\label{dressed-E}
\bs E^{(\bs\upupsilon, \u)} 
=
\bs\upupsilon^* \left( 
\bs E 
+ 
dE \big( 
\iota_{\bs{d\upupsilon}\circ\bs \upupsilon\-} \phi; \phi
\big)
+
 dE \big(\bs d\u\, \u\-  - \u\, \mathfrak L_{\bs{d\upupsilon} \circ \bs\upupsilon\-} \u\-
 ; \phi \big)\,
\right).
\end{equation}
This reproduces and  generalises results of  \cite{Francois2021, Francois-et-al2021, Francois2023-a}, as it  specialises to general-relativistic and gauge field theories respectively as
\begin{equation}
\label{dressed-E-special}
\begin{aligned}
\bs E^{\bs\upupsilon} 
&=
\bs\upupsilon^* \left( 
\bs E 
+ 
dE \big( 
\iota_{\bs{d\upupsilon}\circ\bs \upupsilon\-} \phi; \phi
\big)
\right), \\
\bs E^{ \u} 
&=
\bs E 
+
 dE \big(\bs d\u \, \u\-  
 ; \phi \big). 
\end{aligned}
\end{equation}

Equation \eqref{dressed-E} is our most important equation.
It implies that the field equations $\bs E^{(\bs\upupsilon, \u)} = E\big(\bs d\phi^{(\bs\upupsilon, \u)}; \phi^{(\bs\upupsilon, \u)}\big)=0$ satisfied by the physical relational variables $\phi^{(\bs\upupsilon, \u)}$ are functionally identical as those $\bs E=E(\bs d\phi; \phi)=0$ satisfied by the bare variables $\phi$.  
It is of foundational importance as it explains why it has been possible to successfully apply gRGFT \emph{before} solving the issue of its fundamental physical d.o.f. and observables.
Indeed, what we actually confront to observations are the field equations $\bs E^{(\bs\upupsilon, \u)}=0$ of  the dressed/relational theory $L^{(\bs\upupsilon, \u)} =L(\phi^{(\bs\upupsilon, \u)})$  -- and \emph{not} a ``gauge-fixed" version of it, as is often claimed -- 
but due to the fundamental principles underlying gRGFT, the latter are formally equivalent to their bare version $\bs E=0$ using  bare fields $\phi$ -- which are only ``partial observables" in the terminology of  \cite{Rovelli2002, Rovelli2014}.
Both are even equal or in strict covariance relation when the contributions of the variation $(\bs d\bs\upupsilon, \bs d\u)$ of the ``coordinatising" dressing field can be neglected -- or when boundary terms are neglected. 

A final crucial observation to be made about the dressed field equations \eqref{dressed-E} is that they describe a manifestly $\big(\!\Diff(M)\ltimes \Hl\big)$-invariant and fully relational dynamics, where physical d.o.f. evolves w.r.t. each other, \emph{without} a predetermined time variable. 
Which means not only that there are no  issues with determinism, the generalised point-coincidence argument being automatically implemented,
but also that there is no so-called ``problem of~time". 
The~latter, also called the ``frozen formalism problem" \cite{Isham1992, AndersonE2012}, may be formulated as the worry that, since proper time evolution on $M$ is a special case of diffeomorphism transformation, ``time evolution is pure gauge in general-relativistic physics" so that physical observables have no dynamics. 
The issue arises only when overlooking that $M$ is not spacetime, and forgetting the relational core of general-relativistic physics. 
This ``problem of time" is dissolved in the relational formulation of gRGFT, or rather never arises in the first place.
We will further elaborate on the importance of this fact in a forthcoming part of this series dealing with relational quantization \cite{JTF-Ravera2024gRGFTquantum} --  closely followed by  a separate paper where we present a relational formulation of non-relativistic Quantum Mechanics \cite{JTF-Ravera2024NRrelQM}.   

\medskip

In the following last section, we present two  applications of the above relational formalism: 
The case of scalar coordinatisation of GR, and the case of Einstein gravity coupled to electromagnetism (EM) and a $\CC$-scalar field.

\subsubsection{Relational Einstein equations}  
\label{Examples} 

\paragraph{General Relativity with scalar matter}

We consider GR -- with cosmological constant $\Lambda$ -- coupled to matter described phenomenologically as a set of scalar fields. We separate the discussion of the kinematics from that of the dynamics so to highlight that the DFM/relational reformulation takes place at the kinematical level first. 
\medskip

\noindent
\underline{Kinematics}: 
The field space  is $\Phi=\{\phi\}=\{g, \upvarphi\}$, where $g$ is a metric field on $M$ and $ \upvarphi : U\subset M \rarrow N=\RR^n$. 
It means, we consider the  connection  to be Levi-Civita, $\Gamma=\Gamma(g)$, i.e.  there is no torsion $T=0$.
The action of $\Diff(M)\ltimes \Hl$ is 
\begin{align}
  R_{(\uppsi, \upgamma)}\, \phi 
  =
  R_{(\uppsi, \upgamma)} \, \big(g, \upvarphi\big)
  = 
  \big( \uppsi^*g, \uppsi^*\upvarphi \big)
  =
  \big( \uppsi^*g, \upvarphi \circ \uppsi \big). 
\end{align}
Naturally in this model there is no gauge  group $\Hl$, so the structure group of $\Phi$ is actually just $\Diff(M)$. 

The first task at hand is to identify a $\Diff(M)$-dressing field. 
Considering the \emph{section} of $\upvarphi$, $\overline{\upvarphi} : N=\RR^n \rarrow U\subset M$ s.t. $\upvarphi \circ \overline{\upvarphi} =\id_N$ -- i.e. the ``right inverse", so $\overline{\upvarphi} = \upvarphi\-$ -- we may define the dressing field 
\begin{equation}
\begin{aligned}
&\bs\upupsilon=\bs\upupsilon(\upvarphi)  \defeq \overline \upvarphi : N \rarrow M, \\
\text{s.t. }\quad & R^\star_{(\uppsi, \upgamma)} \bs\upupsilon(\upvarphi)
=
\bs\upupsilon\big(R_{(\uppsi, \upgamma)}\upvarphi \big) 
=
\bs\upupsilon\big(\upvarphi \circ \uppsi\big)
\defeq
\uppsi\- \circ \overline\upvarphi
\rdefeq
\uppsi\- \circ \bs\upupsilon(\upvarphi).
\end{aligned}
\end{equation}
Therefore, under $\bs\Diff_v(\Phi)\simeq C^\infty\big(\Phi, \Diff(M) \big)$ the dressing field transforms as 
$\bs\upupsilon^{(\bs\uppsi, \bs\upgamma)}= \bs\uppsi\- \circ \bs \upupsilon$, as is expected. 
It~allows to define dressed regions of spacetime:  
\begin{align}
  U^{\bs\upupsilon}\defeq \bs\upupsilon\-(U), \quad \text{s.t.} \quad
  \big(U^{\bs\upupsilon}\big)^{\bs\uppsi} = U^{\bs\upupsilon}. 
\end{align}
This gives a $\Diff(M)$- \emph{and} $C^\infty\big(\Phi, \Diff(M) \big)$-invariant relational definition of physical spacetime $M^{\bs\upupsilon}$ via  matter $\upvarphi$ as a reference physical system, as is expected from the point-coincidence argument. 
The dressed field space is then 
\begin{align}
\label{dressed-field-space-GR-scalar}
\Phi^{\bs\upupsilon}
=
\{\phi^{\bs\upupsilon}\} 
=
\{ g^{\bs\upupsilon}, \upvarphi^{\bs\upupsilon}\}
=
\{ {\bs\upupsilon}^*g, \id_N\}. 
\end{align}
The $C^\infty\big(\Phi, \Diff(M) \big)$-invariant dressed metric $g^{\bs\upupsilon}$ encodes the geometric properties of spacetime $M^{\bs\upupsilon}$.
It can be understood as the  bare metric $g$ ``written in" (``dressed by") the coordinate system supplied by the matter distribution~$\upvarphi$: writing in abstract index notation we have
\begin{align}
\label{dressed-metric-index}
(g^{\bs\upupsilon})_{ab} = \frac{\d x^{\,\mu}}{\d\bs\upupsilon^a} \frac{\d x^{\,\nu}}{\d\bs\upupsilon^b}\, g_{\mu\nu}.
\end{align}
Clearly, $\upvarphi^{\bs\upupsilon}=\id_N$ simply expresses the fact of coordinatising matter distribution w.r.t itself, i.e. that it is (invariantly) at rest in its own reference frame. 
\medskip

\noindent
\underline{Dynamics}: The (bare) Lagrangian of the theory $L$ is $\Diff(M)$-equivariant and  (trivially) $\Hl$-invariant,
\begin{align}
\label{Bare-lagrangian-GR-scalar}
L(g, \upvarphi) 
=
L_\text{\tiny{EH+$\Lambda$}}(g)  +   L_\text{\tiny{Matter}}(g, \upvarphi)
=
\tfrac{1}{2\kappa }\vol_g  \big(\mathsf R(g) - 2 \Lambda \big)  +   L_\text{\tiny{Matter}}(g, \upvarphi),
\end{align}
where $\kappa=\tfrac{8 \pi G}{c^4}$ is the gravitational coupling constant,  $\vol_g = \sqrt{|g|} d^nx$ is the volume form induced by $g$, and $\mathsf R(g)$ is the Ricci scalar: in abstract index notation $\mathsf R(g) \defeq g^{\,\mu \nu} R_{\mu\nu}$, or $\mathsf R(g) \defeq g\- \cdot \text{Ricc}$, with $\text{Ricc}$ the Ricci tensor. 
The~field equations associated to the variation w.r.t. $g$ are
\begin{equation}
\label{Einstein-field-eq}
\begin{aligned}
\bs E &= E(\bs d g; \{g,\upvarphi\})
=
\bs d g \cdot \big( G(g) + \Lambda g - \kappa T(g, \upvarphi) \big) \vol_g, \\[.5mm]
\text{s.t. } \quad 
\bs E&=0 \quad \Rightarrow \quad  G(g) +\Lambda g = \kappa T(g, \upvarphi),
\end{aligned}
\end{equation}
where $G(g) \defeq \text{Ricc} -\tfrac{1}{2} \mathsf R g$ is the Einstein tensor, and the energy-momentum tensor associated to the matter distribution $\upvarphi$ is defined by $\bs d g \cdot T(g, \upvarphi) \vol_g \defeq -\tfrac{1}{2} \bs d  L_\text{\tiny{Matter}}$.
Since $\upvarphi$ is not a fundamental field, i.e. non-variational, the equation of motion of one of its constituting particles with 4-velocity $v$ is given by the geodesic equation $\nabla^g v = dv+ \Gamma(g) v=0$.
Clearly $R^\star_{(\uppsi, \upgamma)} \bs E = \uppsi^* E$, so satisfies hypothesis \eqref{equiv-E-theta} as one would expect in gRGFT. 
By application of \eqref{GT-E}-\eqref{GT-E-special}, the $\bs\Diff_v(\Phi)\simeq C^\infty\big(\Phi, \Diff(M) \big)$-transformation of $\bs E$ is
\begin{equation}
 \label{GT-E-GR-scalar}
\begin{aligned}
&\bs E^{\bs\uppsi} 
=
\bs\uppsi^* \big(\bs E + dE( \iota_{\bs d\bs\uppsi \circ \bs\uppsi\- }g ; \{g, \upvarphi \}) \big),&& \\[1mm]
\Rightarrow \quad &\bs d g^{\bs\uppsi} \cdot \big( G^{\bs\uppsi} + \Lambda g^{\bs\uppsi} - \kappa T^{\bs\uppsi} \big) \vol_{g^{\bs\uppsi}}
=
\bs \uppsi^* &&\hspace{-4mm}\left[
\bs d g \cdot \big( G(g) + \Lambda g - \kappa T(g, \upvarphi) \big) \vol_g 
\right.\\
&\ && +
\left.
d \left( 
\iota_{\bs d\bs\uppsi \circ \bs\uppsi\- }g\cdot \big( G(g) + \Lambda g - \kappa T(g, \upvarphi) \big) \vol_g
\right)
\right].
\end{aligned}
\end{equation}
Which implies that if $G+\Lambda g = \kappa T$ then $G^{\bs\uppsi}+\Lambda g^{\bs\uppsi} = \kappa T^{\bs\uppsi}$, for any  $\phi$-\emph{dependent diffeomorphism} $\bs\uppsi$.
This shows that the covariance group of GR is much bigger than $\Diff(M)$ as required by hypothesis, and as usually understood:
it includes $\phi$-dependent diffeomorphisms $C^\infty\big(\Phi, \Diff(M)\big)$ -- as already observed, and first noticed by \cite{Bergmann1961, Bergmann-Komar1972}. 
\medskip

We can now write down the relational reformulation of GR via DFM. 
The dressed Lagrangian is 
\begin{equation}
\label{Dressed-lagrangian-GR-scalar}
\begin{aligned}
L^{\bs\upupsilon}(g, \upvarphi)
\defeq
{\bs\upupsilon}^*L(g, \upvarphi)
=
L(g^{\bs\upupsilon}, \upvarphi^{\bs\upupsilon}) 
&=
L_\text{\tiny{EH+$\Lambda$}}(g^{\bs\upupsilon})  +   L_\text{\tiny{Matter}}(g^{\bs\upupsilon}, \upvarphi^{\bs\upupsilon}) \\
&=
\tfrac{1}{2\kappa }\vol_{g^{\bs\upupsilon}}  \big(\mathsf R(g^{\bs\upupsilon}) - 2 \Lambda \big)  +   L_\text{\tiny{Matter}}(g^{\bs\upupsilon}, \upvarphi^{\bs\upupsilon}),
\end{aligned}
\end{equation}
with corresponding \emph{dressed Einstein equations}
\begin{equation}
\label{Dressed-Einstein-field-eq}
\begin{aligned}
\bs E^{\bs\upupsilon} &= E(\bs d g^{\bs\upupsilon}; \{g^{\bs\upupsilon},\upvarphi^{\bs\upupsilon}\})=\bs d g^{\bs\upupsilon} \cdot \big( G(g^{\bs\upupsilon}) +\Lambda g^{\bs\upupsilon} - \kappa T(g^{\bs\upupsilon}, \upvarphi^{\bs\upupsilon}) \big) \vol_g, \\[.5mm]
\text{s.t. } \quad 
\bs E^{\bs\upupsilon}&=0 \quad \Rightarrow \quad  G^{\bs\upupsilon} +\Lambda g^{\bs\upupsilon} = \kappa T^{\bs\upupsilon}.
\end{aligned}
\end{equation}
These are the strictly $\Diff(M)$- \emph{and} $C^\infty\big(\Phi, \Diff(M)\big)$-invariant relational Einstein field equations, which have a well-posed Cauchy problem. 
We stress that $\bs E^{\bs\upupsilon}$ are \emph{not} a gauge-fixed version of the bare Einstein equations $\bs E$, as argued at the end of section \ref{Relational field theory}. 
Their explicit form in terms of the bare field equations $\bs E$ is read from \eqref{GT-E-GR-scalar} and applying the DFM rule of thumb:
\begin{equation}
\begin{aligned}
&\bs E^{\bs\upupsilon} 
=
\bs\upupsilon^* \big(\bs E + dE( \iota_{\bs d\bs\upupsilon \circ \bs\upupsilon\- }g ; \{g, \upvarphi \}) \big),&& \\[1mm]
\Rightarrow \quad &\bs d g^{\bs\upupsilon} \cdot \big( G^{\bs\upupsilon} +\Lambda g^{\bs\upupsilon} - \kappa T^{\bs\upupsilon} \big) \vol_{g^{\b\upupsilon}}
=
\bs \upupsilon^* &&\hspace{-4mm}\left[
\bs d g \cdot \big( G(g) +\Lambda g - \kappa T(g, \upvarphi) \big) \vol_g 
\right.\\
&\ && +
\left.
d \left( 
\iota_{\bs d\bs\upupsilon \circ \bs\upupsilon\- }g\cdot \big( G(g) +\Lambda g - \kappa T(g, \upvarphi) \big) \vol_g
\right)
\right].
\end{aligned}
\end{equation}
It shows the equivalence between the bare Einstein equations $G+\Lambda g = \kappa T$ and the relational Einstein equations $G^{\bs\upupsilon}+\Lambda g^{\bs\upupsilon} = \kappa T^{\bs\upupsilon}$. 
We also see that if the variation of the coordinatising (dressing) field $\bs d \bs\upupsilon$ is neglected, the two are in simple covariance relation: $\bs E^{\bs \upupsilon} =\bs \upupsilon^* E$.
This, as we observed on more than one occasion, explains why  confrontation of GR with observations has been so successful early on, even though the problem of identifying its fundamental physical d.o.f. and observables has remained a longstanding open issue.
It is thus not a gauge-fixed version of GR that passes observational tests, as is sometimes claimed, but its relational dressed version \eqref{Dressed-lagrangian-GR-scalar}-\eqref{Dressed-Einstein-field-eq}. 

\medskip
This model can be 
considered to apply to the historical version of GR where gravity is coupled to matter which is described heuristically, or effectively, as a fluid (gas, particles, dust, etc.). 
The scalar fields $\upvarphi^a$ could then represents the 4-velocity field of matter.
This encompasses and extends various ``scalar coordinatisations" of GR, such as \cite{Rovelli1991, Brown-Kuchar1995, Rovelli2002b}.
A variant of it may be applied to pure gravity: then the dressing field is metric dependent, $\bs\upupsilon=\bs\upupsilon(g)$, and supplies a $g$-dependent $\Diff(M)$-invariant definition of space time regions $U^{\bs\upupsilon}$. Or, seen otherwise, a
$g$-dependent coordinate system.
This encompasses e.g. the classic works \cite{Komar1958, Bergmann1961, Bergmann-Komar1972} where $\bs\upupsilon$  is built from $n$ ($=4$) independent (non-vanishing) scalar invariants, like the Kretschmann-Komar invariants -- compare e.g. \eqref{dressed-metric-index} to eq.(2.2) \cite{Komar1958}.

\medskip
Finally, let us highlighting the fact that the relational formulation  suggests  a natural heuristic argument for the necessity of quantization of gravity.  
An often repeated, such heuristics relies on the dynamics: In view of
the bare Einstein equations \eqref{Einstein-field-eq}, $G(g) +\Lambda g = \kappa T(g, \upvarphi)$, the left-hand side contains only metric d.o.f., while the right-hand side contains also the d.o.f. of matter. 
Now, as  the argument goes, if the matter d.o.f. are quantized, so that the energy-momentum becomes an operator $\h T$ with vacuum expected value $\langle \h T \rangle = T$,  one expects that the left-hand side should also contain quantized metric d.o.f. $\h g$ s.t. $\langle\, \h g\,\rangle =g$, and so that  a ``quantum Einstein equation" $( \h G  + \Lambda \h g \,)\,|\Psi \rangle = \kappa \h T\, |\Psi \rangle$ may 
hold. 

In view of the relational Einstein equation 
\eqref{Dressed-Einstein-field-eq},  $G^{\bs\upupsilon} +\Lambda g^{\bs\upupsilon} = \kappa T^{\bs\upupsilon}$, this argument is flawed.
Indeed, the matter d.o.f. involved in the dressing field $\bs\upupsilon$ are seen to actually contribute, like the metric d.o.f., to both sides of the field equations. 
Actually, on the basis of the relational formulation, a stronger, more compelling argument can be made, that does not rely on the classical on-shell dynamics, but rather holds \emph{at the kinematical level}. 
Indeed,  the very definition of the physical relational metric 
\eqref{dressed-field-space-GR-scalar}-\eqref{dressed-metric-index} involves the matter d.o.f., hence if the latter are quantized one expects to naturally obtain a notion of quantized gravitational d.o.f., so that $g^{\bs\upupsilon} \mapsto \h{g^{\bs\upupsilon}}$ and 
$\langle  \h{g^{\bs\upupsilon}}\rangle
=
g^{\bs\upupsilon}$.
Furthermore,  by the same argument, one may obtain a notion of quantized (relationally defined) spacetime regions $U^{\bs\upupsilon} \mapsto U^{\h{\bs\upupsilon}}$, 
that \emph{does not} imply a quantization of the underlying manifold $M$ -- which  is unobservable. This kinematical argument can be summarised via the following sketch:

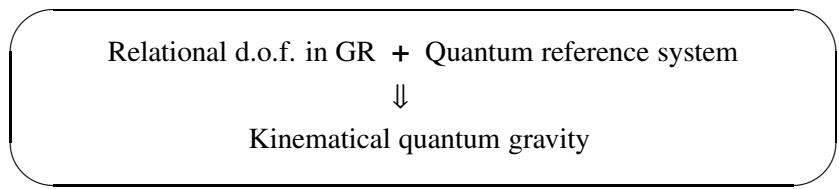
\begin{center}
\ovalbox{\parbox{\dimexpr\linewidth-40\fboxsep-200\fboxrule\relax}{\centering 
\vspace{0.3cm}
 Relational d.o.f. in GR $\  \bs +\ $ Quantum reference system \\ 
\vspace{0.1cm}
\hspace{-6mm}$\Downarrow$ \\
\vspace{0.1cm}Kinematical quantum gravity
\vspace{0.3cm}
}} 
\end{center}

\noindent
We will revisit and expand on these observations in a forthcoming paper \cite{JTF-Ravera2024gRGFTquantum}. We now turn to the second example of gravity coupled to EM and a charged complex scalar matter field.

\paragraph{Einstein-$\CC$-Maxwell model}

We shall now consider GR plus cosmological constant coupled to electromagnetism and a charged complex scalar matter field $\upphi$.
\medskip

\noindent
\underline{Kinematics}: 
The field space  is $\Phi=\{\phi\}=\{g, A, \upphi\}$, where $g$ is a metric field, $A$ is a $\mathfrak u(1)$-valued EM potential, and $\upphi$ is a $\CC$-valued scalar field. It means here $\Hl=\U(1)$.
The EM field strength is $F=dA$, and the minimal coupling between the EM potential and the matter field is $D\upphi=d\upphi + A\upphi$.
The action of $\Diff(M)\ltimes \U(1)$ is 
\begin{align}
\label{action-grp-GR-EM}
  R_{(\uppsi, \upgamma)}\, \phi 
  =
  R_{(\uppsi, \upgamma)} \, \big(g, A, \upphi\big)
  = 
  \big( \uppsi^*g, \,\uppsi^*(A^\upgamma), \, \uppsi^*(\upphi^\upgamma) \big)
  =
  \big( \uppsi^*g, \,  \uppsi^*(A+ \upgamma\-d\upgamma),\,  \uppsi^*( \upgamma\-\upphi ) \big). 
\end{align}
In this model, it is possible to dress in steps, for $\U(1)$ first, and then for $\Diff(M)$, as indicated in section \ref{Residual transformations, and composition of dressing operations}. 
\medskip

The first step in thus to identify a $\U(1)$-dressing field. It is easily done: 
Using a polar decomposition of the matter field $\upphi = \uprho \exp{i\, \theta}$, where $\uprho=|\upphi|^{\sfrac{1}{2}}$ is the modulus and $\theta$ is the phase, it is clear by  \eqref{action-grp-GR-EM}  that we may define
\begin{equation}
\begin{aligned}
\label{U(1)-dressing}
&\u=\u(\upphi)\defeq \exp{i\, \theta}, \\[.5mm]
\text{ s.t. } \quad 
&R^\star_{(\uppsi, \upgamma)} \u(\upphi)
=
\u \big( R_{(\uppsi, \upgamma)} \upphi\big)
\defeq
\uppsi^*\big(\upgamma\- \exp{i\, \theta}\big) 
\rdefeq
\uppsi^*\big(\upgamma\- \u(\upphi)\big).
\end{aligned}
\end{equation}
By the dressing property $R^\star_\upgamma \u = \upgamma\-\u$, we define the $\U(1)$-invariant fields
\begin{align}
\label{dressed-field-space-1-GR-EM}
\Phi^\u
=
\{\phi^\u\} 
=
\{ g^\u, \, A^\u,\, \upphi^\u\}
\defeq
\{g,\,  A+ \u\-d\u,\, \uprho \},
\end{align}
with $\upphi^u = \u\- \upphi= \uprho$ by definition. 
The variable $A^\u$ implements the idea that the gauge invariant internal d.o.f. of the EM potential are coordinatised w.r.t. that  of the scalar matter field (i.e. its phase $\theta$), in accordance with  the insight gained from the internal point-coincidence argument from section \ref{Relationality in  general-relativistic gauge field theory}. 
Meanwhile $\upphi^\u=\rho$ signifies the self-coordinatisation of the internal d.o.f. of matter, i.e. it has constant $0$ phase   w.r.t. itself: this is the internal version of ``being at rest" in its own reference frame, analogue to the case of the real scalar field in the previous example.

The dressed EM field strength is $F^\u\defeq dA^\u =\u\-F \u=F$, and happens to coincide with the bare EM field strength, which as is well-known is the only (local) gauge-invariant field in the bare formulation of classical electromagnetism. 
Also, observe that the dressed covariant derivative is $(D\upphi)^\u=\u\- (D\upphi)= D^\u \upphi^u = D^\u \uprho = d\uprho + A^\u\uprho$, which represents the minimal coupling between the invariant d.o.f. of the EM potential and the matter field. 
It shows that in the relational formulation it is possible to write the coupling of invariantly defined fields -- which is partially in tension with some comments around the main thesis proposed in \cite{Rovelli2014}.
As a side comment, we remark that the variables $\{A^\u, \uprho\}$ allow in particular to model the Aharanov-Bohm effect in an invariant \emph{and local} way, i.e. as arising from the pointwise (field-local) interaction between physical fields, described by $D^\u \uprho = d\uprho + A^\u\uprho$.  
See e.g. \cite{Wallace2014} or \cite{Berghofer-et-al2023} Chap.5, and also \cite{JTF-Ravera2024c}. 
\medskip

By the compatibility condition $R^\star_\uppsi \u = \uppsi^* \u$, the $\U(1)$-invariant  (``internally relational") variables have well-defined $\Diff(M)$-transformations: the action of $\Diff(M)$ on $\Phi^\u$ being as expected 
\begin{align}
R_\uppsi \phi^\u
=
R_\uppsi \, \big( g, \, A^\u,\,  \uprho \big)
=
\big( \uppsi^*g, \, \uppsi^*(A^\u),\,  \uppsi^*\uprho \big). 
\end{align}
This makes it natural to look for a $\Diff(M)$-dressing field. 
The invariant EM potential is $A^\u={A^\u}_\mu \,dx^{\,\mu}$, and its components form a scalar field $\mathsf A^\u \defeq {A^\u}_\mu:M \rarrow N=\RR^n$.
We consider its section $\overline{\mathsf A^\u}: N=\RR^n \rarrow M$, which is s.t. $\mathsf A^\u \circ \overline{\mathsf A^\u} =\id_N$ (a right inverse).
We then define the $\Diff(M)$-dressing field
\begin{equation}
\begin{aligned}
&\bs\upupsilon
=
\bs\upupsilon(A^\u)
=
\bs\upupsilon(A, \upphi)
\defeq \overline{\mathsf A^\u} : N \rarrow M, \\
\text{s.t. }\quad 
& R^\star_{(\uppsi, \upgamma)} \bs\upupsilon(A, \upphi)
=
R^\star_{(\uppsi, \upgamma)} \bs\upupsilon(A^\u)
=
\bs\upupsilon\big(R_{(\uppsi, \upgamma)}A^\u \big) 
=
\bs\upupsilon\big( \uppsi^*(A^\u) \big)
\defeq
\uppsi\- \circ \overline{\mathsf A^\u}
\rdefeq
\uppsi\- \circ \bs\upupsilon(A, \upphi).
\end{aligned}
\end{equation}
Therefore, under $\bs\Diff_v(\Phi)\simeq C^\infty\big(\Phi, \Diff(M) \big)$ this dressing field transforms as 
$\bs\upupsilon^{(\bs\uppsi, \bs\upgamma)}= \bs\uppsi\- \circ \bs \upupsilon$, as is expected. 
It~allows to define dressed regions of spacetime: 
\begin{align}
\label{spacetime-regions-GR-EM}
  U^{\bs\upupsilon}\defeq \bs\upupsilon\-(U), \quad \text{s.t.} \quad
  \big(U^{\bs\upupsilon}\big)^{(\bs\uppsi, \bs\upgamma)} = U^{\bs\upupsilon}. 
\end{align}
As expected from the point-coincidence argument, this gives a $\big(\!\Diff(M)\ltimes \U(1)\big)$- \emph{and} $C^\infty\big(\Phi, \Diff(M)\ltimes \U(1) \big)$-invariant relational definition of spacetime, with  the electromagnetic sector field content $\{A, \upphi\}$, or more precisely the invariant physical electromagnetic d.o.f. $A^\u$, being the reference physical system.

The final dressed field space is then
\begin{align}
\label{dressed-field-space-GR-EM}
\Phi^{(\bs\upupsilon, \u)}
=
\{\phi^{(\bs\upupsilon, \u)}  \}
=
\{ g^{\bs\upupsilon},\,  A^{(\bs\upupsilon, \u)}, \, \uprho^{\bs\upupsilon}\}
=
\{ {\bs\upupsilon}^*g, {\bs\upupsilon}^*(A^\u), \, {\bs\upupsilon}^*\uprho\}. 
\end{align}
These are living on the \emph{physical} spacetime, whose regions are defined by \eqref{spacetime-regions-GR-EM}. 
For example, the scalar matter field is s.t. $\uprho^{\bs\upupsilon} = \rho \circ \bs\upupsilon : U^{\bs\upupsilon} \rarrow \RR$.
This  again illustrates the point-coincidence argument, and the co-definition of the physical field d.o.f., as the values of $\uprho^{\bs\upupsilon}$ are given \emph{at} values of $A^{(\bs\upupsilon, \u)}$ (via $\upupsilon$), this coincidence of values  defining, ``tagging" or identifying,  a physical spacetime point. The minimal coupling of both fields is given by the dressed covariant derivative 
$D^{(\bs\upupsilon, \u)}\uprho^{\bs\upupsilon} =d\uprho^{\bs\upupsilon} + A^{(\bs\upupsilon, \u)} \uprho^{\bs\upupsilon}$, which is then actually an \emph{invariant} derivative.
Likewise, the $C^\infty\big(\Phi, \Diff(M)\ltimes \U(1) \big)$-invariant dressed field $g^{\bs\upupsilon}$ describes the metric properties of the physical, relationally defined, spacetime $M^{(\bs\upupsilon, \u)}$.
It can be seen as the  bare metric $g$ ``written" in the coordinate system supplied by the (invariant) electromagnetic field content $A^\u \sim (A, \upphi)$: 
in abstract index notation 
\begin{align}
\label{dressed-EMscal-metric-index}
(g^{\bs\upupsilon})_{ab} = \frac{\d x^{\,\mu}}{\d\bs\upupsilon^a} \frac{\d x^{\,\nu}}{\d\bs\upupsilon^b}\, g_{\mu\nu}, \quad \text{ with } \quad \bs\upupsilon
=
\bs\upupsilon(A^\u)
=
\bs\upupsilon(A, \upphi).
\end{align} 
 From the invariant metric $g^{\bs\upupsilon}$ one derives the invariant (physical) Levi-Civita connection field $\Gamma^{\bs\upupsilon}= \Gamma(g^{\bs\upupsilon})$, thus the invariant covariant derivative $\nabla^{g^{\bs\upupsilon}}$, which defines the physical parallel transport and geodesic on spacetime $M^{(\bs\upupsilon, \u)}$. The invariant Riemann tensor of spacetime is $\text{Riem}^{\bs\upupsilon}=\text{Riem}(g^{\bs\upupsilon})$, fom which one derive the invariant Ricci tensor $\text{Ricc}^{\bs\upupsilon}$ and Ricci scalar $\mathsf R(g^{\bs\upupsilon})=(g^{\bs\upupsilon})\-\cdot \text{Ricc}^{\bs\upupsilon}$. 
 The physical Einstein tensor is thus 
 $G^{\bs\upupsilon} = G(g^{\bs\upupsilon}) = \text{Ricc}^{\bs\upupsilon} - \tfrac{1}{2} \mathsf R^{\bs\upupsilon} g^{\bs\upupsilon}$.

This achieves our analysis of the relational kinematics of the theory. 
We now consider its dynamics.
\medskip

\noindent
\underline{Dynamics}: The (bare) Lagrangian of the theory $L$ is
\begin{equation}
\begin{aligned}
\label{Bare-lagrangian-GR-scalar-EM}
L(g, A, \upphi) 
&=
L_\text{\tiny{EH+$\Lambda$}}(g)  +   L_\text{\tiny{EM}}(g, A) +
L_\text{\tiny{KG}}(g, A, \upphi)\\
&=
\tfrac{1}{2\kappa }\vol_g  \big(\mathsf R(g) - 2 \Lambda \big)  
\ +\ 
\tfrac{1}{2} \, F *_g\!F 
\ + \ 
\tfrac{1}{2} \left(  \langle D\upphi, *_g D\upphi \rangle + m^2\langle \upphi, *_g\upphi  \rangle \right),
\end{aligned}
\end{equation}
where $\langle\  ,\,\rangle: \CC \times \CC \rarrow \CC$, $(v, w) \mapsto \langle v  ,w\rangle \defeq v^\dagger w$, with $v^\dagger$ the  conjugate of $v$, 
and $*_g: \Omega^p(U) \rarrow \Omega^{n-p}(U)$ is the Hodge dual operator. 
The field equation 1-form is
\begin{equation}
\label{Einstein-EM-field-eq}
\begin{aligned}
\bs E 
=
E(\bs d \phi; \phi)
&=
E(\bs dg; \phi) + E(\bs d A; \phi) + E(\bs d\upphi; \phi)\\
&=
\bs d g \cdot \left( G(g) +\Lambda g - \kappa T(g, A, \upphi) \right) \vol_g 
\ \ +\ \  \bs d A \,  \left( d *_g\!F - J(A, \upphi) \right)
\ \ + \ \ \langle \bs d \upphi, D*_g\! D \upphi + m^2 \upphi\rangle,
\end{aligned}
\end{equation}
where $T$ and $J$ are, respectively, the  energy-momentum tensor of the EM field and charge matter field, and the EM current. Explicitly,
\begin{equation}
\begin{aligned}
 \label{EM-tensor-current}   
T(g, A, \upphi)  =&\, T_{\mu \nu}(g, A) + T_{\mu\nu}(g, \upphi) = \left( F_{\mu \alpha} F_{\nu \beta} \, g^{\alpha \beta} - \tfrac{1}{4} g_{\mu \nu} \, F \cdot F \right) \vol_g +\left( \partial_\mu \upphi \, \partial_\nu \upphi - \tfrac{1}{2} g_{\mu\nu} \, \partial \upphi \cdot \partial \upphi \right) \vol_g, \\[1mm]
J(A, \upphi)  \defeq&\, \tfrac{1}{2} \left( \langle *_g D\upphi , \upphi \rangle - \langle \upphi, *_g D\upphi \rangle \right). 
\end{aligned}
\end{equation}
Then, $\bs E=0$ implies
\begin{align}
\label{Einstein-EM-field-eq-2}
 G(g) +\Lambda g = \kappa T(g, A, \upphi), \quad
 d *_g\!F = J(A,\upphi), 
 \quad
 \big(D*_g\!D\, + m^2 \big)\upphi=0.
\end{align}
These are respectively the Einstein equation, the Maxwell equation, and the Klein-Gordon equation.

It is easy to check that $R^\star_{(\uppsi, \upgamma)} \bs E = \uppsi^*(\bs E^\upgamma) = \uppsi^*\bs E$, satisfying the general  hypothesis \eqref{equiv-E-theta} of gRGFT.
One may also verify that, applying \eqref{GT-E}-\eqref{GT-E-special}, the $\bs\Diff_v(\Phi)\simeq C^\infty\big(\Phi, \Diff(M)\ltimes \U(1) \big)$-transformation  $\bs E^{(\bs\uppsi, \bs\upgamma)}$
are s.t. $\bs E=0\ \Rightarrow \bs E^{(\bs\uppsi, \bs\upgamma)}=0$. 
Which means that the covariance group of the theory is much bigger than $\Diff(M)\ltimes \U(1)$, and encompass $\phi$-dependent transformations: its full covariance group is $C^\infty\big(\Phi, \Diff(M)\ltimes \U(1) \big)$.
An observation that generalises the one made in \cite{Bergmann1961, Bergmann-Komar1972}.
\medskip

We now write down the relational reformulation of the theory via DFM. 
The dressed Lagrangian reads
\begin{align}
\label{Dressed-lagrangian-GR-EM-scalar}
L^{(\bs\upupsilon,\u)}(g, A, \upphi)
\defeq&\,
{\bs\upupsilon}^*L^{\u}(g, A, \upphi)=
L(g^{\bs\upupsilon}, A^{(\bs\upupsilon,\u)},\uprho^{\bs\upupsilon})\\
=&\,
L_\text{\tiny{EH+$\Lambda$}}(g^{\bs\upupsilon})  +   L_\text{\tiny{EM}}(g^{\bs\upupsilon}, A^{(\bs\upupsilon,\u)}) +
L_\text{\tiny{KG}}(g^{\bs\upupsilon}, A^{(\bs\upupsilon,\u)}, \uprho^{\bs\upupsilon}) \notag\\
=&\,
\tfrac{1}{2\kappa }\vol_{g^{\bs\upupsilon}}  \big(\mathsf R(g^{\bs\upupsilon}) - 2 \Lambda \big)  
\ +\ 
\tfrac{1}{2} \, F^{(\bs\upupsilon,\u)} *_{g^{\bs\upupsilon}}\!F^{(\bs\upupsilon,\u)} 
\ + \ 
\tfrac{1}{2} \left(  \langle D^{(\bs\upupsilon,\u)} \uprho^{\bs\upupsilon}, *_{g^{\bs \upupsilon}} D^{(\bs\upupsilon,\u)}\uprho^{\bs\upupsilon} \rangle + m^2\langle \uprho^{\bs\upupsilon}, *_{g^{\bs\upupsilon}}\uprho^{\bs \upupsilon}  \rangle \right), \notag
\end{align}
with corresponding dressed field equation 1-form
\begin{equation}
\begin{aligned}
\bs E^{(\bs \upupsilon,\u)} 
&=
E(\bs dg^{\bs \upupsilon}; \phi^{(\bs \upupsilon,\u)}) + E(\bs d A^{(\bs \upupsilon,\u)}; \phi^{(\bs \upupsilon,\u)}) + E(\bs d\uprho^{\bs \upupsilon}; \phi^{(\bs \upupsilon,\u)})\\
&=
\bs d g^{\bs \upupsilon} \cdot \left( G(g^{\bs \upupsilon}) +\Lambda g^{\bs \upupsilon} - \kappa T(g^{\bs \upupsilon}, A^{(\bs \upupsilon,\u)}, \uprho^{\bs \upupsilon}) \right) \vol_{g^{\bs \upupsilon}}  
\ \ +\ \ 
\bs d A^{(\bs \upupsilon,\u)} \,  \left( d *_{g^{\bs\upupsilon}}\!F^{(\bs \upupsilon,\u)} - J(A^{(\bs \upupsilon,\u)}, \uprho^{\bs\upupsilon}) \right) \\
& \quad + \langle \bs d \uprho^{\bs\upupsilon}, D^{(\bs \upupsilon,\u)}*_{g^{\bs\upupsilon}}\! D^{(\bs \upupsilon,\u)} \uprho^{\bs\upupsilon} + m^2 \uprho^{\bs\upupsilon}\rangle,
\end{aligned}
\end{equation}
so that $\bs E^{(\bs\upupsilon,\u)}=0$ implies
\begin{align}
\label{Dressed-Einstein-EM-scalar-field-eq}
  G^{\bs\upupsilon} +\Lambda g^{\bs\upupsilon} = \kappa T^{(\bs\upupsilon,\u)}, \quad
 d *_{g^{\bs\upupsilon}}\!F^{(\bs\upupsilon,\u)} = J(A^{(\bs\upupsilon,\u)},\uprho^{\bs\upupsilon}), 
 \quad
\big(D^{(\bs\upupsilon,\u)}*_{g^{\bs\upupsilon}}\!D^{(\bs\upupsilon,\u)}\, + m^2 \big)\uprho^{\bs\upupsilon}=0.
\end{align}
The latter are the strictly $\big(\!\Diff(M)\ltimes \U(1)\big)$- \emph{and} $C^\infty\big(\Phi, \Diff(M)\ltimes \U(1)\big)$-invariant relational Einstein, Maxwell, Klein-Gordon field equations, which all have a well-posed initial value (Cauchy) problem. 
One may check, using \eqref{dressed-E}-\eqref{dressed-E-special} to express $\bs E^{(\bs\upupsilon,\u)}$ in terms of the bare  $\bs E$, that $\bs E=0 \Rightarrow \bs E^{(\bs\upupsilon,\u)}=0$.
This is the reason why the bare formalism gives sensible physical results, as 
it is actually the relational version  \eqref{Dressed-lagrangian-GR-EM-scalar}-\eqref{Dressed-Einstein-EM-scalar-field-eq} of the theory which is confronted to experimental tests, not a gauge-fixed version of it as is often said -- we indeed insist that $L^{(\bs\upupsilon,\u)}$ and $\bs E^{(\bs\upupsilon,\u)}$ are \emph{not} gauge-fixed versions of the bare $L$ and $\bs E$. 
\medskip
 
Finally, let us observe that this relational formulation provides a straighforward heuristic argument for the naturality of quantization of both the EM and the gravitational field.

Such an argument, for the electromagnetic field, is a dynamical one and  relies  on the bare Maxwell equation $d*_g\!F=J(A, \upphi)$: if the matter d.o.f. --  both internal and external -- on the right-hand side are quantized, one would expect that 
the EM d.o.f. on the left-hand side are also quantized. 
But looking at the 
$\U(1)$-invariant, \emph{internally
relational} Maxwell equation $d*_g\!F^\u=J(A^\u, \uprho)$, we see that actually the matter d.o.f. contribute on both sides, via $\u=\u(\upphi)$. 
The relational formulation actually allows to make the argument at the kinematical level: 
In view of the $\U(1)$-invariant relational variable $A^\u= A + \u\-d\u$, with $\u=\u(\upphi)$, if $\upphi$ is quantized, it seems unavoidable that the invariant EM potential $A^\u$, and the EM field strength $F^\u$,  should be too. 
This argument can be thus summarised as follows:
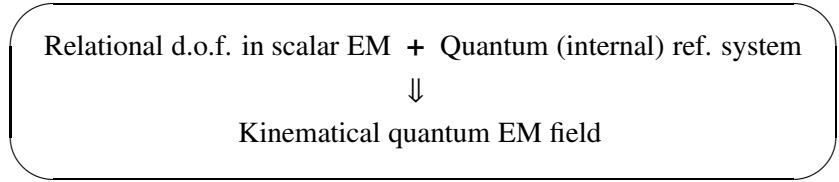
\begin{center}
\ovalbox{\parbox{\dimexpr\linewidth-40\fboxsep-200\fboxrule\relax}{\centering 
\vspace{0.3cm}
 Relational d.o.f. in scalar EM $\  \bs +\ $ Quantum (internal) ref. system \\ 
\vspace{0.1cm}
\hspace{-2mm}$\Downarrow$ \\
\vspace{0.1cm}Kinematical quantum EM field
\vspace{0.3cm}
}} 
\end{center}

Similarly, and as  seen in the previous example, looking at the bare Einstein equations 
$G(g) +\Lambda g = \kappa T(g, A, \upphi)$, 
it may seem that quantization of the d.o.f. of $\upphi$ on the right-hand side would suggest the quantization of the the d.o.f. of $g$, while leaving $A$ a priori unaffected. One would need to appeal to the bare Maxwell equation to hint at quantization of $A$. 
But inspection of the relational Einstein equations 
$G(g^{\bs\upupsilon}) +\Lambda g^{\bs\upupsilon} = \kappa T(g^{\bs\upupsilon}, A^{(\bs\upupsilon, \u)}, \uprho^{\bs\upupsilon})$
shows that the matter d.o.f \emph{and} the EM d.o.f. contribute to both sides via $\bs\upupsilon=\upupsilon(A^\u)=\bs\upupsilon(A, \upphi)$. Also, just by looking at the right-hand side, it already appears that quantization of $\upphi$ would  kinematically imply that of $A^{(\bs\upupsilon, \u)}$, without using the relational Maxwell equation $d*_{g^{\bs\upupsilon}}\!F^{\bs\upupsilon}=J(A^{(\bs \upupsilon,\u)},\uprho^{\bs\upupsilon})$.

This indicates that the relational formulation via DFM makes for a solely kinematical argument: 
Simply by looking at the relational variables $A^{(\bs\upupsilon, \u)} = \bs\upupsilon^*(A^\u)$ and  $g^{\bs\upupsilon}=\bs\upupsilon^* g$, with $\u=\u(\upphi)$ and $\bs\upupsilon=\bs\upupsilon(A^\u)=\bs\upupsilon(A, \phi)$,
it appears that quantization of the d.o.f. of $\upphi$ suggests that of the physical EM field $A^{(\bs\upupsilon, \u)}$, as well as that of the physical gravitational field $g^{\bs\upupsilon}$. 
We summarise this by
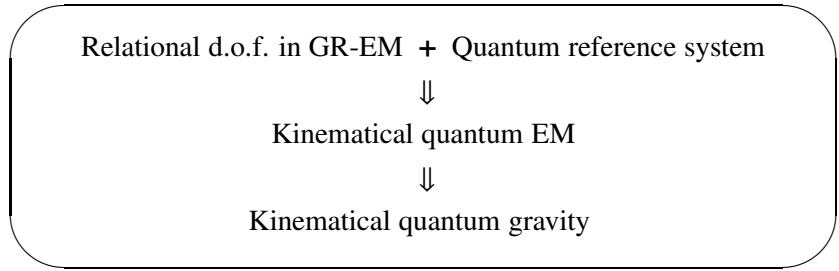
\begin{center}
\ovalbox{\parbox{\dimexpr\linewidth-40\fboxsep-200\fboxrule\relax}{\centering 
\vspace{0.3cm}
 Relational d.o.f. in GR-EM $\  \bs +\ $ Quantum reference system \\ 
\vspace{0.1cm}
\hspace{1mm}$\Downarrow$ \\
\vspace{0.1cm}Kinematical quantum EM\\
\vspace{0.1cm}
\hspace{1mm}$\Downarrow$ \\
\vspace{0.1cm}Kinematical quantum gravity
\vspace{0.3cm}
}} 
\end{center}
The topic of relational quantization, expanding on these observations and going beyond these heuristic arguments, will be addressed in \cite{JTF-Ravera2024gRGFTquantum}.

\section{Conclusion}  
\label{Conclusion}  

In this paper we have developed a relational formulation for general-relativistic gauge field theory (gRGFT) based on the dressing field method (DFM). 
To do so we exploited the bundle geometry of field space $\Phi$, highlighting in particular the structures encoding the physical d.o.f., i.e. the moduli space $\M$ and what we called the associated bundle of regions $\b{\bs U}(M) \defeq \Phi \times \bs U(M)/\sim$.
The latter in particular encodes the \emph{physical} spacetime regions, as understood via the point-coincidence argument. 
A key result stemming from understanding the geometry of the field space of gRGFT is \eqref{GT-E},
\begin{center}
\doublebox{\,$
\bs E^{(\bs\uppsi, \bs\upgamma)} 
=
\bs\uppsi^* \left( 
\bs E 
+ 
dE \big( 
\iota_{\bs{d\uppsi}\circ\bs \uppsi\-} \phi; \phi
\big)
+
 dE \big(\bs{d\upgamma} \, \bs\upgamma\-  - \bs\upgamma\, \mathfrak L_{\bs{d\uppsi} \circ \bs\uppsi\-} \bs\upgamma\-
 ; \phi \big)\,
\right)$\,}
\end{center}
which shows that, while the covariance group of gRGFT is usually understood -- or defined -- to be  $\Diff(M)\ltimes \Hl$, it actually is the much bigger group 
$C^\infty\big(\Phi, \Diff(M)\ltimes \Hl \big)$ of \emph{field-dependent transformations}. This  generalises the observation by \cite{Bergmann1961, Bergmann-Komar1972}.

The DFM is then, in a nutshell, a way to realise  \emph{basic} objects on $\Phi$ and $\Phi \times \bs U(M)$, which can be naturally interpreted as relational coordinatisations on  $\M$ and $\b{\bs U}(M)$. 
The relational formulation via DFM therefore technically implements the core insight of gRGFT stemming from the generalised hole and point-coincidence arguments, as described in section \ref{Relationality in  general-relativistic gauge field theory} -- see also \cite{JTF-Ravera2024c}. 
An essential result of this paper is \eqref{dressed-E}  giving the \emph{relational field equations} of gRGFT,
\begin{center}
\doublebox{\,$
\bs E^{(\bs\upupsilon, \u)} 
=
\bs\upupsilon^* \left( 
\bs E 
+ 
dE \big( 
\iota_{\bs{d\upupsilon}\circ\bs \upupsilon\-} \phi; \phi
\big)
+
 dE \big(\bs d\u\, \u\-  - \u\, \mathfrak L_{\bs{d\upupsilon} \circ \bs\upupsilon\-} \u\-
 ; \phi \big)\,
\right)$\,}
\end{center}
which are strictly $C^\infty\big(\Phi, \Diff(M)\ltimes \Hl \big)$-invariant, and thus have a well-posed Cauchy problem. 
The relational equations $\bs E^{(\bs\upupsilon, \u)} =0$ are the ones tacitly used in practical experimental situations.
The fact that, as the result shows, they are ``equivalent"  to the bare field equations $\bs E=0$,  explains why gRGFT was successfully compared to experiments much before  the problem of identifying the fundamental physical d.o.f. and observables was solved.

Another fundamental result of the relational reformulation via DFM is the definition of $C^\infty\big(\Phi, \Diff(M)\ltimes \Hl \big)$-invariant \emph{dressed regions} $U^{\bs\upupsilon(\phi)}$, which formally implement the point-coincidence argument and provides a definition of spacetime regions by its field content $\phi$. Hence one obtains a relational definition of spacetime $M^{(\bs\upupsilon, \u)}$. 
From this, we observed, follows that the so-called ``boundary problem", i.e. the often repeated claim that ``boundaries break $\Diff(M)$ and/or $\Hl$ symmetry", is erroneous: a physical relational boundary $\d U^{\bs\upupsilon} \subset M^{(\bs\upupsilon, \u)}$ is of necessity $\big(\!\Diff(M)\ltimes \Hl\big)$-invariant, and even invariant under the full covariance group $C^\infty\big(\Phi, \Diff(M)\ltimes \Hl \big)$ of gRGFT. 
Such ``boundary problems" dissolve in the relational understanding of physics. 
We shall investigate further the consequences of this in a forthcoming paper \cite{JTF-Ravera2024bdy}.

We considered a natural relational reformulation of GR, featuring the \emph{relational Einstein equations}
\begin{center}
\doublebox{\,$
G^{\bs\upupsilon} +\Lambda g^{\bs\upupsilon} = \kappa T^{(\bs\upupsilon,\u)}$\,}
\end{center}
which encompass all manners of ``scalar coordinatisation" such as \cite{Rovelli1991, Brown-Kuchar1995, Rovelli2002b} and \cite{Komar1958, Bergmann1961, Bergmann-Komar1972}. 
This  formulation of GR has as many natural applications:  black holes physics,
gravitational waves physics, and cosmological perturbation theory\footnote{Where we may make contact with e.g.
\cite{Giesel-et-al2010, Giesel-et-al2018}.}, etc.;
We may also assess the degree to which the relational formulation influences dark matter models.
In the above relational Einstein equations, it should be stressed that the matter d.o.f., like those of the (invariant) metric $g^{\bs\upupsilon}$, appear on both sides, allowing for a new take on the heuristics motivating the necessity of quantum gravity.  

\medskip

This brings to the next phase of our program dealing with relational path integral quantization, or relational Quantum Field Theory (rQFT), which will be explored in \cite{JTF-Ravera2024gRGFTquantum}. 
As we have shown in section \ref{Relational formulation via dressing}, the relational reformulation of a theory with bare action $S$ failing to be $\Hl$-invariant implements an automatic  mechanism of cancellation  of
\emph{classical} $\Hl$-anomaly, 
where a \emph{twisted dressing field} plays the role of Wess-Zumino term. 
We expect that the same will hold in the QFT context. 
We also highlighted that anomalies, either classical or quantum, are to be understood via the \emph{twisted connections} on field space $\Phi$. 
It is all but certain that the geometry of twisted connections will play a key role in relational quantization. 

Furthermore, as a separate item of our program, in \cite{JTF-Ravera2024gRGFTp3} we will also investigate the \emph{relational covariant phase space} formalism of gRGFT. 
This may be the starting point of a (formal) relational geometric quantization, where again we expect twisted geometry to play a non-trivial role.

\section*{Acknowledgment}  

J.F. is supported by the Austrian Science Fund (FWF), \mbox{[P 36542]} and by the OP J.A.C. MSCA grant, number CZ.02.01.01/00/22\_010/0003229, co-funded by the Czech government Ministry of Education, Youth \& Sports and the EU. 
L.R. acknowledges partial financial support from INFN during her visits at MUNI in which this project started and part of this work was developed. She thanks the Dept. of Physics and the Dept. of Mathematics and Statistics at the Faculty of Science of MUNI for the kind hospitality during her stay.

\appendix

\section{Semi-direct product structure of $\Aut(P)$}  
\label{Semi-direct product structure of Aut(P)}  

As is well-known, the group of vertical automorphisms $\Aut_v(P)$ of  a principal bundle $P$ is a normal subgroup of its automorphisms group $\Aut(P)$: $\Aut_v(P) \triangleleft \Aut(P)$.
Their quotient is thus a group, isomorphic to the group of diffeomorphisms of the orbit space $P/H=M$, $\Aut(P)/\Aut_v(P) \rdefeq \overline{\Diff} \simeq \Diff(P/H)=\Diff(M)$. 
One shows that $\Aut(P)$ has a natural structure of \emph{inner} semi-direct product: 
\begin{equation}
\label{Semi-dir-Aut}
\begin{aligned}
\Aut(P) &= \overline{\Diff} \ltimes \Aut_v(P), \\
\psi&=\Big(\b \uppsi, \eta \Big), \\[1mm]
\text{with product } \quad
\psi' \circ \psi &= \left( {\b\uppsi'} \circ \b\uppsi, \ \eta' \!\circ \text{Conj}(\b\uppsi')\, \eta \right),
\end{aligned}
\end{equation}
with the group morphism $\text{Conj} : \overline{\Diff} \rarrow \Aut\big(\Aut_v(P) \big)$, 
$\b\uppsi \mapsto \text{Conj}(\b\uppsi)$, and $\text{Conj}(\b\uppsi)\,\eta \defeq \b\uppsi \circ \eta \circ {\b\uppsi}\-$. 
Indeed, writing an automorphsism as $\psi = \eta \circ \b\uppsi$, as read from the following graph 
\begin{equation}
\label{Aut-graph}
\begin{tikzcd}
P \arrow[bend left,"\psi"]{rr} \arrow[r,swap,"\b\uppsi"] \arrow[d] & 
P \arrow[r,swap,"\eta"] \arrow[d] &
P \arrow[d]
\\
M \arrow[r,swap,"\uppsi"] & M \arrow[equal]{r} & M 
\end{tikzcd}
\end{equation}
we get the composition
\begin{equation}
\label{SDeq1}
\begin{aligned}
\psi' \circ \psi &= \big(\eta' \circ \b\uppsi' \big) \circ \big(\eta \circ \b\uppsi \big) \\
 &=  \big(\eta' \circ \b\uppsi' \big) \circ \big(\eta \circ {\b\uppsi}'^{-1} \circ \b\uppsi' \circ \b\uppsi \big)\\
 &= \left( \eta' \circ 
 \Big( \b\uppsi'  \circ \eta \circ  {\b\uppsi}'^{-1} \Big) 
 \right) \circ 
 \Big( \b\uppsi' \circ \b\uppsi \Big).
\end{aligned}
\end{equation}
We have then that a form $\phi \in \Omega^\bullet(P)$ transform under $\Aut(P)$ as $\psi^*\phi = \b\uppsi^* (\eta^*\phi)$, where $\eta^*\phi$ defines the $\Aut_v(P)\simeq\H$-gauge transformation of $\phi$, while $\b\uppsi^*\phi$ defines its $\overline{\Diff}$-transformation. 

The alternative decomposition of an automorphism, given by the following graph:
\begin{equation}
\label{Aut-graph-pullback}
\begin{tikzcd}
P \arrow[bend left,"\psi"]{rr} \arrow[r,swap,"\eta"] \arrow[d] & 
P \arrow[r,swap,"\b\uppsi"] \arrow[d] &
P \arrow[d]
\\
M \arrow[equal]{r} & M \arrow[r,swap,"\uppsi"] & M 
\end{tikzcd}
\end{equation}
gives rise to the semi-direct product rule
\begin{align}
\psi' \circ \psi &= \left( {\b\uppsi'} \circ \b\uppsi, \  \!\text{Conj}(\b\uppsi\-)\, \eta'\, \circ \eta \right),
\end{align}
from the composition 
\begin{equation}
\begin{aligned}
\psi' \circ \psi 
&= \big(  \b\uppsi' \circ \eta' \big) \circ \big(\b\uppsi \circ \eta \big) \\
 &=  \big( \b\uppsi' \circ   \b\uppsi \circ {\b\uppsi}^{-1}  \circ \eta' \big) \circ \big( \b\uppsi \circ \eta \big)\\
 &=\Big( \b\uppsi' \circ \b\uppsi \Big) \circ \left( \Big( {\b\uppsi}^{-1}  \circ \eta' \circ \b\uppsi \Big) \circ \eta \right).  
\end{aligned}
\end{equation}
This decomposition may be seen  as a special case of the canonically split of bundle morphisms arising from the pullback bundle construction: Given a bundle $Q\rarrow N$ and a diffeomorphism $\uppsi: N \rarrow M$, one may define the pullback bundle $\uppsi^*P \rarrow N$. 
Then, there is a bundle morphism $\b\uppsi: \uppsi^*P \rarrow P$ covering $\uppsi$, i.e. s.t. $\b \pi ( \b\uppsi) = \uppsi$. 
By the (categorical) universality property of the pullback, for
a bundle morphism $\psi: Q \rarrow P$ covering $\uppsi$, there is a unique bundle morphism $\eta: Q \rarrow \uppsi^*P$ covering $\id_N$. 
This is synthesised in the following graph:
\begin{equation}
\label{Morph-graph-pullback}
\begin{tikzcd}
Q \arrow[bend left,"\psi"]{rr} \arrow[r,swap,"\eta"] \arrow[d] & 
\uppsi^* P \arrow[r,swap,"\b\uppsi"] \arrow[d] &
P \arrow[d]
\\
N \arrow[equal]{r} & N \arrow[r,swap,"\uppsi"] & M 
\end{tikzcd}
\end{equation}
On the other hand, the natural semi-direct structure \eqref{Semi-dir-Aut}-\eqref{Aut-graph} may be understood to follow as a special case of the split of a bundle morphism given by the graph
\begin{equation}
\label{Morph-graph-2}
\begin{tikzcd}
Q 
\arrow[bend left,"\psi"]{rr} \arrow[r,swap,"\b\uppsi"] \arrow[d] & 
 P \arrow[r,swap,"\eta"] \arrow[d] &
P \arrow[d]
\\
N \arrow[r,swap,"\uppsi"] & M \arrow[equal]{r} & M 
\end{tikzcd}
\end{equation}
We may fit the two options into a single  graph:
\begin{equation}
\label{Morh-graph-3D}
\begin{tikzcd}[ampersand replacement=\&]
    \& Q \arrow{rr}{\eta} \arrow{ddr}
         \arrow[swap, near end]{drrr}{\b\uppsi} \arrow[dashed,no head, "{\psi}" description]{drrrrr}
    \&  
    \& \uppsi^*P \arrow[dotted]{ddl} \arrow[rightharpoonup]{drrr}{\b\uppsi}
    \& 
    \&  
    \& \\
    \& 
    \&   
    \& 
    \& P \arrow[swap,rightharpoondown]{rr}{\eta} \arrow{ddr} 
    \& 
    \& P \arrow{ddl}\\
    \& 
    \& N  \arrow{drrr}{\uppsi}
    \& 
    \& 
    \& 
    \& \\
    \& 
    \&   
    \&  
    \& 
    \& M 
    \&  
\end{tikzcd}
\end{equation}
The decomposition \eqref{Morph-graph-2} is the one we refer to in section \ref{Composition of dressing operations} to split an $\Aut(P)$-dressing into an $\Aut_v(P)$-dressing and a $\overline{\Diff}$-dressing, as it is the one respecting the necessary compatibility conditions \eqref{CompCond1}-\eqref{CompCond2}.

\subsection{Semi-direct structure of the local symmetry group}
\label{Semi-direct structure of the local symmetry group}

From \eqref{SDeq1} follows a semi-direct product on $\overline{\Diff} \ltimes \H \simeq \overline{\Diff}\ltimes \Aut_v(P)=\Aut(P)$. Indeed, one finds that the $\Aut_v(P)$ part of $\psi'\circ \psi$ is
\begin{equation}
\begin{aligned}
\label{aut-v-part-comp}
\Big[\eta' \circ 
 \Big( \b\uppsi'  \circ \eta \circ  {\b\uppsi}'{}^{-1} \Big)\Big](p)
 &= \eta' \circ \b\uppsi' \left( {\b\uppsi'}{}\-(p)\, \gamma\Big( {\b\uppsi'}{}\-(p) \Big)\right)\\
&= \eta' \left( \b\uppsi' \big( {\b\uppsi'}{}\-(p)\big) \, \gamma\big( {\b\uppsi'}{}\-(p) \big)\right)\\
&=\eta'(p) \, \gamma\big( {\b\uppsi'}{}\-(p) \big)\\
&= p\, \gamma'(p)\, \gamma\big( {\b\uppsi'}{}\-(p) \big)\\
&= p\, \Big(\gamma'\!\cdot \big({\b\uppsi'}{}\-{}^* \gamma \big) \Big)(p).
\end{aligned}
\end{equation}
One has thus the product structure
\begin{align}
\label{semi-dir-prod-Diff-H}
\big( \b\uppsi', \gamma' \big) \cdot \big(\b\uppsi, \gamma \big)
=\Big(\b\uppsi'\circ \b\uppsi,\, \gamma'\!\cdot \big({\b\uppsi'}{}\-{}^* \gamma \big) \Big).
\end{align}
This directly induces the local version on ($U\subset$) $M$, i.e. the semi-direct product of $\Diff(M) \ltimes \H_\text{\tiny loc}$:
\begin{align}
\label{semi-dir-DiffM-H}
\big( \uppsi' , \upgamma' \big) \cdot \big( \uppsi, \upgamma \big) = \left( \uppsi' \circ \uppsi , \, \upgamma' \!\cdot\big({\uppsi'}{}\-{}^* \upgamma \big)  \right),
\end{align}
with $\H_\text{\tiny loc}$ defined in \eqref{Hloc-def}.
This is directly relevant to (local) field theory as discussed in section \ref{Local field space}.
The inverse element is $\big(\uppsi, \upgamma \big)\-=\big( \uppsi\-, \uppsi^*\upgamma\- \big)$.

From this one may derive the adjoint action of $\Diff(M) \ltimes \Hl$ on its Lie algebra $\diff(M) \oplus\text{Lie}\Hl$:
For an element $(X, \lambda) \defeq  \left(\tfrac{d}{d\tau} \uppsi_\tau \big|_{\tau=0}, \tfrac{d}{ds}\upgamma_s \big|_{s=0} \right)$
of the Lie algebra, we get
\begin{align}
\Ad_{(\uppsi, \upgamma)} (X, \lambda) 
\defeq& \tfrac{d^2}{d\tau ds}\ \text{Conj}(\uppsi, \upgamma)\, \big(X, \lambda \big)
\,\big|_{\tau=0,\, s=0}  \notag\\
=&\, \tfrac{d^2}{d\tau ds}\ 
(\uppsi, \upgamma)\circ (\uppsi_\tau, \upgamma_s) \circ (\uppsi\-, \uppsi^*\upgamma\-) 
\,\big|_{\tau=0,\, s=0} \notag\\
=&\, \tfrac{d^2}{d\tau ds}\ 
(\uppsi, \upgamma)\circ \Big(\uppsi_\tau \circ \uppsi\-, \ \upgamma_s \cdot \uppsi_\tau^{-1*} \uppsi^* \upgamma\- \Big)
\,\big|_{\tau=0,\, s=0} \notag\\
=&\, \tfrac{d^2}{d\tau ds}\
\Big( \uppsi\circ \uppsi_\tau \circ \uppsi\- , \ \upgamma\cdot \uppsi^{-1*}\upgamma_s \cdot \uppsi^{-1*} \uppsi_\tau^{-1*} \uppsi^* \upgamma\- \Big)
\,\big|_{\tau=0,\, s=0} \notag\\
=&\, \tfrac{d^2}{d\tau ds}\
\Big( \uppsi\circ \uppsi_\tau \circ \uppsi\-, \ \upgamma\cdot \uppsi^{-1*}\upgamma_s \cdot (\uppsi \circ \uppsi_\tau \circ \uppsi\-)^{-1*} \upgamma\- \Big)
\,\big|_{\tau=0,\, s=0} , \notag\\[1.5mm]
\Ad_{(\uppsi, \upgamma)} (X, \lambda)
=&\, \Big(\,\uppsi_* \,X \circ \uppsi\-, \ \Ad_\upgamma (\,\uppsi^{-1*}\lambda) - \upgamma\cdot \mathfrak{L}_{\uppsi_* X \circ \uppsi\-} \upgamma\- \,\Big).   \label{SD-loc-adj-action}
\end{align}
The identity \eqref{Psi-relatedness} is used in the last step. Similarly, one gets
\begin{align}
\Ad_{(\uppsi, \upgamma)\-} (X, \lambda) 
:=&\, \tfrac{d^2}{d\tau ds}\ 
(\uppsi, \upgamma)\- \circ (\uppsi_\tau, \upgamma_s) \circ (\uppsi, \upgamma) 
\,\big|_{\tau=0,\, s=0} \notag\\
=&\, \tfrac{d^2}{d\tau ds}\ 
 \Big(\uppsi\- \circ \uppsi_\tau \circ \uppsi , \ \uppsi^* \upgamma\- \cdot \uppsi^* \upgamma_s \cdot \big(\uppsi\- \circ \uppsi_\tau \big)^{-1 *} \, \upgamma \Big) \,\big|_{\tau=0,\, s=0} \notag\\
=&\, \tfrac{d^2}{d\tau ds}\
\Big(\uppsi\- \circ \uppsi_\tau \circ \uppsi , \ \uppsi^* \big(\upgamma\- \cdot \upgamma_s \cdot \uppsi^{-1 *}_\tau \, \upgamma \big) \Big)
\,\big|_{\tau=0,\, s=0} \notag\\[2mm]
=&\, \left\{
\begin{matrix}
\, \hspace{-20mm} \Big( \, \uppsi\-_* \, X \circ \uppsi , \ \uppsi^* \big( \Ad_{\upgamma\-} \lambda - \upgamma\- \mathfrak{L}_X \, \upgamma \big) \, \Big) , \\[2mm]
\ \ \ \Big(\,\uppsi\-_* \, X \circ \uppsi , \ \Ad_{\uppsi^* \upgamma\-} \big( \uppsi^* \lambda \big) - \uppsi^* \upgamma\- \cdot \mathfrak{L}_{\uppsi\-_* X \circ \uppsi} \uppsi^* \upgamma \,\Big), 
\end{matrix}
\right. 
\label{SD-loc-adj-action-inverse}    
\end{align}
using the fact that $\uppsi^{-1 *}_\tau=\uppsi^*_{-\tau}$ and, in the last line, \eqref{lemma1}.

\section{Lie algebra (anti-)isomorphisms and pushforward by a vertical diffeomorphism of (local) field space}
\label{Lie algebra (anti-)isomorphisms}

\subsection{Lie algebra morphisms}
\label{Lie algebra morphisms}

We prove here that the verticality map $|^v :\aut(P) \rarrow \Gamma(V\Phi)$, $X \mapsto X^v$, is a morphism of Lie algebra. 
We write the flow through $\phi \in \Phi$ of $X^v \in \Gamma(V\Phi)$ as $\tilde \psi_\tau(\phi) := R_{\psi_\tau} \phi := \psi_\tau^* \phi$, with $X =\tfrac{d}{d\tau} \psi_\tau\ \big|_{\tau=0} \in \Gamma(TP)$, so that
\begin{align}
X^v_{|\phi}=\tfrac{d}{d\tau} \tilde \psi_\tau (\phi)\, \big|_{\tau=0} \quad \left(\  =X(\phi)^v \tfrac{\delta}{\delta \phi}, \ \text{written as a derivation of $C^\infty(\Phi)$} \right).
\end{align}
One can thus write the bracket of two vertical vector fields:
\begin{align}
\label{demo1}
[X^v, Y^v]_{|\phi} = {\bs L_{X^v} Y^v}_{|\phi} :=&\, \tfrac{d}{d\tau} (\tilde\psi_\tau\-)_\star Y^v_{|\tilde\psi_\tau(\phi)}\, \big|_{\tau=0} \notag\\
:=&\, \tfrac{d}{d\tau} \tfrac{d}{ds} \left(  \tilde \psi_\tau\- \circ \tilde\eta_s \circ \tilde\psi_\tau \right) (\phi)\, \big|_{s=0}	   \, \big|_{\tau=0} \notag\\
:=&\, \tfrac{d}{d\tau} \tfrac{d}{ds}\,  R_{\psi\-_\tau} \circ R_{\eta_s} \circ R_{\psi_\tau} \phi\, \big|_{s=0}	   \, \big|_{\tau=0}  \notag\\
:=&\, \tfrac{d}{d\tau} \tfrac{d}{ds}\,  R_{(\psi_\tau \circ \eta_s \circ \psi\-_\tau)} \phi\, \big|_{s=0}	   \, \big|_{\tau=0}  \notag\\
:=&\, \tfrac{d}{d\tau} \tfrac{d}{ds}\,  (\underbrace{\psi_\tau \circ \eta_s \circ \psi\-_\tau}_{\text{flow of $-[X,Y]$}})^*\phi\, \big|_{s=0}	   \, \big|_{\tau=0} \notag\\
=&\!:\, \mathfrak L_{-[X, Y]} \, \phi =: (-[X, Y]_{\text{{\tiny $\Gamma(TP)$}}})^v_{|\phi}= ([X, Y]_{\text{{\tiny $\mathfrak{aut}(P)$}}})^v_{|\phi}.
\end{align}
In the $1^{\text{st}}$-$2^{\text{nd}}$ and $5^{\text{th}}$-$6^{\text{th}}$ lines the definition of $\tilde\psi_\tau$ and $\psi_\tau$-relatedness are used, while  in the $1^{\text{st}}$ and $6^{\text{th}}$ lines, we have used the definitions of the Lie derivatives of a vector field on $\Phi$ and of a field on $P$. 
\medskip

We now prove that the map $|^v : \bs{\aut}(P) \rarrow \Gamma_{\text{\!\tiny{inv}}}(V\Phi)$, $\bs X \mapsto \bs X^v$, is a Lie algebra \emph{anti}-morphism. 
The flow through $\phi \in \Phi$ of $\bs X^v \in \Gamma_{\text{\!\tiny{inv}}}(V\Phi)$ as $\tilde{\bs\psi}_\tau(\phi) :=(R_{\bs\psi_\tau})(\phi)= R_{\bs\psi_\tau(\phi)} \phi := (\bs\psi_\tau(\phi))^* \phi$, with  $\bs X =\tfrac{d}{d\tau} \bs\psi_\tau\ \big|_{\tau=0} \in \Gamma(TP)$. So,
\begin{align}
[\bs X^v, \bs Y^v]_{|\phi} = {\bs L_{\bs X^v} \bs Y^v}_{|\phi} :=\, \tfrac{d}{d\tau} (\tilde{\bs\psi}_\tau\-)_\star \bs Y^v_{|\tilde{\bs\psi}_\tau(\phi)}\, \big|_{\tau=0} 
				     = \tfrac{d}{d\tau} \tfrac{d}{ds} \left(  \tilde{\bs\psi}_\tau\- \circ \tilde{\bs\eta}_s \circ \tilde{\bs\psi}_\tau \right) (\phi)\, \big|_{s=0} \big|_{\tau=0} .
\end{align}
The equivariance property of the elements of the gauge group has been used. 
We have then 
\begin{align}
\tilde{\bs\eta}_s \circ \tilde{\bs\psi}_\tau  (\phi) 
&= R_{\bs\eta_s(\tilde{\bs\psi}_\tau (\phi))}\,  \tilde{\bs\psi}_\tau (\phi)
= R_{\bs\eta_s\big (R_{\bs\psi_\tau (\phi)} \phi  \big)}\,  R_{\bs\psi_\tau (\phi)} \phi  \notag\\ 
&= R_{\bs\psi_\tau (\phi)\- \circ \bs\eta_s(\phi) \circ \bs\psi_\tau (\phi) }\,  R_{\bs\psi_\tau (\phi)} \phi
= R_{\bs\eta_s(\phi) \circ \bs\psi_\tau (\phi) }\, \phi \notag \\
&= \left( R_{\bs\eta_s \circ \bs\psi_\tau }\right) \phi.
\end{align}
So, 
\begin{align}
 \left(  \tilde{\bs\psi}_\tau\- \circ \tilde{\bs\eta}_s \circ \tilde{\bs\psi}_\tau \right) (\phi) =  \tilde{\bs\psi}_\tau\- \big( \tilde{\bs\eta}_s \circ \tilde{\bs\psi}_\tau  (\phi)\big) 
 &= R_{\bs\psi_\tau \-\big( R_{\bs\eta_s(\phi) \circ \bs\psi_\tau (\phi) }\, \phi  \big)   } \, R_{\bs\eta_s(\phi) \circ \bs\psi_\tau (\phi) }\, \phi \notag \\
&=R_{[\bs\eta_s(\phi) \circ \bs\psi_\tau (\phi)]\- \circ  \bs\psi_\tau (\phi)\- \circ [\bs\eta_s(\phi) \circ \bs\psi_\tau (\phi)]}\, R_{\bs\eta_s(\phi) \circ \bs\psi_\tau (\phi)}\, \phi \notag\\
&= R_{ \bs\psi_\tau (\phi)\- \circ \bs\eta_s(\phi) \circ \bs\psi_\tau (\phi)}\, \phi. 
\end{align}
Finally, 
\begin{align}
\label{demo2}
[\bs X^v, \bs Y^v]_{|\phi} &=\tfrac{d}{d\tau} \tfrac{d}{ds} \,  R_{ \bs\psi_\tau (\phi)\- \circ \bs\eta_s(\phi) \circ \bs\psi_\tau (\phi)}\, \phi  \, \big|_{s=0}\big|_{\tau=0}  \notag\\
&= \tfrac{d}{d\tau} \tfrac{d}{ds}\,  (\underbrace{\bs\psi_\tau (\phi)\- \circ \bs\eta_s(\phi) \circ \bs\psi_\tau (\phi) }_{\text{flow of $[\bs X,\bs Y]$}})^*\phi\, \big|_{s=0}	   \, \big|_{\tau=0} \notag\\
&=:\, \mathfrak L_{[\bs X, \bs Y]} \, \phi =: ([\bs X, \bs Y]_{\text{{\tiny $\Gamma(TP)$}}})^v_{|\phi}= (-[\bs X, \bs Y]_{\text{{\tiny $\aut(P)$}}})^v_{|\phi},
\end{align}
which ends the proof of the assertion. 
Observe that the above computations hold the same for both $\aut_v(P)$ \cite{Francois2021, Francois-et-al2021} and $\diff(M)$ \cite{Francois2023-a}.

\subsection{Pushforward by a vertical diffeomorphism of field space}
\label{Pushforward by a vertical diffeomorphism of field space}

We consider a generic vector field $\mathfrak X \in \Gamma(T\Phi)$ with flow $\vphi_\tau$ and a vertical diffeomorphism $\Xi \in \bs\Diff_v(\phi)$ to which corresponds $\bs \psi \in C^\infty\big(\Phi, \Aut(P)\big)$. 
The pushforward of $\mathfrak X_{|\phi} \in T_\phi\Phi$ by $\bs\psi$ is
 \begin{align}
 \Xi_\star \mathfrak X_{|\phi} = \tfrac{d}{d\tau} \,  \Xi \big( \vphi_\tau(\phi) \big)  \,\big|_{\tau=0} = \tfrac{d}{d\tau} \,  R_{\bs\psi \left( \vphi_\tau(\phi) \right)}  \vphi_\tau(\phi)   \, |_{\tau=0}  
&=  \tfrac{d}{d\tau} \,   R_{\bs\psi \left( \vphi_\tau(\phi) \right)}  \phi   \,\big|_{\tau=0} + \tfrac{d}{d\tau} \,  R_{\bs\psi (\phi)}  \vphi_\tau(\phi)   \, |_{\tau=0}  \notag\\
&=  \tfrac{d}{d\tau} \,   R_{\bs\psi \left( \vphi_\tau(\phi) \right)}  \phi   \,\big|_{\tau=0} + R_{\bs\psi (\phi)\star}  \mathfrak X_{|\phi}.
 \end{align}
In the last equality the definition of the pushforward by the right action of $\bs\psi (\phi) \in \Aut(P)$ is used. 
The remaining term is manifestly a vertical vector field. The question is to find the element of $\aut(P)$ that generates it,   knowing  that it must be anchored at the point $\Xi(\phi)=R_{\bs\psi(\phi)} \phi=\bs\psi(\phi)^*\phi$ of field space:
 \begin{align}
 \tfrac{d}{d\tau} \,   R_{\bs\psi \left( \vphi_\tau(\phi) \right)}  \phi   \,\big|_{\tau=0}  = 
 \tfrac{d}{d\tau} \,   R_{\bs\psi (\phi) \circ \bs\psi(\phi)^{-1} \circ \bs\psi \left( \vphi_\tau(\phi) \right)}  \phi   \,\big|_{\tau=0} = 
  \tfrac{d}{d\tau} \,   R_{\bs\psi(\phi)^{-1} \circ \bs\psi \left( \vphi_\tau(\phi) \right) }  \, R_{ \bs\psi (\phi)} \phi   \,\big|_{\tau=0}.   \notag
 \end{align}
So, $\bs\psi(\phi)^{-1} \circ \bs\psi \left( \vphi_\tau(\phi)\right)$ is the flow of the  vector field on $\Aut(P)$ we are looking for. 
Now,  on the one hand we have
 \begin{align}
  \tfrac{d}{d\tau} \,  \bs\psi\big( \vphi_\tau(\phi)  \big)  \,\big|_{\tau=0}  = \bs d \bs\psi_{|\phi} \big( \mathfrak X_{|\phi} \big) = \bs\psi_\star \mathfrak X_{|\phi}  \in T_{\bs\psi(\phi)}\Aut(P),\notag
  \end{align}
  since $\bs\psi : \Phi \rarrow \Aut(P)$. 
  On the other hand, the Maurer-Cartan form -- given by the left translation --  on $\Aut(P)$ is
  \begin{align}
   L_{\psi^{-1} \star} : T_\psi \Aut(P) \rarrow&\  T_{\id_P}\Aut(P)=\aut(P)\simeq \Gamma(TP),  \notag\\
   			\mathcal X_{|\psi}  \mapsto &\   L_{\psi^{-1} \star} \mathcal X_{|\psi} \defeq (\psi^{-1})_* \mathcal X_{|\psi}. 
  \end{align}
  So we have
  \begin{align}
     L_{\bs\psi(\phi)^{-1} \star} : T_{\bs\psi(\phi)} \Aut(P) \rarrow&\  \aut(P)\simeq \Gamma(TP),  \notag\\
	       \big[  \bs d \bs\psi_{|\phi} \big( \mathfrak X_{|\phi} \big) \big]_{|\bs\psi(\phi)}  \mapsto& \   L_{\bs\psi(\phi)^{-1}\star} \big[  \bs d \bs\psi_{|\phi} \big( \mathfrak X_{|\phi} \big) \big]_{|\bs\psi(\phi)} = (\bs\psi(\phi)^{-1})_* \big[  \bs d \bs\psi_{|\phi} \big( \mathfrak X_{|\phi} \big) \big]_{|\bs\psi(\phi)} \notag\\
&\ =L_{\bs\psi(\phi)^{-1} \star}  \tfrac{d}{d\tau} \,  \bs\psi\big( \vphi_\tau(\phi)  \big)  \,\big|_{\tau=0}  
=  \tfrac{d}{d\tau} \,  \bs\psi(\phi)^{-1}\circ \bs\psi\big( \vphi_\tau(\phi)  \big)  \,\big|_{\tau=0}. 	
  \end{align}
  Thus, $(\bs\psi(\phi)^{-1})_* \big[  \bs d \bs\psi_{|\phi} \big( \mathfrak X_{|\phi} \big) \big]$ is the generating vector field of $P$ we were searching for. We then get
   \begin{align}
 \tfrac{d}{d\tau} \,   R_{\bs\psi \left( \vphi_\tau(\phi) \right)}  \phi   \,\big|_{\tau=0}  = 
  \tfrac{d}{d\tau} \,   R_{\bs\psi(\phi)^{-1} \circ \bs\psi \left( \vphi_\tau(\phi) \right) }  \, R_{ \bs\psi (\phi)} \phi   \,\big|_{\tau=0} =
  \left\{    (\bs\psi(\phi)^{-1})_* \big[  \bs d \bs\psi_{|\phi} \big( \mathfrak X_{|\phi} \big) \big]  \right\}^v_{|R_{\bs\psi(\phi)} \phi}.
 \end{align}
Hence, finally we get
 \begin{align}
 \Xi_\star \mathfrak X_{|\phi} &= R_{\bs\psi(\phi) \star} \mathfrak X_{|\phi} + \left\{   \bs\psi(\phi)\-_* \bs d \bs\psi_{|\phi}(\mathfrak X_{|\phi})   \right\}^v_{|\Xi(\phi)}  \notag\\
 					    &=R_{\bs\psi(\phi) \star} \left( \mathfrak X_{|\phi} +  \left\{  \bs d \bs\psi_{|\phi}(\mathfrak X_{|\phi})  \circ \bs\psi(\phi)\-  \right\}^v_{|\phi} \right),
\end{align}
where in the second line the property \eqref{Pushforward-fund-vect} of the fundamental vertical vector fields under pushforward by the right action of the structure group has been used. 	 
This result is essential to compute geometrically the vertical and gauge transformations of  forms on $\Phi$.    

\subsection{Pushforward by a vertical diffeomorphism of local field space}
\label{Pushforward by a vertical diffeomorphism of local field space}

We here derive the local version of the above result, important for the standard formulation of field theory.
Despite some repetition, we deem it pedagogically useful. 
The local field space $\Phi$ is now that of local representatives of fields on ($U\subset$) $M$, and its structure group is $\Diff(M)\ltimes \Hl$. 
A vertical diffeomorphism $\Xi \in \bs\Diff_v(\Phi)$ is generated by $(\bs\uppsi, \bs\upgamma) \in C^\infty\big(\Phi, \Diff(M)\ltimes \Hl \big)$.
As above, we~consider the pushforward along $\Xi$ of a generic vector field $\mathfrak X \in \Gamma(T\Phi)$ with flow $\vphi_\tau$:
 \begin{align}
 \Xi_\star \mathfrak X_{|\phi} = \tfrac{d}{d\tau} \,  \Xi \big( \vphi_\tau(\phi) \big)  \,\big|_{\tau=0} =
 \tfrac{d}{d\tau} \,  R_{(\bs\uppsi, \bs\upgamma) \circ \left( \vphi_\tau(\phi) \right)}\  \vphi_\tau(\phi)   \, \big|_{\tau=0}  
&=  \tfrac{d}{d\tau} \,   R_{(\bs\uppsi, \bs\upgamma) \circ\left( \vphi_\tau(\phi) \right)}  \phi   \,\big|_{\tau=0} + R_{\big(\bs\uppsi(\phi), \bs\upgamma(\phi)\big)\star}  \mathfrak X_{|\phi}.
 \end{align}
The first term is clearly a vertical vector field, which must be anchored at the point  $\Xi(\phi)=R_{\big(\bs\uppsi(\phi), \bs\upgamma(\phi)\big)} \phi$.
One needs only to find its  $\diff(M)\oplus\text{Lie}\Hl$-generating element. 

For notational convenience, let us write $(\bs\uppsi, \bs\upgamma)_\tau \defeq(\bs\uppsi, \bs\upgamma) \circ\left( \vphi_\tau(\phi) \right)$, 
and $\big(\bs\uppsi(\phi), \bs\upgamma(\phi)\big)=(\bs\uppsi, \bs\upgamma)$.
We have
\begin{align}
 \tfrac{d}{d\tau} \,   R_{(\bs\uppsi, \bs\upgamma)_\tau}  \phi   \,\big|_{\tau=0}  
 =  \tfrac{d}{d\tau} \,   R_{(\bs\uppsi, \bs\upgamma)^{-1} \cdot (\bs\uppsi, \bs\upgamma)_\tau }  \, R_{(\bs\uppsi, \bs\upgamma)} \phi   \,\big|_{\tau=0},   \notag
 \end{align}
where $(\bs\uppsi, \bs\upgamma)^{-1} \cdot (\bs\uppsi, \bs\upgamma)_\tau$ is the flow of the  vector field on $\Diff(M)\ltimes \Hl$ we are seeking. 
This is a curve through the identity $(\id_M, \id_{\Hl})$ in the group. 
Given that the left action, by an inverse element, on $\Diff(M)\ltimes \Hl$ is 
\begin{align}
  L_{(\uppsi, \upgamma)\-}   (\uppsi', \upgamma')\defeq (\uppsi, \upgamma)\- \cdot (\uppsi', \upgamma') = (\uppsi\-, \uppsi^*\upgamma\-) \cdot (\uppsi', \upgamma') = \big(\uppsi\- \circ \uppsi', \uppsi^*(\upgamma\- \upgamma')\big) ,
\end{align}
the Maurer-Cartan form  is
  \begin{align}
   L_{(\uppsi, \upgamma)^{-1} \star} : T_{(\uppsi, \upgamma)} \big(\!\Diff(M)\ltimes \Hl \big)
    \rarrow&\  T_{(\id_M, \id_{\Hl})}\big(\!\Diff(M)\ltimes \Hl \big)=\diff(M)\oplus\text{Lie}\Hl   \notag\\
   T_\uppsi \Diff(M) \oplus T_\upgamma \Hl \rarrow & T_{\id_M} \Diff(M) \oplus T_{\id_{\Hl}} \Hl  \\
   \big( \mathcal X_{|\uppsi},  \mathcal Y_{|\upgamma} \big) \mapsto&\   L_{(\uppsi, \upgamma)^{-1} \star} \big( \mathcal X_{|\uppsi},  \mathcal Y_{|\upgamma} \big)
   \defeq  (\uppsi, \upgamma)^{-1} \cdot \big( \mathcal X_{|\uppsi},  \mathcal Y_{|\upgamma} \big)=
   \left(\uppsi^{-1}_* \mathcal X_{|\uppsi}, \uppsi^* \big( \upgamma\- \mathcal Y_{|\upgamma}\big) \right). \notag
  \end{align}
Now, writing $(\bs\uppsi, \bs\upgamma)_\tau = (\bs\uppsi_\tau, \bs\upgamma_\tau)$, we have that
\begin{equation}
\begin{aligned}
  \tfrac{d}{d\tau} \,  (\bs\uppsi, \bs\upgamma)^{-1} \cdot (\bs\uppsi, \bs\upgamma)_\tau  \,\big|_{\tau=0}  
&=  \tfrac{d}{d\tau} \, L_{(\bs\uppsi, \bs\upgamma)^{-1} } (\bs\uppsi,\upgamma)_\tau  \,\big|_{\tau=0} 
  = L_{(\bs\uppsi, \bs\upgamma)^{-1} \star} \tfrac{d}{d\tau}\, \big( \bs\uppsi_\tau, \bs\upgamma_\tau \big)\, \big|_{\tau=0} \\
&= L_{(\bs\uppsi, \bs\upgamma)^{-1} \star}\,  \left(\bs d \bs\uppsi_{|\phi} (\mathfrak X_{|\phi}) , \bs d\bs \upgamma_{|\phi} (\mathfrak X_{|_\phi}) \right) \\
&= (\bs\uppsi, \bs\upgamma)^{-1} \cdot  \left(\bs d \bs\uppsi_{|\phi} (\mathfrak X_{|\phi}) , \bs d\bs \upgamma_{|\phi} (\mathfrak X_{|_\phi}) \right)\\
&= \left( \bs\uppsi\-_* \big[ \bs d \bs\uppsi_{|\phi} (\mathfrak X_{|\phi})\big], \bs\uppsi^*\, \bs\upgamma\- \big[\bs d\bs \upgamma_{|\phi} (\mathfrak X_{|_\phi})  \big] \right)
\in \diff \oplus \text{Lie}\Hl ,
\end{aligned}
\end{equation} 
with $\bs d \bs\uppsi_{|\phi} (\mathfrak X_{|\phi}) \in T_{\bs\uppsi} \Diff(M)$ and $\bs d \bs\upgamma_{|\phi} (\mathfrak X_{|\phi}) \in T_{\bs\upgamma} \Hl$.
So finally, 
\begin{equation}
\label{Pushforward-X-loc-1}
\begin{aligned}
\Xi_\star \mathfrak X_{|\phi}
& = R_{(\bs\uppsi, \bs\upgamma ) \star}\mathfrak X_{|\phi} 
+ \left\{\big[(\bs\uppsi, \bs\upgamma)^{-1}\! \cdot  \big(\bs d \bs\uppsi , \bs d\bs \upgamma\, \big)\big]_{|\phi} (\mathfrak X_{|\phi}) \right\}^v_{|\Xi(\phi)} \\
& = R_{(\bs\uppsi, \bs\upgamma ) \star}\mathfrak X_{|\phi} 
+ \left\{
\big(
\bs\uppsi\-_* \bs d\bs \uppsi, 
\bs\uppsi^*\, (\bs\upgamma\- \bs d\bs \upgamma)
\big)_{|\phi} (\mathfrak X_{|\phi})
\right\}^v_{|\Xi(\phi)}. \\
\end{aligned}
\end{equation}
\medskip 

There is an alternative form of the above result, which is most useful in concrete computations. It relies on the pushforward of fundamental vector fields \eqref{pushf-fund-vec-loc}: we have indeed that
\begin{align}
\left\{(\bs\uppsi, \bs\upgamma)^{-1}\! \cdot  \big(\bs d \bs\uppsi , \bs d\bs \upgamma\, \big)_{|\phi} (\mathfrak X_{|\phi}) \right\}^v_{|\Xi(\phi)} 
&= R_{(\bs\uppsi, \bs\upgamma) \star} \left\{\Ad_{(\bs\uppsi, \bs\upgamma)} \left(
\big[(\bs\uppsi, \bs\upgamma)^{-1}\! \cdot  \big(\bs d \bs\uppsi , \bs d\bs \upgamma\, \big)\big]_{|\phi} (\mathfrak X_{|\phi}) 
\right) \right\}^v_{|\phi}.
\end{align}
We compute the $\diff(M)\oplus\text{Lie}\Hl$-valued 1-form on the right to be
\begin{align}
\Ad_{(\bs\uppsi, \bs\upgamma)} \left(
(\bs\uppsi, \bs\upgamma)^{-1}\! \cdot  \big(\bs d \bs\uppsi , \bs d\bs \upgamma\, \big) 
\right) 
&=\Ad_{(\bs\uppsi, \bs\upgamma)} \Big(
\bs\uppsi\-_*\bs{d\uppsi}, \ \bs\uppsi^*(\bs\upgamma\- \bs{d\upgamma})\Big) \notag\\
&=\left( 
\bs\uppsi_*\big(  \bs\uppsi\-_*\bs{d\uppsi} \big)\circ \bs\uppsi\-,
\Ad_{\bs\upgamma}\big( 
\bs\uppsi^{-1*}\big[ \bs\uppsi^*(\bs\upgamma\- \bs{d\upgamma}) \big]
\big)
- \bs\upgamma \, \mathfrak L_{\bs\uppsi_* ( \bs\uppsi\-_*\bs{d\uppsi}) \circ \bs\uppsi\-} \bs\upgamma\-
\right)  \notag\\
&= \Big( 
\bs{d\uppsi} \circ \bs\uppsi\-,\ 
\bs{d\upgamma} \, \bs\upgamma\-  - \bs\upgamma\, \mathfrak L_{\bs{d\uppsi} \circ \bs\uppsi\-} \bs\upgamma\-
\Big). 
\end{align}
This result we denote by
\begin{align}
\label{linear-right-action-loc}
    \big( \bs d \bs \uppsi, \bs d \bs \upgamma \big) \cdot \big( \bs \uppsi,\bs \upgamma \big)^{-1} \equiv \left(\bs d \bs \uppsi \circ \bs \uppsi\- , \bs d \bs \upgamma \,\bs \upgamma\- - \bs \upgamma \, \mathfrak{L}_{\bs d \bs \uppsi \circ \bs \uppsi\-} \bs \upgamma\- \right).
\end{align}
This is justified by the fact that it can be understood via the linearisation of the right action
\begin{align}
R_{(\uppsi, \upgamma)\-} (\uppsi', \upgamma') 
= (\uppsi', \upgamma') \cdot (\uppsi, \upgamma)\-
=(\uppsi', \upgamma') \cdot (\uppsi\-, \uppsi^*\upgamma\-)
=\big( \uppsi' \circ \uppsi\-, \upgamma'\, {\uppsi'}^{-1*}(\uppsi^*\upgamma\-) \big).
\end{align}
So, one has that
\begin{align}
\label{right-action-tangent}
R_{(\uppsi, \upgamma)\- \star} : T_{(\uppsi, \upgamma)} \big(\!\Diff(M)\ltimes \Hl \big) &\rarrow T_{(\id_M, \id_{\Hl})}\big(\!\Diff(M)\,\ltimes&&\hspace{-9mm} \Hl \big) \rdefeq \diff(M)\oplus\text{Lie}\Hl, \notag \\
\big( \mathcal X_{|\uppsi},  \mathcal Y_{|\upgamma} \big) &\mapsto 
R_{(\uppsi, \upgamma)\- \star}\big( \mathcal X_{|\uppsi},  \mathcal Y_{|\upgamma} \big)
=&&\hspace{-10mm}
\big( \mathcal X_{|\uppsi},\,  \mathcal Y_{|\upgamma} \big) \cdot (\uppsi, \upgamma)\- \notag \\
&\phantom{\mapsto 
R_{(\uppsi, \upgamma)\- \star}\big( \mathcal X_{|\uppsi},  \mathcal Y_{|\upgamma} \big)}
\defeq&&\hspace{-10mm}\Big( \mathcal X_{|\uppsi} \circ \uppsi\-, \,  \mathcal Y_{|\upgamma} \upgamma\- - \upgamma\, \mathfrak{L}_{\mathcal X_{|\uppsi} \circ \uppsi\-} \upgamma\- \Big).  
\end{align}
Thus, we get the final alternative result
\begin{equation}
\label{Pushforward-X-loc-2}
\begin{aligned}
\Xi_\star \mathfrak X_{|\phi}
& =   R_{(\bs\uppsi, \bs\upgamma ) \star}\left(
\mathfrak X_{|\phi} + 
\left\{
 \Big(\bs d \bs\uppsi , \bs d\bs \upgamma\, \Big)_{|\phi} (\mathfrak X_{|\phi})
 \cdot \big(\bs\uppsi, \bs\upgamma \big)\-
\right\}^v_{|\phi} 
\right)\\
&=  R_{(\bs\uppsi, \bs\upgamma ) \star}\left(
\mathfrak X_{|\phi} + 
\left\{\Big( 
\bs{d\uppsi} \circ \bs\uppsi\-,\ 
\bs{d\upgamma} \, \bs\upgamma\-  - \bs\upgamma\, \mathfrak L_{\bs{d\uppsi} \circ \bs\uppsi\-} \bs\upgamma\-
\Big)_{|\phi}(\mathfrak X_{|\phi})
\right\}^v_{|\phi} 
\right).
\end{aligned}
\end{equation}

{
\normalsize 
 \bibliography{Biblio9.5}
}

\end{document}